\documentclass[a4paper,11pt]{book}
\usepackage{amsmath}
\usepackage{palatino}
\usepackage{german}
\usepackage{amsfonts}
\usepackage{graphicx}
\usepackage{tensind}
\usepackage{slashed}
\usepackage{longtable}
\usepackage{color}
\usepackage{ifthen}
\usepackage{listings}
\usepackage{verbatim}
\usepackage{misctools}
\usepackage{feynmp}
\usepackage{units}
\usepackage{tikz}
\usepackage{listings}

\usepackage{fancyhdr}
\usepackage[noweb]{ocamlweb}
\renewcommand{\ocwindent}[1]{\dimen0=#1\noindent\kern0.3\dimen0}

\setboolean{preview}{false}

\newcommand{\threeshl}{Three Site Model}

\newcommand{\singleplotwidth}{6cm}

\newcommand{\doubledplotwidth}{5.4cm}
\newcommand{\doubleplotwidth}{4.6cm}
\newcommand{\dbltableplotheight}{6.8cm}

\newcommand{\ilum}{\int\!\!\LL}

\DeclareMathOperator{\atan}{atan}

\begin{document}

\begin{fmffile}{fmf}

% start plain pages
{
\pagestyle{plain}
\renewcommand{\thepage}{}

\selectlanguage{german}
\begin{center}\Huge
LHC Phenomenology\\
of the\\
Three-Site Higgsless Model
\end{center}
\vspace{1.5cm}
\begin{center}
Dissertation zur Erlangung des\\
naturwissenschaftlichen Doktorgrades\\
der Julius-Maximilians-Universit"at W"urzburg
\end{center}
\vspace{1cm}
\centerline{\includegraphics[width=6cm]{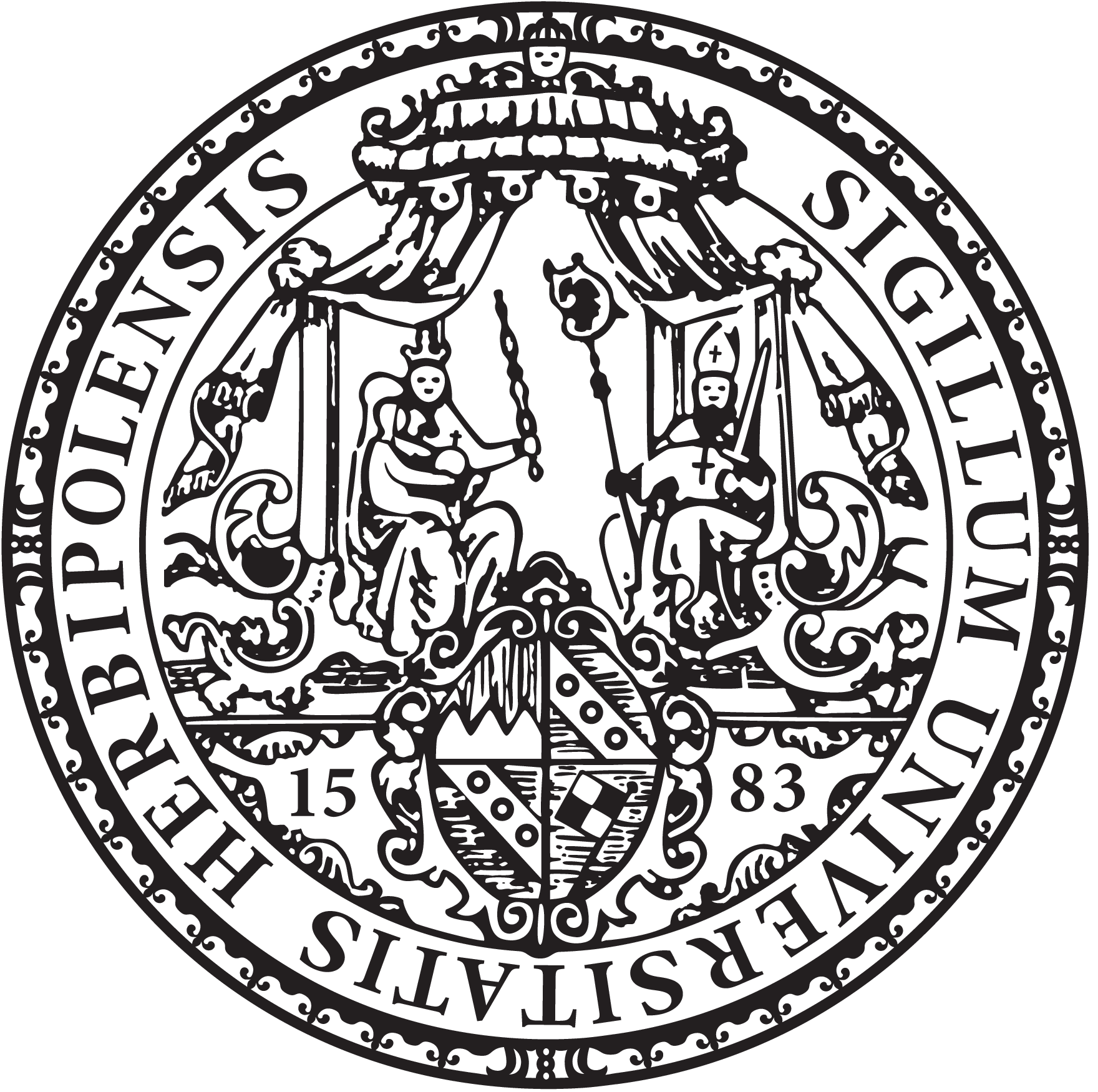}}
\vspace{1cm}
\begin{center}
{\small vorgelegt von}\\[4mm]
{\large Christian Speckner}\\[4mm]
aus W"urzburg
\end{center}
\vspace{1.5cm}
\centerline{W"urzburg, 2009}
\selectlanguage{english}

\newpage
\mbox{}
\newpage

\selectlanguage{german}
\begin{center}\Huge
LHC Phenomenology\\
of the\\
Three-Site Higgsless Model
\end{center}
\vspace{1.5cm}
\begin{center}
Dissertation zur Erlangung des\\
naturwissenschaftlichen Doktorgrades\\
der Julius-Maximilians-Universit"at W"urzburg
\end{center}
\vspace{1cm}
\centerline{\includegraphics[width=6cm]{./siegel.ps}}
\vspace{1cm}
\begin{center}
{\small vorgelegt von}\\[4mm]
{\large Christian Speckner}\\[4mm]
aus W"urzburg
\end{center}
\vspace{1.5cm}
\centerline{W"urzburg, 2009}
\selectlanguage{english}

\newpage

\selectlanguage{german}
\mbox{}\\[9cm]
Einereicht am:\hspace{5mm}\underline{\hspace{3cm}}\\[2mm]
bei der Fakult"at f"ur Physik und Astronomie\\[2cm]
1. Gutachter:\hspace{5mm}\underline{\hspace{5cm}}\\[2mm]
2. Gutachter:\hspace{5mm}\underline{\hspace{5cm}}\\[2mm]
3. Gutachter:\hspace{5mm}\underline{\hspace{5cm}}\\[2mm]
der Dissertation\\[1cm]
1. Pr"ufer:\hspace{5mm}\underline{\hspace{5cm}}\\[2mm]
2. Pr"ufer:\hspace{5mm}\underline{\hspace{5cm}}\\[2mm]
3. Pr"ufer:\hspace{5mm}\underline{\hspace{5cm}}\\[2mm]
im Promotionskolloquium\\[1cm]
Tag des Promotionskolloquiums:\hspace{5mm}\underline{\hspace{3cm}}
\selectlanguage{english}
\newpage

}
% end plain pages

% Roman page numbers
{
\setcounter{page}{1}
\renewcommand{\thepage}{\roman{page}}

\selectlanguage{german}
\chapter*{Zusammenfassung}

Das Standardmodell der Teilchenphysik, welches in den sp"aten sechziger Jahren in seiner heutigen
Form eingef"uhrt wurde, hat sich "uber die letzten 40 Jahre hinweg in zahllosen Experimenten als eine
au"serordentlich genaue Beschreibung der Physik auf kleinsten Skalen erwiesen. Alle von dem Modell
vorhergesagten Fermionen und Vektorbosonen wurden mittlerweile entdeckt, ihre Eigenschaften
experimentell vermessen, und die Vorhersagen des Modells auf Basis dieser Daten
haben sich auf Schleifen- und sogar Mehrschleifenniveau als g"ultig erwiesen.

Der letzte noch nicht entdeckte Baustein des Modells ist das Higgs-Boson. Da von den bisher
durchgef"uhrten Experimenten im wesentlichen der Skalenbereich unterhalb von $\unit[100]{GeV}$
abgedeckt wurde, ist dies jedoch noch kein Problem und immer noch mit der Theorie konsistent. Der LHC
(der in der nahen Zukunft anlaufen sollte) wird den experimentell zug"angliche Skalenbereich
voraussichtlich um eine Gr"o"senordnung auf mehrere $\unit{TeV}$ erweitern. Sollte das Higgs auch
bei diesen Energien nicht gefunden werden, so w"are das Standardmodell falsifiziert und 
aus den Grundlagen der Quantenmechanik w"urde folgen, da"s entweder bei etwa $\unit[1]{TeV}$ die
St"orungstheorie zusammenbrechen oder neue Physik in Erscheinung treten m"u"ste, welche die
Perturbativit"at der Theorie gew"ahrleistet --- anderenfalls w"urde bei hohen Energien die
Unitarit"at des Zeitentwicklungsoperators verlorengehen und die statistische Interpretation der
Quantenmechanik zusammenbrechen.

Obwohl viele Modelle neuer Physik das Konzept der Symmetriebrechung durch fundamentale Skalarfelder
vom Standardmodell "ubernehmen und lediglich die Details des entsprechenden Sektors modifizieren,
gibt es auch eine gro"se Gruppe von Szenarien, welche keine derartigen Felder enthalten. Unter
diesen Higgslosen Modellen (und m"oglicherweise auch durch die ber"uhmte AdS/CFT-Korrespondenz
verkn"upft) sind die wohl bekanntesten Beispiele Technicolor-Modelle sowie extradimensionale
Higgslose Szenarien. Beiden Klassen von Modellen gemeinsam ist das Auftreten neuer Resonanzen im Spektrum
oberhalb von etwa $\unit[100]{GeV}$, deren Austausch perturbative Unitarit"at bei hohen
Skalen gew"ahrleistet. Diese neuen Resonanzen stellen jedoch ebenfalls eine ernstzunehmende Gefahr
f"ur solche Szenarien dar, da ihr Austausch zus"atzliche Beitr"age zu den bei LEP / LEP-II sehr
genau vermessenen Pr"azisionsobservablen liefert. Falls keine speziellen Vorkehrungen getroffen
werden, haben diese Beitr"age die Tendenz, derartige Modelle bereits auf der Basis existierender Daten
auszuschlie"sen.

In den letzten Jahren wurden extradimensionale Modelle vorgeschlagen, in welchen die Fermionen des
Standardmodells in der Extradimension delokalisiert sind, wodurch die Pr"azisionsobservablen durch
Justage der Kopplungen an die neuen Resonanzen korrekt reproduziert werden k"onnen. Derartige
Modelle sind ein gangbarer Weg, die elektroschwache Symmetrie zu brechen und die
perturbative Unitarit"at an der $\unit{TeV}$-Skala aufrecht zu erhalten, ohne ein fundamentales Higgs-Feld
zu postulieren. Allerdings sind extradimensionale Modelle (von Trivialf"allen abgesehen) nicht renormierbar
und nur unterhalb einer Cutoff-Skala g"ultig, und die meisten neuen Resonanzen liegen oberhalb dieses
Cutoffs. Eine \glqq ehrliche\grqq\ Erweiterung des Standardmodells sollte lediglich die Struktur
unterhalb dieses Cutoffs enthalten und den extradimensionalen Mechanismus zur Symmetriebrechung und
Verz"ogerung der Unitarit"atsverletzung implementieren, ohne Annahmen "uber die Physik jenseits der
Cutoff-Skala zu machen.

Das \glqq Three-Site Higgsless Model\grqq\ ist eine minimale Implementation dieser Idee. Obwohl dieses Modell
durch extradimensionale Higgslose Modelle motiviert werden kann, enth"alt es lediglich eine
Generation zus"atzlicher Resonanzen, welche vollst"andig unterhalb des Cutoffs liegt und die
Verletzung der Unitarit"at auf $2-\unit[3]{TeV}$ hinausschiebt. Der nicht mit dem Standardmodell
"ubereinstimmende Teil des Spektrums besteht aus einem kompletten Satz von Partnern f"ur alle
Standardmodellteilchen mit Ausnahme des Gluons und des Photons. Eine Analyse der bestehenden
experimentellen Einschr"ankungen zeigt, da"s das Modell die Pr"azisionsobservablen korrekt
reproduzieren
kann, falls die Kopplungen zwischen den schweren Eichbosonpartnern und den Fermionen des
Standardmodells sehr klein (etwa $1\%$ des Isospinkopplung) und die Fermionpartner
mit Massen $\ge\unit[1.8]{TeV}$ verh"altnism"a"sig schwer sind.

In dieser Doktorarbeit wurde die LHC-Ph"anomenologie dieses Szenarios untersucht. Zu diesem Zwecke
wurden die Kopplungen und Breiten aller neuen Teilchen berechnet und das Modell in den
Monte-Carlo-Generator WHIZARD / O'Mega implementiert. Diese Implementation wurde verwendet, um die
Produktion der Fermion- und Eichbosonpartner auf Partonniveau in verschiedenen Kan"alen zu
simulieren, welche sich f"ur die Entdeckung am LHC eignen k"onnten. In dieser Arbeit
werden die Ergebnisse zusammen mit ein Einf"uhrung in das Modell sowie einer Diskussion der
Modelleigenschaften pr"asentiert.

Obwohl ihre fermiophobe Natur die Entdeckung der schweren Eichbosonen an Teilchenbeschleunigern grunds"atzlich
erschwert, zeigt sich, da"s der LHC die entsprechenden Resonanzen finden kann sowie sogar einige
R"uckschl"usse auf die St"arke der fermiophoben Kopplungen (was ein wesentlicher Test der Konsistenz
eines solchen Szenarios w"are) zulassen sollte. Bei der Berechnung der Breite der schweren Fermionen
stellt sich heraus, da"s zu der gro"sen Masse auch relative Breiten von $10\%$ und mehr kommen, so
da"s diese Teilchen sich eher schlecht f"ur eine direkte Entdeckung am LHC eignen. Trotzdem zeigen
die Simulationen da"s, hinreichend viel Zeit, Geduld sowie ein gutes Verst"andnis von Detektor und Hintergrund
vorausgesetzt, eine direkte Entdeckung zumindest in einem Teil des Parameterraums m"oglich ist.

\selectlanguage{english}
\chapter*{Abstract}

The Standard Model of particle physics, conceived in the late 1960s, has been confirmed as an
extremely accurate description of microscopic physics in a multitude of experiments conducted over
the last 40 years. Over time, all fermions and vector bosons predicted by the model have been
discovered, their properties have been measured, and the predictions of the model based upon these
properties have shown to be accurate to the one loop and even multiloop level.

The last piece of the Standard Model which hasn't been discovered yet is the Higgs boson. As
experiments have essentially only been probing scales below $\unit[100]{GeV}$, this is not
necessarily a problem and still consistent with the theory. However, the LHC (which should be
commencing operation anytime soon) will be hopefully extending the energy scale accessible to
experiments by a an order of magnitude to several $\unit{TeV}$s. If the Higgs is not discovered in
this energy range, then the Standard Model would be falsified and the very fundamentals of quantum
mechanics would tell us that either perturbation theory must break down at about $\unit[1]{TeV}$
or that new physics must enter the stage in order for the theory to remain perturbative
--- otherwise, unitarity would be lost at high
energies and the probabilistic interpretation of quantum mechanics would break down.

While many models of new physics retain the concept of fundamental scalars in the spectrum being
responsible for the symmetry breaking and just modify the details of the implementation, there is
also a large group of scenarios which do not contain any such fields at all. Among these Higgsless
models (and probably also connected by the celebrated AdS/CFT correspondence), the arguably most
prominent examples are technicolor and extra dimensional Higgsless models. In both classes of
models, new resonances appear in the spectrum above $\unit[100]{GeV}$, the exchange of which retains
perturbative unitarity at high scales. However, at the same time, the presence of these new resonances
also turns out to be a severe threat to such scenarios, their exchange leading to additional
contributions to the precision observables measured in the LEP / LEP-II experiments to very high accuracy and
tending to exclude such models if no special care is taken.

In the last years, extra dimensional models have been proposed which can evade these
constraints by delocalizing the Standard Model fermions within the extra dimension, thus allowing to
tune the couplings to the new resonances in order to avoid these constraints. This way, such models
are a viable method of breaking the electroweak symmetry and retaining perturbative
$\unit{TeV}$ scale unitarity without introducing a fundamental Higgs field. However, extra
dimensional models (excluding trivial cases) are intrinsically nonrenormalizable
and valid only below a cutoff scale,
with most of the new resonances lying in fact above the cutoff.
Conceptionally, a honest extension of the Standard Model should only contain the structure below
this cutoff, incorporating the extra dimensional mechanism of breaking the symmetry and delaying
unitarity violation without making assumptions on the high energy physics above the cutoff scale.

The Three-Site Higgsless Model is a minimal implementation of this idea. While it can be motivated
by extra dimensional Higgsless models of electroweak symmetry breaking, it in fact contains only
one set of extra resonances which lies below the cutoff, delaying unitarity violation to
$\approx 2-\unit[3]{TeV}$. The non-Standard Model part of the spectrum consists of a set of heavy
partners for all Standard Model particles with the exception of photon and gluon. The analysis of
the experimental constraints reveals that, while the model is consistent with the precision
observables, the couplings between the new heavy gauge bosons and the Standard Model fermions have
to be exceedingly small ($\approx 1\%$ of the isospin gauge coupling) while the new fermions are
constrained to be rather heavy with masses above $\unit[1.8]{TeV}$.

In this thesis, we explored the LHC phenomenology of this scenario. To this end, we calculated the
couplings and widths of all the new particles and implemented the model into the Monte-Carlo
eventgenerator and WHIZARD / O'Mega. With this implementation, we simulated the parton-level
production of the gauge boson and fermion partners in different channels possibly
suitable for their discovery at the LHC. The results are presented in this work together with an
introduction to the model and a discussion of the properties and couplings of the model.

We find that, while the fermiophobic nature of the new heavy gauge bosons does make them
intrinsically difficult to observe at a collider, the LHC should be able to establish the
existence of both resonances and even give some hints about the properties of their couplings which
would be a vital test of the consistency of such a scenario. For the heavy fermions, we find that
their large mass is accompanied by relative widths of more than $10\%$, making them ill-suited for a
direct discovery at the LHC. Nevertheless, our simulations reveal that there is a part of parameter
space where, given enough time, patience and a good understanding of detector and backgrounds, a direct
discovery might be possible.

\tableofcontents

\newpage
}
% end Roman page numbers.

\setcounter{page}{1}

% A labeled equation with the label optionally printed on the right.
\newenvironment{lequation}[1]%
{\ifthenelse{\boolean{preview}}{\marginpar{\mbox{}\\[0.5ex]\color{green}label:\\#1}}{}\begin{equation}\label{#1}}%
{\end{equation}}

\newcommand{\warning}[2]{\ifthenelse{\boolean{preview}}%
{{\color{red}#1}\marginpar{\color{red}%
\begin{minipage}{0.2\textwidth}\raggedright#2\end{minipage}}}%
{#1}}

\newcommand{\myurl}{\texttt{http://theorie.physik.uni-wuerzburg.de/{\textasciitilde}cnspeckn/}}

\chapter{Introduction}

\begin{quote}\itshape
Of course our model has too many arbitrary features for these predictions to be taken very
seriously...
\end{quote}
\hfill\begin{minipage}{0.7\textwidth}\small\raggedleft
(Steven Weinberg, ``A theory of leptons'')
\end{minipage}
\\[5mm]
This quote taken from \cite{Weinberg:1967tq} on what would later become the Standard Model of elementary
particle physics may well be one of the most spectacular cases of false modesty in the history of
science. Only six years
after Steven Weinberg wrote these memorable lines on the then speculative unification of
electromagnetic and charged current interactions in a spontaneously broken
$\sun{2}_\text{L}\times\un{1}_\text{Y}$ gauge theory, the discovery of the neutral current
interaction at the Gargamelle bubble chamber in 1973 at CERN suggested that there was more to this
model than its authors originally had dared to hope.

Attempting to include the three known quarks into this model of weak interactions and explain the
smallness of flavor violation, Glashow,
Iliopoulos and Maiani postulated an additional quark (the charm) in 1970 \cite{Glashow:1970gm}, while
in 1972 Kobayashi and Maskawa suggested another doublet of quarks (top and bottom) in order to
explain the violation of CP symmetry observed in the hadron sector \cite{Kobayashi:1973fv}. Again,
nature chose to prove the theorists right, and in 1974, the charm quark was discovered in form of
the $J/\Psi$ charmonium state, while the $\Upsilon$ bottomonium state was discovered in 1977.

By the end of the 1970s, the Nobel prize committee in Stockholm deemed the evidence for the unified
theory of electroweak interactions to be convincing enough to win
Glashow, Salam and Weinberg the 1979 Nobel prize in physics, even before the postulated $W$ and $Z$
bosons were discovered as resonances at the UA1 and UA2 detectors in 1983, with the 1984
physics Nobel prize going to Rubbia and van der Meer for this discovery.

Up to this day, the success story of the Standard Model has continued
with the impressive confirmation of the one loop structure of the model by the LEP / LEP-II
experiments, the discovery of the top quark in 1995 at the Tevatron and the discovery of tau neutrino
in 2000. Indeed, quite contrary to Weinbergs initial feelings, every single prediction of
the Standard Model has been confirmed in experiment, and all significant discoveries which are
usually called ``new physics'' (e.g. neutrino oscillations) can be easily accommodated within the
model.

The last remaining specimen from the Standard Model particle zoo is the Higgs boson. If it exists,
then it must have managed to escape detection both at LEP / LEP-II and Tevatron, and together with
indirect bounds on its mass coming from a global fit to the Standard Model, this implies that it
most probably is lurking around $\unit[120]{GeV}$ and in any case must be lighter than several
hundred $\unit{GeV}$ where the LHC together with the ATLAS and CMS experiments would
surely find it. After all the past success of the model, it is arguably not the discovery but the absence of
a Higgs resonance at the LHC which would be exciting news from the land of particle physics.

Still, there is much truth in Weinbergs statement (after all, it comes from a Nobel laureate), and
the Standard Model does exhibit many features that deserve to be called arbitrary and seem to be
unfitting for a deep model of nature, e.g. the large number of free parameters (especially the
Yukawa sector), the seemingly magical cancellation of gauge anomalies which is essential for the
celebrated renormalizability of the model or the three generations of fermions, each of which is an
exact copy of the others, differing only in mass.

Also, at the very least, the Standard Model does not describe dark matter and is incompatible
with general relativity, and this fact alone implies that it can be only an effective description
of nature. If we accept this as fact and still want to keep the Standard Model as-is up to the
Planck scale, then we must answer the question why nature would adjust the Higgs mass at the
matching scale to ridiculous precision in order to achieve the moderate value we observe.
This is the famous Hierarchy Problem which would leave an ugly aftertaste if the Higgs were the only
``new'' piece of nature waiting to be discovered at the LHC.

So, what happens if the LHC fails to discover the Higgs and puts an end to the success of the
Standard Model? If there is no Higgs boson, then (unless our understanding of quantum field theory
is fundamentally flawed) consistency conditions require that either new physics enters the game
below $\unit[1]{TeV}$ or that the model becomes strongly interacting at this scale. In this sense,
the LHC cannot fail in its mission: Higgs or not, something yet undiscovered most certainly has to
lurk within its reach.

Over the last decades, particle physicists have invested a lot of time and thought into finding out
what such a discovery might look like. The resulting models can be divided into two categories:
renormalizable models which (even if they may not be candidates for a fundamental theory of nature
due to their physics content) can be valid to arbitrarily high scales, and effective field theories
which are only valid below a cut-off scale and which do not claim to cover physics at scales
significantly higher than those probed by the LHC.

Models which fall into the first category are often motivated in a top-down fashion by theoretical
concepts of how nature might organize itself at hight scales, for example supersymmetry or
technicolor, while effective models are usually built in a more pragmatic way as an extension
of the low-energy Standard Model physics with little or no assumptions on the high energy
structure of nature.

Either way, it turns out that building a model which is in agreement with all experimental
constraints collected over the last decades is quite challenging. In particular, any model of
electroweak symmetry breaking different from the usual Higgs mechanism
tends to change the structure of the leptonic current-current
interactions which have been very precisely mapped out in the LEP / LEP-II experiments. The major part
of the resulting constraints can be summarized in only three numbers which, however, have a
devastating effect on many models of electroweak symmetry breaking. In particular, technicolor
models and models of Higgsless symmetry breaking from compact extra dimension tend to violate these precision
constraints badly, and extra care has to be taken if models of this kind are to remain
viable candidates for a description of nature\footnote%
{
Of course, if the celebrated AdS/CFT correspondence is correct, then the difficulties technicolor
and 5D theories are faced with are connected by this duality, and solutions in one formalism induce
solutions in the other framework.
}.

In the specific case of compact extra dimensions, all particles usually come with a tower of
partners of
increasing mass which correspond to the higher excitations of the 5th momentum component (exceptions
are particles which are explicitly localized on four-dimensional submanifolds by constuction), the
so called Kaluza-Klein (KK) modes.
Depending on the conditions enforced on the fields on the boundary of the extra dimensions, massless
modes can be forbidden for the towers, thus facilitating the breaking of electroweak gauge symmetry
without introducing a Higgs.

As stated above, if a model of new physics is to remain perturbative at the $\unit{TeV}$ scale, then
it must contain some agent which retains perturbative unitarity at high scales, this part being played
by the Higgs in the Standard Model. In Higgsless extra-dimensional models, this is achieved via the
exchange of higher KK modes, which requires nonzero couplings between the Standard Model particles and
their KK partners. However, these couplings also imply new contributions to the
current-current interactions which lead to the aforementioned conflict with the LEP observables.
Luckily, these constraints can be evaded by a moderate tuning of the wavefunctions of the Standard
Model fermions, and this way, extra dimensions remain a potential candidate for the mechanism of
electroweak symmetry breaking.

However, if we accept this fine-tuning of the parameters in order to comply with the precision
observables, extra dimensional models contain another conceptional flaw. Although they are built in
a top-down way with the high energy structure of spacetime in mind, the resulting models are in fact
nonrenormalizable with a UV cutoff scale which is typically of the order of $\unit[5-10]{TeV}$. Therefore,
most of the new physics in these models is meaningless as it lies above the cutoff, and an effective
field theory which only describes the KK modes below the cutoff would be a more honest way to
describe the remaining sector which is not dominated by other high energy physics.

The Three-Site Higgsless Model \cite{Chivukula:2006cg},
the LHC phenomenology of which is the topic of this thesis, is
an implementation of this idea. While it is clearly inspired by such Higgsless extra
dimensional models of electroweak symmetry breaking, it only describes the Standard Model particles
and the first generation of KK excitations, making no assumptions on the underlying high energy
physics. The precision constraints severely restrict the
couplings between the new heavy particles, and we will find that, albeit the structure of the model
is simple compared to other, more ambitious extensions of the Standard Model, these restrictions
lead to a quite interesting and rich LHC phenomenology.

In chapter \ref{chap-1}, we motivate the model in a bottom-up approach, starting with the
assumption of unitarity being maintained by $W$ and $Z$ partners in the absence of a Higgs boson
and show that, while the model doesn't cure the arbitrariness in the Standard Model,
most of the new structure is a straightforward consequence of this assumption.
The explicit construction of the model using a more ideological top-down approach inspired by
extra dimensions is shown in chapter \ref{chap-2}, and in chapter \ref{chap-3} we will move on to
discussing the bounds on the model, its parameter space, the couplings and the widths of the new
particles. Chapter \ref{chap-4} elaborates on the tools we have been using for the phenomenology
study, the results of which are presented in chapters \ref{chap-5} -- \ref{chap-7}.

\chapter{The Model --- Bottom-up Approach}
\label{chap-1}
\begin{quote}\itshape
I have a cunning plan.
\end{quote}
\hfill\begin{minipage}{0.7\textwidth}\small\raggedleft
(Archetypical quote from Baldrick in ``Blackadder'')
\end{minipage}
\\[5mm]
The \threeshl\ is only one example out of a multitude of models for physics beyond the
Standard Model that have been proposed. Before examining the phenomenology of any such model in
detail, we should have a clear notion why looking at this particular kind of new physics is
worthwhile.

The purpose of this chapter is to give a bottom-up type motivation for the \threeshl\
as a straightforward option for pushing the unitarity bounds of the Standard Model once the
notion of a fundamental Higgs field that unitarizes scattering amplitudes is discarded.

Although this chapter contains some general remarks on quantum field theories, familiarity
of the reader with the formalism and with the Standard Model ist assumed. These topics can be found covered
in-depth in many textbooks, e.g. \cite{peskin,itzykson,weinberg1,weinberg2}.

\section{Restrictions on Quantum Field Theories}

\subsubsection{Setting the stage}
\label{chap-1-1}

According to our current understanding of nature, the world at very small scales is properly
described in the language of relativistic quantum field theories. To define such a theory, we need
to specify the field content of the theory as well as the Lagrangian.

The field content tells us what different kinds of particles are described by the theory. The fields
transform in different representations of the Poincar\'e group  which are distinguished by
spin and mass of the field.

The Lagrangian $\LL$ of the theory is a Lorentz invariant local function of the fields and their derivatives and
determines the time evolution of the states described by the theory. It is commonly assumed that
$\LL$ either is a polynomial of the fields and their derivatives or at least can be expanded in a
series in a meaningful way.

The most useful known way to extract predictions out of this kind of description of nature is the
use of perturbation theory to describe scattering experiments. Such experiments are modeled in an
idealized way as a world that only consists of two particles with well-defined momenta in the
asymptotic past, which then interact and scatter to finally evolve into a set of particles
with well defined momenta in the distant future. This description relies on the hypothesis that there
are such asymptotically free states and that all interactions between the particles are negligible
at asymptotic times (``adiabatic switching''). The time evolution operator that mediates the
transition from the free particle states in the asymptotic past (``in states'') to those in the
asymptotic future (``out states'') is called the $S$ matrix operator.

To describe the scattering process at finite times, the Lagrangian is split into a
quadratic propagation part $\LL_\text{free}$ and an interaction part $\LL_\text{int}$
that consists of monomials of higher order in the fields\footnote%
{
Because of the Lorentz invariance of the Lagrangian, linear terms in the fields are only allowed for
scalars and generate a shift in the field vacuum expectation value in this case.
}
. It is then assumed that the interaction part is just a small perturbation on top of the free
propagation, and time dependent perturbation theory is applied to calculate the transition amplitudes
order for order as an expansion in the coupling constants appearing as prefactors at the monomials in
$\LL_\text{int}$. This formalism then leads to the well-known expansion of scattering amplitudes in
terms of Feynman diagrams.

Although this setup suggests a lot of freedom in the construction of a quantum field theory, there are
some properties any sensible theory has to fulfill.
The severity of these constraints depends on whether we want our
theory to be acceptable as a fundamental description of nature or whether we just want it to be an
effective description of our world that breaks down at some energy scale.

\subsubsection{Renormalization and renormalizability}

Any physical model comes with a number of free parameters that have to be fixed
before the theory can be used to quantitatively predict experimental results. One
criterion for a sensible model is the existence of predictions that can be checked
to potentially falsify the theory, and therefore, these free parameters must be
such that they can be determined with a finite number of measurements.

When we set out to calculate higher orders in the perturbation expansion, we encounter loop diagrams
that translate to unbounded integrals over four-momenta. Quite contrary to the assumption that higher
order contributions are small compared to the leading order, the vast majority of these integrals is
infinite. To study these infinities, it is useful to bound the integration domain and cut
off the high energy contributions at some energy scale using a regularization scheme (e.g. a
na"ive cutoff, Pauli-Villars, dimensional regularization etc.). Calculations in the
regularized theory then give finite predictions for observables which now carry an explicit
dependence on the parameters and on the cutoff scale (in which they are divergent).

Interestingly, there is a class of theories in which the reexpression of the free parameters
in terms of a finite set of measurable defining observables\footnote%
{
An introduction of a finite number of new free couplings that undergo the same treatment might also be
necessary.
}
leads to a cancellation of the divergences. In this case, the cutoff scale can be taken to infinity,
leading to well-defined, finite observables that now only depend on a finite set of numbers that must
be measured to completely fix the theory. This process is called renormalization, and theories which
can be treated this way are called renormalizable.

Of course, there is a lot of freedom in the choice of defining observables. Different
observables will lead to different convergence properties of the perturbation series. In particular,
for a given observable, a clever choice of the scale $\mu$ at which the defining observables are measured can
significantly reduce the contributions of higher orders; this corresponds to a resummation of the
perturbation series.

In the limit of infinite cutoff scale, the parameters which are expressed through the defining
observables are divergent. Therefore, the above renormalization trick can be equivalently achieved
by splitting the parameters $g$ into a finite part $g_R$ and an infinite part $\delta g$
which is called a counterterm. The finite part is fixed by measurement while
the counterterm is chosen according to some prescription such that it cancels the infinities. If
the splitting prescription depends on some renormalization scale $\mu$ like in
$\overline{\text{MS}}$, then a change of this scale shifts contributions between $g_R$ and $\delta
g$, changing the value of $g_R$. This dependence of the $g_R$ on $\mu$ is called the
renormalization group flow of the parameters, and a suitable choice of $\mu$ again allows to resum the
perturbation series and reduce the contributions from higher orders.

The whole process of
renormalization can be interpreted as absorbing the effects of physics far above the renormalization
scale into an (albeit infinite) redefinition of the couplings.

\subsubsection{Effective field theories}

Renormalizable field theories can be taken to be valid at arbitrarily high energies and therefore are
candidates for a fundamental theory of nature. However, with any theory, the claim of it being
fundamental is a very ambitious one and, furthermore, the renormalizability condition turns out to be
very restrictive on the types of allowed Lagrangians.

However, if we drop the requirement of the theory being fundamental and instead think of it in terms of an
effective field theory that is only valid below some scale $\Lambda$, then there is nothing bad
in the idea of having a cutoff scale to make our integrals finite.
Quite contrary, as the theory is only valid up to $\Lambda$, we should only include intermediate
states below $\Lambda$ anyway. Effects that come from
high energy physics should be included as effective couplings in our low energy theory. The cutoff
explicitly enters our calculations and again influences the convergence of the perturbation series.

If we change the cutoff scale $\Lambda$, we can calculate how we have to adjust the couplings in
front of the operators to retain the predictions. This leads to a renormalization group flow of the
couplings that can again be used to resum parts of the perturbation series and this way allow for a
meaningful perturbative calculation of observables at different scales. The downside is that, as
opposed to renormalizable theories, we must allow for infinitely many couplings as they will be
introduced by the renormalization group flow anyway.

To keep our theory predictive, we must assume some kind of ordering scheme where
the higher order operators are suppressed by some scale $\Lambda_\text{NP}$ and therefore are negligible at
low energies. If we construct an effective theory e.g. by integrating out heavy particles from a
renormalizable theory, the new operators describing the effects of the degrees of freedom that have
been removed obey exactly such an ordering scheme, so this is not an unnatural assumption. This way
the theory can be predictive at low energies even though the Lagrangian describing
it contains arbitrarily many operators.

If we want to calculate observables at some scale $\Lambda$, we must evolve the couplings to this
scale in order to perform the perturbative calculation. The higher order interactions increase
if we evolve to higher scales, and at some scale $\Lambda_0$, the ordering scheme will break down.
At this scale the very latest, we have no control over the contributions from higher order operators
anymore, and the effective theory is not predictive anymore. It then has
to be replaced by another theory that might contain new degrees of freedom and which can either be
a fundamental, renormalizable theory or again an effective theory.

This way, if we don't require our theory to be fundamental, renormalizability is not necessary for
the theory to make sense. We only have to be aware that the energy range in which the theory is
valid is limited and that we have to replace it by something else above $\Lambda_0$. Not insisting
on renormalizability allows for a much larger class of theories which can be effective
descriptions of nature even if quantum field theories should turn out not to be the suitable language for the
``theory of everything''. Even better, the scale $\Lambda_0$ as well as a self-consistent ordering
scheme for the higher dimension operators can be estimated from the parameters of
the theory by simple dimensional arguments without requiring any knowledge about the underlying high
energy physics; this procedure is called NDA \cite{Manohar:1983md} (``na"ive dimensional analysis'').

The remainder of this chapter will be dealing with effective field theories, so
renormalizability will not be an issue as long as we stay in the energy range where these theories
are valid.

\subsubsection{Unitarity --- limits on the validity of the perturbation expansion}

Unrelated to the issue of renormalizability, there is another criterion for the consistency of field
theories. Because it is the limit of a time evolution operator, the axioms of quantum mechanics require the $S$
matrix to be unitary. This is the prerequisite for the probabilistic interpretation of quantum
mechanics.

The diagonal elements of the $S$ matrix contain contributions from the case
of direct propagation of the initial into the final state without any scattering. As we are usually
not interested in these contributions, the cross section is conveniently defined as a flux normalization
factor times the modulus of an $T$ matrix element, $T$ being defined as
\begin{equation}\label{equ-1-1-tdef} S = iT + \mathbb{I} \end{equation}
The unitarity condition on $S$ then implies a relation for $T$:
\begin{lequation}{1-1-tunit}
SS^\dagger = \mathbb{I} \quad\longrightarrow\quad i\left(T^\dagger - T\right) = TT^\dagger
\end{lequation}
Taking the matrix element between two states $\ket{a}$ and $\ket{b}$ and inserting a complete set of
basis vectors, we obtain
\begin{lequation}{1-1-tunit-states}
i\left(\braXket{b}{T}{a}^* - \braXket{a}{T}{b}\right) = \sum_{\ket{c}}
\braXket{a}{T}{c}\braXket{b}{T}{c}^*
\end{lequation}

From \eqref{1-1-tunit-states}, important consequences for matrix elements can be derived.
Restricting the equation to the forward scattering case $\ket{a}=\ket{b}$ yields
\begin{lequation}{1-1-otheorem}
2\Im\braXket{a}{T}{a} = \sum_{\ket{c}}\abs{\braXket{a}{T}{c}}^2
\end{lequation}
This is the famous optical theorem which basically states that the imaginary part of the forward
scattering amplitude of some state is equal to the probability for this state scattering into
anything. If $\ket{a}$ is an eigenstate of $T$ with eigenvalue $T\ket{a}=t\ket{a}$ and norm
$\braket{a}{a} = 1$, \eqref{1-1-otheorem} further reduces to
\[ 2\Im t = \abs{t}^2 \]
which is solved by\footnote%
{
The same result can also be directly obtained from the definition of $T$ in terms of the unitary $S$
matrix \eqref{equ-1-1-tdef}.
}
\begin{lequation}{1-1-argand}
t = i + e^{i\delta}
\end{lequation}
\eqref{1-1-argand}  restricts the eigenvalue $t$ to a ``Argand circle'' in the complex plane with
center $i$ and radius $1$.

Due to rotational invariance, two particle angular momentum eigenstates below the inelastic
threshold (the regime in which the total energy is not sufficient to create any new particles) are
eigenstates of the $T$ matrix. If we examine the cross section between incident momentum
eigenstates, then these states can be decomposed in a relativistic partial wave
expansion \cite{Jacob:1959at}, and \eqref{1-1-argand} implies relations for the partial wave amplitudes.
If we cross the inelastic threshold \eqref{1-1-argand} changes into an inequality, and the
amplitudes are constrained to lie within the Argand circle rather than on it. For a more complete
discussion of unitarity and partial wave amplitudes see \cite{Jacob:1959at,itzykson} 

If the Lagrangian is hermitian, then the $S$ matrix will be unitary by definition\footnote%
{
Exceptions can arise if the quantization is done incorrectly and states are inconsistently removed
from the Hilbert space, e.g. due to improper quantization of gauge theories.
}.
However, as \eqref{1-1-tunit} -- \eqref{1-1-argand} are nonlinear in the $T$ matrix elements, they
don't hold order by order if the $T$ matrix elements are expanded in a perturbation series\footnote%
{
Of course, the equations themselves can be expanded to hold order by order, but in this case,
different orders in the expansion of the matrix elements will mix.
}.
Therefore, at a fixed order in perturbation theory, the partial wave amplitudes can violate the
unitarity criteria. Worse, if an amplitude grows with energy in perturbation theory, then it will
leave the Argand circle at some scale and higher orders must become increasingly large to compensate
and restore consistency to the theory.

In particular, if the tree level amplitude for some scattering process grows with energy, then there
will be a scale where the tree level scattering probability exceeds one and at which perturbation theory
stops to be a trustworthy calculational instrument and breaks down. If we write down an effective
field theory, this unitarity cutoff will limit the range of applicability of the theory
independently from the NDA cutoff.

\section{The Fermi Model}
\label{chap-1-2}

At ``low'' energies much smaller than the $W$ mass, our world can be well described by the Fermi model.
At this scale, the observable fundamental degrees of freedom are\footnote%
{
As color and QCD are not related in any way to the line of reasoning aimed at in this chapter,
all colored particles will be ignored for the remainder.
}:
\begin{itemize}
\item Three generations of leptons $e^-, \mu^-, \tau^-$ which differ only by their mass and which
are described by Dirac spinors $L_j$.
\item The associated three generations of neutrinos $\nu_e, \nu_\mu, \nu_\tau$ which we will
approximate to be massless and which are described by Dirac spinors $N_j$.
\item The photon which mediates the electromagnetic interaction and which is described by the vector
field $A^\mu$.
\end{itemize}

The Fermi Lagrangian $\LL_\text{Fermi}$ can be decomposed as
\[ \LL_\text{Fermi} = \LL_\text{QED} + \LL_4 \]
The first part is the usual QED Lagrangian $\LL_\text{QED}$ given by
\begin{lequation}{1-2-lqed}
\LL_\text{QED} = \sum_{j = 1}^3\left(\overline{L}_j\left(i\slashed{\partial} + m_j\right)L_j +
i\overline{N}_j\slashed\partial N_j\right) - eA_\mu J_Q^\mu - \frac{1}{4}F^{\mu\nu}F_{\mu\nu}
\end{lequation}
with the electromagnetic gauge coupling $e$, the electromagnetic current
\begin{lequation}{1-2-curem}
J^\mu_Q = \sum_{j=1}^3 \overline{L}_j\gamma^\mu L_j
\end{lequation}
and the electromagnetic field strength tensor
\[ F^{\mu\nu} = \partial^\mu A^\nu - \partial^\nu A^\mu \]
Defining the left- and right-handed fermion fields $\Psi_L$ and $\Psi_R$ via \eqref{conv-chiral},
we can introduce the charged and neutral currents
\begin{lequation}{1-2-curweak}\begin{aligned}
J^+_\mu &= \frac{1}{\sqrt{2}}\sum_{j=1}^3 \overline{L}_{L,j}\gamma_\mu N_{L,j} \quad,\quad
\\
J^-_\mu &= J^{+\dagger}_\mu
\\
J^0_\mu &= \sum_{j=1}^3\left(\overline{L}_j\gamma_\mu\left(\sin^2\theta_W -
\frac{1}{2}\Pi_-\right) L_j + \frac{1}{2}\overline{N}_{L,j}\gamma_\mu N_{L,j}\right)
\end{aligned}\end{lequation}
(with the Weinberg angle $\theta_W$) and write the second piece $\LL_4$ as
\begin{lequation}{1-2-l4pt}
\LL_4 = -4\sqrt{2}G_F\left(J^+_\mu J^{-\mu} + J^0_\mu J^{0\mu}\right)
\end{lequation}
where the Fermi coupling $G_F$ has mass dimension $\left[G_F\right]=-2$.

\eqref{1-2-lqed} contains the bilinear parts of the Lagrangian that generate the propagators as well
as the three point gauge couplings of the photon to two leptons, while \eqref{1-2-l4pt} encodes the
charged and neutral current four point couplings among four fermions. Apart from the lepton
masses, the only free parameters are the electromagnetic gauge coupling $e$, the Fermi coupling
$G_F$ and the Weinberg angle $\theta_W$. Once these
parameters are fixed, the Fermi model does a very good job at describing the low energy
phenomenology of our world.

However, while pure QED is renormalizable, the four point interactions
make the Fermi model a nonrenormalizable model that is only valid up to a UV cutoff scale
$\Lambda_\text{UV}$. To estimate this scale, consider the one loop correction to the
charged current four point interaction\footnote%
{
The $t$ channel contribution can be avoided by considering the four point function for two
different flavors (e.g. $e^-\bar{\nu}_e\rightarrow\mu^-\bar{\nu}_\mu$).
}%
:
\[
\parbox{17mm}{\begin{fmfgraph}(17,14)
\fmfleft{i2,i1}\fmfright{o2,o1}\fmf{fermion}{i2,v,i1}\fmf{fermion}{o1,v,o2}
\fmfblob{8thick}{v}\end{fmfgraph}}
\quad=\quad
\parbox{20mm}{\begin{fmfgraph}(17,14)\fmfleft{i2,i1}\fmfright{o2,o1}
\fmf{fermion}{i2,v,i1}\fmf{fermion}{o1,v,o2}\fmfdot{v}\end{fmfgraph}}
\quad+\quad
\parbox{28mm}{\begin{fmfgraph}(28,14)\fmfleft{i2,i1}\fmfright{o2,o1}\fmf{fermion}{i2,v1,i1}
\fmf{fermion}{o1,v2,o2}\fmf{fermion,left=0.6,te=0.5}{v1,v2,v1}\fmfdot{v1,v2}\end{fmfgraph}}
\quad+\quad\ldots
\]
Na"ive powercounting reveals a quadratic divergence in the one loop contribution, and we can
estimate the leading contribution by
\[
8{G_F}^2N_f\int \frac{d^4p}{\left(2\pi\right)^4}\;\tr\frac{\slashed{p}\slashed{p}}{p^4} =
\frac{4}{\pi^2}{G_F}^2N_f\int_0^{\Lambda_\text{UV}}dp\;p =
\frac{2}{\pi^2}{G_F}^2\Lambda_\text{UV}^2N_f
\]
where $N_f$ is the number of flavors which can run in the loop. Considering only the leptons we
have $N_f=3$, and
demanding that the one loop correction must not exceed the tree level contribution if perturbation
theory is to make sense then leads to a estimate for the cutoff scale
\[ \Lambda_\text{UV} = \pi\sqrt\frac{\sqrt{2}}{N_fG_F} \approx \unit[630]{GeV} \]
Of course, if we had also included the quarks into our version of Fermi theory, they would appear
within the loop, and the cutoff would be lowered further to
$\Lambda_\text{UV}\approx\unit[315]{GeV}$.

However, even when setting aside all issues related to renormalization, a straightforward calculation of the amplitude
for $ff\longrightarrow ff$ type processes reveals terms induced by the four point
couplings which grow quadratically with energy. Partial wave analysis yields a
scale of $\approx\unit[600]{GeV}$ \cite{quigg,Ohl:1999qm} at which the $s$ wave amplitude exceeds the unitarity bound. At
this scale the very latest, perturbation theory can't be trusted anymore. If our world is to be
described by a perturbative quantum field theory, then new physics must come into play below this
scale and unitarize the scattering amplitudes or at least delay the scale of unitarity violation.

\section{The Standard Model}
\label{chap-1-3}

Indeed, experimentalists didn't have to go to scales of $\unit[600]{GeV}$ to discover new physics
that supersedes the Fermi model: in the early eighties, the $W^\pm$ and $Z$ bosons were discovered at
the Super Proton Synchrotron SPS at CERN and at Fermilab with masses of $m_W\approx\unit[80]{GeV}$
and $m_Z\approx\unit[91]{GeV}$ \cite{Amsler:2008zzb}.

Together with the photon, the $W^\pm$ and $Z$ are understood as the gauge bosons of a
$\sun{2}_L\times \mathbf{U}(1)_Y$ gauge group which unifies electromagnetism and weak interactions.
In the following, we will denote the $\sun{2}_L$ gauge fields as $W^\mu=W^\mu_k\tau_k$ (with the
$\sun{2}$ generators $\tau_k$, see \eqref{conv-su2gen}) and the $\mathbf{U}(1)_Y$
gauge field as $B^\mu$; the gauge couplings will be called $g$ and $g^\prime$ respectively.
Introducing the field strength tensors
\[ F_W^{\mu\nu} = \partial^\mu W^\nu - \partial^\nu W^\mu - ig\komm{W^\mu}{W^\nu} \quad,\quad
F_B^{\mu\nu} = \partial^\mu B^\nu - \partial^\nu B^\mu \]
the gauge kinetic Lagrangian can be written as
\[ \LL_\text{gauge} = -\frac{1}{2}\tr F_W^{\mu\nu}F_{W\mu\nu} -
\frac{1}{4}F_B^{\mu\nu}F_{B\mu\nu}\]

As the $W$ and $Z$ are massive, the gauge symmetry must be broken, and the only known mechanism
which achieves this while still allowing for a consistent quantization is spontaneous symmetry
breaking. Without any
assumptions on the underlying dynamics, the breaking can be parameterized by the introduction of
a $\sun{2}$ valued field\footnote%
{
More accurately, it is the rescaled field $\frac{1}{v}\Sigma$ which is valued in $\sun{2}$.
}
\[ \Sigma = ve^{i\frac{\phi_k}{v}\tau_k} \]
where $v$ is the vacuum expectation value\footnote%
{
Note that this definition of $v$ differs by a factor of $\sqrt{2}$ from that usually
used found textbooks.
}
of $\Sigma$
\[ \left<\Sigma\right> = v\mathbb{I} \]
This field is assigned the transformation behavior
\[ \Sigma\quad\xrightarrow{\;\sun{2}_L\times\mathbf{U}(1)_Y\;}\quad
e^{i\lambda_{L,k}\tau_k}\Sigma e^{-i\lambda_Y\tau_3} \]
with the gauge transformation parameters $\lambda_{L,k},\lambda_Y$. With this charge assignment,
the covariant derivative of $\Sigma$ reads (c.f. \eqref{conv-codev})
\[ D^\mu\Sigma = \partial^\mu\Sigma - igW^\mu_k\tau_k\Sigma + ig^\prime B^\mu\Sigma\tau_3 \]

The minimal gauge invariant Lagrangian required to make $\Sigma$ a dynamic field is
\begin{lequation}{1-3-lsigma}
\LL_\Sigma = \tr\left(D^\mu\Sigma\right)\left(D_\mu\Sigma\right)^\dagger
\end{lequation}
Expanding \eqref{1-3-lsigma} in the component fields $\phi_i$, we find that the lowest order (aka
the vacuum expectation value of $\Sigma$) leads to mass terms for the gauge bosons which can be
written as
\begin{lequation}{1-3-gbmterm}
\frac{v^2}{2}\left(g^2\sum_{k=1}^{2}W_k^\mu W_{k,\mu} + \left(W_{3,\mu},B_\mu\right)
\begin{pmatrix}g^2\vs{2ex} & -gg^\prime \\ -gg^\prime\vs{2ex} & {g^\prime}^2\end{pmatrix}
\cvect{W_{3,\mu}\vs{2ex} \\ B_\mu\vs{2ex}}\right)
\end{lequation}
\eqref{1-3-gbmterm} is diagonalized by the mass eigenstates
\begin{equation}\label{equ-1-3-meigen}
W^\pm_\mu = \frac{W_{1,\mu} \mp i W_{2,\mu}}{\sqrt{2}} \quad,\quad
Z_\mu = \frac{gW_{3,\mu} - g^\prime B_\mu}{\sqrt{g^2 + {g^\prime}^2}} \quad,\quad
A_\mu = \frac{g^\prime W_{3,\mu} + gB_\mu}{\sqrt{g^2 + {g^\prime}^2}}
\end{equation}
and yields the $Z$ and $W$ masses as eigenvalues
\begin{equation}
m_W = gv \quad,\quad m_Z = \sqrt{g^2 + {g^\prime}^2}v \label{equ-1-3-mgb}
\end{equation}

Classically, the $\Sigma$ field can be put to $v\mathbb{I}$ at every point in spacetime by a gauge
transformation, absorbing the three corresponding degrees of freedom into the gauge fields.
This identifies the three $\phi_k$ fields as the Goldstone boson fields which are
required by the Goldstone theorem \cite{Goldstone:1961eq} and which are ``eaten'' by the heavy gauge
bosons to constitute the longitudinal degrees of freedom. As the $\phi_k$ transform in a nonlinear
realization of the gauge group, $\Sigma$ is called a nonlinear sigma field.

To reproduce the physics of the Fermi model, we have to specify the representation in which the
leptons transform. The left-handed leptons of each generation are grouped in isospin doublets
\[ \Psi_{L,j} = \cvect{N_{L,j}\vs{2ex} \\ L_{L,j}\vs{2ex}} \]
that transform in the fundamental representation of the $\sun{2}_L$, the right-handed leptons are
singlets under this group factor. The hypercharge transformation properties are derived from the
electric charge of the fermions:
\[ \Psi_L\xrightarrow{\;\mathbf{U}(1)_Y\;} e^{i\lambda_Y(Q - \tau_3)}\Psi_L \quad,\quad
\Psi_R \xrightarrow{\;\mathbf{U}(1)_Y\;} e^{i\lambda_Y Q} \Psi_R \]
with the electric charge $Q$. These assignments completely fix the covariant derivative and the
conserved currents which couple to the gauge bosons.
Defining the electromagnetic gauge coupling $e$ and the Weinberg angle $\theta_W$ as
\begin{equation}\label{equ-1-3-cpldef}
e = \frac{gg^\prime}{\sqrt{g^2 + {g^\prime}^2}} \quad,\quad \sin\theta_W =
\frac{g^\prime}{\sqrt{g^2 + {g^\prime}^2}}
\end{equation}
the resulting fermionic Lagrangian including the kinetic terms can be written as 
\begin{multline}\label{1-3-lferm}
\LL_\text{fermion} = i\sum_{j=1}^3\left(\overline{L}_k\slashed\partial L_k +
\overline{N}_k\slashed\partial N_k\right) + \\
\frac{e}{\sin\theta_W}\left(W^+_\mu J^{+\mu}
+ W^-_\mu J^{-\mu}\right) + \frac{2e}{\sin2\theta_W} Z^\mu J_{0\mu} - eA^\mu J_{Q\mu}
\end{multline}
The conserved currents appearing in \eqref{1-3-lferm} are the very same currents already defined
in \eqref{1-2-curem} and \eqref{1-2-curweak}.

In the Fermi model, the fermion mass terms can be inserted into the Lagrangian directly. However, with
the above $\sun{2}_L\times\mathbf{U}(1)_Y$ charge assignments, such terms would now violate gauge
invariance. Fortunately, the nonlinear sigma field $\Sigma$ allows to generate the necessary terms in
a gauge invariant way from Yukawa couplings\footnote%
{
For the quark masses, the mass matrix in \eqref{1-3-lyuk} is replaced by $\frac{1}{v}\diag\left(m_u,
m_d\right)$.
}%
:
\begin{lequation}{1-3-lyuk}
\LL_\text{Yukawa} = \sum_{j=1}^3 \Psibar_{L,j}\Sigma\begin{pmatrix}0 \vs{2ex} & 0 \\ 0 \vs{2ex} &
\frac{m_j}{v}\end{pmatrix}\cvect{N_{R,j}\vs{2ex} \\ L_{R,j}\vs{2ex}}
\end{lequation}

Putting together all pieces of the puzzle, the full Lagrangian of the theory reads
\begin{lequation}{1-3-ltot}
\LL = \LL_\text{gauge} + \LL_\text{fermion} + \LL_\text{Yukawa} + \LL_\Sigma
\end{lequation}
The resulting model is the Standard Model, but without any assumptions on the dynamics of the
electroweak symmetry breaking and therefore without the Higgs field.

The Standard Model without a Higgs is nonrenormalizable due to the dimension $6+$ operators
appearing in the expansion of the $\Sigma$ fields and therefore must be treated as an effective
field theory. Even without knowing the physics responsible for the symmetry breaking, NDA\footnote%
{
\eqref{equ-1-3-nda} differs from \cite{Manohar:1983md} by a factor of $2$ due to a difference in
convention.
}
can
be used to get an upper bound on the UV cutoff \cite{Manohar:1983md}:
\begin{equation}\label{equ-1-3-nda}
\Lambda_\text{NDA} = 8\pi v = 8\pi \frac{m_W}{g} \approx \unit[3.1]{TeV}
\end{equation}

How does the transition from the Fermi model to the Higgsless Standard Model affect tree level
unitarity? In the Fermi model, the divergent parts violating unitarity in the $ff\longrightarrow ff$
amplitude stem from the four point fermion coupling. However, in the Higgsless Standard Model, these
couplings are replaced by the exchange of the $W^\pm$ and $Z$ bosons:
\[\parbox{17mm}{\begin{fmfgraph}(17,14)\fmfleft{i2,i1}\fmfright{o2,o1}\fmf{fermion}{i1,v1,i2}
\fmf{fermion}{o1,v1,o2}\fmfdot{v1}\end{fmfgraph}}
\quad\longrightarrow\quad
\parbox{28mm}{\begin{fmfgraph}(28,12)\fmfleft{i2,i1}\fmfright{o2,o1}
\fmf{fermion}{i1,v1,i2}\fmf{fermion}{o1,v2,o2}\fmf{wiggly}{v1,v2}\fmfdot{v1,v2}
\end{fmfgraph}}
\quad+\quad
\parbox{16mm}{\begin{fmfgraph}(16,20)\fmfleft{i2,i1}\fmfright{o2,o1}
\fmf{fermion}{i1,v1,o1}\fmf{fermion}{i2,v2,o2}\fmf{wiggly}{v1,v2}\fmfdot{v1,v2}
\end{fmfgraph}}\]
At low energies, the $W^\pm$ and $Z$ can be integrated out by replacing the propagators by
$-\frac{1}{m^2}$ to obtain the Fermi theory as an
effective theory. However, above the $W^\prime$ mass scale, the propagators and the amplitude fall off as
$\frac{1}{p^2}$. This damping removes the quadratic growth in energy from the
partial wave amplitudes and eliminates the unitarity violating terms.

The entrance of the $W^\pm$ and $Z$ bosons on the stage restores perturbative unitarity to the
amplitudes for four fermion processes. However, the exorcism of unitarity violation is not
complete --- perturbative unitarity is still violated by amplitudes containing external heavy
vectors. To understand this, recall the high energy limit of the longitudinal polarization
vector
\[ \epsilon_L^\mu \xrightarrow{\;E\rightarrow\infty\;} \frac{k^\mu}{m} \]
At high energies, $\epsilon_L$ behaves like the four-momentum and therefore, the amplitude for the
scattering of longitudinal vector bosons
\[\MM_{V_LV_L\rightarrow V_LV_L} = \quad
\parbox{28mm}{\begin{fmfgraph}(28,12)\fmfleft{i2,i1}\fmfright{o2,o1}
\fmf{wiggly}{i1,v1,i2}\fmf{wiggly}{o1,v2,o2}\fmf{wiggly}{v1,v2}\fmfdot{v1,v2}
\end{fmfgraph}}
\quad+\quad
\parbox{16mm}{\begin{fmfgraph}(16,20)\fmfleft{i2,i1}\fmfright{o2,o1}
\fmf{wiggly}{i1,v1,o1}\fmf{wiggly}{i2,v2,o2}\fmf{wiggly}{v1,v2}\fmfdot{v1,v2}
\end{fmfgraph}}
\quad+\quad
\parbox{17mm}{\begin{fmfgraph}(17,14)\fmfleft{i2,i1}\fmfright{o2,o1}
\fmf{wiggly}{i1,v,i2}\fmf{wiggly}{o1,v,o2}\fmfdot{v}
\end{fmfgraph}}\]
might develop a high energy behavior as bad as $E^4$. Explicit calculation \cite{Lee:1977eg}
shows that, while the $E^4$ parts cancel, a quadratic divergence remains which leads to a
violation of tree level unitarity at approximately $\unit[1]{TeV}$, well below the NDA cutoff.
Therefore, new physics that unitarizes the scattering amplitudes should be expected below
$\unit[1]{TeV}$ if the theory is to remain perturbative.

\section{The Higgs}
\label{chap-1-4}

The question of unitarity at the $\unit{TeV}$ scale is not the only terra incongnita not covered by the
Higgsless version of the Standard Model. In particular, we didn't make any assumptions on the
physics responsible for the dynamics of electroweak symmetry breaking. As simplicity always has been
an effective guiding principle in physics, it is not unreasonable to assume that these questions
have the same answer and that the physics responsible for breaking the electroweak symmetry also
unitarizes the scattering amplitudes. This is exactly what happens in the Standard Model:
a scalar Higgs field is introduced which
dynamically develops the vacuum expectation value which breaks the electroweak symmetry and which
at the same time unitarizes the scattering amplitudes.

The Higgs can be incorporated into the Higgsless Standard Model presented in the last section by
replacing the vacuum expectation value of the $\Sigma$ field with a dynamical field $H(x)$
\[ \Sigma = ve^{\frac{i}{v}\phi_k(x)\tau_k} \quad\longrightarrow\quad
\Sigma = H(x)e^{\frac{i}{v}\phi_k(x)\tau_k} \]
and adding the potential 
\begin{lequation}{1-4-vHiggs} \LL_H = -V_H(x) = \frac{\mu^2}{2}\tr\Sigma\Sigma^\dagger -
\frac{\lambda}{4}\tr\left(\Sigma\Sigma^\dagger\right)^2 =
\mu^2 H^2 - \frac{\lambda}{2}H^4 \end{lequation}
with the ``mass'' $\mu$ and the coupling constant $\lambda$ to the Lagrangian.
This potential is the famous Mexican
Hat potential whose nontrivial minimum constitutes the tree level vacuum expectation of the Higgs
\[ v = \left<H\right> = \frac{\mu}{\sqrt{\lambda}} \]
Shifting $H(x)$ in order to perform the perturbation expansion around the ground state, we finally
obtain $\LL_H$ and $\Sigma$ in terms of the physical Higgs field
\[ \Sigma =  \left(v + h(x)\right)e^{i\frac{\phi(x)}{v}\phi_k(x)\tau_k} \]
\begin{lequation}{1-4-vHiggs2}
\LL_H = \frac{\mu^4}{2\lambda} - \mu^2h^2 - 2\mu\sqrt{\lambda}h^3 - \frac{\lambda}{2}h^4
\end{lequation}
Inserting the modified $\Sigma$ into \eqref{1-3-ltot} and adding \eqref{1-4-vHiggs2} finally leads
to the
usual Standard Model. The nonlinear representation of the Higgs and Goldstone fields may look
unfamiliar compared to the linear doublet representation usually used in textbooks, but both
formulations can be transformed into each other by means of a nonlinear field redefinition and
therefore are equivalent \cite{Coleman:1969sm}.

What has changed with the introduction of the Higgs? In the Higgsless version of the Standard Model,
the symmetry breaking was just parameterized by the introduction of the Goldstone bosons, but the
physics responsible for the breaking was not included and the vacuum expectation value $v$ was
fixed by hand. In the version of the model including the Higgs, the vacuum expectation
value arises dynamically from the Higgs potential, and the observable consequence is the appearance
of a physical scalar in the particle spectrum.

The Higgs couples to every massive field and, in particular, new contributions to the scattering
amplitude for longitudinal gauge bosons arise
\[
\parbox{28mm}{\begin{fmfgraph}(28,14)\fmfleft{i2,i1}\fmfright{o2,o1}
\fmf{wiggly}{i1,v1,i2}\fmf{wiggly}{o1,v2,o2}\fmf{dashes}{v1,v2}\fmfdot{v1,v2}
\end{fmfgraph}}
\quad+\quad
\parbox{16mm}{\begin{fmfgraph}(16,20)\fmfleft{i2,i1}\fmfright{o2,o1}
\fmf{wiggly}{i1,v1,o1}\fmf{wiggly}{i2,v2,o2}\fmf{dashes}{v1,v2}\fmfdot{v1,v2}
\end{fmfgraph}}
\]
As already mentioned above, explicit calculation shows that these new diagrams completely cancel the divergent
pieces and, if the Higgs is not too heavy, perturbative unitarity is restored at
arbitrarily high scales \cite{Lee:1977eg}.

In addition to resolving the unitarity problem, the inclusion of the Higgs makes the Standard Model
a renormalizable theory \cite{'tHooft:1972fi}, removing the remaining obstacle in interpreting it as
a fundamental theory valid at arbitrarily high energies\footnote%
{%
Even if the model is renormalizable and tree-level unitary at all energies, Landau poles in the
renormalization group flow might limit the range of applicability of the model, so this statement is
a bit oversimplified.
}%
. From this, one might draw the conclusion that the Higgs is the preferred candidate for the
mechanism of electroweak symmetry breaking.

However, the Standard Model does not include gravity, and therefore, it cannot be a
fundamental theory of nature but must be superseded by new physics at some (although possibly high)
scale. Once we have accepted this fact, the Higgs looses much of its
appeal through the so-called Hierarchy Problem. Consider the one loop corrections to the Higgs mass
\begin{lequation}{1-4-mhcor}
\parbox{20mm}{\begin{fmfgraph}(20,10)\fmfleft{i}\fmfright{o}\fmf{dashes}{i,v1}\fmf{dashes}{v2,o}
\fmf{dashes,left,te=.5}{v1,v2,v1}\fmfdot{v1,v2}\end{fmfgraph}}
\;+\;
\parbox{20mm}{\begin{fmfgraph}(20,10)\fmfleft{i}\fmfright{o}\fmf{dashes}{i,v1}\fmf{dashes}{v2,o}
\fmf{wiggly,left,te=.5}{v1,v2,v1}\fmfdot{v1,v2}\end{fmfgraph}}
\;+\;
\parbox{20mm}{\begin{fmfgraph}(20,10)\fmfleft{i}\fmfright{o}\fmf{dashes}{i,v,o}\fmfdot{v}
\fmf{wiggly}{v,v}\end{fmfgraph}}
\;+\;
\parbox{20mm}{\begin{fmfgraph}(20,10)\fmfleft{i}\fmfright{o}\fmf{dashes}{i,v,v,o}\fmfdot{v}\end{fmfgraph}}
\;+\;
\parbox{20mm}{\begin{fmfgraph}(20,10)\fmfleft{i}\fmfright{o}\fmf{dashes}{i,v1}\fmf{dashes}{v2,o}
\fmf{fermion,left,te=.5}{v1,v2,v1}\fmfdot{v1,v2}\end{fmfgraph}}
\end{lequation}
Apart from the first two diagrams, all pieces of \eqref{1-4-mhcor} contain quadratic divergences inducing
a quadratic running of the Higgs mass. Therefore, if the Higgs is to have
a mass $m_H$ at low scales, then the mass must be tuned with an accuracy of
approximately $\frac{m_H}{\Lambda_\text{UV}}$  at the scale $\Lambda_\text{UV}$ where the Standard Model is matched
to some more fundamental description of nature.

In order to preserve perturbative unitarity, the
Higgs mass must be considerably smaller then $\unit[1]{TeV}$. If we insert the Planck scale as $\Lambda_\text{UV}$,
this implies a fine tuning to a rather ridiculous precision of roughly $10^{-14}\%$!
While this extreme sensitivity to the underlying high energy physics is not technically inconsistent,
it arguably seems unfitting for a theory of nature which is deep enough to be valid over $19$ orders
of magnitude.
Apart from the Hierarchy Problem, there is also a more mundane issue concerning the Higgs boson:
no trace of it has been yet discovered in collider experiments and, looking at current
exclusion plots (see e.g. \cite{Alcaraz:2007ri}) for the Higgs mass, the air is growing thin.

Considering the effective nature of the Standard Model in conjunction with the Hierarchy Problem and the
reluctance of the Higgs to show up in experiments, the concept of a fundamental Higgs certainly
drops in grace, and other mechanisms for electroweak symmetry breaking and/or maintaining
perturbative unitarity become attractive.

\section{An Alternative}
\label{chap-1-5}

If the amplitudes for the scattering of longitudinal gauge bosons are to fulfill the tree level
unitarity requirement, then their high energy behavior must be modified. This can
be achieved either through modifying the propagators of the particles or by introducing new particles
into the model, the exchange of which cancels the dangerous growth with energy.

Of course, a scalar is not the only possible particle suitable for achieving this goal;
an alternative ansatz would be the introduction of
new vector bosons. These new vectors would have to be massive and rather heavy in order to
have accomplished the escape from detection until now. However, the only known case of massive spin $1$
particles that can be consistently quantized is that of gauge bosons of a spontaneously broken
symmetry. So, let us introduce an additional $\sun{2}$ gauge group with new heavy gauge bosons
which we will call $W^\prime$ and $Z^\prime$. To parameterize the symmetry breaking, we will also
need a new nonlinear sigma field $\Sigma^\prime$ which describes the three new Goldstone bosons that
arise from the symmetry breaking.

If the new particles are to unitarize the scattering amplitudes, then they must couple to the
$W^\pm$ and $Z$ bosons. However, the couplings of gauge bosons are completely determined by gauge
invariance, and if we have only kinetic and\footnote%
{
As this section focuses on the ideas behind the model, the discussion is qualitative and the
details are postponed to the next chapter.
}
mass terms, then we will only get couplings of the types
\[
\begin{fmfgraph}(17,14)\fmfleft{i2,i1}\fmfright{o}\fmf{wiggly}{i1,v}\fmf{wiggly}{i2,v}
\fmf{wiggly}{o,v}\fmfdot{v}\end{fmfgraph}
\qquad\qquad
\begin{fmfgraph}(17,14)\fmfleft{i2,i1}\fmfright{o}\fmf{dbl_wiggly}{i1,v}\fmf{dbl_wiggly}{i2,v}
\fmf{dbl_wiggly}{o,v}\fmfdot{v}\end{fmfgraph}
\qquad\qquad
\begin{fmfgraph}(17,14)\fmfleft{i2,i1}\fmfright{o2,o1}\fmf{wiggly}{i1,v}\fmf{wiggly}{i2,v}
\fmf{wiggly}{o1,v}\fmf{wiggly}{o2,v}\fmfdot{v}\end{fmfgraph}
\qquad\qquad
\begin{fmfgraph}(17,14)\fmfleft{i2,i1}\fmfright{o2,o1}\fmf{dbl_wiggly}{i1,v}\fmf{dbl_wiggly}{i2,v}
\fmf{dbl_wiggly}{o1,v}\fmf{dbl_wiggly}{o2,v}\fmfdot{v}\end{fmfgraph}
\]
(assigning double lines to the new heavy particles), but no couplings that mix the Standard Model
gauge bosons and the new heavy ones.

A way to lift this constraint is the introduction of terms into the mass matrix
which mix the Standard Model gauge bosons and the new ones. As a results, the
physical fields arise as mixtures of the old and the new fields
\[
\cvect{W^\pm_\mu\vs{4ex} \\ W^{\pm\prime}_\mu\vs{4ex}} =
U \cvect{\frac{W_{1\mu}\pm iW_{2\mu}}{\sqrt{2}}\vs{4ex} \\
\frac{W_{1\mu}^\prime\pm iW_{2\mu}^\prime}{\sqrt{2}}\vs{4ex}} \quad,\quad
\cvect{A_\mu\vs{3ex} \\ Z_\mu\vs{3ex} \\ Z_\mu^\prime\vs{3ex}} =
V \cvect{B_\mu\vs{3ex} \\ W_{3\mu}\vs{3ex} \\ W_{3\mu}^\prime\vs{3ex}}
\]
with orthonormal mixing matrices $U,V$. This mixing induces the desired couplings, resulting in
contributions to the longitudinal gauge boson scattering amplitude
\[
\parbox{28mm}{\begin{fmfgraph}(28,14)\fmfleft{i2,i1}\fmfright{o2,o1}
\fmf{wiggly}{i1,v1,i2}\fmf{wiggly}{o1,v2,o2}\fmf{dbl_wiggly}{v1,v2}\fmfdot{v1,v2}
\end{fmfgraph}}
\quad+\quad
\parbox{16mm}{\begin{fmfgraph}(16,20)\fmfleft{i2,i1}\fmfright{o2,o1}
\fmf{wiggly}{i1,v1,o1}\fmf{wiggly}{i2,v2,o2}\fmf{dbl_wiggly}{v1,v2}\fmfdot{v1,v2}
\end{fmfgraph}}
\]
which can potentially restore perturbative unitarity.

The mixing between the gauge bosons also induces couplings of the $W^\prime$/$Z^\prime$ to
the standard model fermions, leading to contributions to the charged and neutral current
four point function
\[
\parbox{28mm}{\begin{fmfgraph}(28,14)\fmfleft{i2,i1}\fmfright{o2,o1}
\fmf{fermion}{i1,v1,i2}\fmf{fermion}{o1,v2,o2}\fmf{dbl_wiggly}{v1,v2}\fmfdot{v1,v2}
\end{fmfgraph}}
\quad+\quad
\parbox{16mm}{\begin{fmfgraph}(16,20)\fmfleft{i2,i1}\fmfright{o2,o1}
\fmf{fermion}{i1,v1,o1}\fmf{fermion}{o2,v2,i2}\fmf{dbl_wiggly}{v1,v2}\fmfdot{v1,v2}
\end{fmfgraph}}
\]
As new physics would also appear in loop contributions to the charged and neutral current
correlators, this kind of processes was studied very precisely by the LEP-II experiments, resulting
in an astonishing agreement with the one loop predictions of the Standard Model and leading to very
stringent bounds on any kind of new physics which contributes to this kind of processes. The major
part of this so-called electroweak precision data can be distilled into
the famous $\alpha S$, $\alpha T$ and $\alpha U$
parameters defined in \cite{Peskin:1991sw,Peskin:1990zt} or equivalently the three
$\epsilon_{1/2/3}$ parameters defined in \cite{Altarelli:1990zd,Altarelli:1991fk}.

If no special care is taken to avoid these constraints, most models of new physics that introduce
copies of the $W$ and $Z$ which couple to Standard Model fermions are excluded. In our construction
so far, these couplings are completely determined by the mixing matrices $U$ and $V$, and as we have
no further influence on the resulting couplings, we will almost certainly violate the electroweak
precision bounds. Therefore, once we have introduced the $W^\prime$ and $Z^\prime$ this way, we have
to think of a trick to gain control over the undesired couplings.

Of course, we could try to charge the Standard Model fermions under the new $\sun{2}$ gauge group,
but the resulting couplings are completely fixed by gauge invariance, giving us no additional
freedom that could be used to avoid the constraints. What could
work, however, this the introduction of a partner fermion for each Standard Model fermion which is heavy,
charged under the new $\sun{2}$ and mixes with its Standard Model partner. The price
would be new set of new heavy fermions $f^\prime$, but we could
use the mixing parameters to control the unwanted fermion couplings and tune them away.

This way, we are rather naturally led to a model which contains a heavy copy of
every Standard Model particle (with the exception of the gluon and the photon).
There is no physical scalar in the spectrum and consequently there is no hierarchy problem.
The task of unitarizing the scattering amplitudes is taken over by the new set of heavy vector
bosons.

This structure of massive replica of the spectrum is a well-known feature of theories with one or
more compact extra dimensions. In such theories, gauge symmetry can be broken without a Higgs by the
introduction of suitable boundary conditions, and the massive gauge bosons that arise this way are
known to delay the scale of unitarity violation \cite{Csaki:2003dt}. Reducing the extra dimension to
a finite point lattice, this can be exploited to build a model which has exactly the spectrum and
features discussed above, the Three-Site Higgsless model \cite{Chivukula:2006cg}. The next chapter will be devoted to
a detailed discussion of the construction and properties of this model.

\chapter{The Model --- Top-down Approach}
\label{chap-2}

\begin{quote}
\hspace{-2ex}Percy: {\itshape Only this morning in the courtyard I saw a horse with two heads and two bodies.}
\newline
\mbox{}\hspace{-2ex}Blackadder: {\itshape Two horses standing next to each other?}
\end{quote}
\hfill\begin{minipage}{0.7\textwidth}\small\raggedleft
(``Blackadder I ---  Witchsmeller Pursuviant'')
\end{minipage}
\\[5mm]

In the last chapter the Three-Site Model was demonstrated to arise in a rather natural way when
attempting to unitarize scattering amplitudes with heavy copies of the $W$ and $Z$ bosons instead of
a Higgs. The actual construction was postponed and will now be carried out using a convenient framework.

The first two sections are devoted to introducing the concepts of extra dimensions and dimensional
deconstruction, while the third then continues with the actual construction of the model.

\section{Extra Dimensions}
\label{chap-2-1}

\subsubsection*{Manifolds and boundary conditions}

The first ingredient to an extra dimensional field theory is, obviously, the spacetime manifold on
which it is formulated. As the theory must match onto our known 4D physics at low energies, this
manifold must contain the usual Minkowski spacetime (we are not going to consider general
relativity), extended by one or more extra dimensions. For the same matching reason, these additional
dimensions must decouple from the low energy phenomenology and are therefore usually chosen to be
compact with some compactification scale $R$ which suppresses the new physics arising from them.

In the following, we will specialize to a flat 5th dimension which (together with the
case of ``warped'' extra dimensions with Randall-Sundrum metric \cite{Randall:1999ee})
is arguably the best-studied
case in particle phenomenology. To complete the basic setup, we also need to specify
the topology of the extra dimension. Common
choices are a circle with radius $R$ or a compact interval $\left[0;2R\pi\right]$. Out of these
two examples, we now specialize to the latter. The boundaries of such an interval are also sometimes
called ``branes'', while the open interval is called ``bulk''.

The second ingredient to the theory is the set of fields that propagate on the manifold. In analogy
to the 4D case, these form representations of the 5D Poincar\'e group\footnote%
{
Usually, the compactification breaks the 5D Poincar\'e group down to the 4D one, resulting in a violation of
5D momentum conservation.
}%
. In 4D field theories, we
require that the fields decay ``fast enough'' for asymptotic times and
distances, and in a 5D theory, we need to specify boundary
conditions on the boundaries of the compact extra dimension. On the interval, these
are usually chosen either as Dirichlet or as Neumann boundary conditions\footnote%
{
Starting from a fifth dimension compactified on a circle, the case of compactification on an
interval with Dirichlet or Neumann boundary conditions can also be obtained in a very
elegant way by orbifolding \cite{Hebecker:2001jb}.
}%
\begin{equation}\label{equ-2-1-bc}\begin{aligned}
&\text{Neumann:}\quad&\left.\partial_y\Phi(x,y)\vs{2.5ex}\right|_{y\in\left\{0,2R\pi\right\}} = 0
\\
&\text{Dirichlet:}\quad&\left.\Phi(x,y)\vs{2.5ex}\right|_{y\in\left\{0,2R\pi\right\}} = 0
\end{aligned}
\end{equation}
It is also possible to mix these conditions and e.g. impose a Dirichlet boundary condition at one
end of the interval and a Neumann one on the other one.

\subsubsection{5D scalar field and Kaluza-Klein expansion}

The simplest possible 5D theory is that of a real scalar field on an interval:
\[ \Phi:(x,y)\in\mathbb{R}^4\times\left[0;2R\pi\right]\longrightarrow \Phi(x,y)\in\mathbb{R} \]
As an example for how to treat such theories, let us consider the Lagrangian for the
flat 5D variant of $\Phi^4$ theory (see appendix \ref{chap-conv} for the conventions regarding Lorentz indices)
\begin{lequation}{equ-2-1-lp4}
\LL_{\Phi^4} = \frac{1}{2}(\partial_a\Phi)(\partial^a\Phi) - \frac{1}{2}M^2\Phi^2 - \frac{g}{4!}\Phi^4
\end{lequation}
The action functional is then defined as the usual integral of the Lagrangian over spacetime, the
$y$ part of which is bounded
\begin{lequation}{chap-2-1-sp4}
S\left[\Phi\right] = \int d^4x\;\int_0^{2R\pi}dy\;\LL_{\Phi^4} =
\int d^4x\int_0^{2R\pi}dy\;\left(\frac{1}{2}(\partial_a\Phi)(\partial^a\Phi)
- \frac{1}{2}M^2\Phi^2 - \frac{g}{4!}\Phi^4 \right)
\end{lequation}
It is noteworthy that, in order for the action \eqref{chap-2-1-sp4} to be classically scale
invariant, the mass dimensions of the scalar and of the coupling must be
\[ [\Phi] = \frac{3}{2} \qquad,\qquad [g] = -1 \]
Variation of the action to obtain the equations of motion proceeds as usual. However, the result
\[
\delta S = \int d^4x\left(\left.\partial_y\Phi\delta\Phi\vs{4ex}\right|_{y=0}^{y=2R\pi} -
\int_0^{2R\pi}dy\;\left(\partial_a\partial^a\Phi + M^2\Phi +
\frac{g}{3!}\Phi^3\right)\delta\Phi\right)
\]
contains a boundary term that results from partial integration. With the boundary conditions
\eqref{equ-2-1-bc} or mixed ones, this term vanishes, and the resulting
Euler-Lagrange equation is
\begin{lequation}{equ-2-1-elgp4}
\left(\partial_a\partial^a + M^2\right)\Phi = \left(\partial^2 - \partial_y^2 + M^2\right)\Phi =
-\frac{g}{3!}\Phi^3
\end{lequation}

It is convenient to partially decompose the solutions to \eqref{equ-2-1-elgp4} as a Fourier series with the
basis functions $f_k(y)$ chosen such that they satisfy the boundary conditions and are eigenfunctions of
$\partial_y^2$:
\begin{lequation}{equ-2-1-kkexp}\begin{aligned}
&\text{Neumann:}\quad& \Phi(x,y) = \sum_{k=0}^\infty\phi_k(x)f_k(y) &=
\sum_{k=0}^\infty\phi_k(x)\frac{1}{\sqrt{R\pi(1+\delta_{k,0})}}\cos\left(\frac{k}{2R}y\right)
\\
&\text{Dirichlet:}\quad& \Phi(x,y) = \sum_{k=1}^\infty\phi_k(x)f_k(y) &=
\sum_{k=1}^\infty\phi_k(x)\frac{1}{\sqrt{R\pi}}\sin\left(\frac{k}{2R}y\right)
\end{aligned}\end{lequation}
Inserting the expansion into the Lagrangian \eqref{equ-2-1-lp4} and defining the effective 4D action
functional and Lagrangian as
\[
\widetilde{S}[\phi] = \int d^4x\; \widetilde\LL_{\Phi^4} \qquad,\qquad
\widetilde\LL_{\Phi^4} = \int_0^{2R\pi}dy\;\LL_{\Phi^4}
\]
we obtain
\begin{lequation}{equ-2-1-lp4kk}
\widetilde\LL_{\Phi^4} =
\frac{1}{2}\sum_{k}\left(\left(\partial_\mu\phi_k\right)^2 -
\left(M^2 + \frac{k^2}{4R^2}\right)\phi_k^2\right) -
\sum_{k_1,k_2,k_3,k_4}g_{k_1,k_2,k_3,k_4}\phi_{k_1}\phi_{k_2}\phi_{k_3}\phi_{k_4}
\end{lequation}
with the dimensionless coupling constants
\begin{equation}\label{equ-2-1-4dcpl}
g_{k_1,k_2,k_3,k_4} = \frac{g}{4!}\int_0^{2R\pi}dy\;f_{k_1}(y)f_{k_2}(y)f_{k_3}(y)f_{k_4}(y)
\end{equation}
This expansion is the so-called Kaluza-Klein (KK) expansion of the theory.

What is the virtue of the KK expansion and of the effective Lagrangian \eqref{equ-2-1-lp4kk}?
While we started with a 5D theory of a single scalar field, we find that the physics can be equivalently
described by an ordinary 4D field theory with an infinite number of 4D scalar fields
$\phi_k$. Each of these fields has an effective mass of
\[ m_k^2 = \frac{k^2}{4R^2} + M^2 \]
making them arbitrarily heavy for $k>0$, the mass increasing with $k$. The 5D Lagrangian
$\LL_{\Phi^4}$ contains a single four point interaction of the $\Phi$ whereas the effective 4D version
contains four point interactions involving all the different $\phi_k$ fields (although many of
the coupling constants vanish).

The physics behind this is easy to understand: in the 5D version of the theory, the 5-momentum
satisfies the mass shell condition
\[ k_a k^a = k_\mu k^\mu - {\left(k^5\right)}^2 = M^2 \]
with the fifth component of the momentum being discrete\footnote%
{
To be a bit more precise, the $f_k$ are not eigenfunctions of $i\partial_y$ but of $\partial_y^2$,
so we should rather be talking about the square of the five momentum and not about the momentum
itself.
}
rather than continuous%
\[ k^5_n = \frac{n}{2R} \]
due to the compactification of the extra dimension.
The straightforward interpretation is to think of $\Phi$ as a 5D field with a discrete fifth
momentum component and mass $M^2$. However, it can be equivalently regarded as a tower of 4D fields
with masses given by $M^2 + \left(k^5\right)^2$. In the same way we could think about a 4D particle as
a continuum of 3D particles, but as the four-momentum is a continuous degree of freedom, this is not
a very useful intuition.

At tree level, we have the liberty of setting the 5D mass $M$ to zero. In the case of Neumann
boundary conditions, there is a $k=0$ mode in the tower of scalars which is massless.
In the case of Dirichlet or mixed
boundary conditions the flat wavefunction $f_0(k)$ is forbidden, and all the $\phi_k$ are
massive, even if no tree level mass is put into the Lagrangian.

Applying na"ive powercounting to the 5D version of the Lagrangian leads to the conclusion that it
must describe a nonrenormalizable theory as can be easily understood by noting that the 5D coupling
$g$ has negative mass dimension. In the 4D theory, the coupling constants at the four point interactions
are dimensionless and therefore, the theory might appear to be renormalizable at a casual glance.
However, in higher order diagrams, all the infinitely many $\phi_k$ can run in the loops leading to a possibly
divergent series over $k$. This enhances the divergences and recovers the nonrenormalizable
character of the theory.

Although we have been examining just the simple case of a 5D $\Phi^4$ theory, we have found a
lot of interesting structure that generalizes to more complicated 5D theories. These theories can
always be described by an effective 4D theory by virtue of the KK expansion. In this process,
every 5D field
gets replaced by a tower of 4D fields of increasing mass. Asymptotically, the masses of the
modes in the tower are equidistantly spaced\footnote%
{
This changes if the metric of is chosen nontrivially, e.g. in Randall-Sundrum type models.
}%
. The presence of a zero mode which is massless in absence of an explicit mass term in the Lagrangian
depends on the boundary conditions. Extra dimensional theories are usually\footnote%
{
Trivial counterexamples exist, e.g. $\phi^3$ theory in $5$ or $6$ dimension (which, however, is a
slightly pathological example as the potential is not bounded from below).
}
nonrenormalizable and
valid only as effective field theories with an ultraviolet cutoff.
Therefore, it is arguably inconsistent to include all modes
in the KK towers in calculations. Instead, the towers should be cut off in some way at high
scales, and if gauge fields are involved, extra care has to be taken to preserve gauge invariance in
this step.

\subsubsection*{Gauge bosons}

A 5D gauge field transforms in the vector representation of the five dimensional Poincar\'e group
and hence has five components $A^a$. However, boundary conditions break this group down to its 4D
subgroup, under which $A^\mu$ transforms as a vector and $A^5$ transforms as a scalar.
Once we perform the KK expansion to get the 4D effective theory, the $A^\mu$ will give us a tower of
4D vector fields, while $A^5$ will become a tower of 4D scalars. In the following, we will
restrict the discussion to the case of an abelian field for simplicity's sake.

Under a gauge transformation with parameter field $\lambda$, the fields $A^\mu$ and $A^5$ transform as
\begin{equation}\label{chap-2-2-gtrans}
A^\mu(x,y) \longrightarrow A^\mu(x,y) + \frac{1}{g}\partial^\mu\lambda(x,y) \quad,\quad
A^5(x,y) \longrightarrow A^5(x,y) - \frac{1}{g}\partial_y\lambda(x,y)
\end{equation}
with the gauge coupling $g$. Similar to the scalar case, we have to choose boundary conditions for the
fields. Let us choose Neumann boundary conditions for $A^\mu$. We can then KK expand the
4D vector field as
\[
A^\mu(x,y) = \sum_{k=0}^\infty A^\mu_k(x)\frac{1}{\sqrt{R\pi(1+\delta_{k,0})}}
\cos\left(\frac{k}{2R}y\right)
\]
Looking at \eqref{chap-2-2-gtrans}, it is easy to see that the parameter field $\lambda$ has to
satisfy the same boundary conditions as $A^\mu$ and therefore can be KK expanded likewise
\[
\lambda(x,y) = \sum_{k=0}^\infty \lambda_k(x)\frac{1}{\sqrt{R\pi(1+\delta_{k,0})}}
\cos\left(\frac{k}{2R}y\right)
\]
This implies that the shift in $A^5$ generated by such a gauge transformation is
\[
A^5(x,y) \longrightarrow A^5(x,y) + \frac{1}{g}\sum_{k=1}^\infty\lambda_k\frac{k}{2R}\frac{1}{\sqrt{R\pi}}
\sin\left(\frac{k}{2R}y\right)
\]
which means that the field $A^5$ must obey Dirichlet boundary conditions and can be decomposed as
\[ A^5(x,y) = \sum_{k=1}^\infty A^5_k(x)\frac{1}{\sqrt{R\pi}}\sin\left(\frac{k}{2R}y\right) \]
Therefore, after the KK decomposition, we obtain an infinite tower of 4D gauge transformations
which leave the theory invariant, the 4D gauge bosons $A^\mu$ and the 4D scalars $A^5$ transforming as
\begin{equation}\label{equ-2-2-aneu}\begin{aligned}
A^\mu_k(x) \longrightarrow &A^\mu_k(x) + \frac{1}{g}\partial^\mu \lambda_k(x) \quad&\qquad k\ge 0\\
A^5_k(x) \longrightarrow &A^5_k(x) + \frac{1}{g}\frac{k}{2R} \lambda_k(x) \quad&\qquad k\ge 1
\end{aligned}\end{equation}

A closer look at \eqref{equ-2-2-aneu} reveals at least two things of special interest.
First, there is a gauge
transformation generated by $\lambda_0$ which only acts on $A^\mu_0$ and doesn't involve any of the
4D scalars $A^5_k$. Second, it is possible to perform gauge transformations which remove all of
the $A^5_k$ and completely absorb them into the $A^\mu_k$, fixing the gauge for all the 4D gauge
fields with the exception of $A^\mu_0$. This suggests an elegant interpretation.

As we chose Neumann boundary conditions for the $A^\mu$,
the fifth component of the 5-momentum $k^5$ vanishes for $A^\mu_0$, leaving this mode as a massless
gauge boson of an unbroken gauge symmetry generated by $\lambda_0$.
All the other modes $A^\mu_k$ are massive and obtain their
longitudinal modes from ``eating'' a $A^5_k$ field in a process very similar to the Goldstone
mechanism in spontaneous symmetry breaking. Due to the mass terms,
the gauge symmetry generated by the $\lambda_k,\;k\ne 0$
is hidden in a simultaneous transformation of the $A^\mu_k$ and the ``Goldstone'' fields $A^5_k$,
which also is reminiscent of spontaneous symmetry breaking. Gauging away the $A^5_k$ corresponds to
unitarity gauge in a spontaneously broken gauge theory.

Let's have a look at the Lagrangian to see whether this assertion is correct. The 5D
gauge kinetic Lagrangian is given by
\begin{equation}\label{equ-2-1-lgauge} \LL_\text{gauge} = -\frac{1}{4}F_{ab}F^{ab} \end{equation}
with the 5D field strength tensor
\[ F^{ab} = \partial^a A^b - \partial^b A^a \]
The Lagrangian \eqref{equ-2-1-lgauge} can be split into 4D part $\LL_{4D}$ and a 5D
part $\LL_{5D}$
\[
\LL_\text{gauge} = \underbrace{-\frac{1}{4}F^{\mu\nu}F_{\mu\nu}}_{\LL_{4D}} +
\underbrace{\frac{1}{2}\left(\partial_y A^\mu + \partial^\mu A^5\right)^2}_{\LL_{5D}}
\]
Performing the KK expansion and defining the KK masses $m_k$ as
\[ m_k = \frac{k}{2R} \]
we obtain the effective 4D Lagrangian
\begin{equation}\label{equ-2-1-lgaugekk}
\widetilde{\LL}_\text{gauge} = -\frac{1}{4}\sum_{k=0}^\infty\underbrace{F_k^{\mu\nu}F_{k,\mu\nu}}_{
\widetilde{\LL}_{k,4D}} + \sum_{k=1}^\infty\underbrace{
\frac{1}{2}\left(-m_k A^\mu_k + \partial^\mu A^5_k\right)^2}_{\widetilde{\LL}_{k,5D}}
\end{equation}
The $\widetilde{\LL}_{k,4D}$ are the kinetic Lagrangians for the 4D gauge fields, while
the $\widetilde{\LL}_{k,5D}$ are easily identified as gauge invariant Stueckelberg
Lagrangians. The Stueckelberg action is a well-known trick for giving mass to $\mathbf{U}(1)$ gauge
bosons by spontaneous symmetry breaking, introducing a scalar, massless Goldstone boson. Therefore,
the above interpretation is quite correct; in the 4D effective theory, the masses of the
$A^\mu_k,\;k\ge1$ fields arise from a mechanism equivalent to spontaneous symmetry breaking,
yielding an infinite tower of massless Goldstone fields.

In addition to the mass terms for $A^\mu_k$, $k\ge1$
and the kinetic terms for $A^5_k$,\ the Lagrangian $\widetilde{\LL}_{k,5D}$ also contains a bilinear term which mixes
$A^\mu_k$ and $\partial^\mu A^5_k$. This can be removed by adding a suitable gauge fixing
term to \eqref{equ-2-1-lgauge} similar to the $R_\xi$ gauges in spontaneous broken gauge theories
\[ \LL_{GF} = \frac{1}{2\xi}\left(\partial_\mu A^\mu - \xi\partial_y A^5\right)^2 \]

If we impose Dirichlet boundary conditions on $A^\mu$, we find a result only
slightly different. This boundary condition implies Neumann boundary conditions for the
4D scalar $A^5$. The massless $A^\mu_0$ vanishes from the spectrum, and at the same
time, a physical scalar $A^5_0$ arises in which is not ``eaten'' by the gauge bosons.
Imposing mixed boundary conditions would forbid the massless $A^\mu_0$ mode as well as the physical
scalar $A^5_0$.

The analysis of nonabelian 5D gauge theories proceeds along the same lines as the abelian case
but is more complicated due to the more complex group structure. This discussion can be found e.g.
in \cite{Muck:2001yv} or \cite{Csaki:2003dt}. The most exciting difference when compared to the abelian
case is the possibility of assigning different boundary conditions to different component gauge
fields.
As an example, consider a 5D $\sun{2}$ gauge field $A^a = A^a_r \tau_r$.
If we assign Dirichlet boundary conditions to the components belonging to the $\tau_1$ and $\tau_2$
generators and Neumann ones to the $\tau_3$ component, then only the $A^a_3$ tower will contain a
massless mode mode in the KK decomposition. Thus, only the $\mathbf{U}(1)$
subgroup generated by $\tau_3$ will make it into the 4D effective theory as an unbroken gauge
symmetry; all other gauge symmetries of the KK fields get spontaneously broken.

Therefore, in a 5D version of the Standard Model, suitable boundary conditions could be used to break
the $\sun{2}_L\times\mathbf{U}(1)_Y$ gauge symmetry down to the $\mathbf{U}(1)_\text{EM}$ subgroup
without introducing a Higgs field. The price to pay would be towers of massless KK partners for all
Standard Model particles. However, such a scenario is excluded by electroweak precision observables.
Still, if we could cut off the KK towers after the first mode, this theory could be a starting point
for the model suggested in chapter \ref{chap-1-5}.

Unfortunately, another feature of the nonabelian result is a more complicated transformation behavior of
the KK modes under gauge transformation involving mixing between different modes \cite{Muck:2001yv}.
In other words, we have to take
all KK modes into account to retain gauge invariance. Simply cutting off the tower
wouldn't do any good but instead destroy gauge invariance! However, there are loopholes: for
example, dimensional deconstruction as
discussed in section \ref{chap-2-2} provides a gauge invariant way of cutting off the KK
towers.

\subsubsection*{Fermions}

To describe 5D fermions, we have to find a five dimensional representation of the Clifford algebra,
i.e. a set of five matrices $\gamma^a$ obeying the anticommutation relations
\begin{equation}\label{equ-2-1-cliff} \akomm{\gamma^a}{\gamma^b} = 2g^{ab} \end{equation}
Fortunately, this search does not lead far from well-known territory: if we take
the usual set of 4D gamma matrices $\gamma^\mu$, then $\gamma^5$ has all the properties we
need. Defining $\gamma^4$ as
\[ \gamma^4 = i\gamma^5 \]
we obtain a set of matrices that obey \eqref{equ-2-1-cliff}. As a result, 5D fermions can be
described by 4 component complex spinors just as the 4D ones.

However, writing down the generators for 5D Lorentz transformations
\[ \sigma^{ab} = \frac{i}{2}\komm{\gamma^a}{\gamma^b} \]
we note that the $\sigma^{5\mu}$ don't commute with the 4D chirality projectors $\Pi_\pm$ (c.f.
\eqref{conv-chiral}). This means that 5D
Lorentz transformations mix left- and right-handed fields, defeating the notion of chirality:
there is no such thing like a chiral fermion in five dimensions. Nevertheless, as 5D Lorentz invariance is
broken down to the 4D subgroup anyway by the boundaries of the interval, it is possible to obtain
a 4D effective theory that makes a distinction between the chiralities by choosing suitable chiral
boundary conditions.

As an example consider a 5D fermion $\Psi$ with left- and right-handed components
$\Psi_{L/R}=\Pi_\pm\Psi$. The massless 5D kinetic Lagrangian reads
\begin{multline}\label{equ-2-1-lferm}
\LL_\Psi = i\Psibar\gamma^a\partial_a\Psi =
i\Psibar\slashed\gamma\Psi - \Psibar\partial_y\gamma^5\Psi = \\
i\left(\Psibar_L\slashed\gamma\Psi_L + \Psibar_R\slashed\gamma\Psi_R\right) -
\Psibar_L\partial_y\Psi_R + \Psibar_R\partial_y\Psi_L
\end{multline}
If we impose Neumann boundary conditions on the left-handed field $\Psi_L$ and Dirichlet
ones on the right-handed field $\Psi_R$ and insert the corresponding
KK expansions into \eqref{equ-2-1-lferm}, we obtain the effective 4D Lagrangian as
\begin{equation}\label{equ-2-1-lfermkk}
\widetilde{\LL}_\Psi = \Psibar_{0,L}\slashed\gamma\Psi_{0,L} + \sum_{k=1}^\infty\left(
\Psibar_{k,L}\slashed\gamma\Psi_{k,L} +
\Psibar_{k,R}\slashed\gamma\Psi_{k,R} - \frac{k}{2R}\left(\Psibar_{k,L}\Psi_{k,R} +
\Psibar_{k,R}\Psi_{k,L}\right)\right)
\end{equation}
From \eqref{equ-2-1-lfermkk} it is then clear that the 4D effective theory contains a left-handed
massless fermion $\Psi_{0,L}$ together with a tower of massive Dirac fermions.

Other types of boundary conditions as well as an explicit 5D mass term change this picture and may
result in more complicated situations. In particular, a mixing between different KK modes is
possible requiring further diagonalization to obtain the mass eigenstates. However, as this is
of no importance for this work, we refrain from delving deeper into these issues and stop the discussion
at this point. Some further remarks on fermions in compactified 5D theories can be found e.g.
in \cite{Sundrum:2005jf}.

\section{Dimensional Deconstruction}
\label{chap-2-2}

Dimensional deconstruction is the reduction of one or more extra dimensions to a finite point
lattice. This procedure is well-known from lattice gauge theory where Minkowski space is
approximated by a four dimensional finite point lattice which allows for the numerical calculation
of correlators by solving the path integral via Monte Carlo integration.

In the context of model building, the deconstruction of an extra dimension opens up a new class of
models which are akin to extra dimensional models but differ in some important aspects. In particular,
deconstruction allows to cut off the KK towers in a gauge invariant fashion and provides a class of
possible UV completions.

\begin{figure}[!tb]
\centerline{\begin{tikzpicture}
\tikzstyle{lattice site} = [shape=circle,ball color=gray!01,minimum width=1cm,shading=ball]
\path[shade=axis,top color=gray!5,bottom color=gray!20]
(-7.5cm,-1.5cm) -- (4.5cm,-1.5cm) -- (7.5cm,1.5cm) -- (-4.5cm,1.5cm) -- cycle;
\foreach \pos in {-5cm,5cm}{
	\draw[xshift=\pos,draw=gray!90,dotted] (-7mm,-7mm) -- (11mm,11mm);
}
\node[style=lattice site] at (5cm,0cm) {$y_N$};
\node[style=lattice site] at (-5cm,0cm) {$y_0$};
\node[style=lattice site] at (-2cm,0cm) {$y_1$};
\node[style=lattice site] at (2cm,0cm) {\parbox{5mm}{\centerline{$y_{N-1}$}}};
\draw[<->,thick] (-43mm,0mm) -- (-27mm,0mm);
\draw[<->,thick] (27mm,0mm) -- (43mm,0mm);
\node at (0mm,0mm) {$//$};
\draw[<-,thick] (-13mm,0mm) -- (-10mm,0mm);
\draw[thick,loosely dotted] (-10mm,0mm) -- (-3mm,0mm);
\draw[thick,loosely dotted] (3mm,0mm) -- (10mm,0mm);
\draw[->,thick] (10mm,0mm) -- (13mm,0mm);
\begin{scope}[xshift=9mm,yshift=9mm]
	\draw[<-,thick] (-50mm,0mm) -- (-10mm,0mm);
	\draw[thick,loosely dotted] (-10mm,0mm) -- (-3mm,0mm);
	\draw[thick,loosely dotted] (3mm,0mm) -- (10mm,0mm);
	\draw[->,thick] (10mm,0mm) -- (50mm,0mm);
	\node at (0mm,0mm) {$//$};
\end{scope}
\foreach\pos in {-3.5cm,3.5cm}{
	\node[anchor=north] at (\pos,0mm) {$d=\frac{2R\pi}{N}$};
}
\node[anchor=south] at (11mm,11mm) {$2R\pi$};
\draw[->] (-7.3cm,-1.4cm) -- (-6.4cm,-5mm);
\draw[->] (-7.3cm,-1.3cm) -- (-6cm,-1.3cm);
\node[anchor=south] at (-62mm,-6mm) {$x^{0,1,2,3}$};
\node[anchor=south east] at (-59mm,-13mm) {$y$};
\end{tikzpicture}}
\caption{Sketch of the deconstruction of a fifth dimension compactified on an interval
$\left[0;2R\pi\right]$ to a lattice of $N+1$ points.}
\label{fig-2-2-dec}
\end{figure}
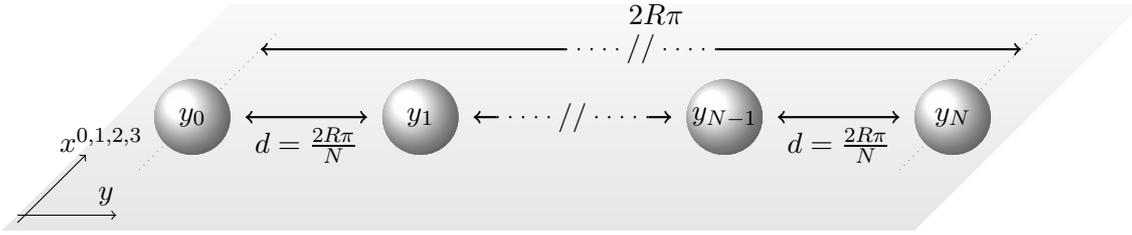
Let us consider again the example of a 5D theory compactified
on an interval which we discussed in the last section
\[ (x,y) \in \mathbb{R}^4\times\left[0;2R\pi\right] \]
and replace the fifth dimension by a discrete lattice of $N+1$ equidistantly spaced sites
$y_0,\dots,y_N$
\[ y_n = nd \quad,\quad d = \frac{2R\pi}{N} \]
This setup is sketched in fig.~\ref{fig-2-2-dec}.

\subsubsection*{Gauge theory revisited}

In order to find out how a 5D gauge field $A^a$ is properly represented in the deconstructed theory, let us
briefly recall some basics about gauge theories. Such a theory is invariant under a
group of spacetime dependent transformations\footnote%
{
For the rest of this section, an unitary representation will be assumed together with the canonical
trace normalization of the Lie algebra generators $T_i$
\[ \tr T_i T_j = \frac{1}{2}\delta_{ij} \]
}%
\[ \phi(x) \longrightarrow U(x)\phi(x) \]
of the fields $\phi$.

This invariance gives rise to the necessity of some prescription for how to compare fields at different points
in space time in a gauge invariant way. Even in a local field theory, this is necessary because the
definition of the partial derivative
\[
\partial^\mu\phi(x) = \lim_{\epsilon\rightarrow 0}\frac{\phi(x+\epsilon n^\mu) -
\phi(x)}{\epsilon}
\]
(with the unit vector $n^\mu$ pointing in direction $\mu$) involves the field at two infinitesimally
separated points and therefore is not gauge covariant
\[ \partial^\mu \left(U(x)\phi(x)\right) \ne U(x)\partial^\mu\phi(x) \]

Such a prescription is given by Wilson lines $W(x_1,x_2)$, operators valued in the gauge group which
transform as
\begin{equation}\label{equ-2-2-transfwilson}
W(x_1,x_2) \longrightarrow U(x_1)W(x_1,x_2)U(x_2)^\dagger
\end{equation}
The gauge field $A^\mu$ is valued in the Lie algebra of the group and generates Wilson lines for infinitesimal
displacements $\epsilon^\mu$
\begin{equation}\label{equ-2-2-defgf}
W(x,x+\epsilon) = \mathbb{I} - ig\epsilon_\mu A^\mu(x) + \order{\epsilon^2}
\end{equation}
with an arbitrary nonzero number $g$ which ends up as the gauge coupling in the covariant
derivative. 
Different finite Wilson lines $W(x_1,x_2)$ can be generated by iterating \eqref{equ-2-2-defgf}
along paths connecting $x_1$ and $x_2$ (see e.g. \cite{peskin} for details).
\eqref{equ-2-2-defgf} implies that the gauge field transforms as
\[ A^\mu(x) \longrightarrow U(x)\left(A^\mu(x) + \frac{i}{g}\partial^\mu\right)U(x)^\dagger \]

Due to their transformation behavior, the Wilson lines can be used to connect different points on
the manifold and define a covariant derivative:
\begin{equation}\label{equ-2-2-defcodev}
D^\mu\phi(x) = \lim_{\epsilon\rightarrow0}\frac{W(x,x+\epsilon n^\mu)\phi(x+\epsilon n^\mu) - \phi(x)}
{\epsilon}
\end{equation}
Inserting \eqref{equ-2-2-defgf} into \eqref{equ-2-2-defcodev} we retrieve the well known expression
for the covariant derivative
\[ D^\mu\phi(x) = \partial^\mu\phi(x) - igA^\mu\phi(x) \]
which by construction is gauge covariant as desired.

The transformation law \eqref{equ-2-2-transfwilson} also implies that Wilson lines can be plugged
together to form new Wilson lines%
\[ W(x_1,x_2)W(x_2,x_3) = W(x_1,x_3) \]
In particular, it follows that the closed Wilson line defined as
\begin{equation}\label{equ-2-2-cwl}
K(x) = W(x,x+\epsilon)W(x+\epsilon,x+\epsilon+\eta)W(x+\epsilon+\eta,x+\eta)W(x+\eta,x)
\end{equation}
with two small displacements $\epsilon^\mu$ and $\eta^\mu$ transforms as
\begin{equation}\label{equ-2-2-tcw}
K(x) \longrightarrow U(x)K(x)U(x)^\dagger
\end{equation}
Plugging \eqref{equ-2-2-defgf} and the Taylor expansion of $A^\mu$
\[ A^\mu(x+\epsilon) = A^\mu(x) + \epsilon_\nu\partial^\nu A^\mu(x) + \order{\epsilon^2}\]
into the definition of $K(x)$ \eqref{equ-2-2-cwl}, we obtain
\[
K(x) = \mathbb{I} - ig\epsilon_\mu\eta_\nu
\underbrace{\left(\partial^\mu A^\nu(x) - \partial^\nu A^\mu(x) -
ig\komm{A^\mu(x)}{A^\nu(x)}\right)}_{F^{\mu\nu}} + \order{\eta^2,\mu^2}
\]
recovering the well known expression for the field strength tensor $F^{\mu\nu}$. The transformation
behavior \eqref{equ-2-2-tcw} must hold in every order in $\epsilon^\mu$ and $\eta^\mu$, and it
follows that
\[ F^{\mu\nu}(x) \longrightarrow U(x)F^{\mu\nu}(x)U(x)^\dagger \]
Therefore, we can use $F^{\mu\nu}$ to write down a gauge invariant Lagrangian that involves
bilinears in the
derivatives of $A^\mu$ and which therefore is suitable to give dynamics to the gauge field
\[ \LL_\text{gauge} = -\frac{1}{2}\tr F^{\mu\nu}F_{\mu\nu} \]

Summing up the discussion: the gauge field acts as connection that relates gauge
transformations at adjacent points on the manifold to each other and therefore allows to compare the
value of fields at different points in a gauge covariant way. This is strongly
reminiscent of Riemannian geometry where the Christoffel symbols provide such a connection w.r.t.
coordinate transformations. The connection can be integrated to provide a parallel transport along
finite paths,
much like the Wilson lines for gauge transformations. Indeed, there is a geometrical description
of gauge theory in terms of fiber bundles \cite{goeckeler}. The field strength tensor is built in a
similar way as the curvature tensor by considering the transport along an infinitesimal
parallelogram.

\subsubsection*{Deconstructing gauge theory}

Let us now carry the above description of gauge invariance over to the deconstructed 5D theory.
Going from the continuous coordinate $y$ to discrete sites $y_n$, we've got a copy of the 4D
gauge group $\mathcal{G}_n$ at every lattice site, and the theory is invariant under transformations
of the fields
\begin{equation}\label{equ-2-2-phidisc} \phi(x,y_n) = \phi_n(x) \rightarrow U_n(x)\phi_n(x)
\end{equation}
If we compare the values of one of the $\phi_n$ at two different
4D coordinates $\phi_n(x_1)$ and $\phi_n(x_2)$, then we can still use 4D gauge fields
$A^\mu_n(x)$ localized at the lattice sites to generate Wilson lines $W_n(x_1,x_2)$ which compensate
for the change in gauge.

However, as far as the fifth
coordinate is concerned, there is no such thing as an infinitesimal displacement in $y$ anymore after
deconstruction. Therefore, the notion of a field $A^5_n(x)$ which connects points infinitesimally
separated in $y$ doesn't make any sense at all.
The smallest displacement in $y$ we can manage in the deconstructed theory is a hop between adjacent lattice
sites. Since this is a finite displacement, we need finite Wilson lines $\Sigma_n(x)$ which connect
the sites and which transform as
\begin{equation}\label{equ-2-2-transfsigma}
\Sigma_n(x) \longrightarrow U_{n-1}(x)\Sigma_n(x) U_n(x)^\dagger
\end{equation}

\begin{figure}[!tb]
\centerline{\begin{tikzpicture}
\tikzstyle{lattice node} = [shape=circle,minimum width=1.3cm,shade=ball,ball color=gray!01]
\draw[thick] (-6cm,0mm) -- (-1cm,0mm);
\draw[thick] (1cm,0mm) -- (6cm,0mm);
\draw[thick,loosely dotted] (-1cm,0mm) -- (-3mm,0mm);
\draw[thick,loosely dotted] (3mm,0mm) -- (1cm,0mm);
\node[style=lattice node] at (-6cm,0mm) {$\begin{matrix}\mathcal{G}_0\end{matrix}$};
\node[style=lattice node] at (-2.5cm,0mm) {$\mathcal{G}_1$};
\node[style=lattice node] at (2.5cm,0mm) {$\mathcal{G}_{N-1}$};
\node[style=lattice node] at (6cm,0mm) {$\mathcal{G}_N$};
\node[anchor=north] at (-6cm,-7mm) {$A^\mu_0(x)$};
\node[anchor=north] at (-2.5cm,-7mm) {$A^\mu_1(x)$};
\node[anchor=north] at (2.5cm,-7mm) {$A^\mu_{N-1}(x)$};
\node[anchor=north] at (6cm,-7mm) {$A^\mu_N(x)$};
\node at (0mm,0mm) {$//$};
\node[anchor=north] at (-42.5mm,-1mm) {$\Sigma_1(x)$};
\node[anchor=north] at (-10mm,-1mm) {$\Sigma_2(x)$};
\node[anchor=north] at (10mm,-1mm) {$\Sigma_{N-1}(x)$};
\node[anchor=north] at (42.5mm,-1mm) {$\Sigma_N(x)$};
\end{tikzpicture}}
\caption{Deconstructing a gauge theory. At each lattice site, there is a 4D copy $\mathcal{G}_n$
of the gauge group $\mathcal{G}$
with the gauge field $A_n^\mu$, the sites being linked with Wilson line fields $\Sigma_n$.}
\label{fig-2-2-gaugedec}
\end{figure}
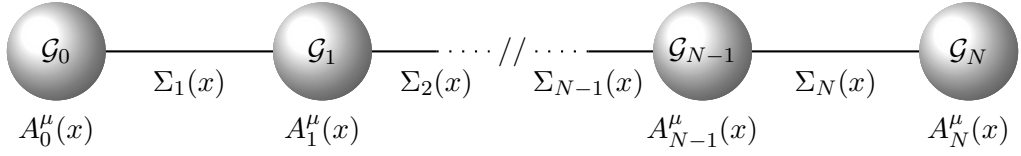
Therefore, in the deconstructed theory, we must replace the 5D gauge field $A^a(x,y)$ with $N+1$ 4D
gauge fields $A^\mu_0(x),\dots,A^\mu_N(x)$ and $N$ Wilson line fields $\Sigma_1(x),\dots,\Sigma_N$.
This setup is shown in fig.~\ref{fig-2-2-gaugedec}.
From these objects, we can build up Wilson lines that connect arbitrary points on the lattice
$(x_1,y_{n_1})$ and $(x_2,y_{n_2})$. The $\Sigma_n$ are valued in the gauge group and can be
parameterized using the group generators $T_k$
\begin{equation}\label{equ-2-2-decsigma}
\Sigma_n(x) = \exp\left(i\frac{\sigma_{nk}(x)}{v}T_k\right)
\end{equation}
The scale $v$ is introduced to make the component fields $\sigma_{nk}$ dimensionful quantities and
will later be fixed by matching the mass spectrum of the deconstructed theory to that of the
continuous one. As the $\Sigma_n$ are valued in the gauge group, they must have a nonvanishing
vacuum expectation value. If the component fields $\sigma_{nk}$ don't develop nontrivial vacuum
expectation values, then \eqref{equ-2-2-decsigma} tells us
\begin{equation}\label{equ-2-2-vacsigma} \left<\Sigma_n\right> = \mathbb{I} \end{equation}

\begin{figure}[!tb]
\centerline{\begin{tikzpicture}
\tikzstyle{lattice site} = [shape=circle,shade=ball,ball color=gray!01,minimum width=1cm]
\path[shade=axis,top color=gray!5,bottom color=gray!20]
(-7.5cm,-1.5cm) -- (4.5cm,-1.5cm) -- (7.5cm,1.5cm) -- (-4.5cm,1.5cm) -- cycle;
\draw[->] (-7.3cm,-1.4cm) -- (-6.4cm,-5mm);
\draw[->] (-7.3cm,-1.3cm) -- (-6cm,-1.3cm);
\node[anchor=south] at (-62mm,-6mm) {$x^{0,1,2,3}$};
\node[anchor=south east] at (-59mm,-13mm) {$y$};
\begin{scope}[xshift=-9mm,yshift=-9mm]
	\draw[dotted] (-4cm,0cm) -- (4cm,0cm);
	\node[style=lattice site] at (-20mm,0mm) {\parbox{5mm}{\centerline{$y_{n-1}$}}};
	\node[style=lattice site] at (20mm,0mm) {$y_n$};
	\node[anchor=south east] at (4cm,0mm) {$x$};
	\node[anchor=south west] at (-4cm,0mm) {$x$};
	\draw[thick,->] (-14mm,0mm) -- (14mm,0mm);
	\node[anchor=north] at (0mm,-0mm) {$\Sigma_{n}(x)$};
\end{scope}
\begin{scope}[xshift=9mm,yshift=9mm]
	\draw[dotted] (-4cm,0cm) -- (4cm,0cm);
	\node[style=lattice site] at (-20mm,0mm) {\parbox{5mm}{\centerline{$y_{n-1}$}}};
	\node[style=lattice site] at (20mm,0mm) {$y_n$};
	\node[anchor=south east] at (4cm,0mm) {$x+\epsilon$};
	\node[anchor=south west] at (-4cm,0mm) {$x+\epsilon$};
	\draw[thick,<-] (-14mm,0mm) -- (14mm,0mm);
	\node[anchor=south] at (0mm,-0mm) {$\Sigma_{n}(x+\epsilon)^\dagger$};
\end{scope}
\draw[thick,->] (15mm,-5mm) -- (25mm,5mm);
\draw[thick,<-] (-25mm,-5mm) -- (-15mm,5mm);
\node [anchor=west] at (20mm,-2mm) {$W_n(x,x+\epsilon)$};
\node [anchor=east] at (-20mm,2mm) {$W_{n-1}(x+\epsilon,x)$};
\end{tikzpicture}}
\caption{Definition of the quantity $K_n$ as the Wilson line along a parallelogram spanning two
adjacent lattice sites and an infinitesimal 4D distance.}
\label{fig-2-2-defk}
\end{figure}
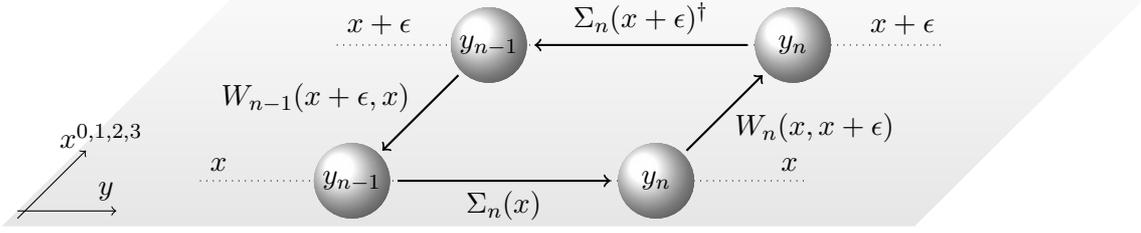
A kinetic term for the $\Sigma_n$ can be obtained by finding the analog to $F^{\mu5}(x,y)$ in the
continuous case. According to our previous discussion, $F^{\mu5}$ is obtained from the Wilson line
along a parallelogram spanning an infinitesimal distance both in a 4D direction and in $y$. As we can
only go finite distances $d$ in the discrete theory, a deconstructed analog is the quantity
\begin{equation}\label{equ-2-2-defk}
K_n(x) = \Sigma_n(x)W_n(x,x+\epsilon)\Sigma_n(x+\epsilon)^\dagger W_{n-1}(x+\epsilon,x)
\end{equation}
which corresponds to an infinitesimal displacement $\epsilon^\mu$ in 4D direction and a finite hop
from $y_{n-1}$ to $y_n$ in the discrete extra dimension (c.f. fig.~\ref{fig-2-2-defk}). Expanding
\eqref{equ-2-2-defk} in a similar fashion as \eqref{equ-2-2-cwl}, we obtain
\[
K_n(x) = \mathbb{I} - \epsilon_\mu\underbrace{\left(\partial^\mu\Sigma_n(x) -
ig\left(A^\mu_{n-1}(x)\Sigma_n(x)
- \Sigma_n(x)A^\mu_n(x)\right)\right)\Sigma_n(x)^\dagger}_{F^\mu_n(x)} + \order{\epsilon^2}
\]
By construction, $F^\mu_n$ must transform as
\[ F^\mu_n(x) \rightarrow U_n(x)F^\mu_n(x)U_n(x)^\dagger \]
and therefore, we can write down a gauge invariant kinetic term for the $\Sigma_n$
\[ \LL_\Sigma = \sum_{n=1}^N v^2\tr F^\mu_n F_{n\mu}^\dagger \]
$\LL_\Sigma$ is chosen such that the component fields $\sigma_{nk}$ are canonically normalized.

As the covariant derivative of the $\Sigma_n$ is fixed by their transformation behavior
\eqref{equ-2-2-transfsigma}
\[ D^\mu\Sigma_n = \partial^\mu\Sigma_n -ig\left(A^\mu_{n-1}\Sigma_n - \Sigma_nA^\mu_n\right) \]
it follows that
\[ F^\mu_n = \left(D^\mu\Sigma_n\right)\Sigma_n^\dagger \]
and finally, we can rewrite the complete Lagrangian for the deconstructed theory as
\begin{equation}\label{equ-2-2-lagdec}
\LL_\text{dec} = -\frac{1}{2}\sum_{n=0}^N \tr F^{\mu\nu}_nF_{n\mu\nu} +
v^2\sum_{n=1}^N \tr \left(D^\mu\Sigma_n\right)\left(D_\mu\Sigma_n\right)^\dagger
\end{equation}

If we expand the $\Sigma_n$ fields in their component fields $\sigma_{nk}$ according to
\eqref{equ-2-2-decsigma} we find that the lowest order (aka the vacuum expectation value of
the $\Sigma_n$) generates mass terms for the component gauge fields $A^\mu_{n,k}$:
\[
\LL_\text{mass} = \frac{v^2g^2}{2} \sum_{n=1}^N\left(
A_{n-1,k}^2 + A_{n,k}^2 - 2A_{n-1,k}^\mu A_{n,k\mu}\right) =
\frac{1}{2}M_{nm}A^\mu_{n,k}A_{m,k\mu}
\]
with the real, symmetric mass matrix
\[
M = v^2g^2\begin{pmatrix}
1			& -1		& 0		& \cdots	& 0		\\
-1			& 2		& -1		& \ddots	& \vdots	\\
0			& -1		& \ddots	& \ddots	&	0		\\
\vdots	& \ddots	& \ddots	& 2		& -1		\\
0			& \cdots	& 0		& -1		& 1		\\
\end{pmatrix}
\]
The eigenvalues of this matrix (see e.g. \cite{Cheng:2001vd} for the eigenvectors) give the gauge
boson masses
\begin{equation}\label{equ-2-2-mdec}
m_n^2 = 4v^2g^2\sin^2\frac{n\pi}{2\left(N+1\right)} \quad,\quad n=0,\ldots,N
\end{equation}
For every component gauge field, the spectrum contains a massless mode together with $N$
massive ones. As we have $N$ sigma fields, we additionally get $N$ scalar fields for every group
generator which is just the amount of Goldstone bosons that are required to constitute
longitudinal modes for the massive vector bosons.

In the limit of large $N$ we can approximate \eqref{equ-2-2-mdec} by
\[ m_n^2 \approx \left(vg\frac{n\pi}{N+1}\right)^2 \]
Therefore, if we fix the scale $v$ as
\[ v = \frac{N+1}{2gR\pi} \approx \frac{1}{gd} \]
we recover the mass spectrum of the continuous case with Neumann boundary conditions from the
deconstructed theory in the large $N$ limit. It seems that our way of deconstructing the theory has
been the correct prescription for a 5D gauge field compactified with Neumann boundary conditions.

\begin{figure}[!tb]
\centerline{\begin{tikzpicture}
\tikzstyle{lattice node} = [shape=circle,shade=ball,minimum width=1.3cm,ball color=gray!01]
\tikzstyle{empty node} = [shape=circle,minimum width=1.3cm,draw,fill=white]
\draw[thick] (-6cm,0mm) -- (-1cm,0mm);
\draw[thick] (1cm,0mm) -- (6cm,0mm);
\draw[thick,loosely dotted] (-1cm,0mm) -- (-3mm,0mm);
\draw[thick,loosely dotted] (3mm,0mm) -- (1cm,0mm);
\node[style=lattice node] at (-2.5cm,0mm) {$\mathcal{G}_1$};
\node[style=lattice node] at (2.5cm,0mm) {$\mathcal{G}_{N-1}$};
\node[anchor=north] at (-2.5cm,-7mm) {$A^\mu_1(x)$};
\node[anchor=north] at (2.5cm,-7mm) {$A^\mu_{N-1}(x)$};
\node[style=empty node] at (-6cm,0mm) {};
\node[style=empty node] at (6cm,0mm) {};
\node at (0mm,0mm) {$//$};
\node[anchor=north] at (-42.5mm,-1mm) {$\Sigma_1(x)$};
\node[anchor=north] at (-10mm,-1mm) {$\Sigma_2(x)$};
\node[anchor=north] at (10mm,-1mm) {$\Sigma_{N-1}(x)$};
\node[anchor=north] at (42.5mm,-1mm) {$\Sigma_N(x)$};
\end{tikzpicture}}
\caption{Deconstructing a gauge theory with Dirichlet boundary conditions. The condition
$\left.A^\mu(x,y)\right|_{y\in\left\{0,2R\pi\right\}}=0$ removes the gauge groups and gauge
fields at the $y_0$ and $y_N$ sites of the lattice (c.f. fig.~\ref{fig-2-2-gaugedec}).}
\label{fig-2-2-dirichlet}
\end{figure}

What about Dirichlet or mixed boundary conditions on $A^\mu(x,y)$? Setting one or all components to
zero at both ends of the interval corresponds to setting the gauge fields at $y_0$ and / or $y_N$ to
zero and basically kills gauge invariance at these sites (c.f. fig.~\ref{fig-2-2-dirichlet}).
However, we have no way of transferring the appropriate boundary conditions on $A^5(x,y)$ to the
deconstructed version of the theory and can only hope that the theory takes care of this itself.

This is indeed the case as can been seen by counting the scalar and vector fields involved: if we remove
the $A_0^\mu$ and $A_N^\mu$ fields, we obtain $N-1$ massive gauge fields for every generator, while
the number of scalars remains $N$. After all the Goldstone bosons have been eaten, a physical scalar
remains for every generator, which we have argued in the last section also happens in the continuous
case with Dirichlet boundary conditions. In the mixed case, we have $N$ massive vectors and $N$
scalars which all get eaten for every generator, also matching the continuous case nicely.

Although deconstruction as presented here does a rather nice job at approximating 5D gauge theories,
the formalism had first been proposed in \cite{ArkaniHamed:2001ca} as a tool for constructing
renormalizable UV completions to 5D theories. At low scales, the deconstructed theory virtually
cannot be distinguished from a true 5D theory. However, while such a theory is inherently
nonrenormalizable and
must break down at some cutoff scale, the deconstructed version can be made renormalizable rather
easily.

For example, note that the $\Sigma_n$ are similar to the $\Sigma$ field introduced in
chapter \ref{chap-1-3} to describe the spontaneous breaking of electroweak gauge symmetry. Therefore,
the deconstructed theory can be UV completed just like the Standard Model by
introducing Higgs fields $H_n$ for each of the $\Sigma_n$ fields.

A different, technicolor-like approach to an UV completion is to describe the $\Sigma_n$
fields similar to the Pions in QCD
as the excitation of a condensate of strongly interacting fermions. At energies above
the confinement scale of this new strong interaction, the $\Sigma_n$ fields would be replaced
by the ``quarks'' and ``gluons'' of the new strongly interacting sector, a theory which is known to
be renormalizable.

For our purposes, it is important to note that deconstructing the theory retains gauge invariance.
Therefore, deconstruction presents a consistent gauge invariant way for cutting off the infinite
KK towers that arise in the continuous case.

\subsubsection*{Matter fields}

The correct prescription for the discretization of matter fields has already been given in
\eqref{equ-2-2-phidisc}: the field $\phi(x,y)$ is replaced by a set of $N+1$ fields $\phi_n(x)$.
The Wilson line fields $\Sigma_n$ then allow for a straightforward discretization of the covariant
derivative\footnote%
{
In fact, different possible discretization prescriptions can be devised, each leading to a slightly
different mass spectrum, but all having the same high energy limit.
} $D^5$
\[
\partial_y\phi(x,y_n) \longrightarrow D_n = \frac{\Sigma_n(x)\phi_n(x) - \phi_{n-1}(x)}{d}
\]
The discretized covariant derivative can be used to build a discretized version of the
Lagrangian for the matter fields, e.g. for a real scalar
\begin{equation}\label{equ-2-2-disclphi}
\LL_\phi = \frac{1}{2}\sum_{n=0}^N\left(D^\mu\phi_n D_\mu\phi_n - m^2\phi^2_n\right) -
\frac{1}{2}\sum_{n=1}^N D_n^\dagger D_n
\end{equation}

Expanding the second term in \eqref{equ-2-2-disclphi} we obtain
\[
D_n^\dagger D_n = \frac{1}{d^2}\left(\phi_n^T\phi_n + \phi_{n-1}^T\phi_{n-1} -
\phi_n^T\Sigma_n^\dagger\phi_{n-1} -\phi_{n-1}^T\Sigma_n\phi_n\right)
\]
We find that the discretized covariant derivative terms 
yield contributions to the diagonal mass terms of the $\phi_n$ as well as Yukawa-type ``hopping'' terms
that couple fields at adjacent lattice sites and contribute off-diagonal elements to the mass
matrix. Diagonalization of the mass matrix then leads to a finite tower of massive modes which
matches the infinite KK tower in the $N\rightarrow\infty$ limit.

If the discretization procedure is applied to fermions, a new issue arises. Plugging in the
discretized derivative and calculating the mass spectrum, an unphysical doubling of modes is
observed: for every mode in the continuous theory, two modes arise in the lattice approximation, one
of which is unphysical. This problem of fermion doubling as well as the solution by adding a
so-called Wilson term to the Lagrangian is well known from lattice gauge theory. A detailed
discussion of the issue in the context of deconstructed QED can be found in \cite{Hill:2002me}.
In the Three-Site Model, however, this problem does not arise due to the extremely small size of the
lattice, so we won't elaborate on this point. Instead, the interested reader is referred to above
reference.

\section{Constructing the Lagrangian}
\label{chap-2-3}

\begin{figure}[!tb]
\centerline{\begin{tikzpicture}
\tikzstyle{node style} = [shape=circle,minimum size=2cm,shade=ball,ball color=gray!01]
\path[draw=black,dotted,fill=gray!20]
	(35mm,0mm) -- (35mm,12mm) arc (0:90:1cm) -- (0mm,22mm) arc (270:90:5mm) --
	(45mm,32mm) arc (90:0:1cm) -- (55mm,-22mm) arc (0:-90:1cm) --
	(-45mm,-32mm) arc (270:90:5mm) -- (25mm,-22mm) arc (-90:0:1cm) -- cycle;
\path[draw=white] (55mm,10mm) -- (55mm,22mm);
\draw[thick] (-4.5cm,0cm) -- (4.5cm,0cm);
\draw[line width=1ex,line cap=round] (-4.5cm,0cm) -- (-4.5cm,-22mm);
\draw[line width=1ex,line cap=round] (0cm,0cm) -- (0cm,-22mm);
\draw[line width=1ex,line cap=round] (0cm,0cm) -- (0cm,22mm);
\draw[line width=1ex,line cap=round] (4.5cm,0cm) -- (4.5cm,22mm);
\draw[thick,dashed] (-4.5cm,-22mm) -- (0cm,22mm);
\draw[thick,dashed] (0cm,-22mm) -- (4.5cm,22mm);
\node[style=node style] at (-4.5cm,0cm)
	{$\begin{matrix}\sun{2}_0 \\ g_0 = g \\ A_0^\mu \end{matrix}$};
\node[style=node style] at (0cm,0cm)
	{$\begin{matrix}\sun{2}_1 \\ g_1 = \tilde{g} \\ A_1^\mu \end{matrix}$};
\node[style=node style] at (4.5cm,0cm)
	{$\begin{matrix}\mathbf{U}(1)_2 \\ g_2 = g^\prime \\ A_2^\mu\end{matrix}$};
\node[anchor=south west] at (-32mm,-1mm) {$\Sigma_1$};
\node[anchor=south west] at (13mm,-1mm) {$\Sigma_2$};
\node[anchor=south east] at (35mm,10mm) {$f_2$};
\node[anchor=south east] at (-10mm,10mm) {$f_1$};
\node[anchor=east] at (-45mm,-17mm) {$\Psi_{0L}$};
\node[anchor=east] at (0mm,-17mm) {$\Psi_{1L}$};
\node[anchor=west] at (0mm,17mm) {$\Psi_{1R}$};
\node[anchor=west] at (45mm,17mm)
	{$\begin{pmatrix}\Psi^u_{2R} \\ \Psi^d_{2R}\end{pmatrix}=\Psi_{2R}$};
\end{tikzpicture}}
\caption{The structure of the Three-Site Higgsless Model in moose notation. See the text for
explanation.}
\label{fig-2-3-moose}
\end{figure}
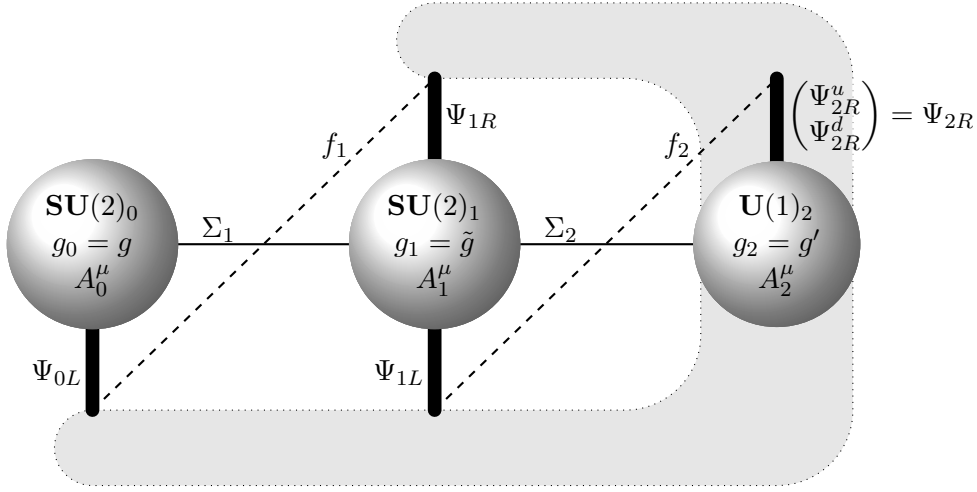
With the introduction of 5D gauge theories and dimensional deconstruction in the last two sections, we now
have the tools at hand for the actual construction of a Lagrangian implementing the Three-Site
Higgsless Model \cite{Chivukula:2006cg} sketched in chapter \ref{chap-1-5}.

In \cite{Csaki:2003dt} it is shown that in 5D gauge
theories the unitarity cutoff is pushed to the NDA cutoff scale by the exchange of KK gauge bosons.
Therefore, a 5D gauge theory is a good starting point for a theory in which unitarity violation is
delayed
by heavy $W^\prime$ and $Z^\prime$ partners. Dimensional deconstruction can then be used to cut off
the higher KK modes in gauge invariant way, retaining only the Standard Model particles and one
generation of KK partners.

Fig.~\ref{fig-2-3-moose} shows the structure of the Three Site Model in moose
notation \cite{Georgi:1985hf}. The gauge group consists of two $\sun{2}$ factors and one
$\mathbf{U}(1)$ factor which can be understood as a 5D $\sun{2}$ gauge theory broken by
boundary conditions to $\mathbf{U}(1)$ on the right brane and deconstructed to three lattice sites.
All three group factors are given their own gauge couplings $g$, $\tilde{g}$ and $g^\prime$
which is required in order to allow the model to be consistent with experimental data and which
corresponds to a $y$ dependent gauge coupling $g(y)$ in the continuous version of the theory.
QCD is left unchanged in the Three-Site Model with only one set of gluon fields $G^\mu$ 
in the spectrum. The gauge kinetic Lagrangian is
\begin{equation}\label{equ-2-3-lgauge}
\LL_\text{gauge} = -\frac{1}{2}\sum_{n=0}^1\tr F^{\mu\nu}_n F_{n\mu\nu}
-\frac{1}{4} F^{\mu\nu}_2 F_{2\mu\nu} - \frac{1}{2}\tr G^{\mu\nu}G_{\mu\nu}
\end{equation}
It should be noted that this setup is identical to the BESS model \cite{Casalbuoni:1985kq}.

Deconstructing the gauge group gives us two nonlinear sigma type Wilson line fields
$\Sigma_1$ and $\Sigma_2$ which connect adjacent lattice sites. Under gauge transformations
generated by parameter fields $\lambda_0 = \lambda_{0k}\tau_k$, $\lambda_1=\lambda_{1k}\tau_k$ and
$\lambda_2$, the Wilson lines transform as
\begin{equation}\label{equ-2-3-transfsigma}
\Sigma_1 \longrightarrow e^{i\lambda_0}\Sigma_1 e^{-i\lambda_1} \qquad,\qquad
\Sigma_2 \longrightarrow e^{i\lambda_1}\Sigma_2 e^{-i\lambda_2\tau_3}
\end{equation}
The transformation behavior \eqref{equ-2-3-transfsigma} fixes the covariant derivative of the
$\Sigma$ fields
\begin{equation}\label{equ-2-3-codevsigma}\begin{aligned}
D^\mu\Sigma_1 &= \partial^\mu\Sigma_1 - igA_0^\mu\Sigma_1 + i\tilde{g}\Sigma_1A_1^\mu \\
D^\mu\Sigma_2 &= \partial^\mu\Sigma_2 - i\tilde{g}A_1^\mu\Sigma_2 + ig^\prime A_2^\mu\Sigma_2\tau_3
\end{aligned}\end{equation}
With the covariant derivatives fixed, we can write down the kinetic term for the $\Sigma_{1/2}$
fields
\begin{equation}\label{equ-2-3-lsigma}
\LL_\Sigma = v^2\sum_{n=1}^2\tr \left(D^\mu\Sigma_n\right)^\dagger\left(D_\mu\Sigma_n\right)
\end{equation}
Expanding \eqref{equ-2-3-lsigma} in terms of the component fields of $\Sigma_{1/2}$ gives mass terms for
the gauge bosons. The mass scale is set by $v$ and fixed by the $W$ and $Z$ masses.

The fermion sector is inspired by deconstructing 5D fermions with chiral boundary conditions.
The Three-Site Model comes with two copies of each Standard Model
fermion, the chiral components of which are distributed according to fig.~\ref{fig-2-3-moose}. Each
Standard Model isospin doublet gives rise to left-handed doublets under the respective $\sun{2}$ at
sites $0$ and $1$, a right-handed $\sun{2}_1$ doublet at site $1$ and two right-handed fermions at
site $2$. As shown in tab.~\ref{tab-2-3-u1charge},
the fermions at site $2$ inherit their $\mathbf{U}(1)_2$ charge from the electromagnetic charge of
the corresponding Standard Model fields, while those at the other two sites inherit the
hypercharge\footnote%
{
While being a slight deviation from straightforward deconstruction as it corresponds to a nonlocal
coupling of the fermion fields to the gauge field on the right brane in the continuous theory, this
setup is necessary in order to reproduce the correct hypercharges of the Standard Model fermions without
introducing additional neutral vectors.
}.
\begin{table}
\centerline{\begin{tabular}{|l||c|c|}
\hline
 & $\Psi_{0L}$ / $\Psi_{1L}$ / $\Psi_{1R}$ & $\Psi_{2R}$ \\\hline\hline
\vs{4ex}Neutrinos & $-\frac{1}{2}$ & $0$ \\\hline
\vs{4ex}Leptons &  $-\frac{1}{2}$ & $-1$ \\\hline
\vs{4ex}Up type Quarks & $\frac{1}{6}$ & $\frac{2}{3}$ \\\hline
\vs{4ex}Down type Quarks & $\frac{1}{6}$ & $-\frac{1}{3}$ \\\hline
\end{tabular}}
\caption{$\mathbf{U}(1)_2$ charge assignments for the fermions on the three site moose.}
\label{tab-2-3-u1charge}
\end{table}

According to the above charge assignments, the covariant derivatives of the fermions (omitting the
gluon contribution in the case of quarks) read
\[\begin{aligned}
&D^\mu\Psi_{0L} = \left(\partial^\mu - igA^\mu_0 - ig^\prime A^\mu_2 Y\right)\Psi_{0L}
\quad&,\quad\quad
&D^\mu\Psi_{1L} = \left(\partial^\mu - i\tilde{g}A^\mu_1 - ig^\prime A^\mu_2 Y\right)\Psi_{1L}
\\
&D^\mu\Psi_{1R} = \left(\partial^\mu - i\tilde{g}A^\mu_1 - ig^\prime A^\mu_2 Y\right)\Psi_{1R}
\quad&,\quad\quad
&D^\mu\Psi_{2R}^{u/d} = \left(\partial^\mu - ig^\prime A^\mu_2 Q\right)\Psi_{2R}^{u/d}
\end{aligned}\]
with the charge operator $Q$ and the hypercharge operator $Y$. The fermion Lagrangian then
reads\footnote%
{In all expressions containing fermions, an implicit sum over all fermion flavors present in the
Standard Model is assumed.
}%
\begin{multline}\label{equ-2-3-lferm}
\LL_f = \sum_{n=0}^1\Psibar_{nL}\slashed{D}\Psi_{nL} + \sum_{n=1}^2\Psibar_{nR}\slashed{D}\Psi_{nR}
\\+
v\left(\lambda \Psibar_{0L}\Sigma_1\Psi_{1R} + \tilde{\lambda}\Psibar_{1L}\Psi_{1R} +
\Psibar_{1L}\Sigma_2
\begin{pmatrix}\lambda^\prime_u & 0 \\ 0 &\lambda^\prime_d\end{pmatrix}\Psi_{2R}
+\text{h.c.}\right)
\end{multline}
In \eqref{equ-2-3-lferm}, the Yukawa couplings between fermions at adjacent lattice sites are left as free
parameters, an effect which can be achieved in the continuous theory by the introduction of brane
kinetic terms. In particular,
the Yukawa couplings $\lambda^\prime_{u/d}$ between $\Psi_{L1}$ and the $\Psi_{R2}^{u/d}$ allow to
accommodate the different masses of up- and down-type fermions observed in nature. The
$\lambda^\prime_{u/d}$ are the only Yukawa couplings that have a nontrivial flavor structure, all
other couplings are taken as flavor universal\footnote%
{
Flavor mixing
is not included in the original model, but CKM-type mixing can be easily achieved through 
the $\lambda^\prime_{u/d}$, retaining the flavor-universal nature of the
other couplings \cite{Chivukula:2006cg}.
}.

With the definitions \eqref{equ-2-3-lgauge}, \eqref{equ-2-3-lsigma} and \eqref{equ-2-3-lferm}, the
complete Three-Site Lagrangian reads
\begin{equation}\label{equ-2-3-l3s}
\LL_\text{3-site} = \LL_\text{gauge} + \LL_\Sigma + \LL_f
\end{equation}
How does this model relate to that sketched in chapter \ref{chap-1-5}? In the limit of large
$\tilde{g}$ and large $\tilde{\lambda}$ we can integrate out the fields at the bulk site
in order to obtain a ``two site model'' with one linear sigma field
\[ \Sigma = \Sigma_1\Sigma_2 \]
Comparison to chapter \ref{chap-1-3} identifies this two site model as the ordinary Standard Model
with $\sun{2}_0$ playing the role of $\sun{2}_L$ and $\mathbf{U}(1)_2$ that of $\mathbf{U}(1)_Y$!
The bulk lattice site adds the additional $\sun{2}$ group factor and the set of fermion partners
suggested in chapter \ref{chap-1-5}, and the desired mixings arise from the terms that
connect the lattice sites via the $\Sigma_{1/2}$ fields, corresponding to the 5D kinetic terms in
the deconstructed setting.

In accordance with \cite{Chivukula:2006cg}, we introduce the following definitions for later convenience
\begin{equation}\label{equ-2-3-abbrev}
\begin{aligned}x&=\frac{g}{\tilde{g}}\vs{5ex} \\ \epsilon_L&=\frac{\lambda}{\tilde{\lambda}}\vs{5ex}
\end{aligned}\qquad\qquad\qquad
\begin{aligned}y&=\frac{g^\prime}{\tilde{g}}\vs{5ex} \\
\epsilon^\prime_f&=\frac{\lambda^\prime_f}{\tilde{\lambda}}\vs{5ex}
\end{aligned}\qquad\qquad\qquad
\begin{aligned}
\vs{5ex}t=\tan\theta=\frac{g^\prime}{g}=\frac{\sin\theta}{\cos\theta}\\\vs{5ex}\end{aligned}
\end{equation}
As argued above (and discussed in more detail in the next chapter),
the Standard Model is recovered in the limit of large $\tilde{g}$ and
$\tilde{\lambda}$, and therefore, we should expect that $x$, $y$, $\epsilon_L$ and
$\epsilon^\prime_f$ are
small quantities if the new structure is indeed a small perturbation on top of the Standard Model.

\chapter{Model Properties}
\label{chap-3}

\begin{quote}\itshape
Who ordered that?
\end{quote}
\hfill\begin{minipage}{0.7\textwidth}\small\raggedleft
(I. I. Rabi on the discovery of the muon on 1936)
\end{minipage}
\\[5mm]
At the end of the last chapter, the Lagrangian of the Three-Site Model as presented in \cite{Chivukula:2006cg}
was assembled. The present chapter is devoted to a detailed discussion of the model. In the first
section, the mass spectrum is calculated. In the second section, the free parameters present in the
model in addition to the Standard Model parameters are identified. A number of experimental
constraints together with the resulting parameter space is discussed. With the third section follows
a brief discussion of the couplings between the Standard Model particles and their new heavy
partners, and the last section is devoted to the widths of the new heavy particles. An autogenerated
sample spectrum is printed in appendix \ref{app-3}.

\section{Masses}
\label{chap-3-1}

The first step in studying the phenomenology of the Three-Site Model is the calculation of the mass
spectrum. The mass terms for the gauge bosons can be readily extracted from \eqref{equ-2-3-lsigma}
\begin{equation}\label{equ-3-1-mgauge}
\LL_\text{mass,gauge} = M^{CC}_{ij} W_i^{+\mu} W_{j\mu}^- + \frac{1}{2}M^{NC}_{ij}B^\mu_i B_{j\mu}
\end{equation}
with the charged gauge bosons
\[
W_n^{\pm\mu} = \frac{A^\mu_{n1} \pm iA^\mu_{n2}}{\sqrt{2}}
\quad\left(n=0,1\right)
\]
the neutral gauge bosons
\[ B^\mu_n = A^\mu_{n3} \quad\left(n=0,1\right)\quad,\quad
B^\mu_2 = A^\mu_2
\]
and the mass matrices
\begin{equation}\label{equ-3-1-mmgb}
M^{CC} = v^2\tilde{g}^2\begin{pmatrix}x^2 & -x \\ -x & 2\end{pmatrix} \quad,\quad
M^{NC} = v^2\tilde{g}^2\begin{pmatrix}x^2 & -x & 0 \\ -x & 2 & -tx \\ 0 & -tx & t^2 x^2\end{pmatrix}
\end{equation}
where we have used the definitions \eqref{equ-2-3-abbrev}.

The fermion mass terms follow from the Yukawa couplings in \eqref{equ-2-3-lferm}
\begin{equation}\label{equ-3-1-mferm}
\LL_\text{mass,fermions} = 
\left(\Psibar_{0L}^f,\Psibar_{1L}^f\right) M_{f} \cvect{\Psi_{1R}^f \\
\Psi_{2R}^f}
\end{equation}
(an implicit sum over all fermions present in the Standard Model is assumed) with the fermion
mass matrices
\begin{equation}\label{equ-3-1-mmf}
M_f = v\tilde{\lambda}\begin{pmatrix}\epsilon_L & 0 \\ 1 & \epsilon^\prime_f\end{pmatrix}
\end{equation}

\subsubsection{Integrating out the bulk: the two-side model}

To get some intuition about the connection between the Standard Model and the Three-Site Model,
let's pick up again the argument presented at the end of chapter \ref{chap-2-3} and integrate out
the bulk lattice site in the limit
\begin{equation}\label{equ-3-1-limit}
\tilde{g}\rightarrow\infty \qquad,\qquad \tilde{\lambda}\rightarrow\infty
\end{equation}
where we keep the other parameters (in particular $\lambda$ and $\lambda^\prime$)
finite in the limiting process such that the limit implies
\[
x=\frac{g}{\tilde{g}}\rightarrow 0 \quad,\quad
y=\frac{g^\prime}{\tilde{g}}\rightarrow 0 \quad,\quad
\epsilon_L = \frac{\lambda}{\tilde{\lambda}}\rightarrow 0 \quad,\quad
\epsilon^\prime_f = \frac{\lambda^\prime_f}{\tilde{\lambda}}\rightarrow 0
\]
for the quantities defined in \eqref{equ-2-3-abbrev}.

The masses of the bulk gauge bosons and of the bulk fermions that are to be integrated out can be
read off from \eqref{equ-3-1-mgauge} and \eqref{equ-3-1-mferm}
\begin{equation}\label{equ-3-1-mbulk}
M_{A_1} = \sqrt{2}\tilde{g}v \qquad,\qquad  M_\text{bulk} = v\tilde{\lambda}
\end{equation}
The resulting model lives on only two lattice sites, connected by one nonlinear sigma field
\[ \Sigma = \Sigma_1\Sigma_2 \]
To stress the connection to the Standard Model, we will use the same identifiers as in chapter
\ref{chap-1-3} for the two-side model fields
\[
W^\mu = A^\mu_0 \quad,\quad B^\mu=A^\mu_2 \quad,\quad \Psi_L = \Psi_{0L}
\quad,\quad \Psi_R = \Psi_{2R}
\]
The covariant derivative of $\Sigma$ is then
\begin{equation}\label{equ-3-1-codevsigma}
D^\mu\Sigma = \partial^\mu\Sigma - igW^\mu\Sigma + ig^\prime B^\mu\Sigma\tau_3
\end{equation}
and the kinetic Lagrangian for $\Sigma$ is
\begin{equation}\label{equ-3-1-lsigma}
\LL_{\Sigma,\text{2S}} = \tilde{v}^2\tr \left(D_\mu\Sigma\right)^\dagger\left(D^\mu\Sigma\right)
\end{equation}
As we want to reproduce the Standard Model in the limit \eqref{equ-3-1-limit},
we can express $\tilde{v}$ through $m_W$ and $g$ by virtue of \eqref{equ-1-3-mgb}
\[ \tilde{v} = \frac{m_W}{g} \]

To find out how $\tilde{v}$ relates to $v$, let's perform the matching to the
Three-Site Model and integrate out the bulk gauge field from the diagram
\begin{equation}
\parbox{45mm}{\fmfframe(5,5)(5,5){\begin{fmfgraph*}(35,17)
\fmfleft{i}\fmfright{o}\fmftop{d1,t1,t2,d2}\fmfbottom{d3,b1,b2,d4}
\fmf{dashes}{t1,v1,b1}\fmf{dashes}{t2,v2,b2}\fmf{wiggly}{i,v1}
\fmf{dbl_wiggly,la=$W_{1}^+$}{v1,v2}\fmf{wiggly}{v2,o}\fmfdot{v1,v2}
\fmfv{la=$v$,la.a=90}{t1}\fmfv{la=$v$,la.a=90}{t2}
\fmfv{la=$v$,la.a=-90,la.d=\thick}{b1}\fmfv{la=$v$,la.a=-90,la.d=\thick}{b2}
\fmfv{la=$W^+$,la.a=180}{i}\fmfv{la=$W^+$,la.a=0}{o}
\end{fmfgraph*}}}
\end{equation}
The vertices can be read off \eqref{equ-3-1-mgauge}, and the resulting change in the Lagrangian is
\[
\Delta\LL_{m,W} =
(-i)\left(-iv^2g\tilde{g}\right)\frac{i}{2v^2\tilde{g}^2}\left(-iv^2g\tilde{g}\right)
W^{+\mu}W_\mu^-= -\frac{1}{2}v^2g^2 W^{+\mu} W_\mu^-
\]
modifying the mass term for $W^\pm_\mu$
\[
\LL_{m,W} = v^2g^2W^{+\mu}W_\mu^- + \Delta\LL_{m,W} = \frac{1}{2}v^2g^2W^{+\mu}W_\mu^-
\]
Comparing this to \eqref{equ-3-1-codevsigma} and \eqref{equ-3-1-lsigma} identifies $\tilde{v}$ as
\begin{equation}\label{equ-3-1-matchv}
\tilde{v}=\frac{v}{\sqrt{2}}
\end{equation}

The factor $\sqrt{2}$ has an interesting effect: plugging $v$ into the formula for the NDA cutoff
\eqref{equ-1-3-nda} in order to estimate the UV cutoff of the Three-Site Model, we obtain
\[
\Lambda_\text{NDA} = 8\pi v = 8\sqrt{2}\pi\tilde{v} = 8\sqrt{2}\pi\frac{m_W}{g} \approx
\unit[4.4]{TeV}
\]
Comparison to \eqref{equ-1-3-nda} shows that the NDA cutoff scale is elevated by a factor $\sqrt{2}$
in the Three-Site Model above that of the Standard Model.

In fermion sector, Yukawa couplings between the left-handed fermions sitting at site $0$ and the
right-handed ones at site $1$ (formerly site $2$) are obtained by integrating out the bulk
fermions from diagrams of the type
\begin{equation}
\parbox{45mm}{\fmfframe(5,5)(5,5){\begin{fmfgraph*}(35,17)
\fmfleft{i2,i1}\fmfright{o2,o1}
\fmf{dashes}{i1,v1}\fmf{fermion}{i2,v1}\fmf{heavy,la=$\Psi_\text{bulk}^f$}{v1,v2}\fmf{dashes}{v2,o1}
\fmf{fermion}{v2,o2}\fmfdot{v1,v2}\fmfv{la=$v$}{i1}\fmfv{la=$v$}{o1}
\fmfv{la=$\Psi_{L}^f$}{i2}\fmfv{la=$\Psi_{R}^f$}{o2}
\end{fmfgraph*}}}
\end{equation}
The vertex factors can be read off from \eqref{equ-3-1-mmf} and inserted to obtain the effective
Yukawa couplings
\begin{equation}\label{equ-3-1-matchyuk}
\LL_\text{Yukawa,2S} = v\left(\Psibar_L^f\epsilon_L\lambda^\prime\Psi_R^f +
\text{h.c.}\right) =
M_\text{bulk}\left(\Psibar_L^f\epsilon_L\epsilon^\prime_f\Psi_R^f +
\text{h.c.}\right)
\end{equation}
with an implicit sum over all Standard Model fermions. We can read off the fermion masses
from \eqref{equ-3-1-matchyuk}
\[ m_f = M_\text{bulk}\epsilon_L\epsilon^\prime_f \]
After choosing the flavor-universal $\epsilon_L$, the $\epsilon^\prime_f$ are to be chosen for each
Standard Model fermion separately to yield the correct Yukawa couplings and masses.

If we would drop the assumption of taking the limit $\tilde\lambda\rightarrow\infty$ such that
$\epsilon^\prime_f\rightarrow 0$ and instead keep the $\epsilon^\prime_f$ finite (which is
perfectly possible by letting $\epsilon_L$ go to zero fast enough), diagrams like
\begin{equation}
\parbox{48mm}{\fmfframe(5,5)(5,7){\begin{fmfgraph*}(35,17)
\fmfleft{i2,i1}\fmfright{o2,o1}
\fmf{dashes}{i1,v1}\fmf{fermion}{i2,v1}\fmf{heavy,la=$\Psi_\text{bulk}^f$}{v1,v2}\fmf{dashes}{v2,o1}
\fmf{fermion}{v2,o2}\fmfdot{v1,v2}\fmfv{la=$v$}{i1}\fmfv{la=$v$}{o1}
\fmfv{la=$\Psi_{R}^f$}{i2}\fmfv{la=$\Psi_{R}^f$}{o2}
\end{fmfgraph*}}}
\label{equ-3-1-wfren}
\end{equation}
would induce finite wave function renormalizations\footnote%
{
Due to the chirality of $\Psi_{R}^f$, the momentum expansion of the diagram
\eqref{equ-3-1-wfren} starts at the order $\slashed{p}$, leading to the change in the
Lagrangian
\[ \Delta\LL = i{\epsilon_f^\prime}^2\Psibar^f_{R}\slashed{\partial}\Psi^f_{R} \]
}
\[ \Psi_{R}^f \;\longrightarrow\; \sqrt{1+{\epsilon^\prime_f}^2}\;\Psi_{R}^f \]
In addition, diagrams of the type
\begin{equation}
\parbox{46mm}{\fmfframe(5,8)(5,8){\begin{fmfgraph*}(30,30)
\fmfleft{i2,i1}\fmfright{o4,o3,o2,o1}\fmf{fermion}{i1,v1}\fmf{phantom}{v1,v2,v3}
\fmf{fermion}{v3,i2}\fmf{dashes}{v1,o1}\fmf{dashes}{v3,o4}\fmffreeze
\fmf{heavy}{v1,v2}\fmf{heavy}{v2,v3}
\fmf{dbl_wiggly}{v2,v4}\fmf{dashes}{o2,v4}\fmf{wiggly}{o3,v4}
\fmfdot{v1,v2,v3,v4}
\fmfv{la=$\Psi_{R}^f$}{i1}\fmfv{la=$\Psi_{R}^f$}{i2}
\fmfv{la=$v$}{o1,o4,o2}\fmfv{la=$W/B$}{o3}
\end{fmfgraph*}}}
\end{equation}
would induce finite couplings e.g. of two $\Psi_R^f$ to a $W^\pm$, leading to deviations
from the Standard Model even in the limit $\tilde{g}\rightarrow\infty\;,\;\tilde{\lambda}\rightarrow\infty$!
Similar arguments apply if $\epsilon_L$ is kept finite, and therefore, the correct limit for
recovering the Standard Model is in fact
\begin{equation}\label{equ-3-1-limit}
x\rightarrow 0 \quad,\quad M_\text{bulk}\rightarrow\infty \quad,\quad
\epsilon_L\rightarrow 0 \quad,\quad \epsilon^\prime_f\rightarrow 0
\end{equation}
while keeping $g$, $g^\prime$ and $v$ finite\footnote%
{
Indeed, while being more intuitive considering the parameterization of the model we will be
developing in the remainder of this chapter, \eqref{equ-3-1-limit} together with the requirement of
keeping $g$, $g^\prime$ and $v$ finite
is just a rephrasing of the limiting conditions stated already at the beginning of
this section.
}.

In the limit \eqref{equ-3-1-limit}, we fully recover the Standard Model and obtain the matching conditions
\eqref{equ-3-1-matchv} and \eqref{equ-3-1-matchyuk}. By comparison to chapter \ref{chap-1-3}
we can immediately write down the spectrum to lowest order in an expansion in
$x$, $\epsilon_L$ and $\epsilon^\prime_f$:
\begin{itemize}
\item The massless photon and gluons.
\item The Standard Model $W^\pm$ and $Z$ bosons with masses
\[
m_W = \frac{vg}{\sqrt{2}} \quad,\quad m_Z = v\sqrt{\frac{g^2+{g^\prime}^2}{2}}
\]
\item The Standard Model fermions with masses given by
\[ m_f = M_\text{bulk}\epsilon_L\epsilon^\prime_f \]
\item Two ${W^\pm}^\prime$ and one $Z^\prime$ which are degenerate with mass
\[ m_{W^\prime} = m_{Z^\prime} = \sqrt{2}\tilde{g}v \]
\item One partner fermion for each Standard Model fermion with mass
\begin{equation} M_\text{bulk} = \tilde{\lambda}v \label{equ-3-1-defmb}\end{equation}
\end{itemize}

\subsubsection*{Exact spectrum}

To get the corrections for finite $x$, $\epsilon_L$ and $\epsilon^\prime_f$
to the above picture, we must explicitly diagonalize the mass matrices \eqref{equ-3-1-mmgb} and
\eqref{equ-3-1-mmf}. This was done analytically with the results then being expanded as a series
in $x$ resp.
$\epsilon_L$ in order to obtain the expressions for masses and wavefunctions presented in this
section.

As the gauge sector of the Three-Site Model corresponds to a deconstructed 5D $\sun{2}$ gauge group
with Neumann boundary conditions for the $\tau_3$ component field and mixed ones for the other two, an
unbroken $\mathbf{U}(1)$ gauge symmetry together with a massless gauge boson is to be expected. Indeed, setting
\[ \lambda_0 = \lambda_1 = \lambda_2\tau_3 \]
in \eqref{equ-2-3-transfsigma}, we find that the vacuum expectation value of the $\Sigma_{1/2}$ is not
changed under the transformation and, therefore, this combination
generates an unbroken $\mathbf{U}(1)$ gauge symmetry. Looking at the fermion charges tab.
\ref{tab-2-3-u1charge} this is readily identified as the electromagnetic
$\mathbf{U}(1)_\text{em}$, and defining the electromagnetic gauge coupling $e$
\begin{equation}\label{equ-3-1-defe}
\frac{1}{e^2} = \frac{1}{g^2} + \frac{1}{\tilde{g}^2} + \frac{1}{{g^\prime}^2}
\end{equation}
the corresponding photon field $A^\mu$ is
\begin{equation}\label{equ-3-1-defa}
A^\mu = \frac{e}{g}B_0^\mu + \frac{e}{\tilde{g}}B_1^\mu + \frac{e}{g^\prime}B_2^\mu
\end{equation}

\begin{table}[!t]
\centerline{\begin{tabular}{|l|l|}
\hline \raisebox{-2ex}[-2ex][0ex]{
$m_W^2 = \frac{g^2v^2}{2}\left(1 - \frac{x^2}{4} + \order{x^6}\right)$} &
$f_0^W = 1 - \frac{x^2}{8} - \order{x^4} $ \vs{4ex} \\
\cline{2-2} &
$f_1^W = \frac{x}{2} + \frac{x^3}{16} - \order{x^5}$\vs{4ex} \\
\hline\hline \raisebox{-2ex}[-2ex][0ex]{
$m_{W'}^2 = 2\tilde{g}^2v^2\left(1 + \frac{x^2}{4} + \order{x^4}\right)$ } &
$f_0^{W'} = -\frac{x}{2} - \frac{x^3}{16} + \order{x^5}$\vs{4ex} \\
\cline{2-2} & $f_1^{W'} = 1- \frac{x^2}{8} - \order{x^4}$ \vs{4ex} \\
\hline\hline \raisebox{-5ex}[-5ex][0ex]{
$ m_Z^2 = \frac{g^2v^2}{2c^2}\left(1 - \frac{x^2\left(c^2-s^2\right)^2}{4c^2} +
\order{x^6}\right)$} &
$f_0^Z = c - \frac{x^2c^3\left(1+2t^2-3t^4\right)}{8} + \order{x^4} $\vs{5ex} \\
\cline{2-2} & $f_1^Z = \frac{xc\left(1-t^2\right)}{2} + \frac{x^3c^3\left(1-t^2\right)^3}{16}
+\order{x^5}$\vs{4ex} \\
\cline{2-2} & $f_2^Z = -s - \frac{x^2sc^2\left(3-2t^2-t^4\right)}{8} + \order{x^4}$\vs{5ex} \\
\hline\hline \raisebox{-5ex}[-5ex][0ex]{
$ m_{Z'}^2 = 2\tilde{g}^2v^2\left(1 + \frac{x^2}{4c^2} + \order{x^4}\right)$} &
$f_0^{Z'} = -\frac{x}{2} - \frac{x^3\left(1-3t^2\right)}{16} + \order{x^5}$\vs{5ex} \\
\cline{2-2} & $f_1^{Z'} = 1 - \frac{x^2\left(1+t^2\right)}{8} + \order{x^4}$\vs{5ex} \\
\cline{2-2} & $f_2^{Z'} = -\frac{xt}{2} + \frac{x^3t\left(3-t^2\right)}{16} + \order{x^5}$\vs{5ex}
\\\hline
\end{tabular}}
\caption{Masses and wavefunctions of the gauge bosons as obtained from diagonalizing the mass
matrices $M^{CC}$ and $M^{NC}$ (c.f. \eqref{equ-2-3-abbrev} for the abbreviations).}
\label{tab-3-1-mgb}
\end{table}
\begin{table}[!t]
\centerline{\begin{tabular}{|l|l|}
\hline \raisebox{-12ex}[-12ex][0ex]{
\parbox{7cm}{
\begin{multline*}
m_f = \frac{M_\text{bulk}\epsilon_L\epsilon^\prime_f}
{\sqrt{1+{\epsilon^\prime_f}^2}}\quad\times\\
\left(1-\frac{\epsilon_L^2}{2\left({\epsilon^\prime_f}^2+1\right)^2} + \order{\epsilon_L^4} \right)
\end{multline*}}}
&
$f^f_{0L} = -1 + \frac{\epsilon_L^2}{2\left(1+{\epsilon^\prime_f}^2\right)^2} +
\order{\epsilon_L^4}$\vs{8ex}
\\\cline{2-2}  & $f^f_{1L} = \frac{\epsilon_L}{1+{\epsilon^\prime_f}^2} +
\frac{\left(2{\epsilon^\prime_f}^2-1\right)\epsilon_L^3}{2\left({\epsilon^\prime_f}^2+1\right)^3} +
\order{\epsilon_L^5}$\vs{8ex}
\\ \cline{2-2} & $f^f_{1R} = -\frac{{\epsilon^\prime_f}}{\sqrt{1+{\epsilon^\prime_f}^2}} +
\frac{\epsilon_L^2{\epsilon^\prime_f}}{\left(1+{\epsilon^\prime_f}^2\right)^\frac{5}{2}} + 
\order{\epsilon_L^4}$\vs{8ex}
\\ \cline{2-2} & $f^f_{2R} = \frac{1}{\sqrt{1+{\epsilon^\prime_f}^2}} +
\frac{\epsilon_L^2{\epsilon^\prime_f}^2}{\left(1+{\epsilon^\prime_f}^2\right)^\frac{5}{2}} +
\order{\epsilon_L^4}$\vs{8ex}
\\ \hline\hline \raisebox{-12ex}[-12ex][0ex]{
\parbox{7cm}{
\begin{multline*}
m_{f^\prime} = M_\text{bulk}\sqrt{1+{\epsilon^\prime_f}^2}\quad\times\\
\left(1 + \frac{\epsilon_L^2}{2\left({\epsilon^\prime_f}^2+1\right)^2} + \order{\epsilon_L^4}\right)
\end{multline*}}} &
$f^{f^\prime}_{0L} = -\frac{\epsilon_L}{1+{\epsilon^\prime_f}^2} - \frac{\left(2{\epsilon^\prime_f}^2
-1\right)\epsilon_L^3}{2\left({\epsilon^\prime_f}^2+1\right)^3} + \order{\epsilon_L^5}$\vs{8ex}
\\ \cline{2-2} & $f^{f^\prime}_{1L} = -1 +
\frac{\epsilon_L^2}{2\left(1+{\epsilon^\prime_f}^2\right)^2} +
\order{\epsilon_L^4}$
\vs{8ex}\\ \cline{2-2} & $f^{f^\prime}_{1R} = -\frac{1}{\sqrt{1+{\epsilon^\prime_f}^2}} 
-\frac{{\epsilon^\prime_f}^2\epsilon_L^2}{\left(1+{\epsilon^\prime_f}^2\right)^\frac{5}{2}}+
\order{\epsilon_L^4}$\vs{8ex}
\\ \cline{2-2} & $f^{f^\prime}_{2R} = -\frac{{\epsilon^\prime_f}}{\sqrt{1+{\epsilon^\prime_f}^2}} +
\frac{{\epsilon^\prime_f}\epsilon_L^2}{\left(1+{\epsilon^\prime_f}^2\right)^\frac{5}{2}}+
\order{\epsilon_L^4}$\vs{8ex} \\\hline
\end{tabular}}
\caption{Fermion masses and wavefunctions expanded in $\epsilon_L$ as obtained from diagonalizing
the fermion mass matrix (c.f. \eqref{equ-2-3-abbrev} for the abbreviations).}
\label{tab-3-1-mf}
\end{table}
The masses and wavefunctions\footnote%
{
The wavefunctions are to be read as e.g.
\[ {Z^\prime}^\mu = \sum_{n=0}^2 f_n B_n^\mu \]
}%
\ of the massive gauge bosons are given in tab.~\ref{tab-3-1-mgb} as an expansion in $x$. As expected,
the masses coincide at lowest order with the above result obtained by integrating out the bulk site.
The most noticeable effect of the corrections is to lift the degeneracy between the $W^\prime$ and
the $Z^\prime$.
Turning our attention to the wavefunctions, we find that the Standard Model $W^\pm$ and $Z$ bosons
are localized at the brane lattice sites with $f_1$ vanishing in the limit $x\rightarrow 0$, while
the ${W^\pm}^\prime$ and $Z^\prime$ are localized at the bulk site with $f_0$ and $f_2$ being of
order $\order{x}$.

The masses and wavefunctions resulting from diagonalizing the fermion mass matrix
\eqref{equ-3-1-mmf} are given in tab.~\ref{tab-3-1-mf} as an expansion in $\epsilon_L$. Again, in
the limit \eqref{equ-3-1-limit} the result from integrating out the bulk site is
recovered, with the corrections lifting the degeneracy between the heavy partner fermions.
As in the case of the gauge bosons, the Standard Model fermions are predominantly
localized at the brane sites, while their heavy partners reside mainly in the bulk with the modes
becoming fully localized in the limit \eqref{equ-3-1-limit}.

\begin{figure}[!tb]
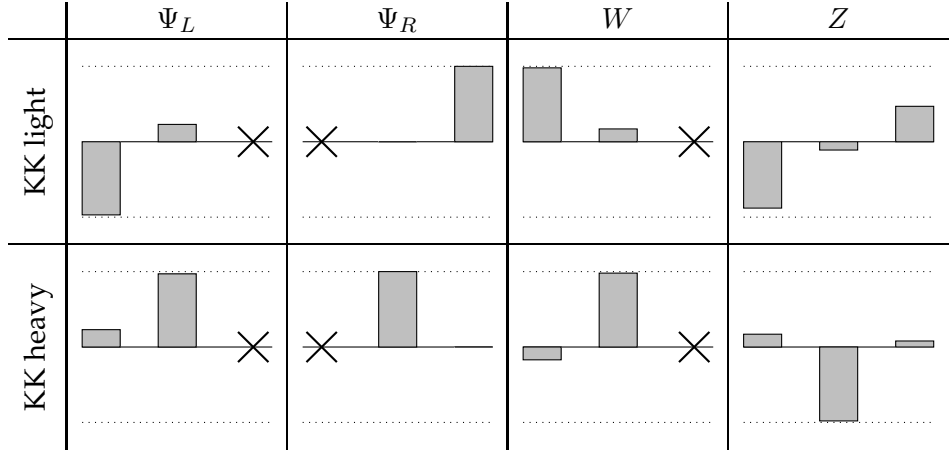

{
\newcommand{\boxhelper}[2]{
\begin{scope}[xshift=#1]
\ifthenelse{\equal{#2}{n}}{
\draw [thick] (-2mm, -2mm) -- (2mm, 2mm);
\draw [thick] (-2mm, 2mm) -- (2mm, -2mm);
}{
\draw [fill=gray!50,draw=black] (-2.5mm, 0mm) -- (-2.5mm, #2) -- (2.5mm, #2) -- (2.5mm, 0mm) -- cycle;
}
\end{scope}
}
\newcommand{\wfuncts}[3]{\parbox{25mm}{
\tikz{
\path (0mm, -10mm) -- (0mm, 10mm);
\draw (-12.5mm, 0mm) -- (12.5mm, 0mm);
\draw [dotted] (-12.5mm, 10mm) -- (12.5mm, 10mm);
\draw [dotted] (-12.5mm, -10mm) -- (12.5mm, -10mm);
\boxhelper{-10mm}{#1}
\boxhelper{0mm}{#2}
\boxhelper{10mm}{#3}
}}}
\tabcolsep2mm
\centerline{
\begin{tabular}{c|c|c|c|c}
& $\Psi_L$ & $\Psi_R$ & $W$ & $Z$ \\\hline
\parbox{2ex}{\rotatebox{90}{KK light}}\vs{15ex} &
\wfuncts{-0.97cm}{0.23cm}{n} & \wfuncts{n}{0cm}{1cm} & \wfuncts{0.98cm}{0.17cm}{n} &
\wfuncts{-0.88cm}{-0.11cm}{0.47cm} \\\hline
\parbox{2ex}{\rotatebox{90}{KK heavy}}\vs{15ex} &
\wfuncts{0.23cm}{0.97cm}{n} & \wfuncts{n}{1cm}{0} & \wfuncts{-0.17cm}{0.98cm}{n} &
\wfuncts{0.17cm}{-0.98cm}{0.08cm}
\end{tabular}
}}
\caption{The wavefunctions $f_n$ of $W$, $Z$ and massless left- and right-handed fermions as well as
those of their heavy KK partners for $m_{W^\prime}=\unit[500]{GeV}$,
$M_\text{bulk}=\unit[3.5]{TeV}$ and $\epsilon_L=0.237$ (ideal delocalization, see section \ref{chap-3-2} below),
the values are taken from app.~\ref{app-3}. The scale is linear, and the dotted lines correspond to $f_n=\pm1$.}
\label{fig-3-1-wfuncts}
\end{figure}
In order to illustrate the formulae in tab.~\ref{tab-3-1-mgb} and tab.~\ref{tab-3-1-mf},
fig.~\ref{fig-3-1-wfuncts} shows
the wavefunctions of $W$, $Z$ and massless left- and right-handed
fermions as well as those of their heavy KK partners, the values being taken from app.~\ref{app-3}.
These profiles are representative for the whole allowed region of parameters space and clearly
exhibit the (de)localization properties of the different modes as discussed above. The wavefunctions
of $t/t^\prime$ and $b/b^\prime$ differ slightly from those shown in fig.~\ref{fig-3-1-wfuncts} due the
comparatively large mass of the respective KK light modes, the most notable difference being a
slight delocalization of the
right-handed components which are (nearly) completely localized for the (almost) massless fermions and
their KK partners.

\section{Parameter space and constraints}
\label{chap-3-2}

While it indeed manages to delay the unitarity cutoff without invoking any scalar fields, the Three-Site
Higgsless Model is not an improvement over the Standard Model as far as the number of free
parameters is concerned. To fix the model and make predictions, all parameters present in the
Standard Model (except for the Higgs mass, obviously) must be specified in addition to a couple of
additional ones.

In the gauge sector we've got four free parameters: the three gauge
couplings $g$, $g^\prime$ and $\tilde{g}$ as well as the scale\footnote%
{
It is possible to choose two different scales $v_1$ and $v_2$ for the link fields $\Sigma_{1/2}$.
The choice $v_1 = v_2 = v$ maximizes the delay of unitarity violation \cite{He:2007ge},
however, this relation is not honored by loop corrections \cite{Abe:2008hb}.
} $v$. After fixing the $W$ and $Z$ masses as well as the electric charge, only
one degree of freedom remains in this sector, which can be chosen as $x$. From the expansions in
tab.~\ref{tab-3-1-mgb} it follows that to leading order in $x$
\begin{equation}\label{equ-3-2-x} \frac{m_W^2}{m_{W^\prime}^2} = \frac{x^2}{4} \end{equation}
Therefore, we can fix $x$ from the mass ratio of $W$ and $W^\prime$, and the limit $x\rightarrow 0$
for recovering the Standard Model is equivalent to $m_{W^\prime}\rightarrow\infty$.

Going over to the fermion sector we have two flavor-universal parameters $\epsilon_L$ and $\tilde{\lambda}$
and one set of parameters $\epsilon^\prime_f$ that are to be chosen separately for
each fermion flavor. The $\epsilon^\prime_f$ are fixed by the
masses of the Standard Model fermions, while $\tilde{\lambda}$ and $\epsilon_L$ remain as free
parameters. We choose to replace $\tilde{\lambda}$ by the heavy fermion mass scale $M_\text{bulk}$
\eqref{equ-3-1-defmb} which has a more intuitive physical meaning.

\subsubsection*{Electroweak precision observables at tree level: ideal delocalization}

In chapter \ref{chap-1-5}, the introduction of heavy fermions was suggested as a loophole for
evading the bounds coming from precision measurements at LEP and LEP-II. In these experiments,
the neutral- and charged current correlators were measured very precisely, revealing an astonishing
degree of agreement with the higher order predictions of the Standard Model and putting tight bounds
on any contributions from new physics.

Under the assumption that all contributions of new physics manifest themselves as modification to
the gauge boson self energies, the resulting constraints can be summarized as bounds on a small number
of parameters. The LEP bounds are commonly parameterized by the $\alpha S$, $\alpha T$ and $\alpha U$
parameters \cite{Peskin:1991sw} or equivalently by the three $\epsilon_{1/2/3}$
parameters \cite{Altarelli:1990zd} which correspond to bounds on the coefficients in an expansion of
the self-energies to order $p^2$.

The data obtained at LEP-II allows for additional
constraints on new physics contributions to the self-energies, corresponding to an expansion to
order $p^4$. Together with the LEP-I constraints, these bounds can be expressed
by a set of 7 parameters introduced in
\cite{Barbieri:2004qk} which also cover the bounds expressed by $\alpha S$, $\alpha T$ and $\alpha
U$.

While these parameters were designed to parameterize deviations from the Standard Model that appear
only as ``oblique corrections'' in the gauge boson self-energies, the authors of \cite{Chivukula:2004af}
show that non-oblique corrections appearing in Higgsless models can also be absorbed into the
extended set of parameters defined in \cite{Barbieri:2004qk}.

An generic analysis of the precision observables in a large class of deconstructed Higgsless models
which also contains the Three-Site Model was published in \cite{Chivukula:2005xm}. There, the
general form of the charged and neutral current scattering amplitudes in the vicinity of the $Z$ and
$W$ poles was derived and it was shown
that a suitable tuning of the fermion wavefunctions dubbed ``ideal delocalization''
can minimize the tree level corrections to the precision
observables. In particular, the $\Delta\rho$ parameter which measures the low energy difference between the
charged and neutral current coupling strengths vanishes, and the tree level contributions to
$\alpha S$ and $\alpha T$ reduce to
\begin{equation}\label{equ-3-2-st}
\alpha S = 4s_W^2c_W^2m_Z^2\sum_n\left(\frac{1}{m_{Z_n}^2} - \frac{1}{m_{W_n}^2}\right)
\quad,\quad
\alpha T = s_W^2m_Z^2\sum_n\left(\frac{1}{m_{Z_n}^2} - \frac{1}{m_{W_n}^2}\right)
\end{equation}
where the sums run over all KK gauge bosons and $c_W$ and $s_W$ are sine and cosine of the Weinberg
angle
\[ c_W = \cos\theta_W = \frac{m_W}{m_Z} \]
For the case of the Three-Site Model, plugging the
expansions of the masses tab.~\ref{tab-3-1-mgb} into \eqref{equ-3-2-st} and using
\eqref{equ-3-2-x} as well as
\begin{equation}\label{equ-3-2-c} c = \frac{m_W}{m_{Z}} + \order{x} \end{equation}
gives the naturally small results
\[
\alpha S = -4\frac{m_W^2}{m_Z^2}\frac{\left(m_Z^2 - m_W^2\right)^2}{m_{W^\prime}^4} +
\order{x^6}
\quad,\quad
\alpha T = -\frac{\left(m_Z^2 - m_W^2\right)^2}{m_{W^\prime}^4} + \order{x^6}
\]

The condition for ideal delocalization is the vanishing of the couplings of the light Standard
Model fermions to the heavy $W$ partners. In the Three-Site Model there is only one $W^\prime$, and the
corresponding coupling is given by the overlap of the wavefunctions
\[
g_{W^\prime ff} =
\frac{1}{\sqrt{2}}\left(gf^{W^\prime}_0f^{f_1}_{0L}\,f^{f_2}_{0L} +
\tilde{g}f^{W^\prime}_1f^{f_1}_{1L}\,f^{f_2}_{1L}\right)
\]
As $\epsilon^\prime_f$ vanishes for massless fermions and the other two parameters of the fermion
sector $M_\text{bulk}$ and $\epsilon_L$ are flavor-universal, all massless fermions have the same
wavefunctions $f^f$. Due the mass matrix being symmetric, the wavefunctions $f^W$ and $f^{W^\prime}$
are orthogonal and, therefore, choosing
\[
\cvect{\left(f^f_{0L}\right)^2\vs{4ex} \\ \left(f^f_{1L}\right)^2\vs{4ex}} \propto
\cvect{\frac{1}{g} f^W_0\vs{4ex} \\ \frac{1}{\tilde{g}} f^W_1\vs{4ex}}
\]
or equivalently
\begin{equation}\label{equ-3-2-ideloc}
\left(\frac{f^f_{1L}}{f^f_{0L}}\right)^2 = x\frac{f^W_1}{f^W_0}
\end{equation}
is sufficient to implement ideal delocalization. Examining the wavefunctions
given in tab.~\ref{tab-3-1-mgb} and tab.~\ref{tab-3-1-mf} reveals that \eqref{equ-3-2-ideloc}
encodes a relationship between $\epsilon_L$ and $x$
\begin{equation}\label{equ-3-2-elx} \epsilon_L^2 = \frac{x^2}{2} + \order{x^2} \end{equation}
which reduces the number of free parameters by one, allowing us to fix the Three-Site Model by
specifying just $m_{W^\prime}$ and $M_\text{bulk}$ in addition to the Standard Model parameters,
(in the case of ideal delocalization). Also, \eqref{equ-3-2-elx} automatically implies
$\epsilon_L\rightarrow 0$ in the limit $x\rightarrow 0$ as required by \eqref{equ-3-1-limit} in
order to recover the Standard Model.

\subsubsection*{Electroweak precision observables at one loop}

\begin{figure}[!tb]
\centerline{\includegraphics[width=\singleplotwidth,angle=270]{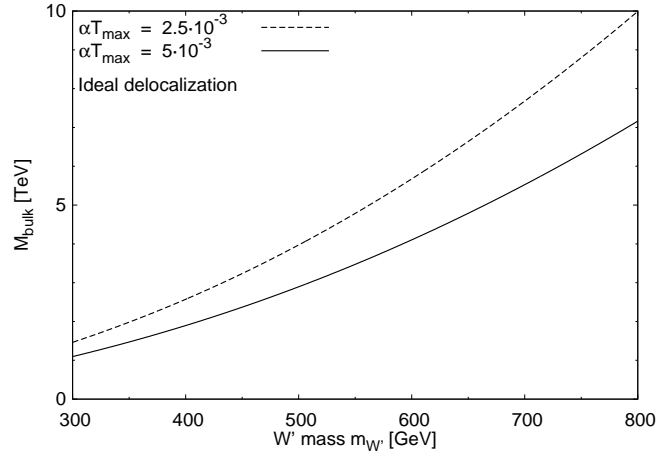}}
\caption{Lower bound on $M_\text{bulk}$ from estimating the $t^\prime$ / $b^\prime$
one loop contribution to $\alpha T$ in the ideally delocalized scenario.}
\label{fig-3-2-bound-mbulk}
\end{figure}
Due to the extremely efficient suppression of the tree level contributions to the precision
observables in the case of ideal delocalization, the loop contributions can be expected to be
dominant. Because the model contains heavy copies of the top and the bottom quarks, new contributions to
$\alpha T$ can arise from diagrams of the type
\[
\begin{fmfgraph*}(40,20)
\fmfleft{i}\fmfright{o}\fmf{wiggly,te=1.5}{i,v1}\fmf{wiggly,te=1.5}{o,v2}
\fmf{heavy,circle,left,la=$t^\prime$}{v1,v2}\fmf{heavy,circe,left,la=$b^\prime$}{v2,v1}
\fmfv{la=$W$}{i,o}\fmfdot{v1,v2}
\end{fmfgraph*}
\]
As $\epsilon_L$ honors isospin, the isospin violating effect of this diagram persists in the limit
$\epsilon_L\rightarrow 0$. The leading contribution to $\alpha T$ in this limit is estimated in
\cite{Chivukula:2006cg} as
\begin{equation}\label{equ-3-2-at}
\alpha T = \frac{1}{32\pi^2}\frac{M_\text{bulk}^2}{v^2}{\epsilon_{t}^\prime}^4
\end{equation}
From tab.~\ref{tab-3-1-mf} we can approximate $\epsilon^\prime_t$ as
\begin{equation}\label{equ-3-2-epspt}
\epsilon^\prime_t \approx \frac{m_t}{M_\text{bulk}\epsilon_L}
\end{equation}
Plugging this into \eqref{equ-3-2-at} with an upper bound $\alpha T_\text{max}$ on $\alpha T$ and using the
condition for ideal delocalization \eqref{equ-3-2-elx}, we obtain a
lower bound on $M_\text{bulk}$
\begin{equation}\label{equ-3-2-bound-mbulk}
M_\text{bulk}\ge \frac{1}{4\pi\sqrt{2\alpha T_\text{max}}}\frac{m_t^2}{v\epsilon_L^2} =
\frac{1}{8\pi\sqrt{2\alpha T_\text{max}}}\frac{m_t^2 m_{W^\prime}^2}{vm_W^2}
\end{equation}
This constraint is shown in fig.~\ref{fig-3-2-bound-mbulk} for two different values of $\alpha
T_\text{max}$
(the exact value depends on the ``reference Higgs boson mass'' chosen for the determination of
$\alpha T$, see \cite{Peskin:1991sw}).

A nice consequence of \eqref{equ-3-2-bound-mbulk} is the fact that it also induces an upper bound
on the $\epsilon^\prime_f$
\[ \epsilon^\prime_f \le 4\pi\sqrt{2\alpha T_\text{max}}\frac{m_fv\epsilon_L}{m_t^2} \]
Together with \eqref{equ-3-2-elx} and \eqref{equ-3-2-x},
this implies that any limit $m_{W^\prime}\rightarrow\infty,M_\text{bulk}\rightarrow\infty$ which
respects the precision constraints leads to $\epsilon^\prime_f\rightarrow0$ and therefore recovers
the Standard Model according to \eqref{equ-3-1-limit}\footnote%
{
Two comments are in order. First, if $M_\text{bulk}$ exceeds the UV cutoff, then the above estimate
of $\alpha T$ from loops with dynamical KK fermions is inconsistent. Instead, an analysis in the effective
theory obtained by integrating out the heavy fermions from an UV-completed version of the model should
be performed. This is carried out in \cite{Chivukula:2006cg} with a toy completion and the result is
shown to be unchanged.

Second, the case of ideal delocalization is excluded by the one-loop result for $\alpha S$ (see
below). However, $g_{W^\prime ff}$ has still to be very close to zero, so the above constraints on
$M_\text{bulk}$ and the $\epsilon^\prime_f$ are still realistic.
}.

\begin{figure}[!tb]
\centerline{\includegraphics[width=6cm]{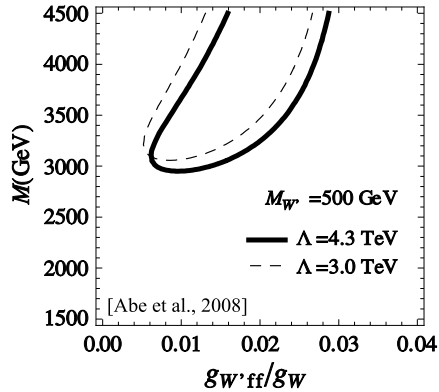}}
\caption{Region in the $M_\text{bulk}$ -- $g_{W^\prime ff}$ plane consistent with $\alpha S$ and
$\alpha T$ at one loop for $m_{W^\prime}=\unit[500]{GeV}$ and different values of the UV cutoff
$\Lambda$; $g_W$ is the isospin gauge coupling. Plot taken from \cite{Abe:2008hb}.}
\label{fig-3-2-abe}
\end{figure}

A complete one loop calculation of $\alpha S$ and $\alpha T$ was started in
\cite{Matsuzaki:2006wn,Sekhar_Chivukula:2007ic} and completed in \cite{Abe:2008hb}. The result is
quite interesting as, although a sizable piece of parameter space remains, the case of ideal
delocalization is excluded. As an example, fig.~\ref{fig-3-2-abe} shows the region in the
$M_\text{bulk}$ -- $g_{W^\prime ff}$ plane which is consistent with $\alpha S$ and $\alpha T$ at $95\%$
confidence level for $m_{W^\prime}=\unit[500]{GeV}$. The resulting $g_{W^\prime ff}$ coupling is
constrained to be smaller than some $2-3\%$ of the isospin gauge coupling, but still finite.

As the Three-Site Model must be treated as an
effective field theory, a logarithmic dependence on the UV cutoff scale $\Lambda$ remains in the result which
fig.~\ref{fig-3-2-abe} demonstrates to be very moderate. Increasing the $W^\prime$ mass moves
the allowed area to higher values of $M_\text{bulk}$ in agreement with the bound
\eqref{equ-3-2-bound-mbulk} obtained from $\alpha T$ for ideal
delocalization. At some value of the $W^\prime$ mass, the bound on $M_\text{bulk}$ crosses the UV
cutoff, indicating the breakdown of the theory that contains the bulk fermions as physical degrees
of freedom. While this is of little consequence to tree level calculations, the effective field
theory analysis performed in \cite{Abe:2008hb} breaks down at this point.

\subsubsection*{Triple gauge boson couplings}

After the discovery of the $W$ and $Z$, LEP-II was the first experiment capable of measuring the
triple gauge boson couplings predicted by nonabelian gauge theory. The results put bounds
on these couplings which differ between the Three-Site Model and the Standard Model.

In the absence of CP-violation, all\footnote%
{
The Hagiwara-Zeppenfeld parameterization excludes operators containing pieces like $\partial_\mu Z^\mu$
whose contributions are suppressed with the electron mass at LEP and which are therefore not
observable in the LEP / LEP-II data.
} possible dimension $4$ operators that generate couplings between two charged and one neutral gauge
bosons can be parameterized in Hagiwara-Zeppenfeld notation \cite{Hagiwara:1986vm}
\begin{multline}\label{equ-3-2-triplegb}
\LL_{3g} = -ie\left(1+\Delta\kappa_A\right)W^+_\mu W^-_\nu A^{\mu\nu}
- ie\left(1+\Delta g_1^A\right)\left(W^{+\mu\nu}W^-_\mu - W^{-\mu\nu}W^+_\mu\right)A_\nu
-\\ ie\frac{c_W}{s_W}\left(1+\Delta\kappa_Z\right)W^+_\mu W^-_\nu Z^{\mu\nu}
-ie\frac{c_W}{s_W}\left(1+\Delta g_1^Z\right)\left(W^{+\mu\nu}W^-_\mu - W^{-\mu\nu}W^+_\mu\right)Z_\nu
\end{multline}
In order to obtain the values of the couplings and the parameters $\Delta\kappa_{A/Z}$ and $\Delta g_1^{Z/A}$, the
mass eigenstates $W,Z$ and $A$ must be expressed through the gauge eigenstates situated at the
lattice sites. The normalization of the wavefunctions together with the photon wavefunction
\eqref{equ-3-1-defa} implies $\Delta\kappa_A=\Delta g_1^A =
0$ by electromagnetic gauge invariance. The remaining two parameters must be
equal and evaluate to
\[
e\frac{c_W}{s_W}\left(1+\Delta\kappa_Z\right) = e\frac{c_W}{s_W}\left(1+\Delta g_1^Z\right) =
gf_0^Z\left(f_0^W\right)^2 + \tilde{g} f_1^Z\left(f_1^W\right)^2
\]
Plugging in the wavefunctions and masses tab.~\ref{tab-3-1-mgb}, we obtain 
\[ \Delta g_1^Z = \Delta\kappa_Z = \frac{x^2}{8c^2} + \order{x^3} \]
Using \eqref{equ-3-2-x} and \eqref{equ-3-2-c} results in a lower bound on the $W^\prime$ mass
\[ m_{W^\prime} = \frac{1}{2}\frac{m_Z^2}{\Delta g_1^Z}
\]
From \cite{lepewwg}, the upper bound on $\Delta g_1^Z$ is $0.028$, and we obtain
\begin{equation}\label{equ-3-2-boundmwp} m_{W^\prime} \ge \unit[380]{GeV} \end{equation}
Note that, together with the above bounds on $M_\text{bulk}$, this implies a lower bound on
$M_\text{bulk}$\footnote%
{
An additional bound on $M_\text{bulk}$ arises from potential dangerous contributions to
$b\rightarrow s\gamma$ originating from the small but finite right-handed $Wtb$ coupling present in
the model. However, this constraint is weaker than \eqref{equ-3-2-bound-mbulk2}, see
\cite{Chivukula:2006cg} for details.
} which we can read off from fig.~\ref{fig-3-2-bound-mbulk} as
\begin{equation}\label{equ-3-2-bound-mbulk2}
M_\text{bulk} \ge \unit[1.8]{TeV} - \unit[2]{TeV}
\end{equation}
(depending on the exact bound on $\alpha T$ assumed, see above).

\section{Couplings}
\label{chap-3-3}

The calculation of couplings in the Three-Site Model follows the pattern already shown for
$g_{W^\prime ff}$ and $g_{WWZ}$ in the last section: the couplings are obtained from reexpressing
the mass eigenstates (which are delocalized over the whole lattice) in terms of the gauge
eigenstates localized at each lattice site, e.g. for couplings of arity three
\[ g_{XYZ} = \sum_n g_n f^X_n f^Y_n f^Z_n \]
In the language of deconstruction, these sums are the deconstructed version of the overlap integrals
which give the couplings in the continuous case (c.f. \eqref{equ-2-1-4dcpl})
\[ g_{XYZ} = \int dy\; g(y)f_X(y)f_Y(y)f_Z(y) \]
(allowing for a possible dependence of the 5D couplings on $y$).

\subsubsection*{Gauge sector}

The only free parameter in the gauge sector of the Three-Site Model not present in the Standard
Model is $x$ and, therefore, all gauge sector couplings must be independent of $M_\text{bulk}$ (and
$\epsilon_L$ in the the case of nonideal delocalization).

As shown at the beginning of the previous section, the Three-Site Model is identical to the Standard
Model without a Higgs in the limit \eqref{equ-3-1-limit}. This implies that couplings
between the Standard Model gauge bosons match those in the Standard Model up to corrections of order
$\order{x^2}$ or higher\footnote%
{
As the sign of the gauge couplings is purely conventional and can be changed by redefining the gauge
field, all couplings have to be either even or odd in $x$ which is the reason why the corrections to
couplings between Standard Model particles start at $\order{x^2}$.
}, where $x$ is constrained to be smaller than $\approx 0.42$ by
\eqref{equ-3-2-boundmwp} and \eqref{equ-3-2-x}.

By the same reasoning, the couplings $g$ and $g^\prime$ at the lattice sites 0 and 2 are equal to
the isospin and hypercharge gauge couplings up to corrections of order $\order{x^2}$.
The coupling $\tilde{g}$ at the bulk site can be written as
\[ \tilde{g} = \frac{g}{x} \approx \frac{g}{2}\frac{m_{W^\prime}}{m_W} \]
and exhibits a linear growth with $m_W$. We can't reasonably expect the delay of
unitarity to work out of the $W^\prime$ is heavier than $\order{\unit[1]{TeV}}$ and, therefore
\[ \tilde{g}\le\left.\tilde{g}\vs{2ex}\right|_{m_{W^\prime}=\unit[1.5]{TeV}} \approx 6.2 < 4\pi \]
This means that the gauge coupling of the $\sun{2}$ at site 1 is moderately strong with radiative corrections
being roughly of order $\frac{\tilde{g}^2}{16\pi^2}\approx 25\%$ in the worst case.

\begin{figure}[!tb]
\centerline{\includegraphics[width=\singleplotwidth,angle=270]{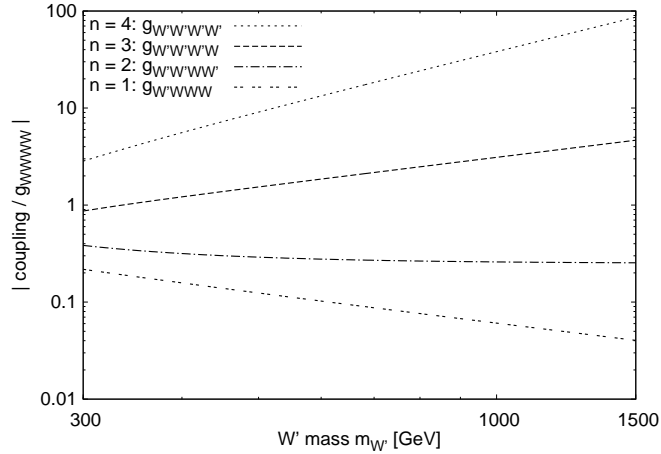}}
\caption{The four-point coupling between different combinations of $W$ and $W^\prime$, double
logarithmic plot to emphasize the leading order dependence on $x$.}
\label{fig-3-3-gwwww}
\end{figure}
In the limit $x\rightarrow 0$, the Standard Model gauge bosons are localized on the
branes with the $f_1$ being of order $x$, while the $W^\prime$ and $Z^\prime$
are localized in the bulk with $f_{0/2}$ of $\order{x}$ (c.f. tab.~\ref{tab-3-1-mgb}).
Naively, one would therefore expect that all couplings involving different KK modes are suppressed by
powers of $x$. However, this is not true because the coupling at the bulk lattice site is
proportional to $x^{-1}$. Taking this into account and
recalling the definition of $x$ \eqref{equ-2-3-abbrev}, it is not difficult to see that
a coupling between gauge bosons involving $n$ $W^\prime$ or $Z^\prime$ (for $n > 0$)
must be of order
$\order{x^{2-n}}$ relative to the corresponding Standard Model coupling. This behavior is
demonstrated in fig.~\ref{fig-3-3-gwwww}.

The only exception from this rule are couplings involving only $W$, $W^\prime$
and photons. In this case, orthonormality of $f^W$ and $f^{W^\prime}$ together with the photon
wavefunction \eqref{equ-3-1-defa} implies that all couplings apart from
\[
g_{WW\gamma} = g_{W^\prime W^\prime\gamma} = e \quad,\quad
g_{WW\gamma\gamma} = g_{W^\prime W^\prime\gamma\gamma} = e^2
\]
vanish.

As QCD is not touched by the additional structure in the Three-Site Model, the couplings between the
gluons are unchanged with respect to the Standard Model.

\subsubsection*{Fermion sector: ideal delocalization}

The fermion sector is more complicated as it depends on $x$ as well as on $M_\text{bulk}$. However,
the Standard Model must be recovered in the limit \eqref{equ-3-1-limit}, so all couplings involving
only KK light particles must equal the respective Standard Model values up to corrections of order
$\order{x^2}$ and $\order{{\epsilon_f^\prime}^2}$ (this also covers the dependence on $\epsilon_L$
as it is proportional to $x$ by ideal delocalization).

In the ideally delocalized scenario, $x$ enters the wavefunctions of the fermions through their dependence
on $\epsilon_L$, while $M_\text{bulk}$ enters through $\epsilon^\prime_f$. For a massless fermion we
have $\epsilon_f^\prime=0$ and therefore, the wavefunctions and couplings of massless fermions only
depend on $x$. In addition, the right-handed zero modes of the massless fermions have to completely
decouple from their left-handed counterparts and therefore are completely localized at site $2$, while
the KK-heavy modes are completely localized in the bulk in this case.

\begin{figure}[!tb]
\centerline{
\includegraphics[width=\doubledplotwidth,angle=270]{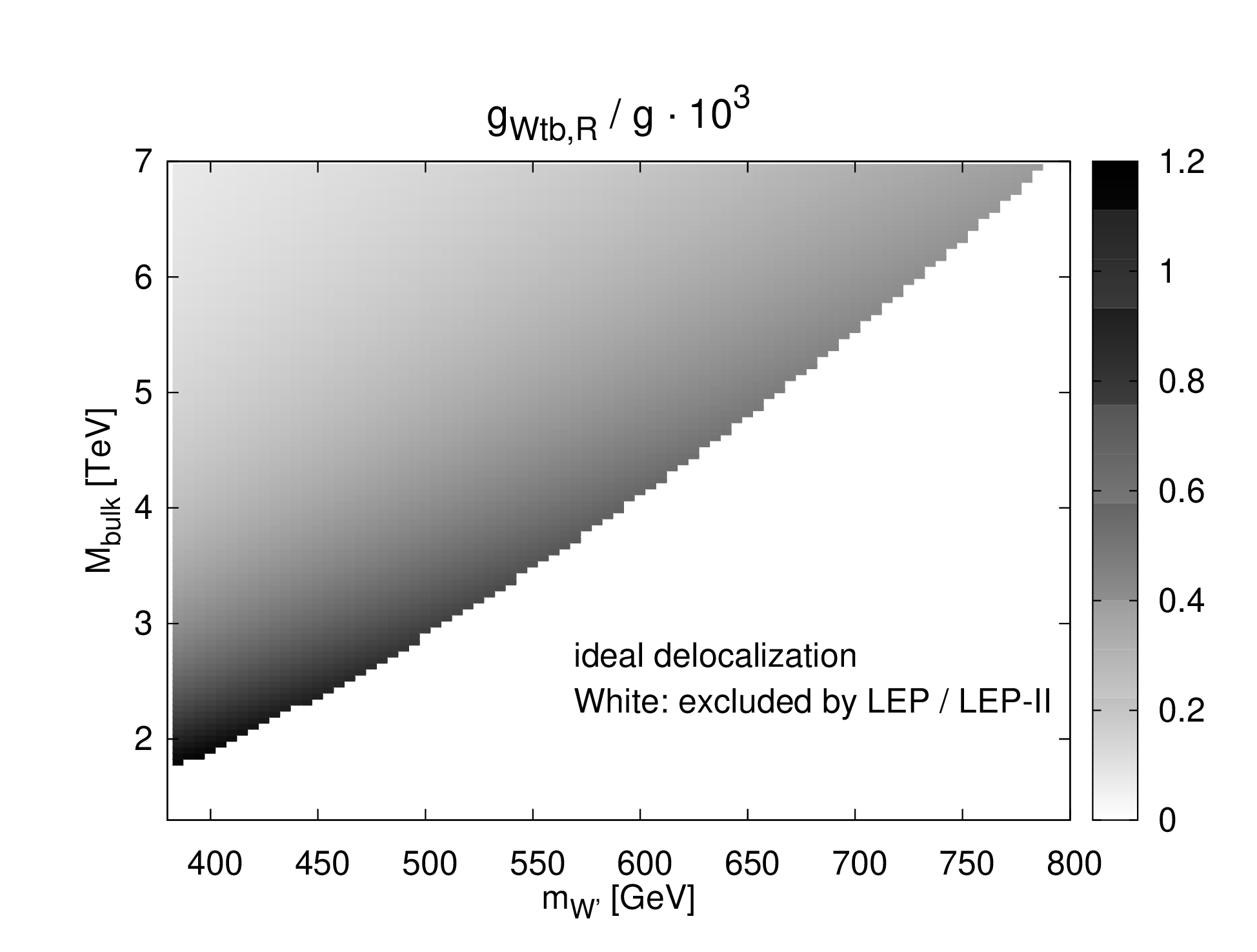}
\includegraphics[width=\doubledplotwidth,angle=270]{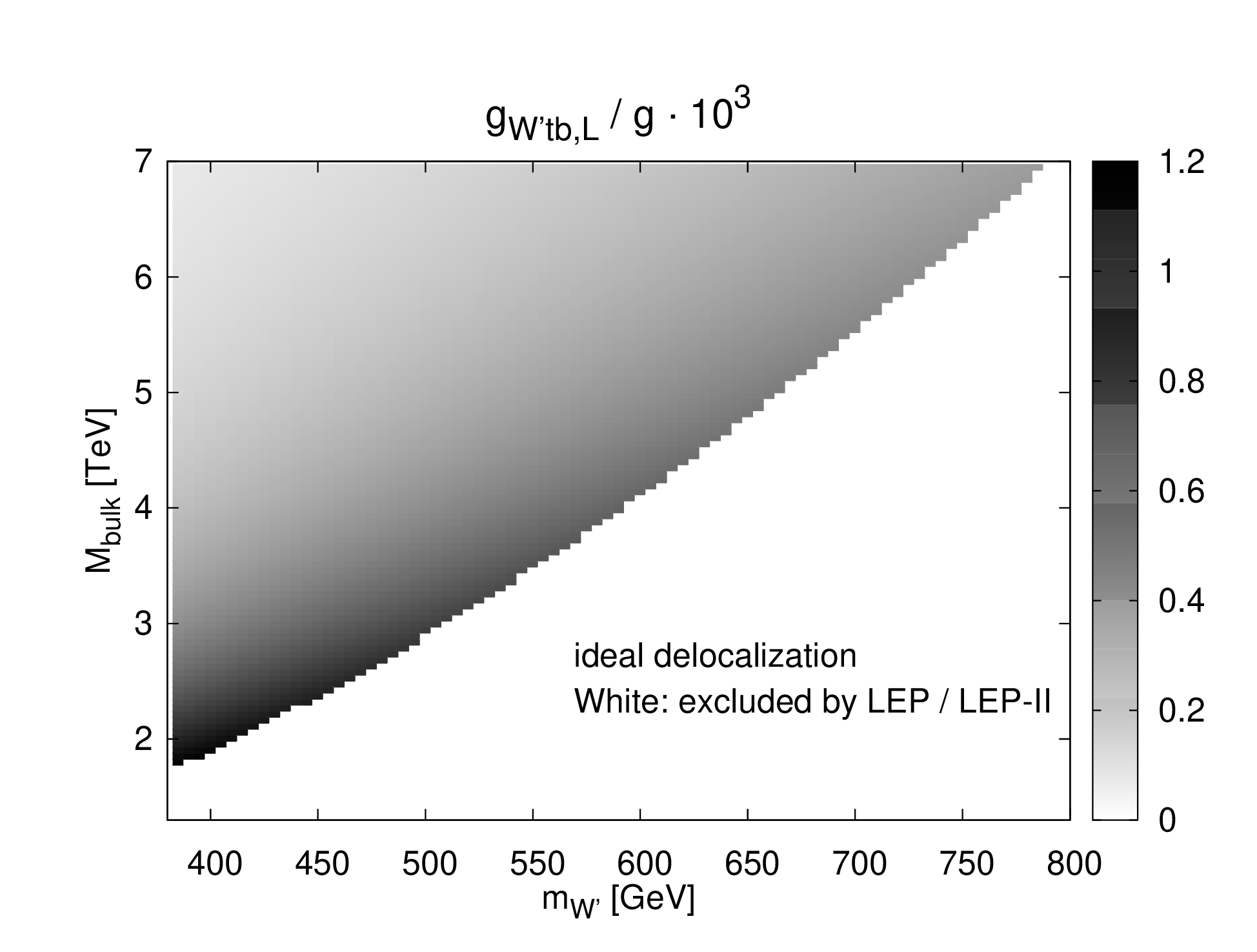}}
\caption{
\emph{Left:} Right-handed $Wtb$ coupling relative to the isospin gauge coupling $g$
(forbidden in the Standard Model)
in the Three-Site Model in as a function of the $m_{W^\prime}$ and $M_\text{bulk}$.\newline
\emph{Right:} Similar plot of the left-handed $W^\prime tb$ coupling which vanishes for massless
fermions due to ideal delocalization.}
\label{fig-3-3-wtbR}
\end{figure}
Consulting tab.~\ref{tab-3-1-mf} shows that the massless KK-light fermions are localized at the branes with
$f^f_1$ suppressed like $\order{\epsilon_L}$ while their KK-heavy partners
are localized in the bulk with $f^f_{0/2}$ similarly suppressed. This implies that the
counting rule derived for the gauge sector also holds for $W$ coupling to massless fermions or their
KK-partners. In the case of massive fermions, $\epsilon_f^\prime$ goes to zero as $M_\text{bulk}$
goes to infinity, recovering the same pattern in this limit and yielding corrections of order
$\epsilon_f^\prime$ for finite $M_\text{bulk}$.

The left-handed couplings between a
$W^\prime$ and two massless fermions are not only suppressed by $\order{x^2}$ but fully vanish
as required by ideal delocalization due to a cancellation between the contributions from different
lattice sites. In the case of massive Standard Model fermions, the cancellation
is not complete, but the resulting coupling is still very small as demonstrated in fig.~\ref{fig-3-3-wtbR}
right by the example of $W^\prime tb$.

As the right-handed KK-light modes of the massive fermions and the $W$ have
a nonvanishing overlap in the bulk, a small right-handed coupling between the $W$ and the massive Standard Model
fermions suppressed like $\order{{\epsilon_f^\prime}^2}$ is present in the Three-Site Model. This
coupling is showcased in fig.~\ref{fig-3-3-wtbR} left for $Wtb$.

The couplings to the $Z$ are more complicated due to the nontrivial charge of the fermions at site
0 and 1 under the $\mathbf{U}(1)$ at site 2. Taking this into account, the left- and right-handed
couplings read
\begin{equation}\label{equ-3-3-gzf}\begin{aligned}
g_{Zf_1f_2,L} =& T_3\left(g\,f^{f_1}_{0L}\,f^{f_2}_{0L}\,f^Z_0 +
\tilde{g}\,f^{f_1}_{1L}\,f^{f_2}_{1L}\,f^Z_1\right) -
Yg^\prime f^Z_2\left(\sum_{n=0}^1 f^{f_1}_{nL}\,f^{f_2}_{nL}\right)
\\
g_{Zf_2f_2,R} =& T_3\tilde{g}\,f^{f_1}_{1R}\,f^{f_2}_{1R}\,f^Z_1 +
Qg^\prime\,f^{f_1}_{2R}\,f^{f_2}_{2R}\,f^Z_2 +
Yg^\prime\,f^Z_2\,f^{f_1}_{1R}\,f^{f_2}_{1R}
\end{aligned}\end{equation}
The first two terms of both couplings come from the wave function overlap and directly fall under the same
counting rule as above in the $\epsilon^\prime_f=0$ case.
The third term comes from the part of the coupling to $B_2^\mu$
which is nonlocal on the lattice and could potentially violate our counting rule.
To see that this is not the case, note that the contributions of these
terms are either of order $\order{x}$
or $\order{1}$, and the latter only if the Standard Model $Z$ is involved. However, in both cases,
the only combination of KK modes that could give a $\order{1}$ contribution is $Zf^\prime f^\prime$ for
which the counting rule also gives $\order{1}$ and therefore, the leading order in $x$ as given
by the rule is unchanged for all possible combinations of modes at the vertex.

\begin{figure}[!tb]
\centerline{\includegraphics[width=\singleplotwidth,angle=270]{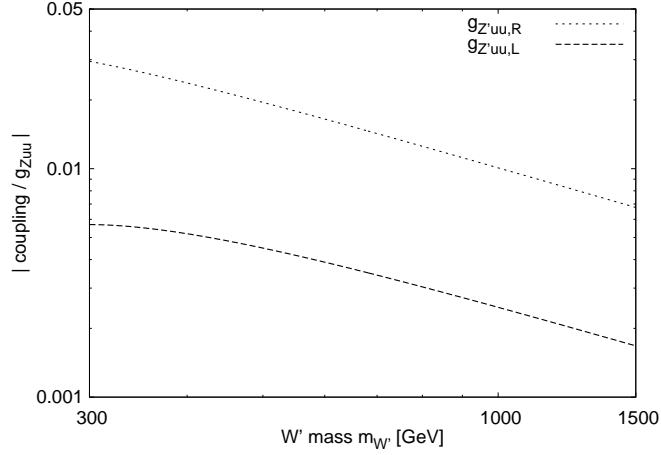}}
\caption{The left- and right-handed $Z^\prime uu$ couplings relative to their $Zuu$ counterpart as a
function of the $W^\prime$ mass $m_{W^\prime}$. Double logarithmic plot.}
\label{fig-3-3-zuu}
\end{figure}
The submatrix of the neutral gauge boson mass matrix $M^{NC}$ (c.f. \eqref{equ-3-1-mmgb}) which links
the gauge fields at sites $0$ and $1$ is identical to the charged gauge boson mass matrix $M^{CC}$.
Therefore, we can expect the wave function of the $Z^\prime$ at these lattice sites to be
approximately proportional to that of the $W^\prime$ (a conjecture which can be confirmed by a look
at the actual wave functions in tab.~\ref{tab-3-1-mgb}). Thus, the cancellation leading to ideal
delocalization also suppresses the left-handed couplings between the $Z^\prime$ and the Standard
Model fermions. This is confirmed by fig.~\ref{fig-3-3-zuu} which shows the right- and left-handed
couplings of $Z^\prime uu$ relative to their $Zuu$ counterparts. While the right-handed coupling is
of order $x$ as expected if there is no special relation between the wave functions at different
lattice sites, the left-handed coupling is suppressed by roughly another order of magnitude by the incomplete
cancellation.

Putting the pieces together, in the limit $\epsilon_f^\prime\rightarrow 0$ (either by vanishing
Standard Model mass or large $M_\text{bulk}$), the couplings of the KK-light fermions to $W$ and
$Z$ are equal to their Standard Model model values up to corrections of $\order{x^2}$, while
couplings involving $n$ ($n\ge 1$) KK-heavy particles go like $\order{x^{2-n}}$\footnote%
{
As a slight exception to this rule, several couplings to right-handed fermions vanish completely due to the exact
localization of the right-handed components of the massless fermions and their KK partners.
}.
As far as the left-handed couplings to the Standard Fermions are concerned, the $W^\prime$ is completely
fermiophobic and the $Z^\prime$ at least close to it.

Due to electromagnetic gauge invariance, the photon doesn't couple to fermions of different mass.
The couplings of a photon to two Standard Model fermions (or both the KK partners) are equal to the
Standard Model couplings. Similarly, the gluons either couple to two KK-light or two KK-heavy
quarks with the couplings borrowed from the Standard Model.

\subsubsection*{Nonideal delocalization}

In \cite{Abe:2008hb} it is shown that, in contrast to the ideally delocalized scenario, a nonvanishing
coupling between the $W^\prime$ and the light fermions is necessary for
compatibility with the precision observables at the one loop level. This coupling $g_{W^\prime ff}$
for which the bounds are derived is defined in the low-energy effective theory obtained by
integrating out the heavy fermions and renormalized at the $W^\prime$ mass scale.

There are two
operators contributing to $g_{W^\prime ff}$ in the one loop analysis. The first one
\[ O_1 = \overline{\Psi}_{0L}\Sigma_1\slashed{A}_1\Sigma_1^\dagger\Psi_{0L} \]
arises from integrating out the bulk fermions from diagrams of the type
\[
\fmfframe(5,2)(5,2){\begin{fmfgraph*}(45,23)
\fmfleft{i2,i1}\fmfright{o2,o1}\fmftop{t1}
\fmf{fermion}{i2,v1}\fmf{scalar}{i1,v1}\fmf{heavy,la=$\Psi_{1R}$,la.si=right}{v1,v2}
\fmf{heavy,la=$\Psi_{1R}$,la.si=right}{v2,v3}
\fmf{scalar}{v3,o1}\fmf{fermion}{v3,o2}\fmffreeze
\fmf{dbl_wiggly}{v2,t1}
\fmfv{la=$\Sigma_1$,la.an=180}{i1}\fmfv{la=$\Psi_{0L}$,la.an=180}{i2}
\fmfv{la=$\Sigma_1$,la.an=0}{o1}\fmfv{la=$\Psi_{0L}$,la.an=0}{o2}
\fmfv{la=$A_1^\mu$,la.an=0}{t1}\fmfdot{v1,v2,v3}
\end{fmfgraph*}}
\]
and corresponds to the Yukawa coupling $\epsilon_L$ in the full model which includes the heavy
fermions. The second operator arises from loop corrections
\[
O_2 = \overline{\Psi}_{0L}\left(D_\nu\left(\Sigma_1 F_1^{\mu\nu}\Sigma_1^\dagger\right)
\right)\gamma_\mu\Psi_{0L}
\]
Although this operator also encodes a coupling between the left-handed Standard Model fermions
and the $W^\prime$, it has a nontrivial momentum structure.
However, a nonlinear field redefinition in the spirit of on-shell effective field
theory \cite{Georgi:1991ch} can be used to apply the equation of motion to $A_1$ and convert $O_2$ to the same
form as $O_1$ at the price of introducing additional higher dimension operators which are
suppressed by a power of the gauge couplings and two powers of $v$. After the
redefinition, $O_2$ also gives a contribution to $g_{W^\prime ff}$ which is accounted for
by $\epsilon_L$ in the full theory.
Therefore, at tree level, the finite $g_{W^\prime ff}$ can be accommodated in the full theory
by tuning $\epsilon_L$ away from the value required by ideal delocalization.

From the fermion and gauge boson wavefunctions tab.~\ref{tab-3-1-mf} and tab.~\ref{tab-3-1-mgb} we
can estimate the value of $\epsilon_L$ required to generate a given $g_{W^\prime ff}$
\[ \epsilon_L \approx \frac{g_{W^\prime ff}}{\tilde{g}} +\frac{x^2}{2} \]
Using the condition for ideal delocalization \eqref{equ-3-2-elx}, we can solve for the relative
change in $\epsilon_L$
\[ \epsilon_L \approx \epsilon_{L,0}\left(1+x\frac{g_{W^\prime ff}}{g}\right) \]
with $\epsilon_{L,0}$ being the value required for ideal delocalization.
According to \cite{Abe:2008hb}, $g_{W^\prime ff}$ is constrained to be smaller than some $2-3\%$ of $g$,
and we find that the required change to $\epsilon_L$ is of the same order.

\begin{figure}[!tb]
\centerline{\includegraphics[width=\singleplotwidth,angle=270]{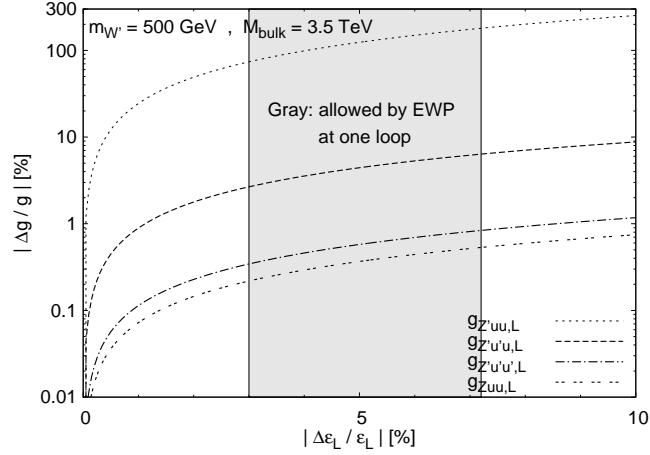}}
\caption{Examples of the change induced in the left-handed $Zuu$ type couplings by  tuning
$\epsilon_L$ always from ideal delocalization.}
\label{fig-3-3-nideloc}
\end{figure}
At leading order, the wavefunctions of the right-handed fermions are independent of $\epsilon_L$ (c.f.
tab.~\ref{tab-3-1-mf}), and therefore, only the left-handed couplings are sensitive to the departure
from ideal delocalization.
Naively, we would expect the resulting relative change in the left-handed couplings to be of the
same order of magnitude as the relative change in $\epsilon_L$. However, this is not true for
the couplings of the Standard Model fermions to the
$W^\prime$ and the $Z^\prime$.

For these couplings, the change of $\epsilon_L$ disrupts the
cancellation among the contributions from different lattice sites, leading to a potentially much
bigger change. This is demonstrated in fig.~\ref{fig-3-3-nideloc} which shows the change induced in
several of the left-handed $Zuu$ type couplings. All couplings change within a few percent or
even significantly less, with the exception of the $Z^\prime uu$ coupling which changes by
nearly $300\%$ when $\epsilon_L$ is changed by $10\%$! However, although the relative change is
sizable, the coupling still remains nearly two orders of magnitude smaller than the isospin gauge
coupling and therefore close to fermiophobic. As the corresponding right-handed couplings to the
$Z^\prime$ are considerably larger and not affected by nonideal delocalization, we shouldn't
expect a big change in most physical observables.

For the $W^\prime$ couplings to Standard Model fermions, the situation is different: although
the coupling is still virtually fermiophobic, the change induced by nonideal delocalization opens up
the possibility of producing this resonance from fermions which is impossible in the case of the
ideally delocalized scenario (c.f. chapter \ref{chap-7}).

The gauge sector as well as the couplings to photons and gluons are not affected by nonideal
delocalization.

\subsubsection*{Goldstone bosons}

From expanding the Wilson lines $\Sigma_{1/2}$ in terms of the component fields and inserting the
expansion into the Three-Site Lagrangian, we would obtain infinitely many couplings involving
arbitrarily many Goldstone bosons. However, choosing unitarity gauge completely removes the Goldstone
bosons from the spectrum, which is what we will be doing for the remainder of this work.

\section{Widths and decay channels}

In order to regularize the mass poles appearing in transition amplitudes, we have to calculate the
widths of the new heavy resonances. In addition, some knowledge on the decay channels and branching
ratios is
required in order to identify suitable final states for the discovery of these particles. In this
section we will only state the results of the calculation, the details of which can be found in
appendix \ref{app-4}.

\subsubsection*{Heavy gauge bosons}

\begin{figure}[!tb]
\centerline{
\includegraphics[width=\doubleplotwidth,angle=270]{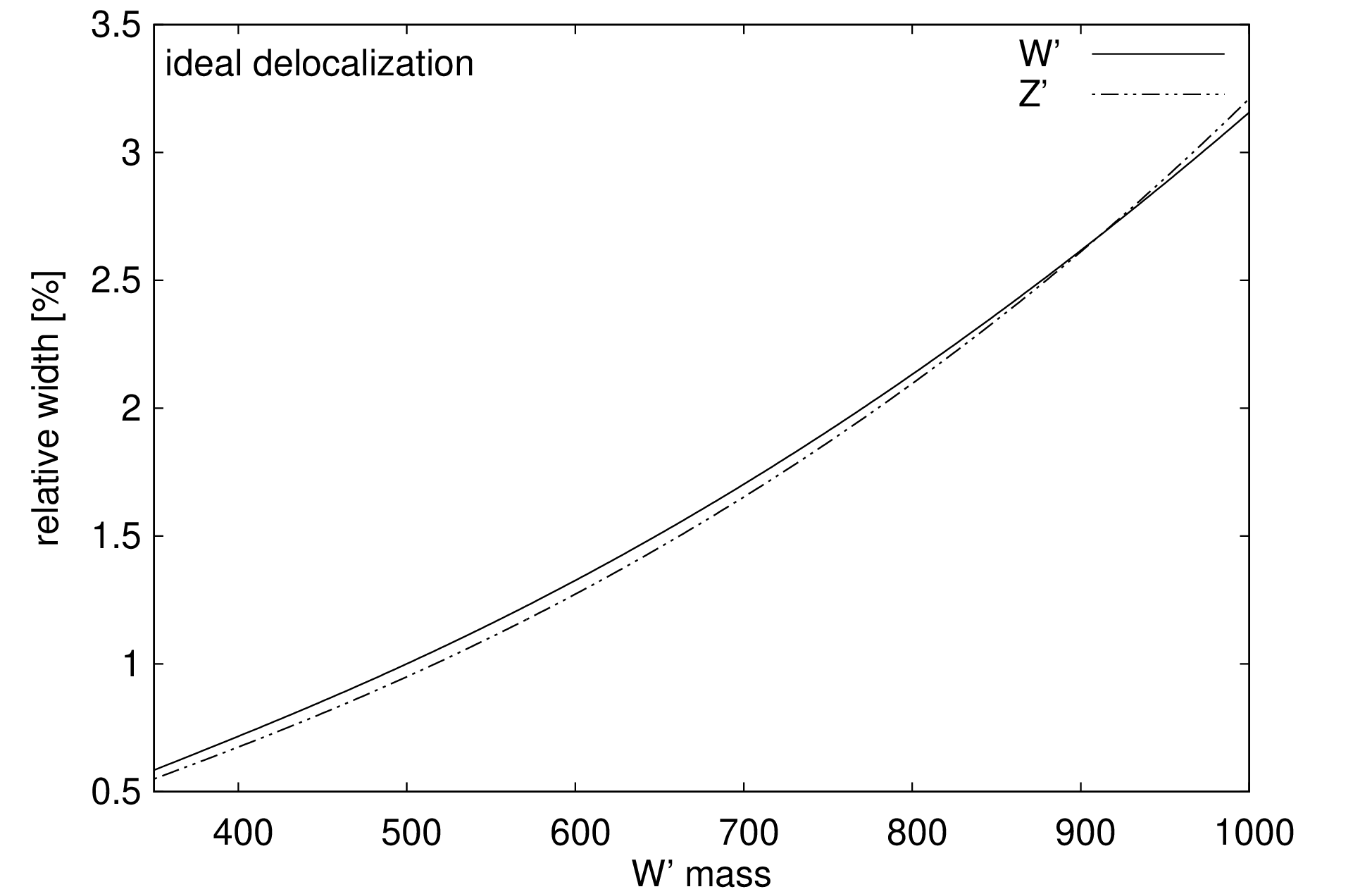}
\quad
\includegraphics[width=\doubleplotwidth,angle=270]{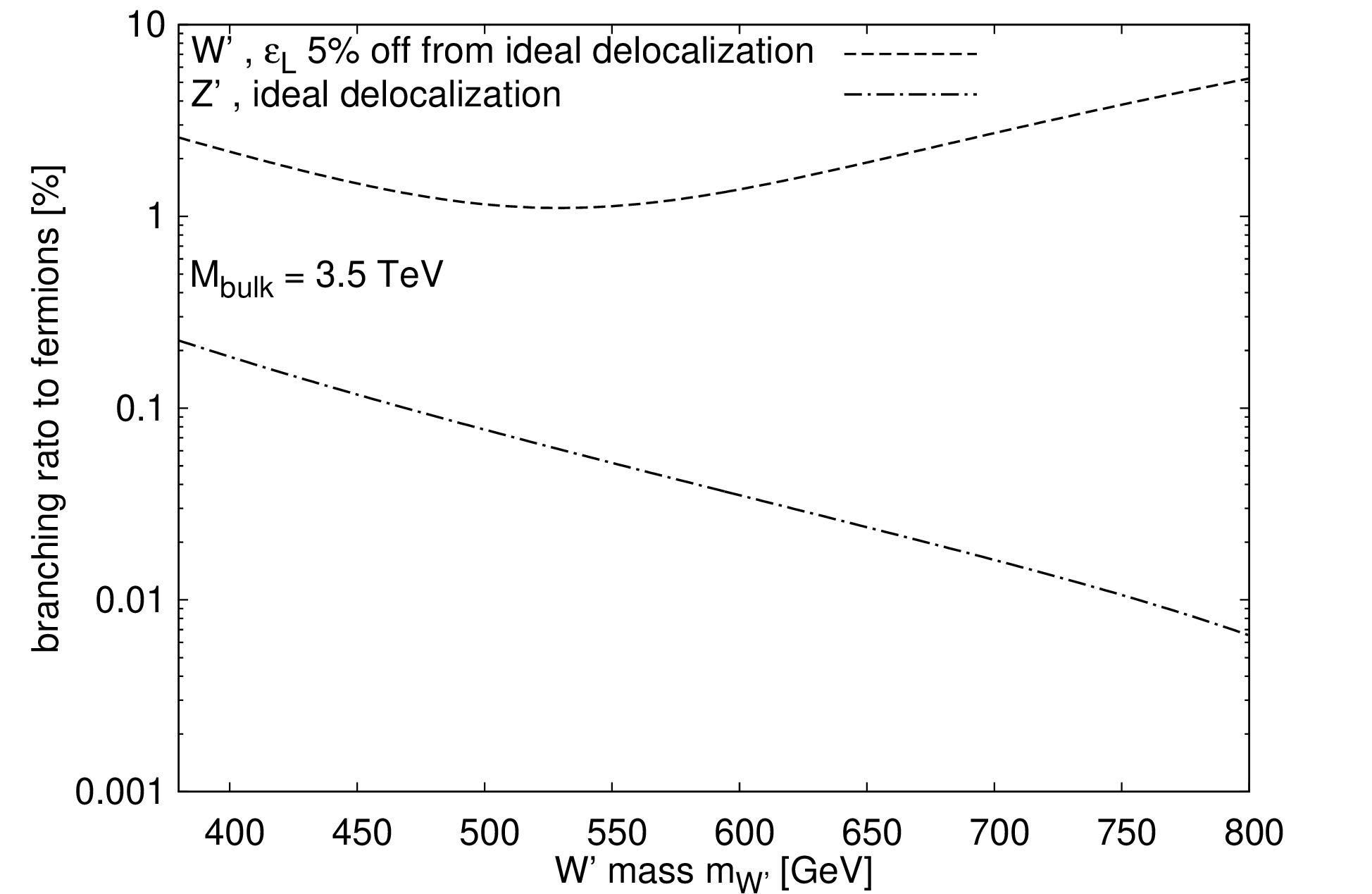}
}
\caption{
\emph{Left:} Relative widths of $W^\prime$ and $Z^\prime$ as a function of $m_{W^\prime}$.
\newline
\emph{Right:} Branching ratio of the heavy gauge bosons going to fermions as a function of
$m_{W^\prime}$.
}
\label{fig-3-4-widthhgb}
\end{figure}
At leading order, the $W^\prime$ and $Z^\prime$ can either decay into two Standard Model gauge
bosons or two Standard Model fermions fermions
\[
\fmfframe(2,2)(2,2){\begin{fmfgraph}(20,15)
\fmfleft{i}\fmfright{o2,o1}\fmf{dbl_wiggly}{i,v}\fmf{wiggly}{o1,v,o2}\fmfdot{v}
\end{fmfgraph}
\qquad\qquad\qquad
\begin{fmfgraph}(20,15)
\fmfleft{i}\fmfright{o2,o1}\fmf{dbl_wiggly}{i,v}\fmf{fermion}{o1,v,o2}\fmfdot{v}
\end{fmfgraph}}
\]
All other potential $1\rightarrow 2$ decay channels are forbidden due to the near-degeneracy of
$W^\prime$ and $Z^\prime$
\[ \frac{m_{W^\prime}}{m_{Z^\prime}} = 1 + \order{x^2} \]
and the bound on the heavy fermion mass scale \eqref{equ-3-2-bound-mbulk}. The resulting relative
widths are shown in fig.~\ref{fig-3-4-widthhgb} left to be very similar for $W^\prime$ and $Z^\prime$ and of
the order of $1-3\%$.

Compared to the fermionic decay channel, the bosonic decay channel receives an enhancement factor
\[  \frac{m_{Z^\prime}^4}{16m_W^2m_Z^2} \]
which is of order $100$ and
which comes from the decay into longitudinal modes. Therefore, although there are more final states
and phasespace available for the fermionic channel, the bosonic final state is highly favored. This
is demonstrated in fig.~\ref{fig-3-4-widthhgb} right which shows the fermionic branching ratio of
$Z^\prime$ and $W^\prime$ as a function of $m_{W^\prime}$ (in the case of the latter, $\epsilon_L$
is tuned away from ideal delocalization by $5\%$). Evidently, the $W^\prime$ decays to more than
$95\%$ into gauge bosons, while the fermionic branching ratio of $Z^\prime$ is even lower than $1\%$.

\subsubsection*{Heavy fermions}

For the heavy fermions there are two two possible decay channels at leading order:
\[
\fmfframe(2,2)(2,2){\begin{fmfgraph}(20,15)
\fmfleft{i}\fmfright{o2,o1}\fmf{heavy}{i,v}\fmf{fermion}{v,o2}\fmf{wiggly}{v,o1}\fmfdot{v}
\end{fmfgraph}
\qquad\qquad\qquad
\begin{fmfgraph}(20,15)
\fmfleft{i}\fmfright{o2,o1}\fmf{heavy}{i,v}\fmf{fermion}{v,o2}\fmf{dbl_wiggly}{v,o1}\fmfdot{v}
\end{fmfgraph}}
\]
The decay into the heavy isospin partners is forbidden due to the near-degeneracy of the heavy
fermions. The $\order{\alpha_s}$ QCD contributions to the decays of the heavy quarks can be expected to be of
order of several percent and may therefore be worthwhile to include. These corrections can be written
diagrammatically as
\begin{multline}\label{equ-3-4-nlo}
\Delta\abs{\MM}^2=2\Re\left(\parbox{20mm}{
\begin{fmfgraph}(20,15)
\fmfset{curly_len}{2mm}
\fmfleft{i1}\fmfright{o2,o1}\fmf{heavy}{i1,v}\fmf{fermion}{v,o2}\fmf{wiggly}{v,o1}\fmfdot{v}
\end{fmfgraph}}\right)\quad\times\\
\left(\parbox{20mm}{
\begin{fmfgraph}(20,15)
\fmfset{curly_len}{2mm}
\fmfleft{i1}\fmfright{o2,o1}\fmf{heavy}{i1,v}\fmf{fermion}{v,o2}\fmf{wiggly}{v,o1}\fmfdot{v}
\fmffreeze\fmf{phantom}{i1,v1,v,v2,o2}\fmffreeze
\fmf{gluon,right,te=0.5}{v1,v2}
\end{fmfgraph}} +\;
\parbox{20mm}{\begin{fmfgraph}(20,15)
\fmfset{curly_len}{2mm}
\fmfleft{i1}\fmfright{o2,o1}\fmf{heavy}{i1,v}\fmf{fermion}{v,o2}\fmfdot{v}\fmf{wiggly}{v,o1}
\fmffreeze\fmf{phantom}{i1,v1,v2,v}\fmffreeze
\fmf{gluon,right}{v1,v2}
\end{fmfgraph}} +\;
\parbox{20mm}{\begin{fmfgraph}(20,15)
\fmfset{curly_len}{2mm}
\fmfleft{i1}\fmfright{o2,o1}\fmf{heavy,te=2}{i1,v}\fmf{fermion}{v,o2}\fmf{wiggly}{v,o1}\fmfdot{v}
\fmffreeze\fmf{phantom}{v,v1,v2,o2}\fmffreeze
\fmf{gluon,right}{v1,v2}
\end{fmfgraph}}
\right)^*+\quad\\
\left|
\parbox{20mm }{\begin{fmfgraph}(20,15)
\fmfset{curly_len}{2mm}
\fmfleft{i1}\fmfright{o4,o3,o2,o1}\fmf{heavy}{i1,v}\fmf{fermion}{v,o4}\fmf{wiggly}{v,o1}
\fmffreeze\fmf{phantom}{v,v1,o4}\fmffreeze\fmf{gluon}{v1,o2}\fmfdot{v}
\end{fmfgraph}}+
\parbox{20mm }{\begin{fmfgraph}(20,15)
\fmfset{curly_len}{2mm}
\fmftop{t4,t3,t2,t1}\fmfleft{i1}\fmfright{o2,o1}\fmf{heavy}{i1,v}\fmf{fermion}{v,o2}\fmf{wiggly}{v,o1}
\fmffreeze\fmf{phantom}{i1,v1,v}\fmffreeze\fmf{gluon}{v1,t2}\fmfdot{v}
\end{fmfgraph}}
\right|^2
\end{multline}
and are included in the calculation of the heavy fermions widths presented here; the
analytical result can be found in appendix \ref{app-4}.

Fig.~\ref{fig-3-4-whf} left shows contour lines of constant relative width
$\Gamma_{e^\prime,\text{rel}}$ for the $e^\prime$ in the $m_{W^\prime}$ -- $M_\text{bulk}$ plane.
Evidently, there is a huge range of possible values from about $10\%$ to more than $100\%$.
However, direct detection as resonances at the LHC will only be possible for moderate
values of $M_\text{bulk}$, say $\le\unit[4]{TeV}$ anyway, and in this part of parameter space, the
relative width does not exceed $30-40\%$.
\begin{figure}[!tb]
\centerline{
\includegraphics[width=\doubleplotwidth,angle=270]{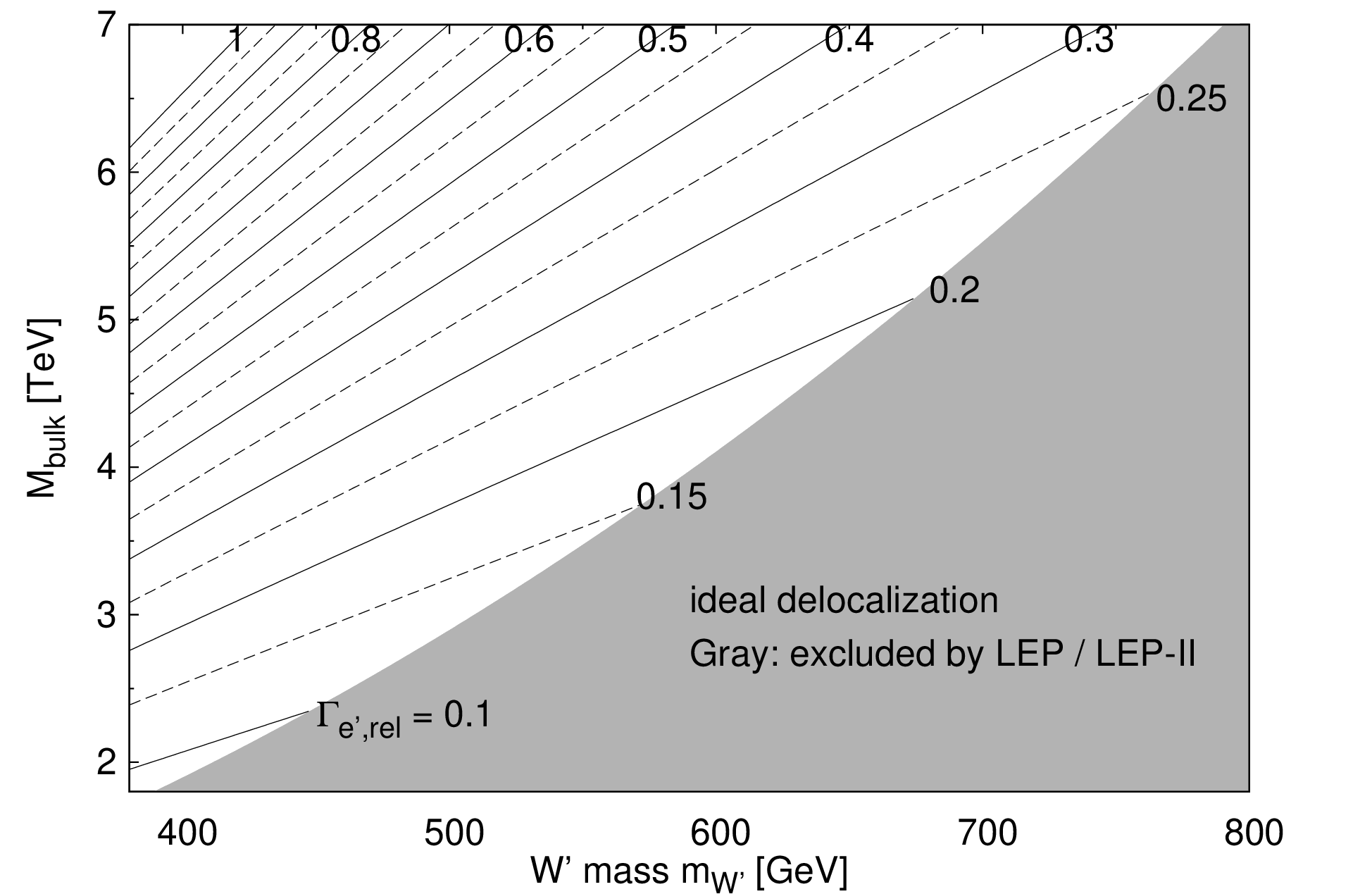}
\includegraphics[width=\doubleplotwidth,angle=270]{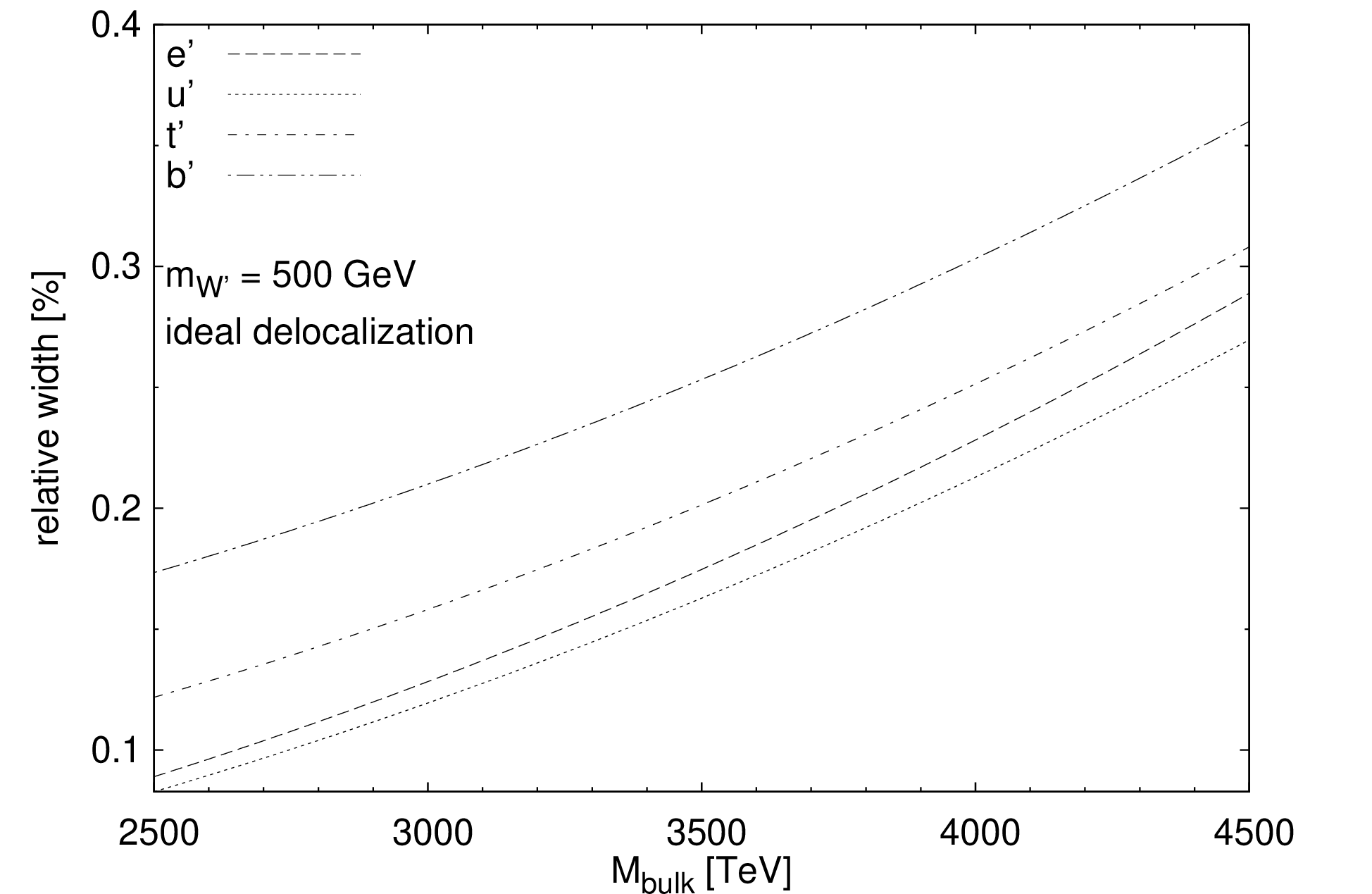}
}
\caption{\emph{Right:} Contour lines of constant relative $e^\prime$ width
$\Gamma_{e^\prime,\text{rel}}=\frac{\Gamma_{e^\prime}}{m_{e^\prime}}$ in the $m_{W^\prime}$ --
$M_\text{bulk}$ plane.\newline
\emph{Left:} Relative widths of $e^\prime$, $u^\prime$, $t^\prime$ and $b^\prime$
(including one loop QCD corrections) as a function of $M_\text{bulk}$.
}
\label{fig-3-4-whf}
\end{figure}

The right-hand plot of fig.~\ref{fig-3-4-whf} shows the relative width of
$e^\prime,u^\prime,b^\prime$ and $t^\prime$ as a function of $M_\text{bulk}$ for
$m_{W^\prime}=\unit[500]{GeV}$. As the tree-level widths of the heavy fermions (with exception of
the $t^\prime$ and $b^\prime$) turn out to be virtually identical, the difference between the
$e^\prime$ and the $u^\prime$ width is solely due to the QCD loop corrections which are
around $7.5\%$ and show only very little variation over the parameter space. The $t^\prime$ is
considerably broader than the partners of the light quarks, which is due to the wavefunctions of the $t$
being fairly nonlocal to accommodate the high top mass, enhancing the couplings between $t^\prime$ and
$t$.

The big surprise, however, is the $b^\prime$ which turns out to be nearly two times as broad as
the partners of the light quarks. The reason for this unexpected effect is the right-handed
$W^{(\prime)}b^\prime t$ coupling which is considerably enhanced by the nonlocal wavefunction of
the right-handed top quark. This can be confirmed by looking at the sample spectrum in appendix
\ref{app-3}.

\begin{figure}[!p]
\centerline{
\includegraphics[width=\doubleplotwidth,angle=270]{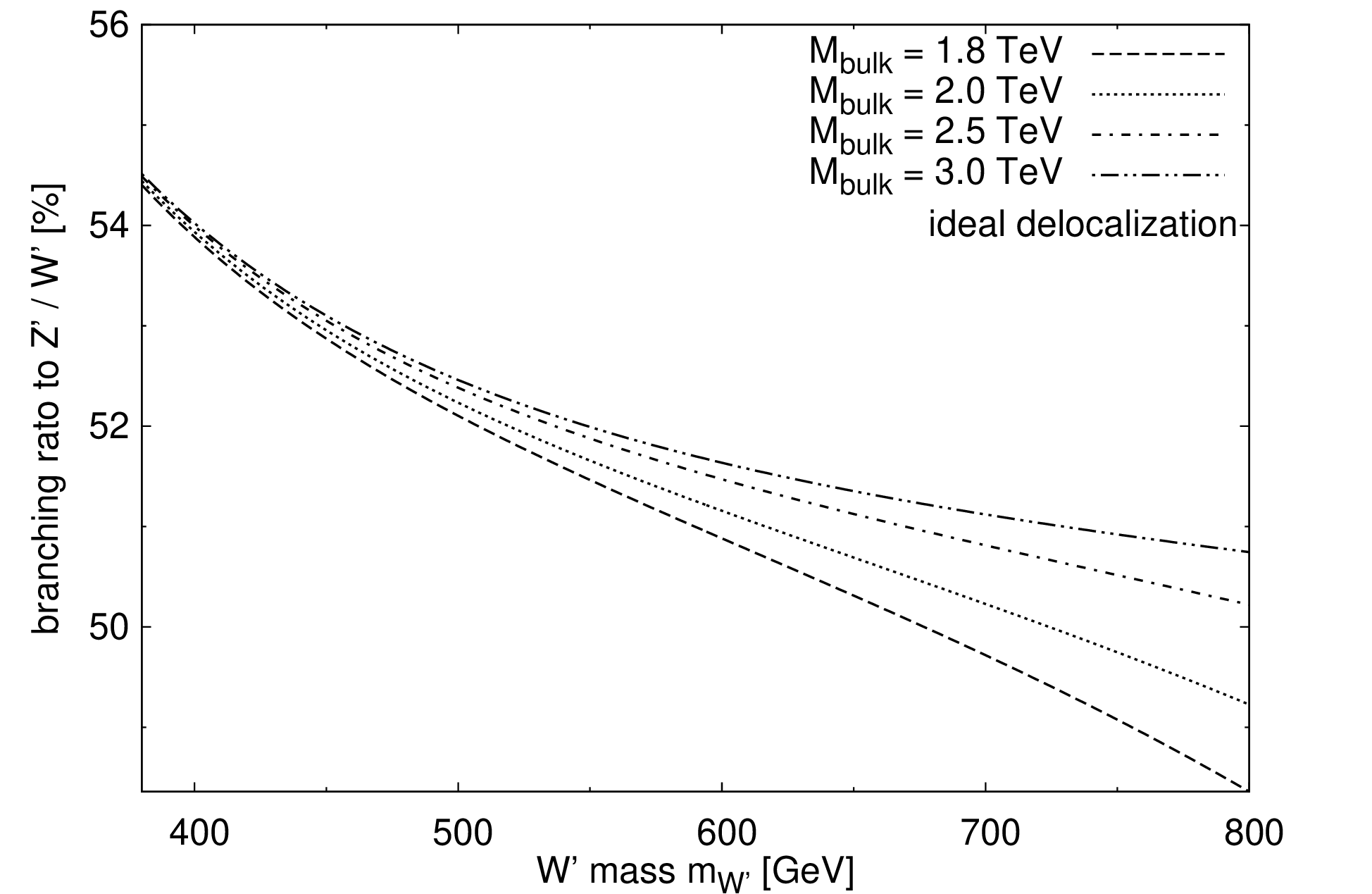}
\includegraphics[width=\doubleplotwidth,angle=270]{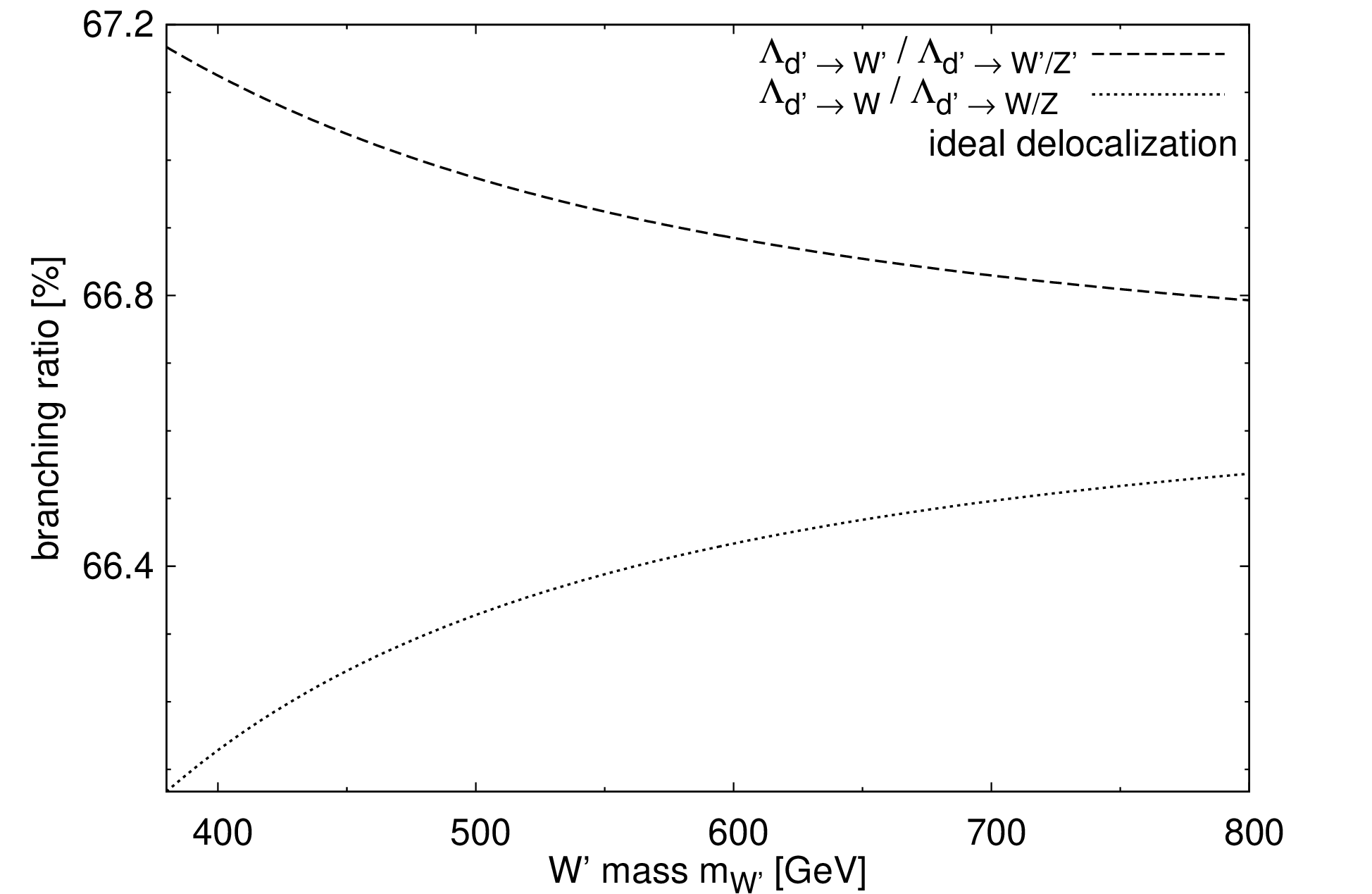}
}
\caption{
\emph{Left:} The branching ratio of the $d^\prime$ into heavy gauge bosons as a function of
$m_{W^\prime}$ for different values of $M_\text{bulk}$.\newline
\emph{Right:} The fraction of the $d^\prime$ taking the light / heavy decay channels going to $W$ /
$W^\prime$.
}
\label{fig-3-4-brhd}
\end{figure}
\begin{figure}[!p]
\centerline{
\includegraphics[width=\doubleplotwidth,angle=270]{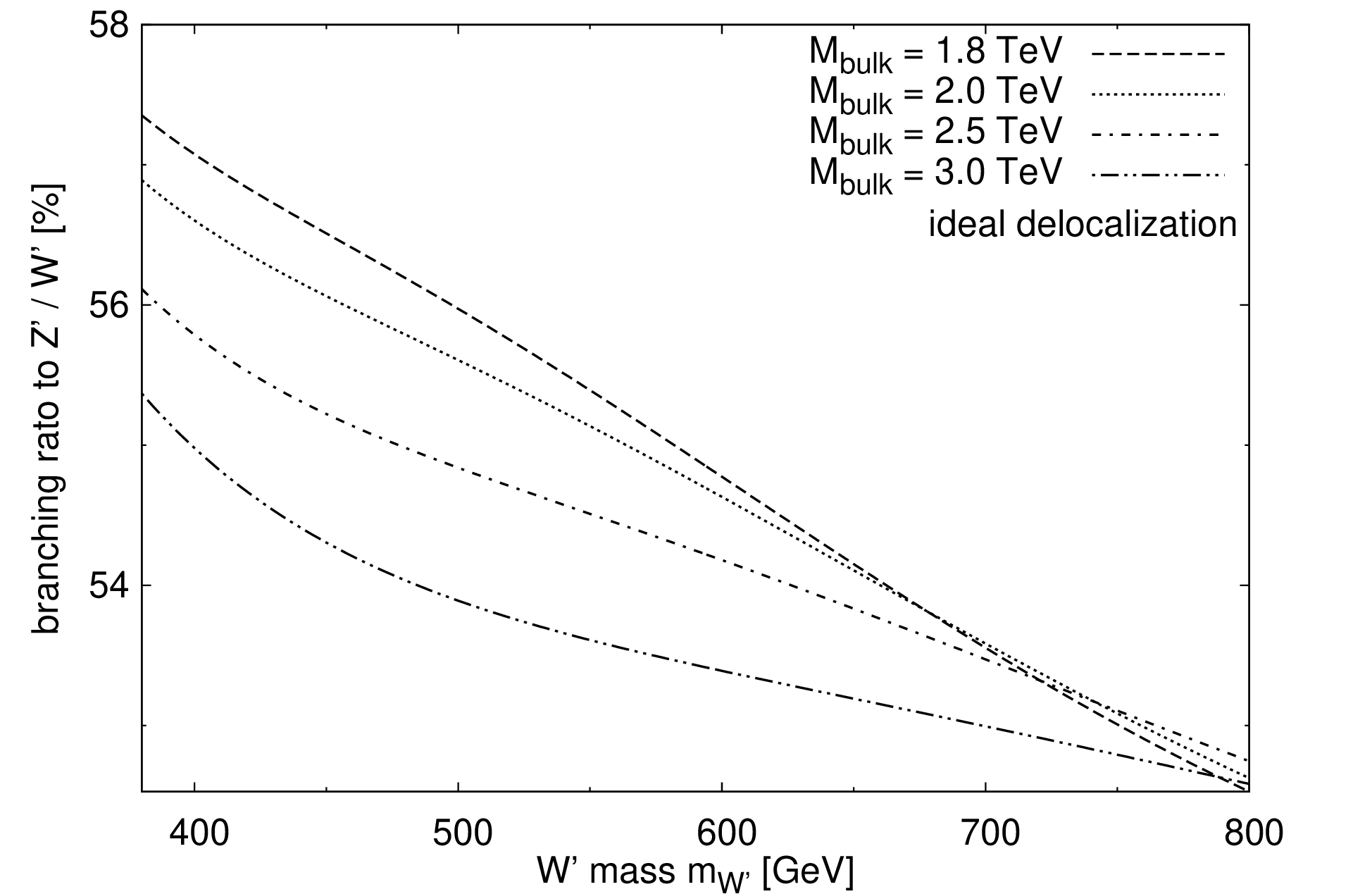}
\includegraphics[width=\doubleplotwidth,angle=270]{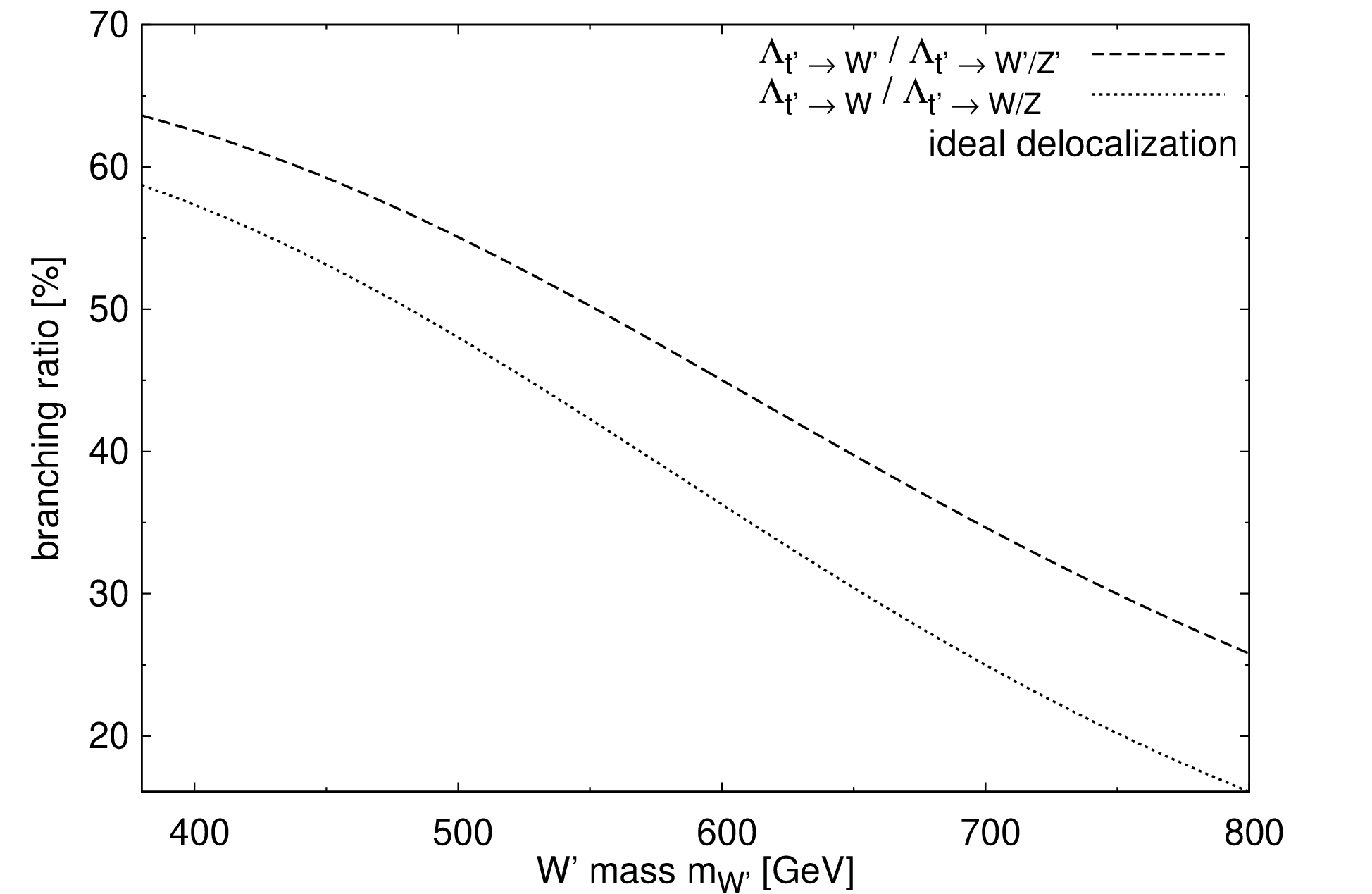}
}
\caption{The same type of plots as fig.~\ref{fig-3-4-brhd} for the $t^\prime$.}
\label{fig-3-4-brht}
\end{figure}
\begin{figure}[!p]
\centerline{
\includegraphics[width=\doubleplotwidth,angle=270]{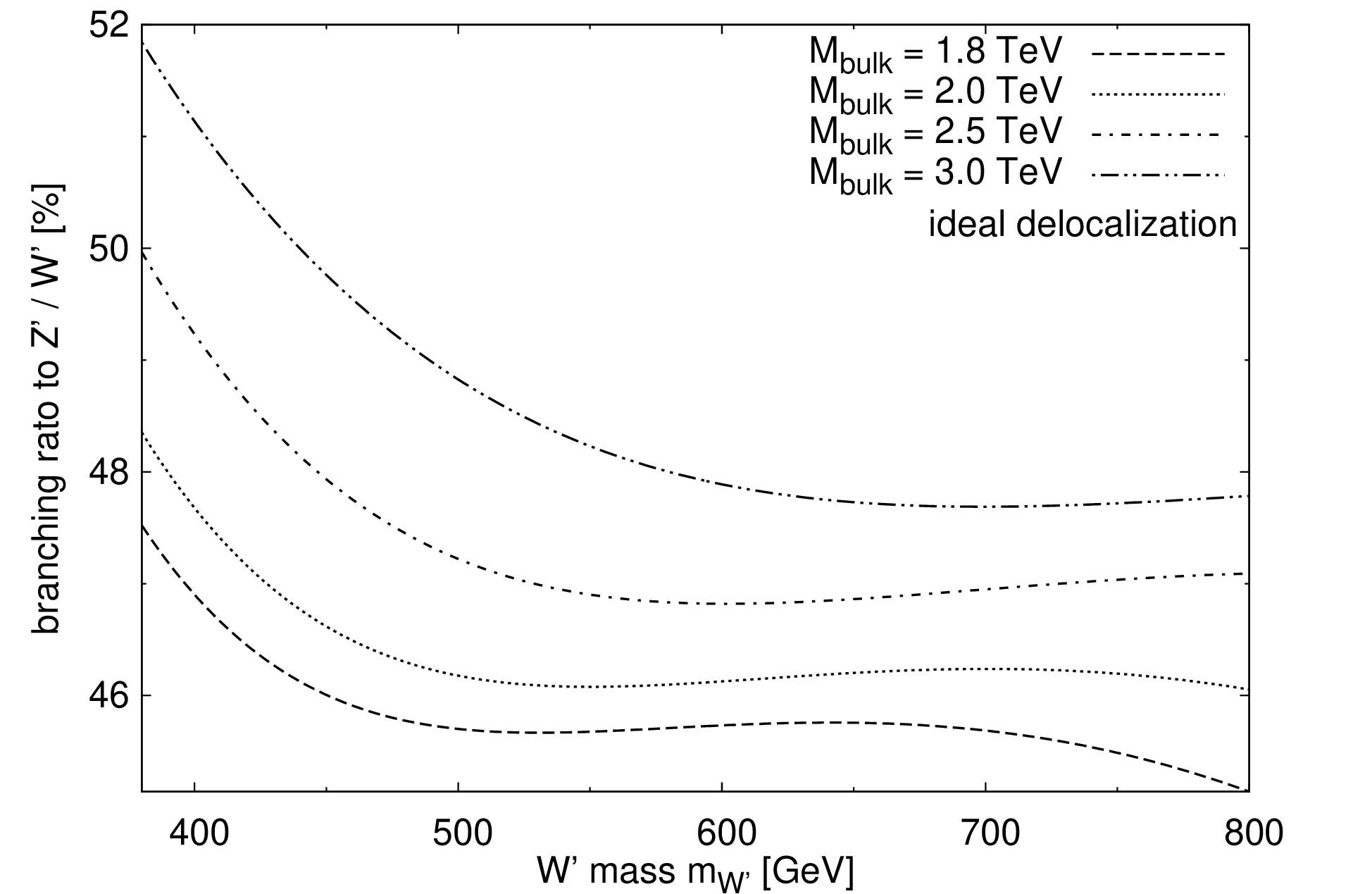}
\includegraphics[width=\doubleplotwidth,angle=270]{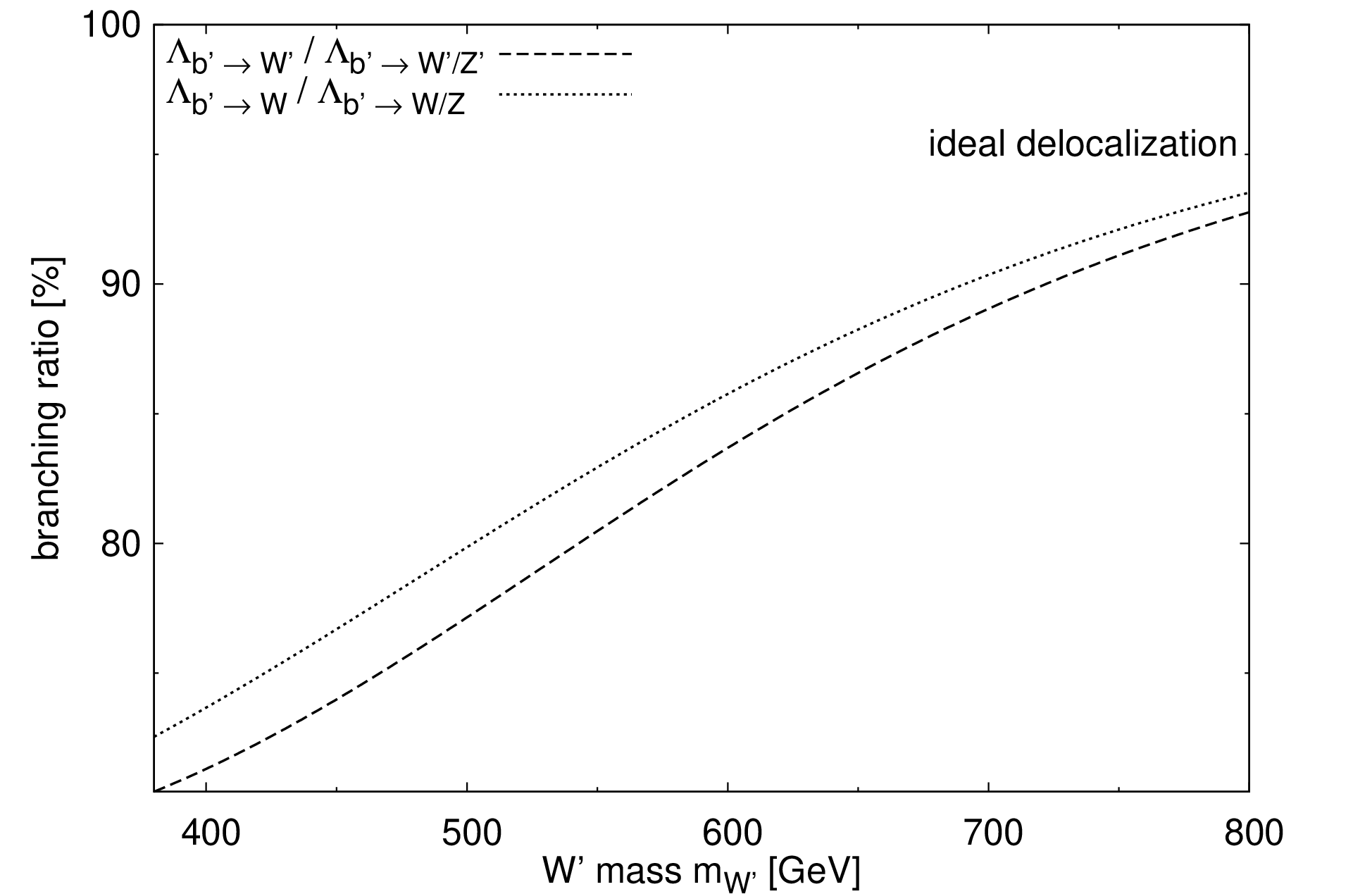}
}
\caption{The same type of plots as fig.~\ref{fig-3-4-brhd} for the $b^\prime$.}
\label{fig-3-4-brhb}
\end{figure}
Concerning the branching ratios, fig.~\ref{fig-3-4-brhd} left shows the fraction of $d^\prime$
decaying to heavy gauge bosons as a function of $m_{W^\prime}$ for different values of
$M_\text{bulk}$.  While this varies a bit over parameter space, it's always around $50\%$.
For $d^\prime$ which decay versus the light
resp. heavy decay channels, fig.~\ref{fig-3-4-brhd} right shows the fractions which goes into a $W$
resp. $W^\prime$, demonstrating these channels to be preferred with branching ratios of around
$65\%$ over the $Z$ / $Z^\prime$ channels. Both results are representative for the other heavy
fermions with little variation over different values of $M_\text{bulk}$,
again excluding the $t^\prime$ and $b^\prime$ due to the high top mass.

For these two, the respective results are shown in fig.~\ref{fig-3-4-brht} and fig.
\ref{fig-3-4-brhb}. Although the functional dependence on $m_{W^\prime}$ and $M_\text{bulk}$ is
different, the branching ratio of both $t^\prime$ and $b^\prime$ into heavy gauge bosons is around
$50\%$ similarly to the partners of the light fermions. However, the branching ratios into $W^{(\prime)}$
and $W$ are very different and exhibit a strong dependence on $m_{W^\prime}$. Further
investigation shows that, while still not drastic, the dependence on $M_\text{bulk}$ is much more
pronounced as for the other KK fermions.

\label{chap-3-4}

\chapter{Tools for Phenomenology}
\label{chap-4}

\begin{quote}\itshape
This admitted, we may propose to execute, by means of machinery, the mechanical branch of these labours,
reserving for pure intellect that which depends on the reasoning faculties.
\end{quote}
\hfill\begin{minipage}{0.7\textwidth}\small\raggedleft
(L. Menabrea, ``Sketch of The Analytical Engine, invented by Charles Babbage'', translated by
Lady A. Lovelace)
\end{minipage}
\\[5mm]
So far, we have dealt with the formal structure of a quantum field theory (the Three-Site Model in
our case) which defines the theory and allows to calculate observables such as scattering
probabilities. To test the theory, we must connect these observables to the quantities that are
measured in collider experiments or, more specific to this work, at the LHC. This is usually done
with the help of computers via Monte Carlo eventgenerators.

In this chapter, we first briefly review the general structure of an event simulation. After that,
the WHIZARD / O'Mega package which has been used for the numerical
simulations presented in this thesis is introduced together with a small review of the challenges
which have to be tackled by Monte-Carlo eventgenerators. In the third section,
the implementation of the Three-Site Model into WHIZARD / O'Mega done as part of this
thesis is discussed. Finally, in the last section, a brief overview over an interface between
FeynRules \cite{Christensen:2008py} and WHIZARD / O'Mega is given.

\section{The structure of an event simulation}
\label{chap-4-1}

The simulation of collision events at particle colliders can be roughly split into three building
blocks. The first is the calculation and integration of the differential cross section which
consists of evaluating the squared matrix elements and performing the phase space integration. The
second block is the actual generation of an ensemble of events that is distributed according to the
differential cross section. The third block consists of simulating the characteristics of the detector
used for the actual observation of the particles produced in the events.

Obviously, the calculation and integration of the differential cross section is the necessary
prerequisite for the generation of events. However, the detector simulation part completely
factorizes from the other two steps. As it also is time consuming and requires an intricate knowledge
of the inner workings of the actual machinery employed in the experiment, it is usually a good idea
to perform studies on the feasibility of detecting new physics without a detailed detector simulation first.
The result then is a machine-independent assessment of the chances for detection which, if it is
positive, can be further improved by adding a detector simulation.

The calculation of the cross section also is an intricate process. The complications start with the
initial state which, at a hadron collider like the LHC, consists of complicated bound states of
quarks and gluons which we cannot describe from first principles. The final state usually consist of leptons,
photons and mesons, and the latter again are bound states of QCD whose formation cannot be
described with perturbation theory.
Luckily, nature has arranged for asymptotic freedom and the factorization theorems of
QCD \cite{Collins:1987pm} which allow us to split this complicated process of
$\text{hadrons}\longrightarrow\text{mesons}+X$ into different regimes which decouple in good
approximation.

More specifically, we may describe the initial hadrons at high energies as ensembles of
non-interacting partons with the probability of finding a given parton carrying a given momentum
fraction being described by parton distribution functions (PDFs) which are universal and independent
of the process under consideration.
In addition, the QCD interactions that lead to the fragmentation of the hard partons and the final
hadronization to the observable bound states can also be approximately factorized.

In the end, as a good
approximation, we can reduce the complicated scattering process to a ``hard'' core process consisting
of two ingoing partons scattering into a handful of hard outgoing partons, photons and leptons.
This then is convoluted with the PDFs to accommodate the hadronic initial state, and
parton showers are added to describe the evolution of the small number of hard partons to a large
number of soft ones. In a final step, empirical models can be applied to describe the formation
of bound states from these soft partons in a process called ``hadronization''.

The only piece of this decomposition which we can calculate in perturbation theory is the hard matrix
element describing the partonic core process (which fortunately usually also is the only part sensitive
to new physics). Algorithms for performing the parton shower can also be
obtained from first principles. However, there is to this day no way of calculating the PDFs from
first principles, and only empirical models exist for the hadronization process.

In this work, we concentrate on the hard process and don't include parton showers, hadronization or a
detector simulation. This is called a ``parton-level'' simulation which essentially relies upon the
assumption that every parton produced in the hard scattering event is distinctly observable
as a jet in experiment. While this is certainly not correct, it is a good assumptions for embarking on
a study of the phenomenology of the Three-Site Model: it is only in the case of significant deviations from
the Standard Model at this level that we can hope to observe something at the LHC.

\section{WHIZARD / O'Mega}
\label{chap-4-2}

In this thesis, the software package WHIZARD / O'Mega \cite{Kilian:2007gr} is used for the
simulation of parton level events. This package actually consists of at least two separate programs,
the optimizing matrix element compiler O'Mega and the core Monte Carlo and infrastructure code of WHIZARD.

\subsubsection*{O'Mega: fast generation of tree level matrix elements}

Easy as it may seem from calculating the matrix elements for $2\rightarrow 2$ by straightforward
application of the Feynman rules, the evaluation of tree level matrix elements actually is very
demanding on computer hardware and requires clever algorithms if it is to run in finite time.

The reason for this is the enormous growth of the number of Feynman diagrams if we increase the
number of external lines. Consider, for example, $\phi^3$ theory\footnote%
{
While this is a pathological example of a field theory due to the indefinite potential, it actually is the
easiest example when it comes to the combinatorics of Feynman diagrams. The only difference to
other, physical models with arity three couplings (like QED) is the ``dressing'' of the
$\phi^3$-topologies with particle flavors which doesn't affect the asymptotic combinatorics.
}.
If we know all the tree level diagrams contributing to the $n$-point correlator of $n$ particles
with momenta $p_1,\ldots,p_n$, we can easily obtain
those for the $n+1$-point function as all possible ways of attaching the $p_{n+1}$ leg to one of the
lines in the $n$-point diagrams, e.g.
\[
\parbox{21mm}{\fmfframe(2,2)(2,2){\begin{fmfgraph*}(17,14)
\fmfleft{i}\fmfright{o2,o1}\fmf{plain}{i,v}\fmf{plain}{o1,v,o2}
\fmfv{la=$p_1$}{i}\fmfv{la=$p_2$}{o1}\fmfv{la=$p_3$}{o2}
\end{fmfgraph*}}}
\quad\longrightarrow\quad
\parbox{32mm}{\fmfframe(2,2)(2,2){\begin{fmfgraph*}(28,14)
\fmfleft{i2,i1}\fmfright{o2,o1}\fmf{dashes}{i1,v1}\fmf{plain}{i2,v1}\fmf{plain}{v1,v2}
\fmf{plain}{o1,v2,o2}\fmfdot{v1}
\fmfv{la=$p_4$}{i1}\fmfv{la=$p_1$}{i2}\fmfv{la=$p_2$}{o1}\fmfv{la=$p_3$}{o2}
\end{fmfgraph*}}}
\quad+\quad
\parbox{20mm}{\fmfframe(2,2)(2,2){\begin{fmfgraph*}(16,20)
\fmfleft{i2,i1}\fmfright{o2,o1}\fmf{dashes}{i1,v1}\fmf{plain}{i2,v2,v1,o1}
\fmf{plain}{v2,o2}\fmfdot{v1}
\fmfv{la=$p_4$}{i1}\fmfv{la=$p_1$}{i2}\fmfv{la=$p_2$}{o1}\fmfv{la=$p_3$}{o2}
\end{fmfgraph*}}}
\quad+\quad
\parbox{20mm}{\fmfframe(2,2)(2,2){\begin{fmfgraph*}(16,20)
\fmfleft{i2,i1}\fmfright{o2,o1}\fmf{dashes}{i2,v1}\fmf{plain}{i1,v2,v1,o2}
\fmf{plain}{v2,o1}\fmfdot{v1}
\fmfv{la=$p_4$}{i2}\fmfv{la=$p_1$}{i1}\fmfv{la=$p_2$}{o1}\fmfv{la=$p_3$}{o2}
\end{fmfgraph*}}}
\]
Therefore, we can calculate the number $N_{n+1}$ of diagrams contributing to the $n+1$-point
function from the number $N_n$ of $n$-point diagrams as
\[ N_{n+1} = L_n\cdot N_n  \]
with the number of lines in any $n$-point diagram $L_n$.
As attaching a leg increases $L_n$ by $2$, we deduce
\[ L_n = 2n - 3 \]
and therefore find
\[
N_{n} = (2n-5)N_{n-1} = (2n-5)(2n-7)\cdot\ldots\cdot1 = \prod_{k=2}^{n-1}(2k-3)
\]
By induction we have
\[ N_n = \left(2n - 5\right)!! = \frac{(2n - 4)!}{2^{n-2}(n-2)!} \ge 2^{n-3}(n-3)! \]
Thus, the number of diagrams grows like factorial , and while we only have three diagrams for
$n=4$, we already have $105$ diagrams for $n=6$ and $10395$ diagrams for $n=8$.

Keeping in mind that the matrix element has to be evaluated at every phase space point
under consideration it is obvious that this factorial growth of the number of Feynman diagrams kills off any attempt
to perform a brute-force, diagram-by-diagram calculation. Things get even
worse if a straightforward analytic calculation is attempted and therefore, traditional tools like
CalcHEP \cite{Pukhov:2004ca} which perform very well for $2\rightarrow 2$ processes have severe
problems handling more than four particles in the final state.

Luckily, this is not the end of the story. If we consider again the addition of an additional leg
to a diagram, we note that recalculating the whole diagram is a waste of resources. Reconsider the
insertion of a line
\[\parbox{35mm}{\begin{fmfgraph*}(35,20)
\fmfleft{i4,i3,i2,i1}\fmfright{o4,o3,o2,o1}
\fmf{plain}{i1,v1}\fmf{plain}{i2,v1}\fmf{plain}{i3,v1}\fmf{plain}{i4,v1}
\fmf{plain}{v1,v2}\fmf{plain}{o1,v2}\fmf{plain}{o2,v2}\fmf{plain}{o3,v2}\fmf{plain}{o4,v2}
\fmfv{de.sh=circle,de.si=10thick,la=$A$,la.d=0,de.fi=empty}{v1}
\fmfv{de.sh=circle,de.si=10thick,la=$B$,la.d=0,de.fi=empty}{v2}
\end{fmfgraph*}}
\quad\longrightarrow\quad
\parbox{35mm}{\begin{fmfgraph*}(35,20)
\fmfleft{i4,i3,i2,i1}\fmfright{o4,o3,o2,o1}
\fmftop{t}
\fmf{plain}{i1,v1}\fmf{plain}{i2,v1}\fmf{plain}{i3,v1}\fmf{plain}{i4,v1}
\fmf{plain}{v1,v2}\fmf{plain}{o1,v2}\fmf{plain}{o2,v2}\fmf{plain}{o3,v2}\fmf{plain}{o4,v2}
\fmffreeze\fmf{phantom}{v1,v3,v2}\fmffreeze\fmf{dashes}{t,v3}\fmfdot{v3}
\fmfv{de.sh=circle,de.si=10thick,la=$A$,la.d=0,de.fi=empty}{v1}
\fmfv{de.sh=circle,de.si=10thick,la=$B$,la.d=0,de.fi=empty}{v2}
\end{fmfgraph*}}
\]
Evidently, we don't have to recalculate the blobs $A$ and $B$,
but just have to cut the line connecting the blobs and reattach them to a
vertex together with the new leg. Also, if $A$ and $B$ are not only pieces of Feynman diagrams but
instead complete correlation functions with all legs amputated except the leg extending to the middle
vertex in the above diagrams (so-called ``one particle off-shell wavefunctions'' 1POWs), we can
calculate a whole class of diagrams in this way. This observation suggests an iterative approach to
the calculation of Feynman amplitudes which is potentially much more efficient than a
straightforward calculation of all Feynman diagrams.

The potential savings in choosing a clever algorithm can be acknowledged by counting the number of
possible 1POWs. In our example of $\phi^3$ theory, these are uniquely enumerated by the
distinct nonzero combinations of momenta which can be formed from external lines, the number
of which
\[ P_n = \binom{n}{1}+\binom{n}{2}+\ldots+\binom{n}{n-1} = (1 + 1)^n -2 = 2^n - 2 \]
only grows exponentially.
Algorithms like the one used in the optimizing matrix element generator O'Mega exploit this redundancy to reduce the
factorial growth in complexity of the calculation to an exponential one \cite{Moretti:2001zz}.

O'Mega represents the amplitude as a directed acyclical graph (DAG) of 1POWs from which it weeds out all
redundancies that would arise from the recalculation of 1POWs.
Finally, this DAG is transferred to a backend which
translates it into a representation suitable for performing numerical calculations. The only
complete backend included in the O'Mega distribution is a FORTRAN 95 backend which emits a process
library coded in FORTRAN 95 that can be called to evaluate the amplitude. The resulting code is
clean, human readable, fast and can be automatically instrumented for numerical checks of gauge
invariance. The matrix element is calculated in a completely numerical way by
numerically multiplying spinors, four-vectors, Lorentz tensors etc. 

\subsubsection*{WHIZARD: phase space, Monte Carlo integration and more}

Once we have a way of calculating the matrix elements, we can go over to integrating them over
phasespace, a step which turns out to be challenging for a number of reasons.

First, the dimensionality is high --- subtracting four degrees of freedom due to momentum conservation and one
due to rotational invariance around the beam axis, we have $3n - 5$ phasespace dimensions for $n$
final state particles plus two additional dimensions for the parton distributions. Therefore, in a
$2\rightarrow4$ process, we already have a $9$-dimensional integral to evaluate. The complexity of
deterministic numerical methods (e.g. Gauss-Kronrod) grows exponentially with the number of
dimensions and therefore, such algorithms are not suitable for this type of integral.

This problem can be overcome by Monte Carlo methods. A simple method for Monte Carlo integration
is drawing $N$ random points $\mathfrak{p}_k$ in the integration domain $\Omega$ and then
approximating the integral of $f$ by
\[ \int_\Omega dx\;f(x) \approx V\left<f\right> = \frac{V}{N}\sum_{k=1}^N f(\mathfrak{p})_k \]
with the volume $V$ of $\Omega$.

If $f$ is a probability density bounded from above by a number $F_0$, then we can use the set of
pairs
\[
\mathcal{W} = \left\{\left(\mathfrak{p}_1,f(\mathfrak{p}_1)\right),
\ldots,\left(\mathfrak{p}_N,f(\mathfrak{p}_N)\right)\right\}
\]
to obtain an ensemble of points distributed according to $f$ by drawing $N$ numbers $z_k$ between $0$ and
$F_0$ and then discarding all points $\mathfrak{p}_k$ for which $f(\mathfrak{p}_k)$ exceeds $z_k$. The
resulting set of $n$ surviving points
\[
\mathcal{U} = \left\{\mathfrak{p}_{k_1},\ldots,\mathfrak{p}_{k_n}\right\} =
\left\{\mathfrak{p}_k\left.\vs{4ex}\right|f(\mathfrak{p}_k) \le z_k\right\}
\]
can be significantly smaller than $\mathcal{W}$ but will obey the desired distribution. The set
$\mathcal{W}$ is usually called a set of ``weighted events'' with statistical weights given by the
$f(\mathfrak{p}_k)$, while $\mathcal{U}$ is called a set of ``unweighted events'' with the processes
leading from $\mathcal{W}$ to $\mathcal{U}$ termed as ``unweighting''. If we want to simulate events
at a particle collider, then it is the unweighted events we are interested in as they give a
realistic estimate of the fluctuations induced by the statistics.

It can be shown that this Monte Carlo integration comes with an error that
scales with $\frac{\alpha}{\sqrt{N}}$
for large $N$ independently of the number of dimensions of the integral.
The coefficient $\alpha$ is determined from the variance functional $\Delta$
\[ \Delta[f] = \int_\Omega dx\;f(x)^2 - \left(\int_\Omega dx\; f(x)\right)^2 \]
Evidently, this is a small number only if the function $f$ is well-behaved over $\Omega$ (aka doesn't
fluctuate much).

However, the differential cross section we want to integrate is far from well-behaved: by the virtue
of the propagators which have mass singularities, it fluctuates wildly, being close to zero in large
regions of phasespace and nearly singular in others. The obvious problems arising from this large
variation can be alleviated by choosing the ensemble of points $\mathfrak{p}_n$ not uniformly
distributed in $\Omega$ but instead with respect to 
a nontrivial probability density $g(x)$. Provided we perform the replacement
\[ f(x)\longrightarrow\frac{f(x)}{g(x)} \]
the Monte-Carlo method presented remains correct, but with the error coefficient $\alpha$ now
determined by $\Delta\left[\frac{f}{g}\right]$.

Choosing a suitable $g$ allows to reduce the Monte Carlo error
drastically at the expense of more calculational time spent on the generation of the
$\mathfrak{p}_i$. Therefore, $g$ should be chosen such that the generation of the points is cheap
enough to retain the speed gained by reducing $\Delta$.
As manual tuning of $g$ usually is too cumbersome, a suitable density is usually
determined in an adaptive process designed to minimize $\Delta$ over a restricted function
space\footnote%
{It is an interesting fact that, for a restricted function space, the $g_\Delta$ which minimizes the variance
functional $\Delta$ usually does not maximize the acceptance in the unweighting step (the percentage
of points in $\mathcal{W}$ that is kept). It is also possible to optimize the acceptance instead of
the variation in the adaption phase, increasing the performance of the unweighting step at the
price of reducing the accuracy of the estimate for the integral.
}.

Monte-Carlo eventgenerators like WHIZARD usually start with an adaption phase during which $g$ is
optimized, then proceed with an integration phase where the integral is obtained and finally an event
generation phase where (unweighted) events are generated. Ideally, new points are generated in all
three phases to avoid pollution of the error estimates by statistical correlations.

In order to choose a suitable class of functions over which the density $g$ is optimized, WHIZARD has
some limited
knowledge of diagrammatics and needs information on the vertices of the model. Prior to the
start of the adaption phase WHIZARD then classifies all singular regions in phasespace that
might contribute to the integral (``integration channels''). This information is passed to the
Monte-Carlo core VAMP \cite{Ohl:1998jn} (a multichannel modification of the VEGAS algorithm \cite{Lepage:1977sw})
which chooses a suitable phasespace map $\phi_k$ for each channel. In these coordinates, a multidimensional
grid function $g_k$ together with a weight $\alpha_n$ is assigned to each channel, and the full
parameterization of $g$ reads
\[ g(x) = \sum_{k=1}^N \alpha_k \left(g_k\circ\phi_k\right)(x) \]
with the sum running over all channels.

In the adaption process, $g$ is optimized by adjusting the grids $g_n$ as well as the weights
$\alpha_n$. After the adaption, the best $g$ obtained in the adaption phase is used to calculate the
Monte Carlo integral of the cross section. Finally, WHIZARD can use the grid and the integral to
generate an ensemble of unweighted events which correspond to a given integrated luminosity.

The emphasis of the WHIZARD package lies on the fast generation of parton level events for BSM
physics. Although its primary source of matrix elements is O'Mega, it can also interface other
matrix element codes\footnote%
{
This functionality is present in the 1.9x branch of WHIZARD but will be dropped from version 2.0
onwards.
}
like MadGraph \cite{Stelzer:1994ta} or CalcHep. It is capable of convoluting the
partonic cross section with PDFs (via PDFLIB \cite{PlothowBesch:1992qj} or
LHAPDF \cite{Whalley:2005nh}) and can interface PYTHIA \cite{Sjostrand:2006za} for parton showering.
The high numerical speed achieved by the combination WHIZARD / O'Mega allows for
the simulation of processes with $6-8$ final state particles which is very hard to achieve with
other tools. Other features include an integrated facility for event analysis which allows to create
histograms on-the-fly from the generated event data.

The package is written in FORTRAN 95 with some perl glue being used to assemble the source and
the matrix element code prior to compilation and evaluation\footnote%
{
This is only true for the 1.9x branch. WHIZARD 2.0 and higher compile the matrix elements at run time
and link them dynamically without relying the perl component which is dropped.
}.
For the exact treatment of color, the color
flow decomposition \cite{Maltoni:2002mq} of Feynman amplitudes is used.

\section{Implementation}
\label{chap-4-3}

In order to study the phenomenology of the Three-Site model with WHIZARD / O'Mega we have
implemented the model into this package. In its current state, the code implements the Three-Site
Model in unitarity gauge without flavor mixing and supports both ideal and
non-ideal delocalization. It has been validated by numerical checks of the Ward identities in the
limit $v\rightarrow 0$ as well as by direct comparison to the CalcHEP implementation used by
\cite{He:2007ge} and (lately) to the FeynRules version done by N. Christensen et al. (see chapter
\ref{chap-4-4}).

The model implementation will be part of a future version of WHIZARD. In the meantime,
the coupling library as well as modified
versions of the O'Mega and WHIZARD packages which include the model can be downloaded from
\newline\centerline\myurl

\subsubsection*{O'Mega}

In order to perform its duties as a matrix element generator, O'Mega needs some information about
the model under consideration. This comes under the guise of a O'Caml module which contains the
necessary definitions and functions needed by O'Mega to operate. We have written such a module
implementing the Three-Site Model.

The implementation offers several options that can be (de)activated at the users leisure to
change the set of particles vertices included. More specifically, it is possible to allow for flavor
off-diagonal couplings (although these are set to zero at the moment, so this would currently just
be a waste of resources), to leave out the heavy fermions and to discard the couplings of the
$W^\prime$ to the leptons and to the first two quark generations.
In addition, it is possible to replace the
unitarity gauge propagators with Feynman gauge ones (this is necessary for using the massless
limit).

Upon compilation of O'Mega, binaries are generated from the model modules which can then be
called to generate code for the desired matrix elements. For the Three-Site Model, different
binaries are generated for the different vertex sets described above both in colored and colorless
versions (the colorless ones completely exclude QCD). A command line switch can be used to replace
the unitarity gauge propagators.

More details on the implementation together with some snippets of the actual code can be found in app.
\ref{app-5-2}.

\subsubsection*{WHIZARD}

In addition to the model implementation in O'Mega, WHIZARD requires a model file which defines the
particles and vertices present in the model together with all parameters that can be changed at run
time (and optionally also a list of secondary parameters calculated from these).

Writing the model file is a straightforward task. The only part which requires some trickery is
the vertex list which WHIZARD requires to be loosely sorted with respect to the mass of the
particles meeting at the vertex. Even if flavor violating couplings are excluded, the Three-Site Model
contains 220 electroweak three-point couplings. To avoid the cumbersome and error-prone task of
collecting and sorting all these couplings per hand, the list in the model file is generated
directly from the O'Mega module by an O'Caml script.

The resulting vertex list may be too long for processes with 6+ particle final states, causing
the phasespace mapping step to take forever or fail altogether. In this case, removing all vertices
containing KK fermions provides a workaround without affecting the quality of the result overmuch.

The parameters\footnote%
{
The actual names of the parameters are documented in the model file.
}
defined in the model file allow to set all Standard Model masses, the $t$, $W$ and
$Z$ widths, the electromagnetic and strong coupling constants, the $W^\prime$ mass and
$M_\text{bulk}$. This is enough to fix the model in the case of ideal delocalization (c.f. chapter
\ref{chap-3-2}). Nonideal delocalization can be activated by toggling a flag\footnote%
{
As WHIZARD only supports real numbers as parameters, a negative number is read as ``false'' and a
positive number as ``true''.
}, making it necessary to supply
a value $\epsilon_L$ to complete the set of model parameters in this case. The
widths of the KK partners and the couplings are automatically calculated from the input parameters
(see below). The inclusion of the QCD corrections into the heavy quark widths can be toggled by a
flag, and another flag can be used to dump the spectrum and couplings (similar
appendix \ref{app-3}) to a file.

Unfortunately, the architecture of WHIZARD in its current form (version 1.9x) requires some patches
to WHIZARD itself to make the model work. However, in a future version of WHIZARD, it will be
possible to use the model without any changes to the actual WHIZARD code, allowing for the
integration of the Three-Site Model into the package. More details on the required changes can be
found in app.~\ref{app-5-3}.

\subsubsection*{Coupling library}

To make the implementation work, a helper is required which calculates all masses, couplings and
widths from the input parameters (c.f. chapter \ref{chap-3}).
This task is hand\-led by a dedicated FORTRAN 90 library.

This library takes the input parameters and uses them to first calculate the masses and
wavefunctions of the particles. The calculation is performed using exact analytic formulae in contrast to
the expansions in $x$ and $\epsilon_L$ given in chapter \ref{chap-3-2}. The wavefunctions are then
used to calculate the couplings as shown in chapter \ref{chap-3-3} (with the exception of the
couplings to the photon which are determined by electromagnetic gauge invariance). After the couplings are
calculated, the library then calculates the widths of the KK particles by looping over all possible
two particle final states, using the analytical formulae given in app.~\ref{app-4-1}.

If activated, the
$\order{\alpha_s}$ corrections to the heavy are also included using the analytic expressions of
app.~\ref{app-4-2}. In this case, the LoopTools library \cite{Hahn:1998yk} is utilized for the
numerical evaluation of the one loop integrals appearing in the virtual corrections and for the dilogarithm
appearing in the three-particle phase space integrals.

In order to validate the O'Mega implementation of the model by numerical checks of the Ward
identities, the library supports a setup in which the full $\sun{2}_0\times\sun{2}_1\times\un{1}_2$
gauge symmetry is left unbroken. In this mode, the model is initialized from $x$, $e$, $t$ and an
arbitrary angle $\phi$. The scale $v$ and the Yukawa couplings
$\lambda,\tilde\lambda,\lambda^\prime$ are set to zero, resulting in vanishing mass matrices. The
gauge couplings are initialized as
\[ g = e\sqrt{1+x^2+t^{-2}} \quad,\quad \tilde{g} = \frac{g}{x} \quad,\quad g^\prime = gt \]
As the mass matrices vanish, we can choose the wavefunctions freely (provided the wavefunction for
KK partners are orthonormal). The $W$ / $W^\prime$ and the fermion wavefunctions are parameterized
by the angle $\phi$
\begin{align*}
f^W = f^{f_L} &=\cvect{\cos\phi \\ \sin\phi} &
f^{W^\prime} = f^{f^\prime_L} &= \cvect{-\sin\phi \\ \cos\phi} \\
f^{f_R} &= \cvect{\sin\phi \\ \cos\phi}&
f^{f^\prime_R} &= \cvect{-\cos\phi \\ \sin\phi}
\end{align*}
The wavefunction of the photon is left unchanged to avoid a change in the couplings to the photon
(which are hardcoded), and the $Z$ and $Z^\prime$ wavefunctions are chosen to form an orthonormal
system together with $f^\gamma$
\[
f^\gamma = e\cvect{\frac{1}{g}\vs{4ex} \\ \frac{1}{\tilde{g}}\vs{4ex} \\ \frac{1}{g^\prime}\vs{4ex}}
\qquad\qquad
f^{Z^\prime} \propto\cvect{-\frac{g}{2}\vs{4ex} \\ \tilde{g}\vs{4ex} \\ -\frac{g^\prime}{2}\vs{4ex}}
\qquad\qquad
f^{Z} \propto \cvect{-\frac{g^\prime}{2\tilde{g}} - \frac{\tilde{g}}{g^\prime}\vs{4ex} \\
\vs{4ex}\frac{g^\prime}{2g} - \frac{g}{2g^\prime} \\
\vs{4ex}\frac{\tilde{g}}{g} + \frac{g}{2\tilde{g}}}
\]
with the normalization factors being suppressed for the sake of readability. This setup tries to avoid
accidental cancellations in the couplings which would impair the effectiveness of the desired consistency
check.

The library is not tied to WHIZARD but can also be used in a standalone fashion. Several
tools are included which generate spectra (like e.g. app.~\ref{app-3}),
density plots (e.g. fig.~\ref{fig-3-3-wtbR}) etc. together with a wrapper that allows to access the
masses, widths and couplings from Mathematica (in fact, many of the plots shown in chapter \ref{chap-3}
were generated from Mathematica this way).

More information about implementation and usage of the library can be found in app.~\ref{app-5-1}
together with some snippets of code; more documentation can be found in the README file and in the
woven source contained in the
distribution which can be downloaded from the URL quoted at the beginning of this chapter.

\section{FeynRules $\rightarrow$ WHIZARD interface driver}
\label{chap-4-4}

The implementation of a new model into an eventgenerator usually is a time-consuming and error-prone
process, and careful testing is necessary to ensure that everything works as expected. The
situation isn't improved by the fact that every eventgenerator has its very own format for
specifying new models, follows its own conventions and has its own deficiencies and weak points which
need to be worked around. Therefore, a generic format for specifying new models models which is as
close as possible to the formal representation by a set of fields and a Lagrangian would be
desirable.

The FeynRules package \cite{Christensen:2008py} is a step in this direction. FeynRules is a
Mathematica package which can be used for the automatic generation of model files for different
Monte Carlo eventgenerators. To this end, the particles in the model are specified in a format which
is derived from FeynArts \cite{Hahn:2000kx}. FeynRules then takes the Lagrangian (which can be entered in a way
which is very close to the usual formal notation) and calculates the Feynman rules from it. The
generated rules can then be passed to an interface driver which generates a model file suitable for
use with the event generator of choice.

This way, the cumbersome task of implementing the model has
to be done just once and can be performed at a fairly abstract level with all the
particularities of the different generators being take care of by the interface drivers.
Having the model available in different eventgenerators is a very important step as all
of them have their own strengths and weaknesses and cross-checking a result over different
generators greatly improves the reliability of a prediction.

So far, the list of eventgenerators officially supported by FeynRules consists of CompHEP / CalcHEP,
Sherpa \cite{Gleisberg:2008ta} and MadGraph. During the preparation of this
thesis, a new interface driver has been developed in corporation with N. Christensen which
generates output for WHIZARD / O'Mega which will be included in a future distribution of FeynRules.

The driver is capable of producing the model module for O'Mega together with the model file and
FORTRAN glue required by WHIZARD. It supports nearly all the features offered by FeynRules,
including the free choice of any $R_\xi$ gauge. The generated code has been successfully validated
for the Standard Model and for the Three-Site Model, validation for the MSSM is in progress.

\chapter{$W^\prime$ Strahlung}
\label{chap-5}

After giving an overview over the model and the tools used for studying it in the last chapters, we
now move on to presenting the results of the actual Monte Carlo simulations performed in this
thesis. In this chapter, we start off with a discussion of the production of $W^\prime$ which are radiated
from a light gauge boson in a process similar to Higgsstrahlung \cite{Ellis:1975ap}. The process is
presented in the first section, the details of the simulation are discussed in the second section
and the result of the simulation is presented in the last part.

\section{The Process}
\label{chap-5-1}

As the Three-Site Model is designed to mimic the Standard Model up to small corrections as far as
the Standard Model particles are concerned, the most prominent signatures of the model at the LHC
should be the resonances of the new heavy particles. While the LHC runs at an energy of $\unit[14]{TeV}$
in the proton center-of-mass system, the energy typically available for the partonic hard process is much
smaller due to the parton distributions and only in the range of several $\unit{TeV}$. Therefore, the
heavy fermions will be hard to detect as resonances due their large mass $\ge\unit[1.8]{TeV}$ (c.f.
chapter \ref{chap-7}), and the $W^\prime / Z^\prime$ are thus the preferred candidates for
direct detection.

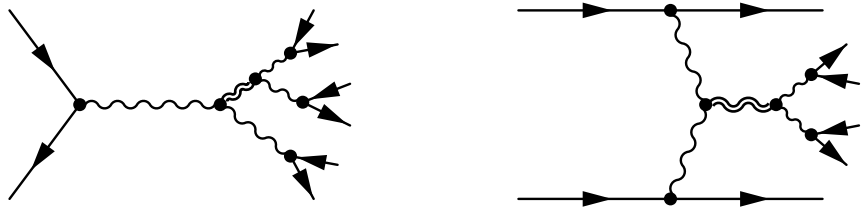
\begin{figure}[!tb]
\centerline{
\parbox{58mm}{\fmfframe(0,0)(8,0){\begin{fmfgraph*}(50,25)
\fmfleft{i2,i1}\fmfright{o6,o5,o4,o3,o2,o1}
\fmf{fermion}{i1,v1,i2}
\fmf{wiggly}{v1,v2}\fmf{phantom}{v2,v3}\fmf{wiggly}{v2,v4}\fmf{fermion}{o1,v3,o2}
\fmf{fermion}{o5,v4,o6}
\fmffreeze
\fmf{dbl_wiggly}{v2,v5}\fmf{wiggly}{v5,v3}
\fmffreeze
\fmf{wiggly}{v5,v6}\fmf{phantom}{v6,o4}
\fmffreeze
\fmf{fermion}{o3,v6,o4}
\fmfdot{v1,v2,v3,v4,v5,v6}
\end{fmfgraph*}}}
\parbox{58mm}{\fmfframe(8,0)(0,0){\begin{fmfgraph*}(50,25)
\fmfleft{i2,i1}\fmfright{o6,o5,o4,o3,o2,o1}
\fmf{fermion}{i1,v1,o1}\fmf{fermion}{i2,v2,o6}
\fmffreeze
\fmf{wiggly}{v1,v3,v2}\fmf{dbl_wiggly}{v3,v4}\fmf{wiggly}{v4,v5}
\fmf{phantom}{v5,o2}\fmf{wiggly}{v4,v6}\fmf{phantom}{v6,o5}
\fmffreeze
\fmf{fermion}{o3,v5,o2}\fmf{fermion}{o4,v6,o5}
\fmfdot{v1,v2,v3,v4,v5,v6}
\end{fmfgraph*}}}}
\caption{Diagrams contribution to the signal for the production of KK gauge bosons in strahlung type
(\emph{left}) and fusion type (\emph{right}) processes.}
\label{fig-5-1-diags}
\end{figure}

The couplings between the heavy gauge bosons and the Standard Model fermions are suppressed due to
their fermiophobic nature, and diagrams containing KK fermions are suppressed with
the bulk mass $M_\text{bulk}$ at the $W^\prime/Z^\prime$ mass scale.
Therefore, the processes that first
spring into mind as promising candidates for producing the heavy gauge bosons on-shell involve the
coupling of the KK particle to a $W$ or $Z$ line. Of this type there are basically two kinds of
processes which are shown in fig.~\ref{fig-5-1-diags}, namely the radiation from SM gauge bosons or
the fusion of two SM gauge bosons.

As shown in chapter \ref{chap-3-4}, the heavy gauge bosons decay to nearly $100\%$ into Standard
Model gauge bosons which then subsequently decay into fermions. Excluding all final states with more
than two jets due to the large QCD backgrounds and all final states with more than one neutrino due
to the missing momentum information, the following final states suitable for $W^\prime /
Z^\prime$ production remain:\\[2ex]
\centerline{\begin{tabular}{|c|c|c|c|}
\hline \multicolumn{2}{|c|}{strahlung} & \multicolumn{2}{|c|}{fusion} \\\hline
$W^\prime$ & $Z^\prime$ & $W^\prime$ & $Z^\prime$ \\\hline\hline
\parbox[t][8ex]{10ex}{$jj+4l$\\$jj+3l+\nu$\\$5l+\nu$} & --- & $jj+3l+\nu$ & ---
\\\hline
\end{tabular}}\\[2ex]
Thus, $W^\prime$ strahlung seems to be the most promising candidate as it can be studied
without involving any neutrinos, the missing momentum information of which would have to be made up
for with tricks like using the transverse mass \cite{Bagger:1995mk} instead of the invariant one or
reconstructing the neutrino momentum from kinematic relations (see chapter \ref{chap-6}).

The only source of notable contributions from the additional structure in
the Three-Site Model to this process is the gauge sector, and therefore, the $W^\prime$ mass
$m_{W^\prime}$ is the only free parameter which has an influence on the
result of the simulation. In particular, we are free to perform the simulations in the ideally
delocalized scenario and can infer the dependence of the result on parameter space solely from
varying $m_{W^\prime}$\footnote%
{
Tuning $\epsilon_L$ away from ideal delocalization in principle allows for an additional
contribution from a $W^\prime$ in the $s$ channel, the coupling of which to the radiated $W^\prime$
is larger than that of the $W$. However, the coupling of the $W^\prime$ to the Standard Model
fermions is still so small in this that the resulting change to the cross section in the peak region
is only about $20\%$ at best.
}.

In the context of this thesis, a simulation of the $W^\prime$ strahlung process via the $jj+4l$
final state was performed using WHIZARD / O'Mega and the implementation of the Three-Site Model
presented in chapter \ref{chap-4}. Unfortunately, after the simulation was started, an already
finished study of $W^\prime$ production via strahlung and fusion was published in \cite{He:2007ge}.
However, the simulation performed in the context of this thesis still is of some value as it
simulates the full six particle partonic final state using unweighted events, while the simulation
performed in above reference uses CalcHEP and only simulates the four particle intermediate state $jjZZ$
using weighted events.

\section{Simulation Details}
\label{chap-5-2}

In order to study the visibility of the $W^\prime$ resonance at the LHC, we have performed a full
parton level simulation of the process $pp\longrightarrow jj+4l$ with
\begin{equation}\label{equ-5-2-passign}
p\in\left\{g,u,d\right\} \quad,\quad j\in\left\{g,u,d,c,s,b\right\} \quad,\quad
l\in\left\{e^-,\mu^-\right\}
\end{equation}
(and any of the respective antiparticles) for a total energy of $\unit[14]{TeV}$ in the
proton CMS. Unweighted events have been generated for an integrated luminosity of
$\ilum = \unit[100]{fb^{-1}}$.
The parton distributions were taken from the CTEQ6M
series \cite{Pumplin:2002vw} and evaluated at the parton CMS energy in a running scale scheme.

All diagrams contributing to the signal are of the form
\[
\fmfframe(3,3)(3,3){\begin{fmfgraph*}(55,28)
\fmfleft{i2,i1}\fmfright{o6,o5,o4,o3,o2,o1}
\fmf{fermion}{i1,v1,i2}
\fmf{wiggly,la=$W$,la.si=left}{v1,v2}
\fmf{phantom}{v2,v3}\fmf{wiggly,la=$Z$,la.si=right,la.di=0.5thick}{v2,v4}
\fmf{fermion}{o1,v3,o2}\fmf{fermion}{o5,v4,o6}
\fmffreeze
\fmf{dbl_wiggly,la=$W^\prime$,la.si=left,la.di=0.5thick}{v2,v5}
\fmf{wiggly,la=$W$,la.si=left,la.di=0.5thick}{v5,v3}
\fmffreeze
\fmf{wiggly,la=$Z$,la.si=right,la.di=thick,la.an=-45}{v5,v6}\fmf{phantom}{v6,o4}
\fmffreeze
\fmf{fermion}{o3,v6,o4}
\fmfdot{v1,v2,v3,v4,v5,v6}
\fmfv{la=$q$}{i1}\fmfv{la=$\bar{q}$}{i2}\fmfv{la=$l^-$}{o6}
\fmfv{la=$l^+$}{o5}\fmfv{la=$l^-$}{o4}\fmfv{la=$l^+$}{o3}\fmfv{la=$j$}{o2,o1}
\end{fmfgraph*}}
\]
(the three-point couplings of the photon don't mix KK modes) and therefore, invariant mass cuts can
be used to reduce the non-resonant contributions to the background. In particular, we have applied a
cut
\[ \unit[60]{GeV}\le m_{jj} \le\unit[100]{GeV} \]
on the invariant mass of the jet pair and a cut
\[ \unit[71]{GeV}\le m_{ll} \le\unit[111]{GeV} \]
on the invariant mass of the lepton pairs, trying both possible combinations in the case of four
leptons of the same generation and discarding events where no unique identification of the $Z$
bosons could be achieved this way.

To further suppress the background, a $p_T$ cut of
\[ p_T\ge\unit[20]{GeV} \]
was imposed on the final state particles. Another cut
\[ -0.99\le\cos\theta\le 0.99 \]
was applied to the polar and intermediate angles of all final state particles, and in order to avoid infrared
singularities, we imposed the condition
\[ E \ge \unit[10]{GeV} \]
on the energy of the incoming partons which corresponds to a small $x$ cut
\[ x\ge 1.4\cdot10^{-3} \]
($x$ being the fraction of the proton momentum which is carried by the parton).

As there is no way of determining which $Z$ originated from the $W^\prime$, we have counted both
possible combinations into the histograms similar to the analysis in \cite{He:2007ge}. Simulations have
been performed in the ideally delocalized scenario at three points in parameter space
\begin{equation}\label{equ-5-2-pspoints}
\left(m_{W^\prime},M_\text{bulk}\right) \in \left\{\left(\unit[380]{GeV},\unit[3.5]{TeV}\right),
\left(\unit[500]{GeV},\unit[3.5]{TeV}\right),\left(\unit[600]{GeV},\unit[4.3]{TeV}\right)\right\}
\end{equation}
where $M_\text{bulk}$ has been varied only for consistency in order to satisfy the
precision constraints as discussed in chapter \ref{chap-3-2}.

\section{Results and Conclusions}
\label{chap-5-3}

\begin{figure}[!tb]
\centerline{
\includegraphics[width=\doubleplotwidth,angle=270]{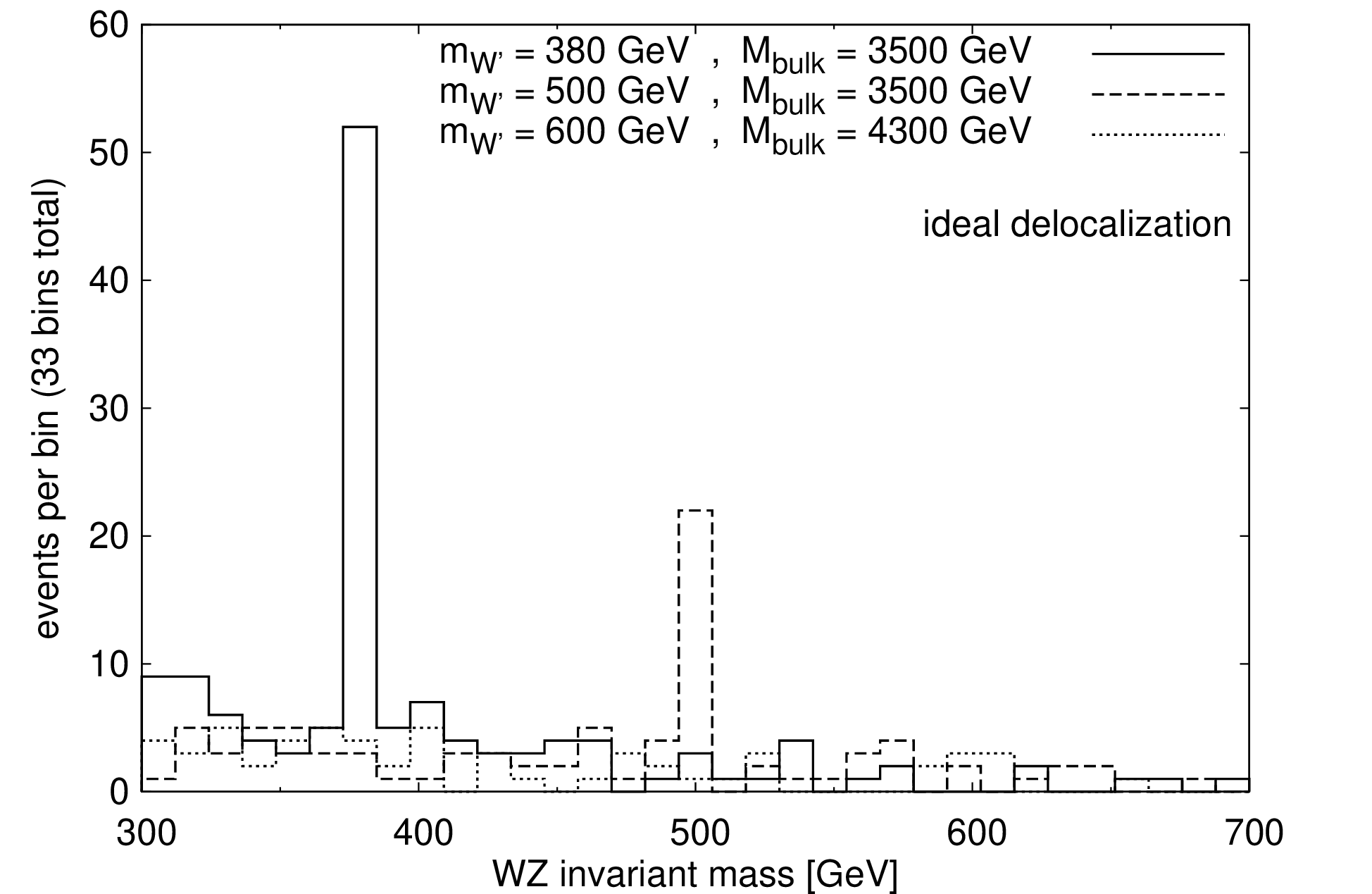}
\includegraphics[width=\doubleplotwidth,angle=270]{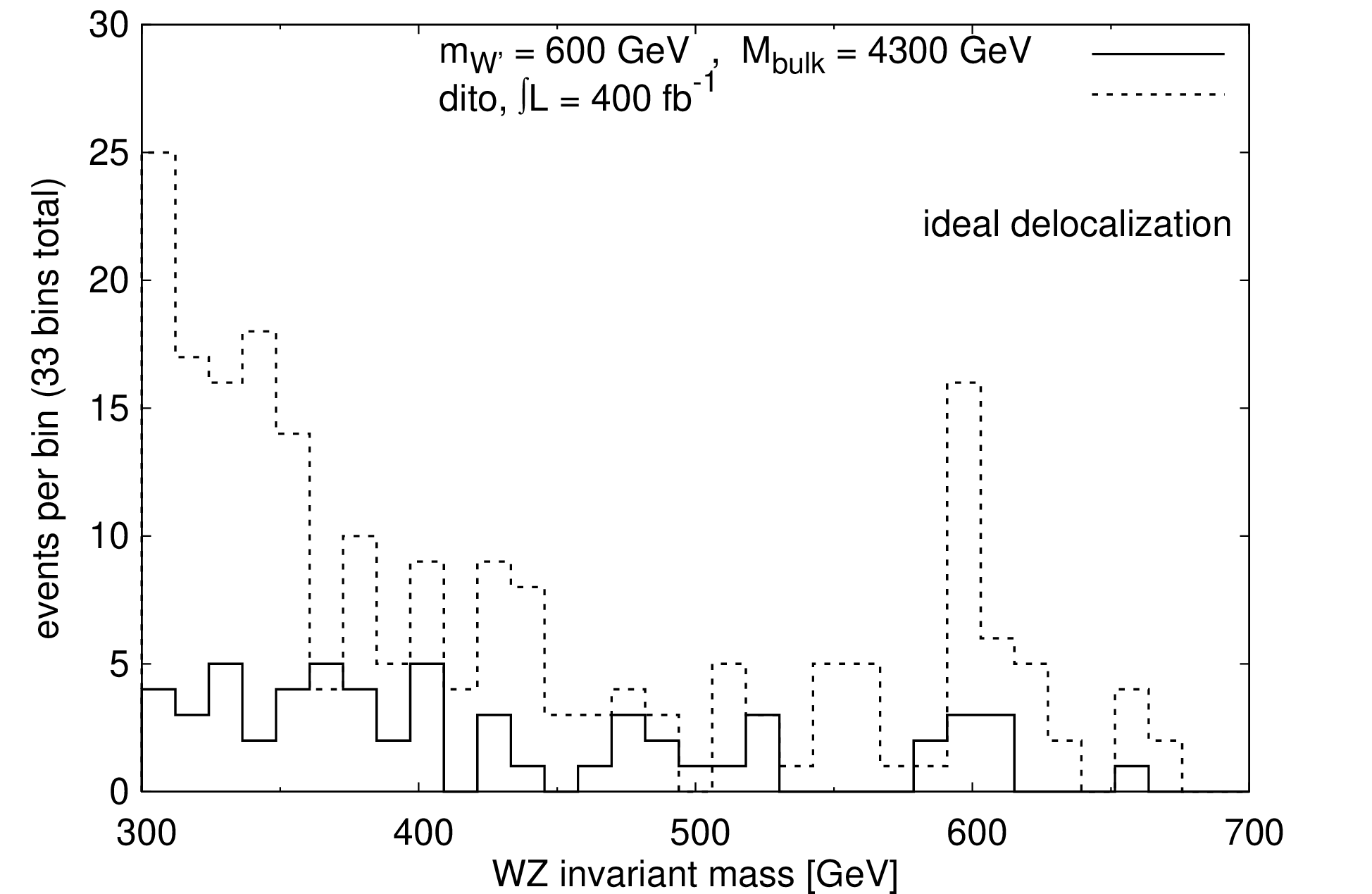}
}
\caption{\emph{Left}: $W^\prime$ resonance in the invariant mass of the $Z/W$ pair for the $W^\prime$
strahlung process; different values of $m_{W^\prime}$ and $\ilum=\unit[100]{fb^{-1}}$.
\newline
\emph{Right}: The $W^\prime$ resonance for $m_{W^\prime}=\unit[600]{GeV}$ and
$\ilum=\unit[100]{fb^{-1}}$, $\unit[400]{fb^{-1}}$.
}
\label{fig-5-3-histhw}
\end{figure}
Fig.~\ref{fig-5-3-histhw} left shows the $W^\prime$ resonance peaks in the invariant mass distribution of
the $W/Z$ pair obtained from the simulation as described above. For $m_{W^\prime}=\unit[380]{GeV}$
and $m_{W^\prime}=\unit[500]{GeV}$, the peaks are clearly visible above the background, while only a
very small bump is present in the data for $m_{W^\prime}=\unit[600]{GeV}$. This bump is reproduced
in fig.~\ref{fig-5-3-histhw} right together with the corresponding distribution for a higher
integrated luminosity of $\unit[400]{fb^{-1}}$. With this increased luminosity, the
$m_{W^\prime}=\unit[600]{GeV}$ resonance can also be clearly discerned.

In order to get a quantitative estimate of the significance of the resonances, let's define the
signal $N_s$ as the number $N$ of events in the $\pm\unit[20]{GeV}$ region around
the peak in the Three-Site Model minus the the same quantity $N_b$ in the Standard
Model.
\[ N_s = N - N_b \]
We can then define the statistical significance of the signal as the deviation from the
Standard Model relative to the standard deviation
\[ s = \frac{N_s}{\sigma_{N_b}} = \frac{N - N_b}{\sqrt{N_b}} \]
which scales like $\sqrt{\ilum}$.

\begin{figure}[!tb]
\centerline{\includegraphics[width=\singleplotwidth,angle=270]{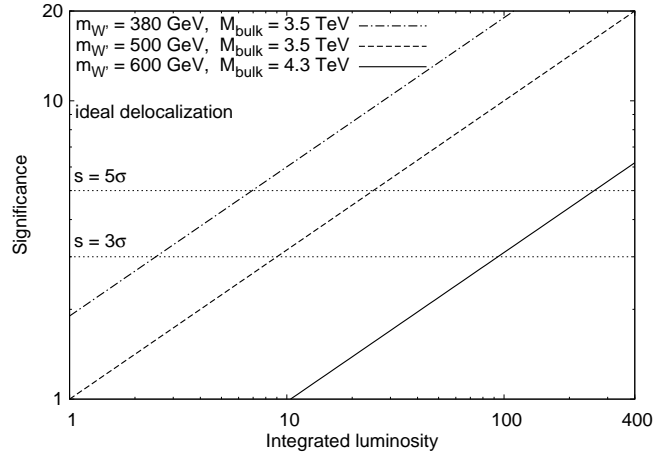}}
\caption{The statistical significance of the $W^\prime$ resonance peaks in $W^\prime$ strahlung as a
function of the integrated luminosity together with $3\sigma$ and $5\sigma$ discovery thresholds.}
\label{fig-5-3-signhw}
\end{figure}
For the determination of the background $N_b$, we have performed a Standard Model simulation of the
process for an integrated luminosity of $\ilum=\unit[4000]{fb^{-1}}$ and downscaled the result by
a factor of $40$ in order to reduce the error on the significance induced by uncertainties in the
background estimate. Fig.~\ref{fig-5-3-signhw} shows the resulting
significance as a function of the integrated luminosity together with the $3\sigma$ and $5\sigma$
discovery thresholds. The result for $m_{W^\prime}=\unit[380]{GeV}$, $\unit[500]{GeV}$
looks encouraging with a $5\sigma$ discovery being possible within the first $\unit[7]{fb^{-1}}$
resp. $\unit[25]{fb^{-1}}$. For $m_{W^\prime}=\unit[600]{GeV}$ however, things are different, and
nearly $\unit[300]{fb^{-1}}$ would be required for a $5\sigma$ discovery in this channel, which is
(hopefully) at least still within LHC running time.

\begin{figure}[!tb]
\centerline{
\includegraphics[width=6.3cm]{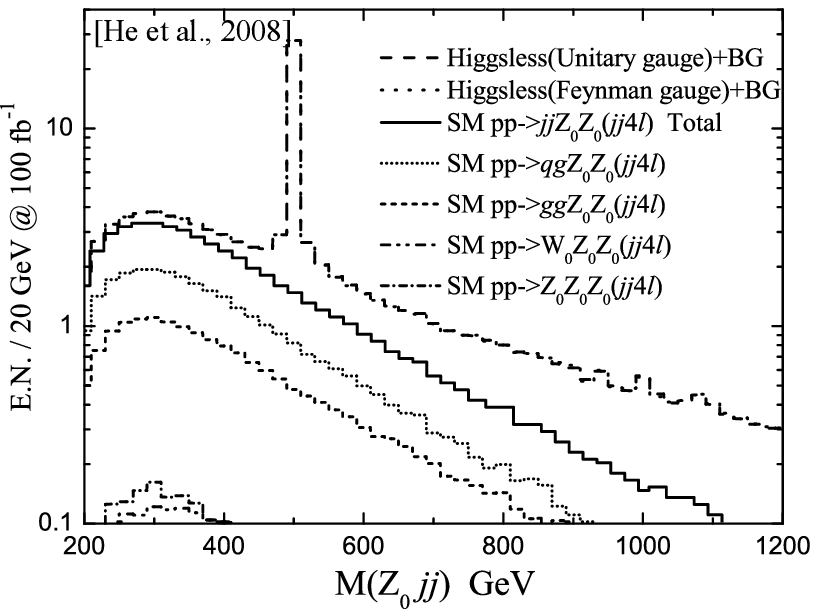}
\hspace{5mm}
\includegraphics[width=6.75cm]{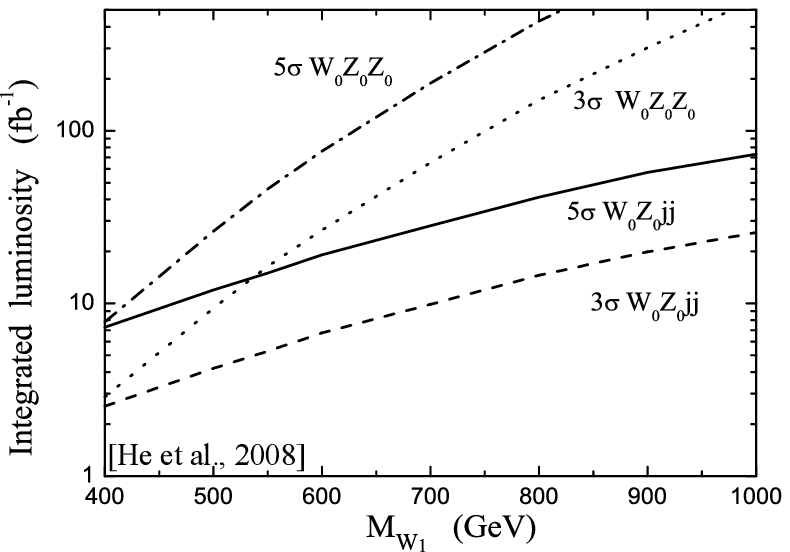}
}
\caption{
\emph{Left}: The $W^\prime$ resonance peak in the $W^\prime$ strahlung process
for $m_{W^\prime}=\unit[500]{GeV}$ as simulated in
\cite{He:2007ge}, plot taken from the same reference.
\newline
\emph{Right}: The integrated luminosity required for a $3\sigma$ resp $5\sigma$ discovery as a
function of the $W^\prime$ mass as obtained in \cite{He:2007ge}, plot taken from the same reference.
$W_0Z_0Z_0$ refers to the $W^\prime$ strahlung process, while $W_0Z_0jj$ refers to the fusion
process (see chapter \ref{chap-5-1}).
}
\label{fig-5-3-he}
\end{figure}

These results should be compared to the results obtained from simulating the $ZZjj$ intermediate
state in \cite{He:2007ge}. To this end, the $W^\prime$ resonance for
$m_{W^\prime}=\unit[500]{GeV}$ published by these authors is reprinted in fig.~\ref{fig-5-3-he} left.
The number of events contained in the peak is about $30$ which is slightly smaller than the same
number in our simulation ($50$ events, c.f. fig.~\ref{fig-5-3-histhw} left). However, as the
CalcHEP implementation of the model used in \cite{He:2007ge} and our WHIZARD / O'Mega implementation
completely agree on the $2\rightarrow 2$ process $pp\rightarrow W^\prime Z$ and have also been validated against
each other in numerous other processes, this discrepance is not overly disturbing and can be
attributed to differences in the cuts, parameters and to the simulation of an intermediate state
in \cite{He:2007ge}.

Reprinted in fig.~\ref{fig-5-3-he} right is the integrated luminosity required for a $3\sigma$ or
$5\sigma$ discovery as a function of the $W^\prime$ mass as estimated in \cite{He:2007ge}, the
$W_0Z_0Z_0$ line referring to the $W^\prime$ strahlung process under consideration. The $5\sigma$
results are $\unit[10]{fb^{-1}}$, $\unit[20]{fb^{-1}}$ for $m_{W^\prime}=\unit[380]{fb^{-1}}$,
$\unit[500]{fb^{-1}}$ and roughly consistent with our results quoted above. However, according to
the authors of \cite{He:2007ge}, a $5\sigma$ discovery for $m_{W^\prime}=\unit[600]{GeV}$ should
be possible within $\unit[100]{fb^{-1}}$ which is much more optimistic than our estimate of
$\unit[300]{fb^{-1}}$.

Unfortunately, \cite{He:2007ge} gives no details about how fig.~\ref{fig-5-3-he} right was obtained and
whether the higher mass region of the curve reflects the results of actual simulations or is an
extrapolation. However, the strong decay of the signal observed in our simulations when going from
$m_{W^\prime}=\unit[380]{GeV}$ to $m_{W^\prime}=\unit[600]{GeV}$ which leads to the high luminosity
required for discovery is completely consistent with the result from performing a simulation of
$pp\rightarrow W^\prime Z$.

Concluding the discussion of our $W^\prime$ strahlung simulation, the discovery of the $W^\prime$ in
this process at the LHC would be possible for the whole segment of parameter space shown to be
compatible with the electroweak precision observables in \cite{Abe:2008hb}. However, depending on
the $W^\prime$ mass, a huge amount of data taking might be necessary before any conclusive result is
within reach. Another disadvantage of this process is that it only probes the electroweak structure
of the Three-Site Model which coincides with other models, e.g. BESS \cite{Casalbuoni:1985kq}. In
order to probe the fermion sector of the Three-Site Model which distinguishes it from other scenarios,
other processes like the $s$ channel production discussed in the next chapter must be explored.

\chapter{KK Gauge Bosons in the $s$ Channel}
\label{chap-6}

In the previous chapter, a study of $W^\prime$ strahlung was presented. The major drawback of
this process is that it only probes the gauge sector and doesn't reveal any information about the
fermion sector. In sharp contrast, the production of the $W^\prime$ and $Z^\prime$ in the $s$ channel is
highly sensitive to the structure of the fermion sector due to the fermiophobic nature of the
couplings between the KK gauge bosons and the light fermions. In this chapter, we discuss the
prospects of exploiting this kind of process at the LHC.

After presenting an overview over the different channels of this type that might be accessible at
the LHC in \ref{chap-6-1}, we move on to the details of our simulations in
\ref{chap-6-2}. In \ref{chap-6-3} and \ref{chap-6-4} respectively, the results
for $Z^\prime$ and $W^\prime$ production are discussed. It turns out that there are contributions
from both
$W^\prime$ and $Z^\prime$ to $jjl\nu$ type final states, and a possible strategy do disentangle
these is presented in \ref{chap-6-5}. The chapter is concluded in \ref{chap-6-6}, and \ref{chap-6-7}
contains a small collection of additional plots which have been omitted from the other sections for the
sake of clarity.

Most of the work presented in this chapter is also covered in \cite{Ohl:2008ri}. The results
presented here at the parton level were supplemented by a detector simulation in the master's thesis
of F. Bach \cite{fabian:master}.

\section{The Processes}
\label{chap-6-1}

The main motivation behind the Three-Site Model is the delay of unitarity violation
without introducing a Higgs while still satisfying the precision constraints imposed by the LEP /
LEP-II data. As discussed in chapters \ref{chap-1} and \ref{chap-2}, this implies strong constraints
on the couplings between the KK gauge bosons (which are responsible for delaying unitarity
violation) and the Standard Model fermions in this scenario. Therefore, if the LHC were to discover
new resonances compatible with a $W^\prime$ and a $Z^\prime$, further tests of the fermion sector
would be necessary in order to probe for such a type of new physics.

While sufficient for
the bare discovery of the $W^\prime$ resonance, the $W^\prime$ strahlung process discussed in
chapter \ref{chap-5} does not probe the structure of the fermion sector and needs to be supplemented
by other channels more sensitive to this part of the Three-Site Model. At a hadron collider,
the simplest process involving the coupling of the $W^\prime/Z^\prime$ to the light fermions is the
production of these particles in the $s$ channel. As the KK gauge bosons decay to nearly $100$
percent into light gauge bosons (c.f. chapter \ref{chap-3-4}), the signal in this kind of process
is provided by diagrams of the type
\[
\fmfframe(0,3)(0,3){\begin{fmfgraph}(45,21)
\fmfleft{i2,i1}\fmfright{o4,o3,o2,o1}
\fmf{fermion}{i1,v1,i2}\fmf{dbl_wiggly}{v1,v2}\fmf{wiggly}{v3,v2,v4}
\fmf{fermion}{o1,v3,o2}\fmf{fermion}{o3,v4,o4}
\fmfdot{v1,v2,v3,v4}
\end{fmfgraph}}
\]
which involve one order of the heavily constrained couplings between light fermions and KK gauge
bosons. Depending on the magnitude of those couplings, the total
invariant mass of such final states should
exhibit a $W^\prime/Z^\prime$ resonance, the magnitude of which depends on the
fermion (de)localization.

\begin{figure}[!tb]
\centerline{\includegraphics[width=\singleplotwidth,angle=270]{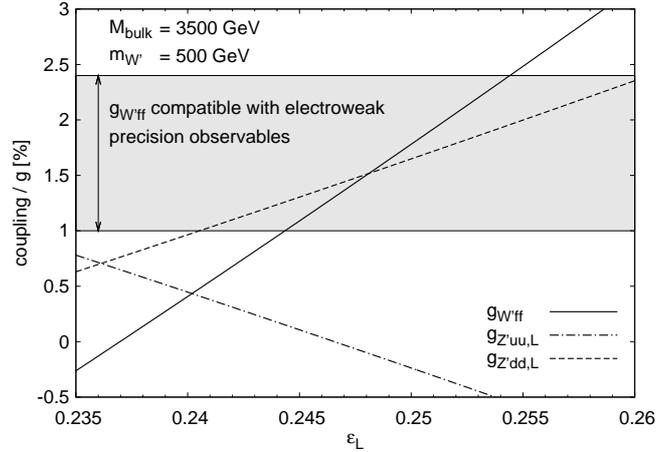}}
\caption{$g_{W'ff}$, $g_{Z'uu,L}$ and $g_{Z'dd,L}$ normalized to the site $0$ gauge
coupling $g$ as a function of the delocalization parameter $\epsilon_L$.
The gray rectangle marks the range for $g_{W^\prime ff}$ allowed by the precision observables as
derived by the authors of~ \cite{Abe:2008hb}.}
\label{fig-6-1-cpl}
\end{figure}
As discussed in chapter \ref{chap-3-3}, the relevant couplings are restricted to be nonvanishing at the
order of several percent of the isospin gauge coupling. Fig.~\ref{fig-6-1-cpl} shows the coupling
of the left-handed\footnote%
{
Recall that in the case of (approximately) massless fermions, the right-handed couplings to the $W^\prime$
vanish and those to the $Z^\prime$ are independent of $\epsilon_L$ (c.f. chapter \ref{chap-3-3}).
}
light quarks to the $W^\prime$ and to the $Z^\prime$ for $m_{W^\prime}=\unit[500]{GeV}$ as a
function of the fermion delocalization parameter $\epsilon_L$ (c.f. chapter \ref{chap-3-2}), the
corresponding plots for $m_{W^\prime}=\unit[380]{GeV}$, $\unit[600]{GeV}$ can be found in fig.
\ref{fig-6-7-cpl}. Remarkably, the left-handed coupling of the $Z^\prime$ to the up quark
$g_{Z^\prime uu,L}$ and that to the down quark $g_{Z^\prime dd,L}$
change in exactly opposite ways under a variation of the delocalization parameter. This is no
accident and can be understood by using the normalization of the fermion wavefunction to recast the
expression for the $Z^\prime ff$ type couplings \eqref{equ-3-3-gzf} as
\begin{equation}\label{equ-6-1-gzf}
g_{Z^\prime ff,L} = \pm\frac{1}{2}\left(gf^{Z^\prime}_0\left(f_{0,L}^f\right)^2
+ \tilde{g}f^{Z^\prime}_1\left(f_{1,L}^f\right)^2\right) + Yg^\prime f^{Z^\prime}_2
\end{equation}
where $Y$ is the hypercharge and the sign depends on the isospin of the fermion (recall that the only
source of
a dependence of \eqref{equ-6-1-gzf} on $\epsilon_L$ is the wavefunction of the left-handed fermion
$f^f_L$).

As the proton contains both up- and down quarks (albeit in the ratio $2:3$) and $g_{Z^\prime
uu,L}$ and $g_{Z^\prime dd,L}$ both start with
a positive value at the point of ideal delocalization, the effect of tuning $\epsilon_L$ away from
nonideal delocalization partially cancels out from the $Z^\prime$ production cross section.
In addition, there are also right-handed couplings of the
same order of magnitude which don't depend on $\epsilon_L$ at all and therefore, we shouldn't expect
$Z^\prime$ production in the $s$ channel to be very
sensitive to $\epsilon_L$. $W^\prime$ production, however, is completely forbidden in the case of
ideal delocalization and therefore highly sensitive to the value of $\epsilon_L$.

The partonic final state of the processes is reached through a cascade of the KK gauge boson going
into two Standard Model bosons which then in turn decay into two fermion pairs. Excluding final
states with four jets due to the QCD background and final states with more than one neutrino due to
the missing momentum information, the following final states remain:
\\[2ex]
\centerline{\begin{tabular}{|c|c|}
\hline
$W^\prime$ & $Z^\prime$
\\\hline\hline
\begin{minipage}[t][8ex]{6ex}\begin{center}$jjl\nu$\\$jjll$\\$lll\nu$\end{center}\end{minipage} &
\begin{minipage}[t]{6ex}\begin{center}$jjl\nu$\end{center}\end{minipage}
\\\hline
\end{tabular}}
\\[2ex]
Evidently, the only such state suitable for the detection of the $Z^\prime$ resonance consists of
two jets, a lepton and a neutrino, while we can choose from three different final states in the
case of the $W^\prime$, $jjl\nu$ featuring the biggest branching ratio and $lll\nu$ the smallest
one.

\section{Simulation Details}
\label{chap-6-2}

If we want to have any chance of observing the $Z^\prime$ in the $s$ channel or to take advantage of
all three final states suitable for the observation of the $W^\prime$, we have to cope with final
states containing one neutrino. Because the neutrino escapes the detector without leaving any traces,
we have no direct information on its momentum which we require in order to measure the total
invariant mass of the final state. Therefore, some trickery is necessary either to reconstruct the
missing information or to live without it.

The problem of missing momentum information is well known e.g. from SUSY decay chains where the
transverse mass \cite{Bagger:1995mk} is usually used instead of the invariant one in order to
observe a resonance without requiring the full momentum information. However, our attempt to apply
the transverse mass to the $Z^\prime/W^\prime$ production processes under consideration was rather
fruitless, the resonance in this observable being almost completely washed out.

Luckily, this is not the only trick available for dealing with the missing neutrino momentum. Even
though we cannot observe the particle itself, we still can infer the projection $\vec{p}_\perp$ of
its momentum on the transverse plane from transverse momentum conservation (assuming that the whole
missing transverse momentum $p_{T,\text{miss}}$
originates from the neutrino). The mass shell conditions of neutrino and $W$ boson (assuming that
the neutrino originates from a $W$ approximately on-shell) then give two additional conditions for
the energy $p_0$ and the longitudinal momentum component $p_L$
\begin{subequations}
\begin{align}
\label{equ-6-2-mshellnu}
  p_0^2 - p_L^2 - \left|\vec p_\perp\right|^2 &= 0 \\
\label{equ-6-2-mshellw}
  p_0 q_0 - p_L q_L - \vec p_\perp \vec q_\perp &= \frac{m_W^2}{2}
\end{align}
\end{subequations}
(with the momentum $q$ of the corresponding lepton and the $W$ mass $m_W$).

\begin{figure}[!tb]
\centerline{\includegraphics[width=\singleplotwidth,angle=270]{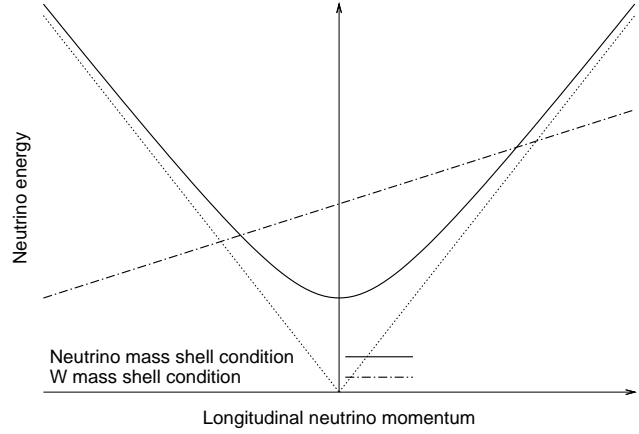}}
\caption{The two curves generated by the mass shell conditions for $W$ and neutrino in the
case of a $W$ decaying to $l\nu$. The two points of intersection give the two possible solutions
for the neutrino momentum.}
\label{fig-6-2-nurec}
\end{figure}
\eqref{equ-6-2-mshellnu} describes a hyperbola in the $p_L$ - $p_0$ plane, while
\eqref{equ-6-2-mshellw} describes a straight line as depicted in fig.~\ref{fig-6-2-nurec}.
These curves can have up to two points of intersection, one of which corresponds to the correct
momentum of the neutrino. The two solutions can be obtained analytically as
\begin{equation*}
p_0 = \frac{q_0^2\left(m_W^2 + 2\vec{p}_\perp\vec{q}_\perp\right) \pm q_L A}
	{2q_0\left(q_0^2  - q_L^2\right)} \quad,\quad
p_L = \frac{q_L\left(m_W^2 + 2\vec{p}_\perp\vec{q}_\perp\right) \pm A}
	{2\left(q_0^2 - q_L^2\right)}\,,
\end{equation*}
with the abbreviation
\[
A = q_0\sqrt{\left(m_W^2 + 2\vec{p}_\perp\vec{q}_\perp\right)^2 +
	4\vec{p}_\perp^2\left(q_L^2 - q_0^2\right)} \,.
\]

It is easy to see from \eqref{equ-6-2-mshellw} that the modulus of the slope of the straight line is
always smaller than $1$
\[ \abs{\frac{q_L}{q_0}} = \frac{\abs{q_L}}{\sqrt{q_L^2 + \abs{\vec{q}_\perp}^2}} < 1 \]
while the asymptotes of the hyperbola \eqref{equ-6-2-mshellnu} have the slopes $\pm1$
\[ p_0 = \sqrt{p_L^2 + \abs{\vec{p}_\perp}^2} \rightarrow \abs{p_L} \]
Therefore, there will be always (excluding the rare case of straight line being a tangent of the
hyperbola) either two points of intersection or none at all, and
this method of reconstructing the neutrino momentum will always yield two solutions, only one of
which will usually be close to the correct value.

To cope with this, one can either try to find a
kinematical criterion for discriminating between the two solutions (at the price of losing part of
the signal) or count them both into the histograms. Of these two approaches, we have
chosen the latter as it does not waste any precious signal events. The price to pay,
however, is a doubling of the background.

Performing the actual simulation reveals that for roughly $10\%$ (a number quite sensitive to the
$W^\prime$ mass used in the reconstruction) of the events, the straight line
doesn't intersect the hyperbola at all but instead lies slightly below the minimum, missing it by a
small margin. Although we have chosen to simply remove these events from the histograms, more
elaborate ways could be devised to deal with this, for example choosing the point on the hyperbola
closest to the straight line or keeping $m_W$ as a free parameter which is determined for each
event such that \eqref{equ-6-2-mshellnu} -- \eqref{equ-6-2-mshellw} have one degenerate solution
and using it as a cut variable. A more in-depth discussion of this issue can
be found in \cite{fabian:master}.

In order to determine the discovery potential of the LHC in these processes at parton level, we have
performed a Monte-Carlo simulation similar to that presented in the last chapter using our
implementation of the model in WHIZARD / O'Mega with $\sqrt{s}=\unit[14]{TeV}$ and an integrated
luminosity of $\ilum=\unit[100]{fb^{-1}}$. We have performed a complete simulation of the partonic
processes $pp\rightarrow X$ with $X$ being one of $jjl\nu$, $jjll$ and $lll\nu$ (using the
assignments \eqref{equ-5-2-passign} and the same parton distributions as in chapter \ref{chap-5}).

\begin{figure}[!tb]
\centerline{
\fmfframe(3,3)(3,3){\begin{fmfgraph*}(45,21)
\fmfleft{i2,i1}\fmfright{o4,o3,o2,o1}
\fmf{fermion}{i1,v1,i2}\fmf{dbl_wiggly,la=$Z^\prime$,la.si=left}{v1,v2}
\fmf{wiggly,la=$W$,la.si=left,la.d=thick}{v2,v3}
\fmf{wiggly,la=$W$,la.si=right,la.d=2thick}{v2,v4}
\fmf{fermion}{o1,v3,o2}\fmf{fermion}{o3,v4,o4}
\fmfdot{v1,v2,v3,v4}
\end{fmfgraph*}}
\hspace{5mm}
\fmfframe(3,3)(3,3){\begin{fmfgraph*}(45,21)
\fmfleft{i2,i1}\fmfright{o4,o3,o2,o1}
\fmf{fermion}{i1,v1,i2}\fmf{dbl_wiggly,la=$W^\prime$,la.si=left}{v1,v2}
\fmf{wiggly,la=$W$,la.si=left,la.d=thick}{v2,v3}
\fmf{wiggly,la=$Z$,la.si=right,la.d=2thick}{v2,v4}
\fmf{fermion}{o1,v3,o2}\fmf{fermion}{o3,v4,o4}
\fmfdot{v1,v2,v3,v4}
\end{fmfgraph*}}
}
\caption{
\emph{Left}: General structure of the signal diagrams for $Z^\prime$ production in the $s$ channel.
\emph{Right}: Dito, but $W^\prime$ production.
}
\label{fig-6-2-diag}
\end{figure}
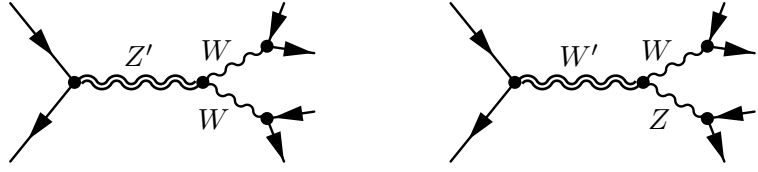
Fig.~\ref{fig-6-2-diag} shows the general structure of the diagrams contribution to the resonance
signal. As the final state fermions are decay products of Standard Model gauge bosons, we can
leverage cuts on the invariant mass $m_{ff}$ of the visible final state fermion pairs in order to reduce
the background\footnote%
{
Note that after the momentum reconstruction, the invariant mass $m_{l\nu}$ of the lepton-neutrino
pairs is equal to $m_W$ per definition, rendering an invariant mass cut on $m_{l\nu}$ pointless.
}.
More specifically, we demanded $m_{ff}$ to lie within
the $\pm\unit[5]{GeV}$ region around the invariant mass of the gauge boson which the fermions
are to reconstruct:
\\[2ex]
\centerline{\begin{tabular}{|c|c|}
\hline
\vs{3.5ex}$pp\rightarrow Z^\prime\rightarrow jjl\nu$ &
$\unit[75]{GeV}\le m_{jj}\le\unit[85]{GeV}$
\\\hline
\vs{3.5ex}$pp\rightarrow W^\prime\rightarrow jjl\nu$ &
$\unit[86]{GeV}\le m_{jj}\le\unit[96]{GeV}$
\\\hline
$pp\rightarrow W^\prime\rightarrow jjll$ &
\vs{7ex}\begin{minipage}{30ex}\begin{center}
$\unit[75]{GeV}\le m_{jj}\le\unit[85]{GeV}$\\
$\unit[86]{GeV}\le m_{ll}\le\unit[96]{GeV}$
\end{center}\end{minipage}
\\\hline
\vs{3.5ex}$pp\rightarrow W^\prime\rightarrow lll\nu$ &
$\unit[86]{GeV}\le m_{ll}\le\unit[96]{GeV}$
\\\hline
\end{tabular}}
\\[2ex]
Note that the narrow cut window of $\pm\unit[5]{GeV}$ is necessary in order to discriminate between
the $W^\prime$ and $Z^\prime$ contributions to $jjl\nu$ (as the masses of $Z^\prime$ and $W^\prime$
are quasidegenerate, the resonance peaks fall together). However, it is quite unclear if ATLAS or
CMS will be able to achieve enough precision in the measurement of the jet momenta for the
separation to work out this way. This issue is discussed at length in section \ref{chap-6-5}.

In the case of three leptons of the same generation in the final state, it is unknown
which of the leptons originates from a decaying $W$. Therefore, we tried both possible
combinations in this case and discarded the event if no unique identification of $Z$ and $W$ could
be achieved this way.

In addition to the invariant mass cuts, we also applied $p_T$ cuts of
\[ p_T \ge \unit[50]{GeV} \]
on the momenta of all visible particles and also on $p_{T,\text{miss}}$ in order to suppress the
background even more. Another cut of
\[ -0.95\le\cos\theta\le0.95 \]
was applied to the polar and intermediate angles of all visible particles, and as in the last
chapter, we demanded
\[ E \ge \unit[10]{GeV} \]
for the incoming partons in order to avoid infrared divergences.

For each final state, we performed simulations for the three parameter space points
\eqref{equ-5-2-pspoints}, choosing three different values for $\epsilon_L$ from the interval allowed
by \cite{Abe:2008hb} in addition to the point of ideal delocalization (c.f. fig.~\ref{fig-6-1-cpl}
and \ref{fig-6-7-cpl}) for each value of $m_{W^\prime}$. As in the case of $W^\prime$
strahlung, the influence of $M_\text{bulk}$ (which was varied only for consistency)
on the processes is much too small to be resolved, with the
results depending virtually only on $m_{W^\prime}$ and the fermion delocalization $\epsilon_L$.

\section{$Z^\prime$ Production}
\label{chap-6-3}

\begin{figure}[!tb]
\centerline{\begin{tabular}{cc}
\includegraphics[width=\doubleplotwidth,angle=270]{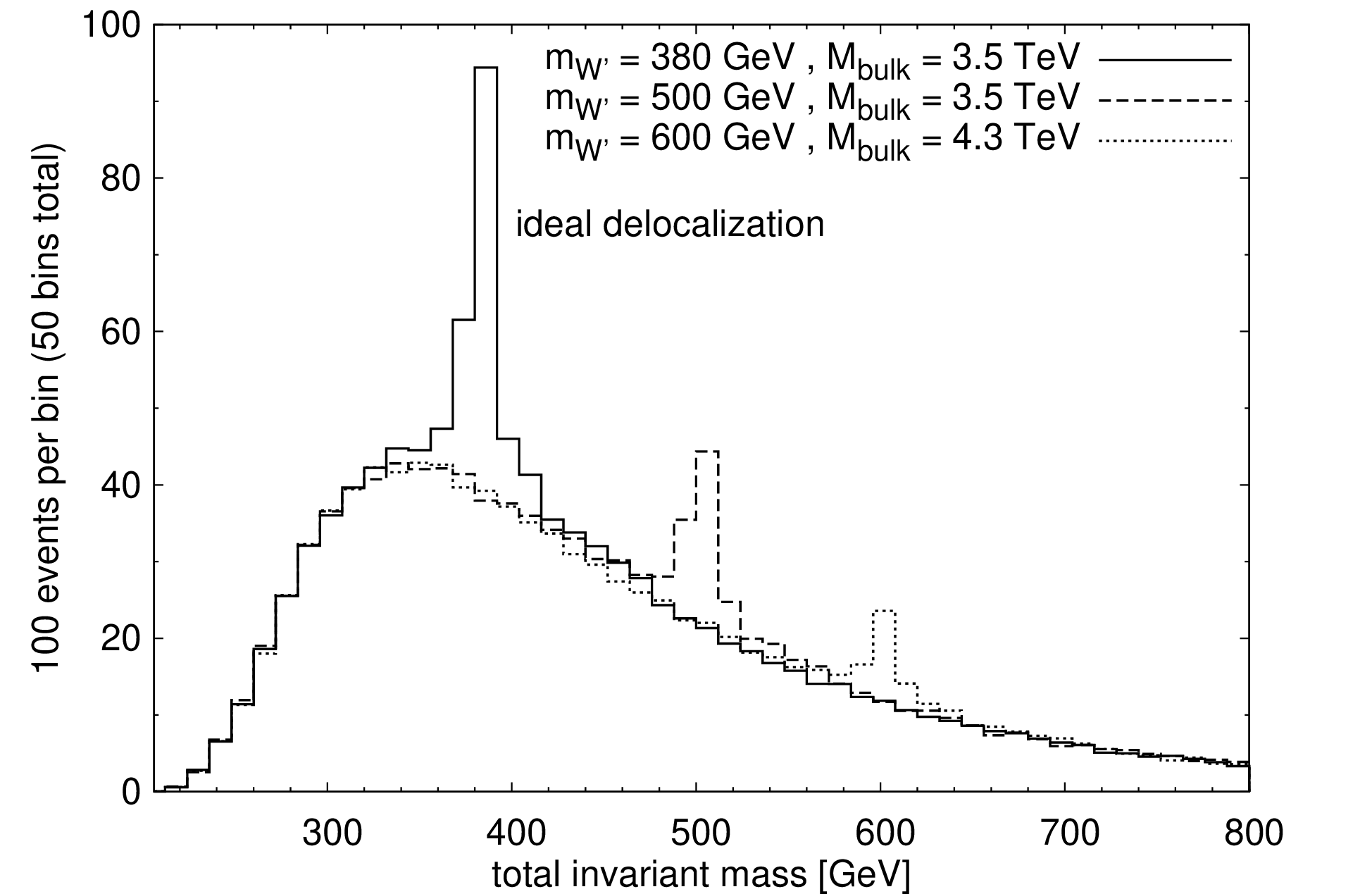}
\includegraphics[width=\doubleplotwidth,angle=270]{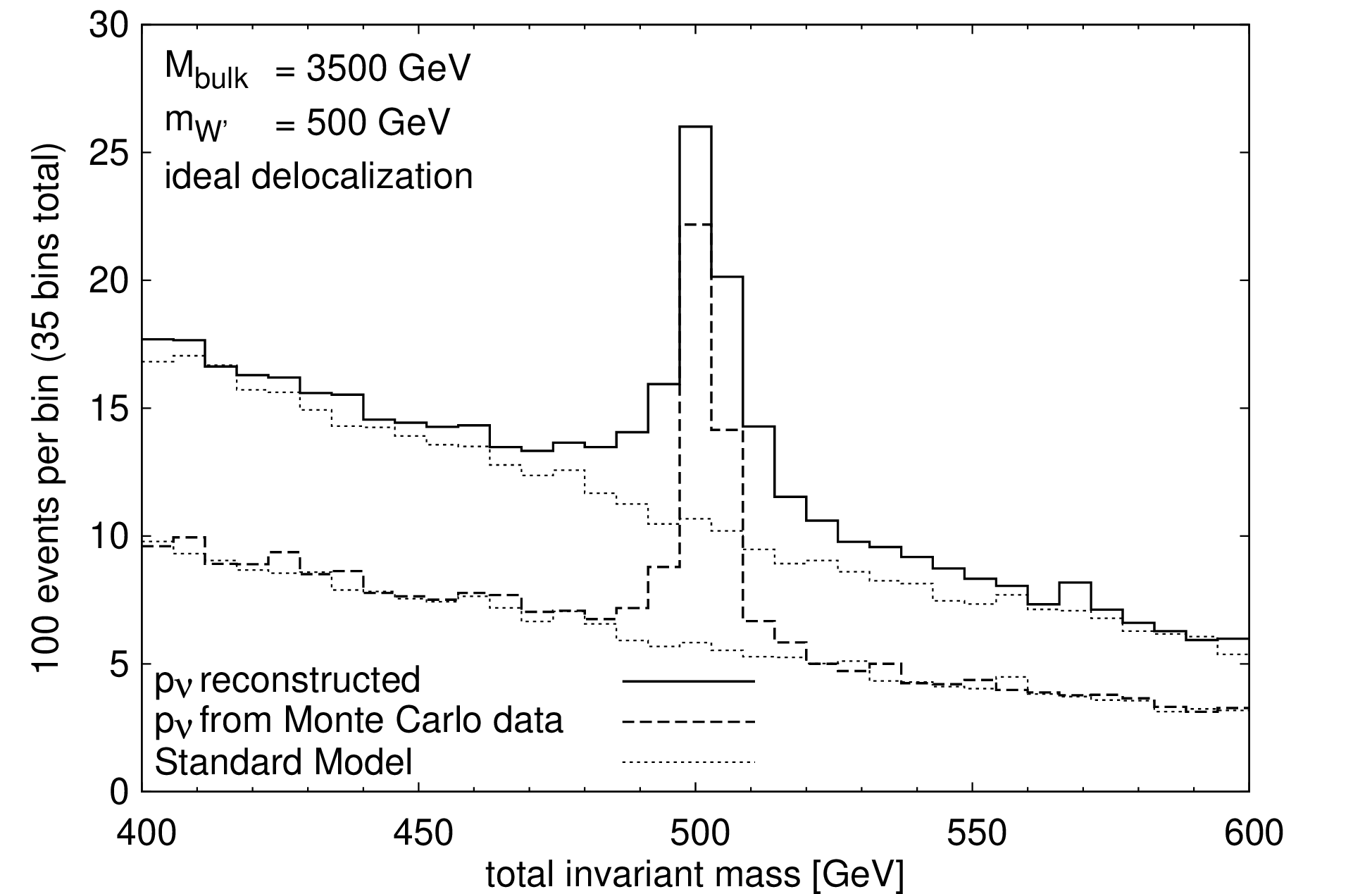} &
\end{tabular}}
\caption{
\emph{Left}: Invariant mass distribution in $pp\rightarrow jjl\nu$ for different values of
$m_{W^\prime}$ and $M_\text{bulk}$ ($Z^\prime$ production).
\newline
\emph{Right}: The $m_{W^\prime}=\unit[500]{GeV}$ peak obtained
from the reconstructed neutrino momenta vs. the corresponding distribution obtained directly from
Monte Carlo data.
}
\label{fig-6-3-nurec}
\end{figure}
Fig.~\ref{fig-6-3-nurec} left shows the invariant mass distributions for $pp\rightarrow
Z^\prime\rightarrow jjl\nu$ for $m_{W^\prime}=\unit[380]{GeV}$, $\unit[500]{GeV}$ and
$\unit[600]{GeV}$ (again, recall that $m_{W^\prime}\approx m_{Z^\prime}$) in the case of ideal
delocalization. For all three values of the mass, the resonance peaks are clearly visible, their
overall size declining with growing $W^\prime$ mass as expected from the behavior of the parton
distribution of the antiquark.

In order to check the reliability of the neutrino momentum reconstruction, fig.
\ref{fig-6-3-nurec} right compares the $m_{W^\prime}=\unit[500]{GeV}$ resonance peak obtained by
reconstructing the neutrino momenta to the distributions created from the exact momenta
taken directly from the eventgenerator.
As expected, the reconstruction roughly doubles the number of background events, while the size of
the peak remains more or less the same. However, the peak in the reconstructed data exhibits
broad ``sidebands'' which closer investigation reveals to originate from the second solutions to
the neutrino momentum in events within the peak.

\begin{figure}[!tb]
\centerline{
\includegraphics[width=\doubleplotwidth,angle=270]{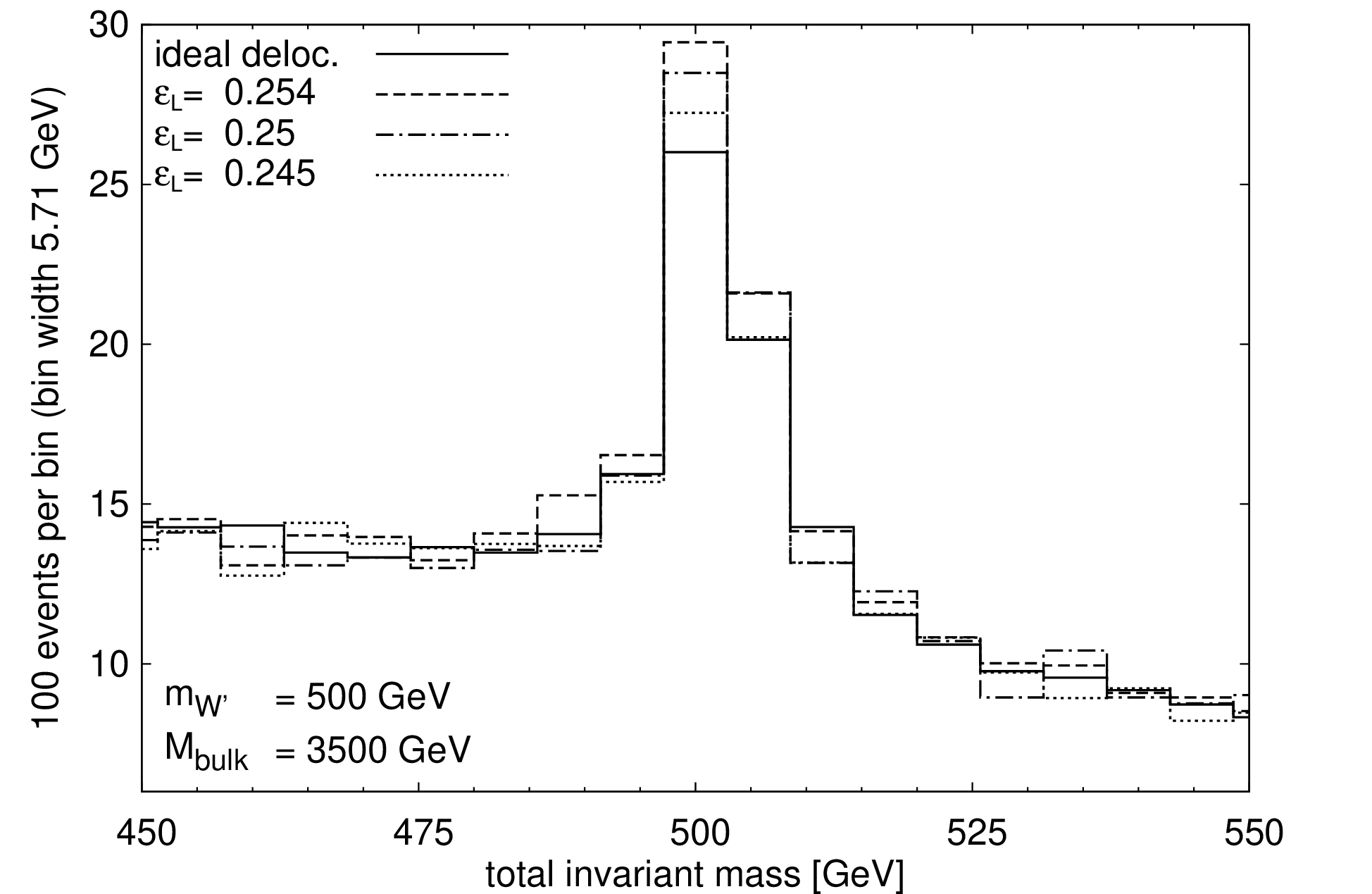}
\includegraphics[width=\doubleplotwidth,angle=270]{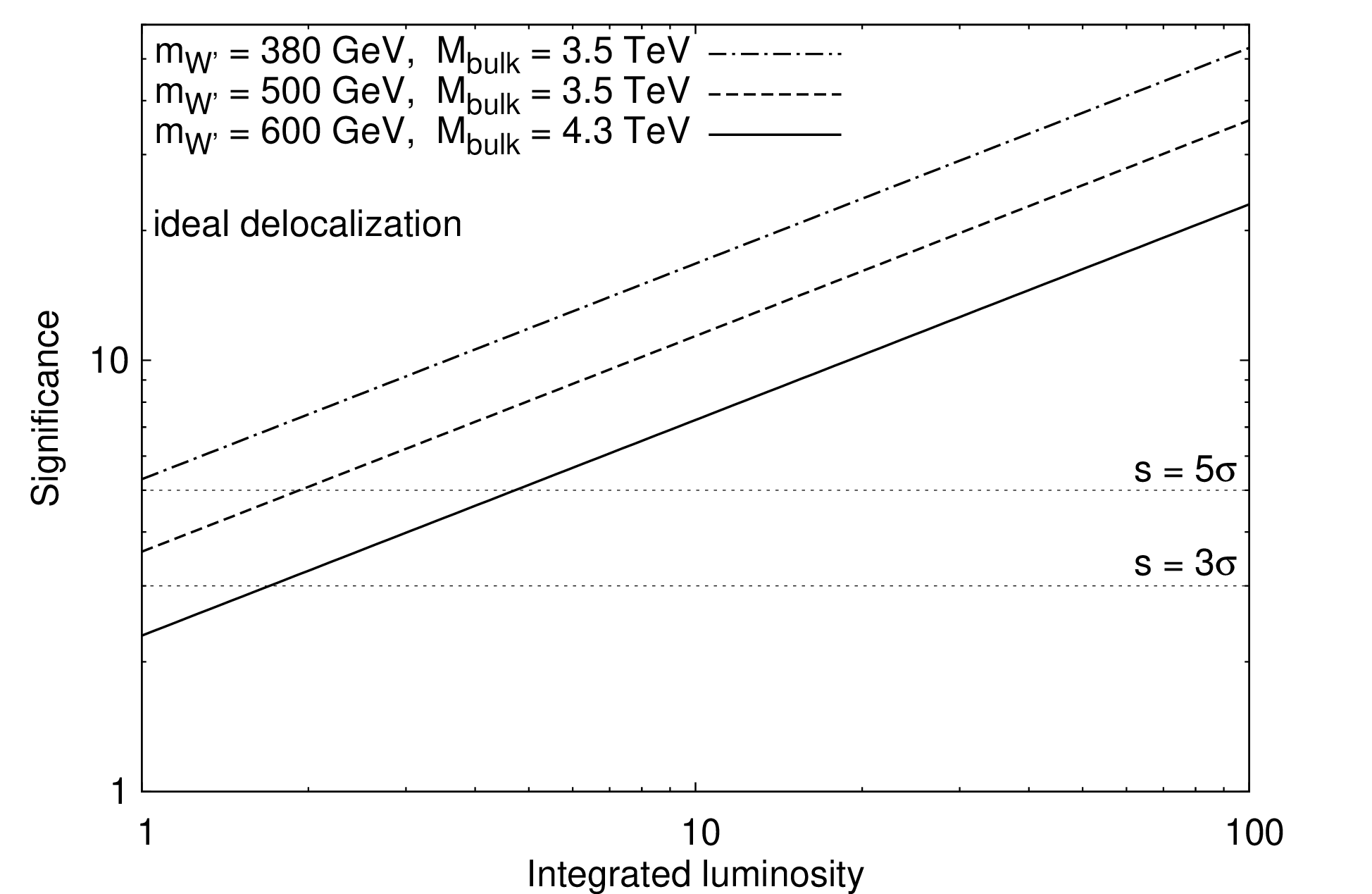}
}
\caption{
\emph{Left}: The effect of tuning $\epsilon_L$ away from
ideal delocalization (cf. fig.~\ref{fig-6-1-cpl}, \ref{fig-6-7-cpl}) on the $Z^\prime$ peak.
\newline
\emph{Right}: The significance as defined in the text as a function of the integrated
luminosity. The dotted lines mark the $3\sigma$ resp. $5\sigma$ discovery thresholds.
}
\label{fig-6-3-hz}
\end{figure}
The histograms in fig.~\ref{fig-6-3-nurec} have been generated for the case of ideal
delocalization in spite of the claim made in \cite{Abe:2008hb} (and discussed in chapter
\ref{chap-3-2})
that this case is in fact excluded by electroweak precision data. Instead,
$\epsilon_L$ must be tuned away from ideal delocalization in order to allow for a small but nonvanishing
$W^\prime ff$ coupling. The resulting variation of the $m_{W^\prime}=\unit[500]{GeV}$ resonance
is shown in fig.~\ref{fig-6-3-hz} left for three
different $\epsilon_L$ chosen from the allowed interval (c.f. fig.~\ref{fig-6-1-cpl}).

As argued in
\ref{chap-6-1}, this is rather small and causes a small
increase of the event count in the peak as compared to the case of ideal delocalization. Therefore, we
will confine ourselves to presenting the results for this scenario, with the effect caused by
varying $\epsilon_L$ being small and pushing in the direction of slightly larger significance.

To get a quantitative handle on the significance of the signal and to estimate the minimal
luminosity necessary for discovering the $Z^\prime$, let's proceed similarly to the analysis in chapter
\ref{chap-6-3} and define the raw signal $N$ to be the
number of events in the $\pm\unit[20]{GeV}$ region around the peak. In order to estimate the
background, we have generated SM events for an integrated luminosity of
$\ilum=\unit[400]{fb^{-1}}$, analyzed this data the in the same way
as the Monte-Carlo data for the Three-Site Model and then downscaled the resulting
distributions by a factor of $4$ to reduce the error coming
from fluctuations in the background. We denote the number of background events in the
$\pm\unit[20]{GeV}$ region around the peak obtained this way by $N_b$.

We then again define the signal $N_s$ as
\[ N_s = N - N_b \]
The determination of the standard deviation of the background is complicated by the fact that the
pairs of entries in the histograms for the different solutions to the neutrino momentum lead to a
nontrivial statistical correlation in the histograms. For simplicity, let's assume that the
background is simply doubled by the momentum reconstruction
\[ N_b = 2N^\prime_b \]
The standard deviation of $\sigma_{N_b}$ of $N_b$ must scale
accordingly
\[ \sigma_{N_b} = 2\sigma_{N^\prime_b} = 2\sqrt{N^\prime_b} = \sqrt{2N_b} \]
resulting in the statistical significance defined as in chapter \ref{chap-5-3} being reduced by a
factor of $\sqrt{2}$
\[ s = \frac{N_s}{\sigma_{N_b}} = \frac{N - N_b}{\sqrt{2N_b}} \]
which nevertheless scales like $\sqrt{\ilum}$.

The significance of the signal in the ideally delocalized scenario thus calculated is
shown in figure~\ref{fig-6-3-hz} right as a function of the integrated luminosity
together with the $5\sigma$ and $3\sigma$ discovery thresholds.
The $5\sigma$ thresholds are approximately $\unit[1]{fb^{-1}}$,
$\unit[2]{fb^{-1}}$, $\unit[5]{fb^{-1}}$ for
$m_{W^\prime}=\unit[380]{GeV}$, $\unit[500]{GeV}$, $\unit[600]{GeV}$, respectively.
Considering the fact that tuning
$\epsilon_L$ into the region allowed by the precision observables does not significantly
change the signal, the Three-Site $Z^\prime$ may be discovered in this process as early as
in the first $\unit[1-2]{fb^{-1}}$ and even in the worst case can be expected to
manifest itself in the first $\unit[10-20]{fb^{-1}}$ of data.

\section{$W^\prime$ Production}
\label{chap-6-4}

\begin{figure}[!p]
\centerline{\begin{tabular}{c||c||c||}
& $m_{W^\prime}\in\left\{\unit[380]{GeV},\unit[500]{GeV},\unit[600]{GeV}\right\}$ &
$m_{W^\prime}=\unit[500]{GeV}$, different values of $\epsilon_L$
\\\hline\hline
\vs{5cm}\parbox{2ex}{\rotatebox{90}{$pp\rightarrow jjl\nu$}} &
\parbox{\dbltableplotheight}{
\includegraphics[height=\dbltableplotheight,angle=270]{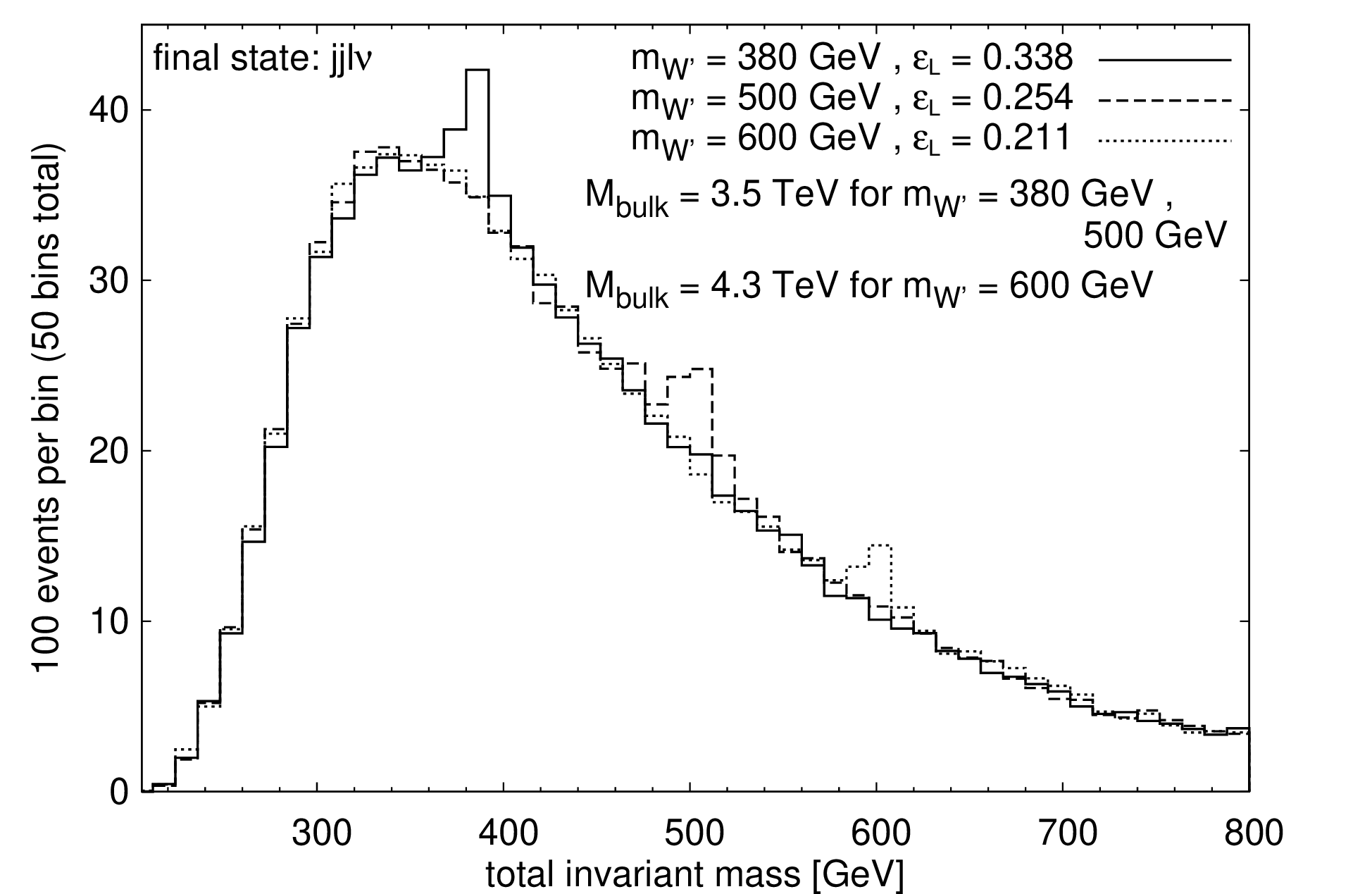}} &
\parbox{\dbltableplotheight}{
\includegraphics[height=\dbltableplotheight,angle=270]{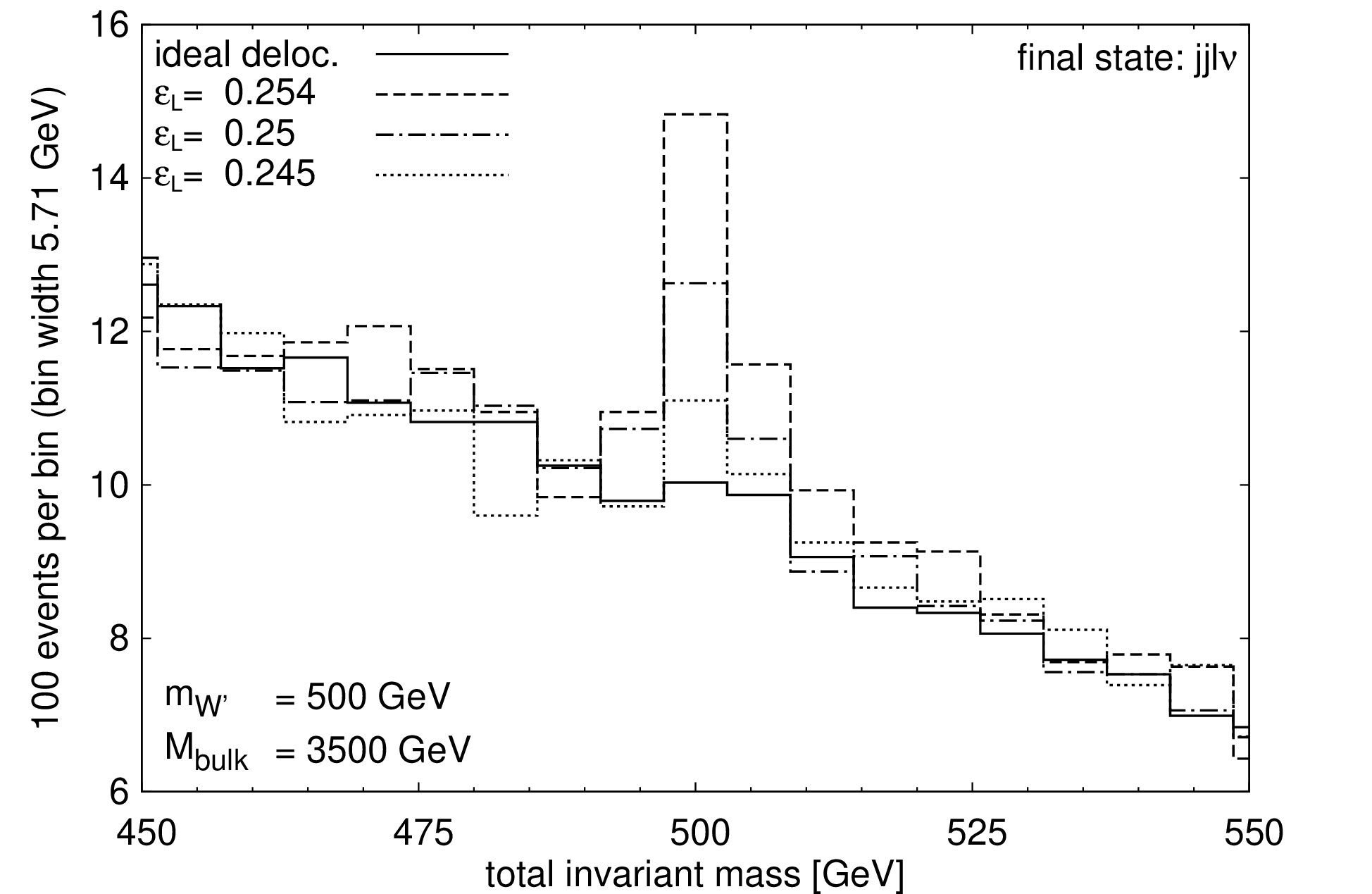}}
\\\hline\hline
\vs{5cm}\parbox{2ex}{\rotatebox{90}{$pp\rightarrow jjll$}} &
\parbox{\dbltableplotheight}{
\includegraphics[height=\dbltableplotheight,angle=270]{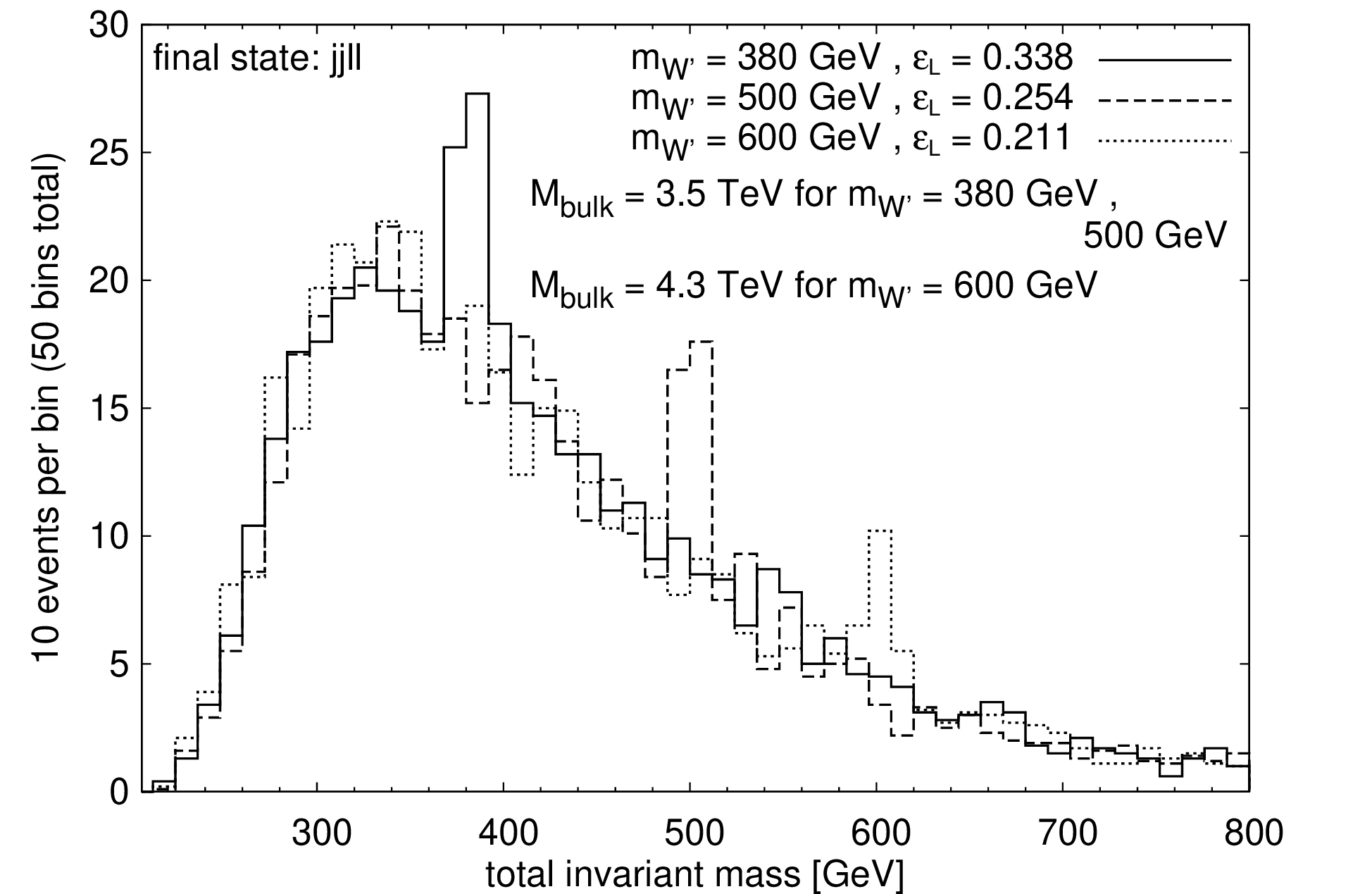}} &
\parbox{\dbltableplotheight}{
\includegraphics[height=\dbltableplotheight,angle=270]{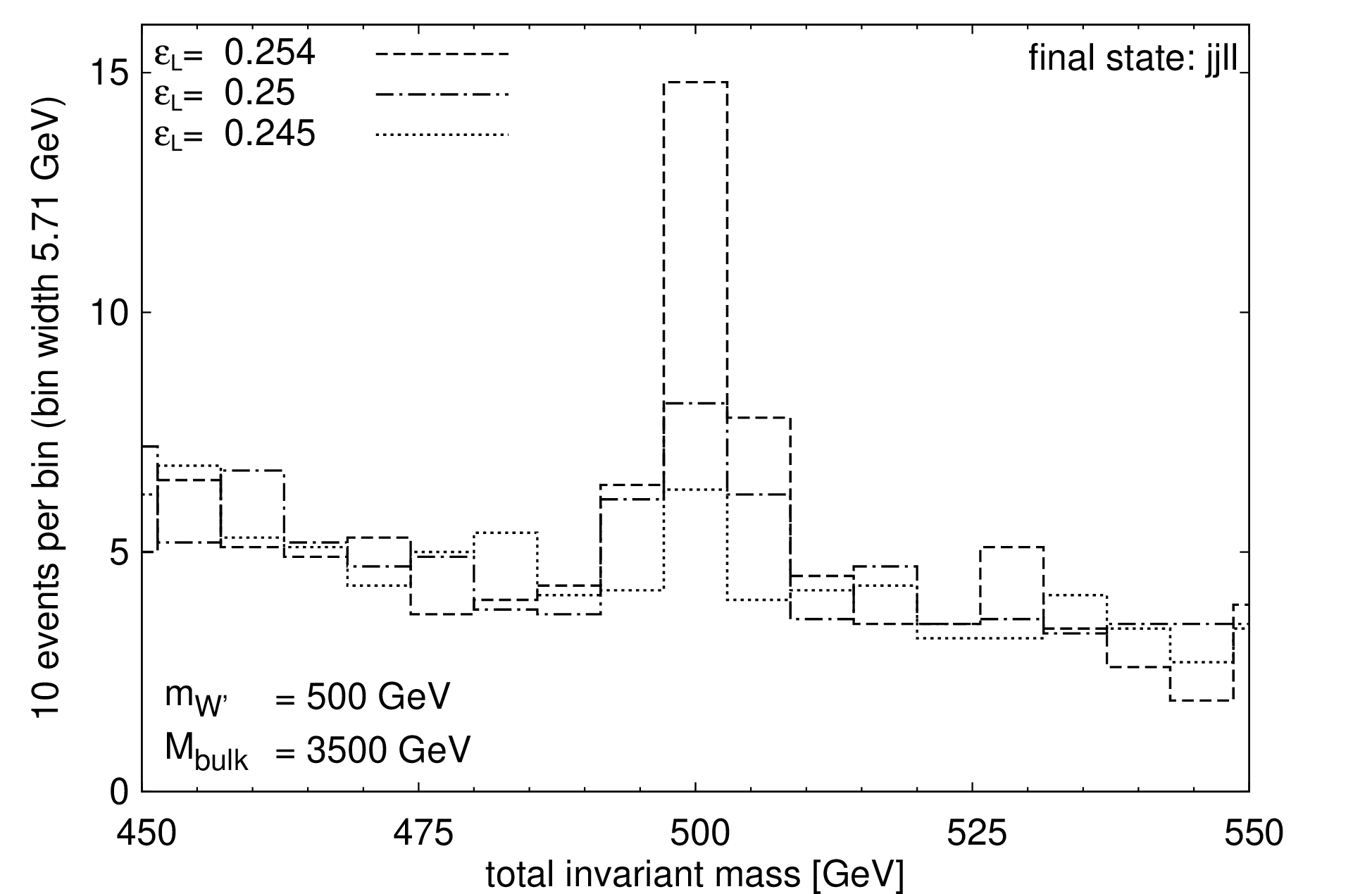}}
\\\hline\hline
\vs{5cm}\parbox{2ex}{\rotatebox{90}{$pp\rightarrow lll\nu$}} &
\parbox{\dbltableplotheight}{
\includegraphics[height=\dbltableplotheight,angle=270]{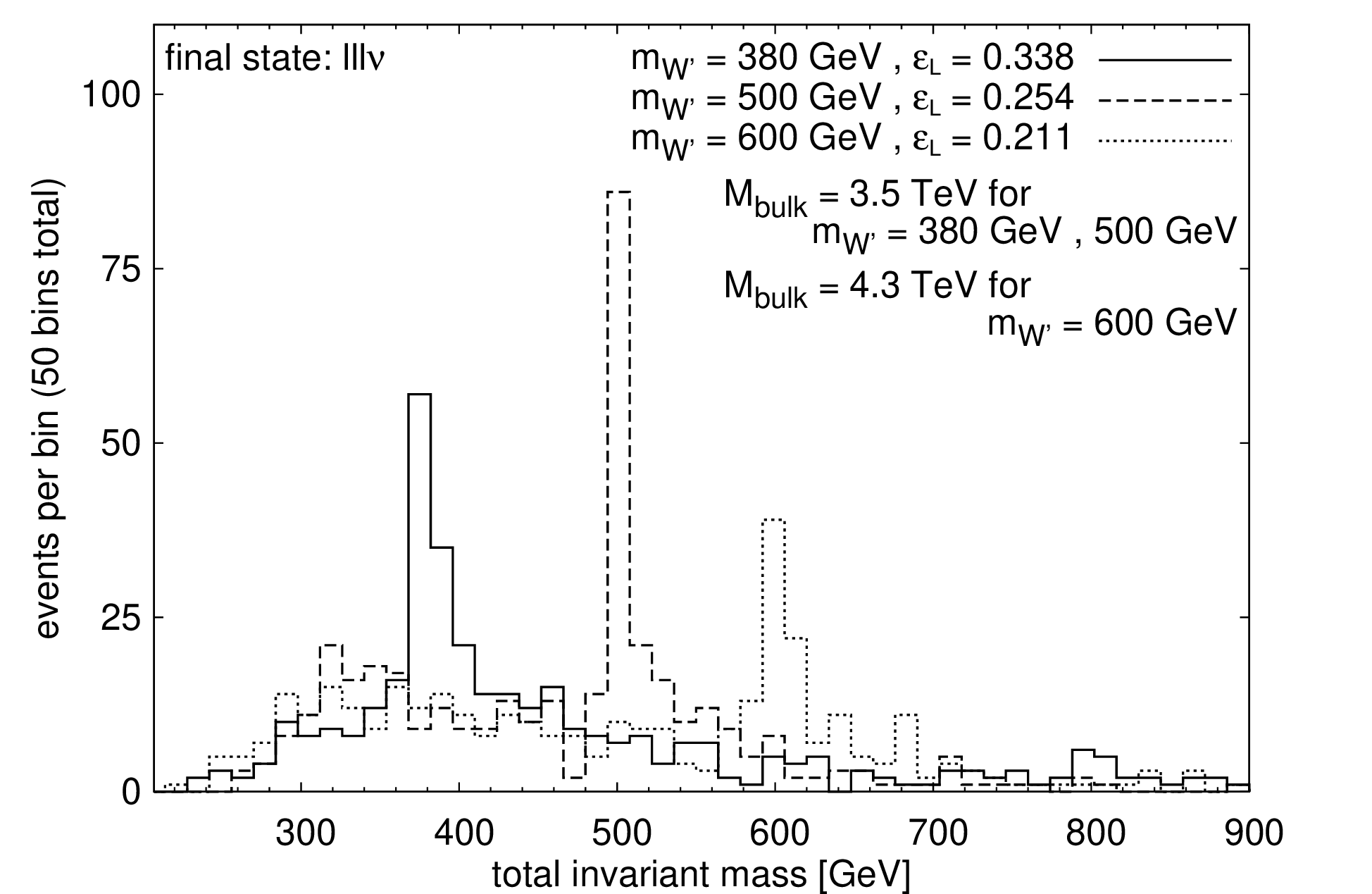}} &
\parbox{\dbltableplotheight}{
\includegraphics[height=\dbltableplotheight,angle=270]{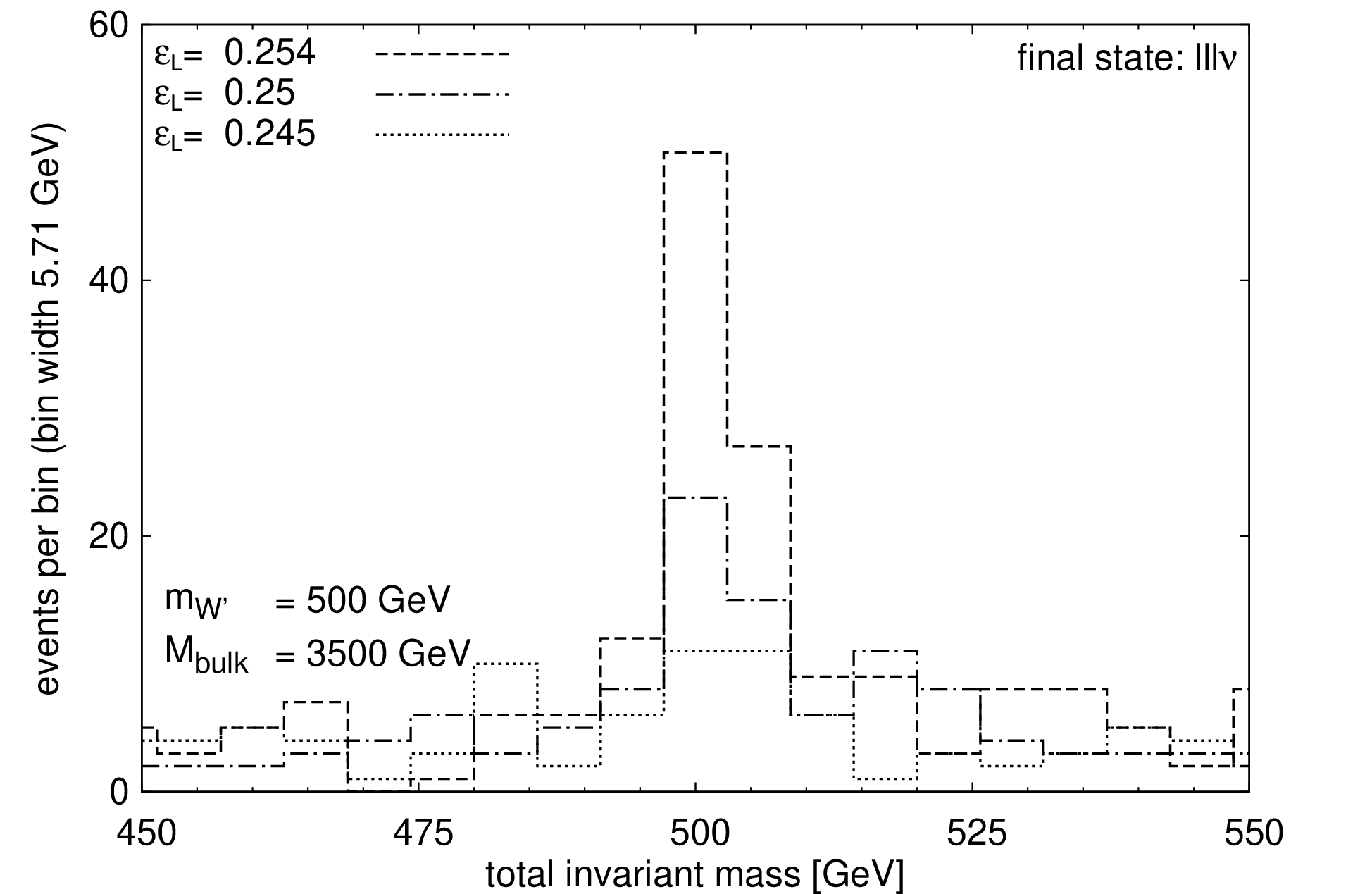}}
\\\hline\hline
\end{tabular}}
\caption{
\emph{Left column:} The $W^\prime$ resonance peaks for the three different final states under
consideration. Different values of $m_{W^\prime}$, near-maximal $g_{W^\prime ff}$.
\newline
\emph{Right column:} Close-up of the $m_{W^\prime}=\unit[500]{GeV}$ resonance peak for
different values of $\epsilon_L$ from the interval allowed by the precision observables.
}
\label{fig-6-4-whist}
\end{figure}

\begin{figure}[!tb]
\centerline{
\includegraphics[width=\doubleplotwidth,angle=270]{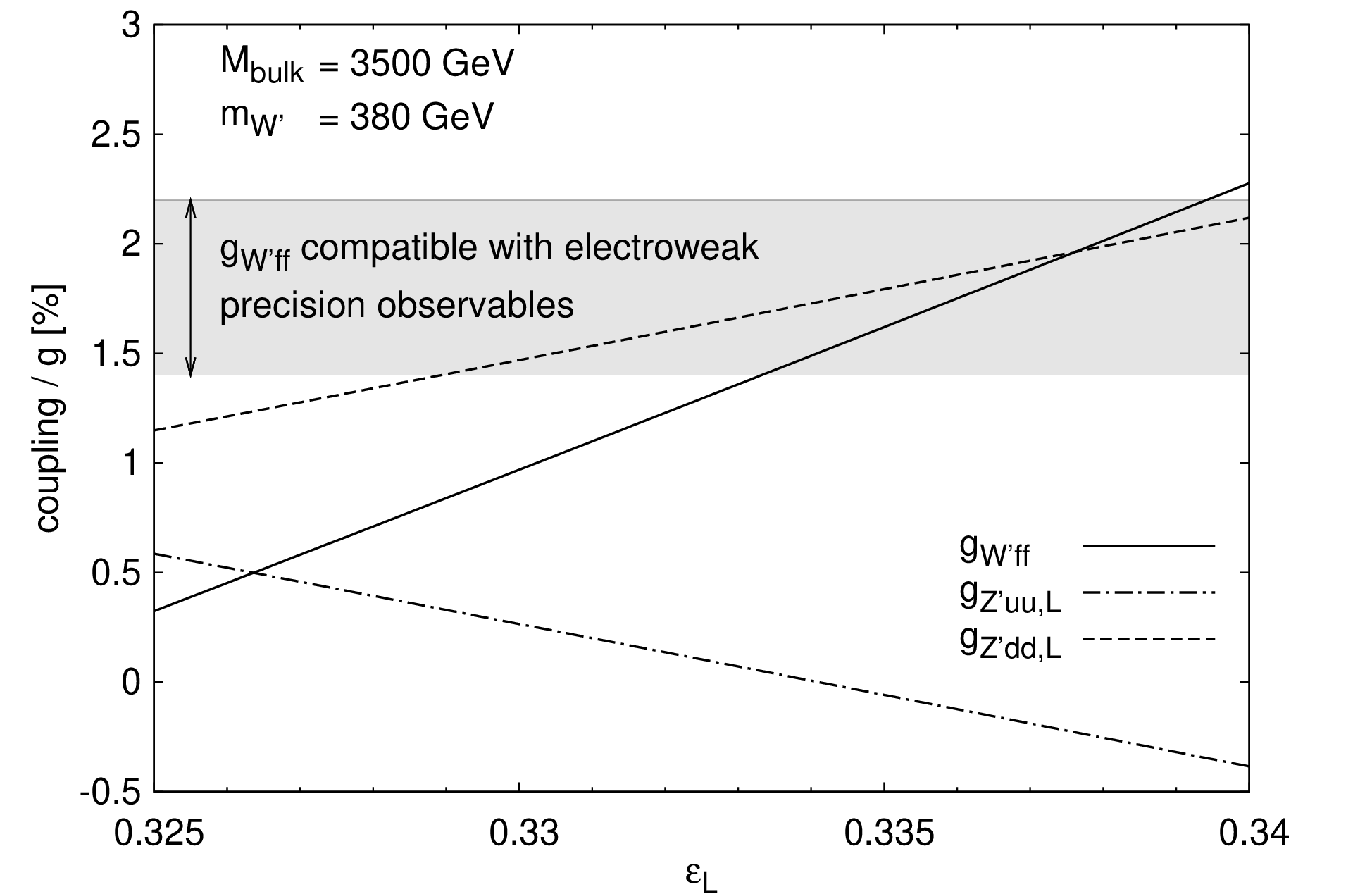}
\includegraphics[width=\doubleplotwidth,angle=270]{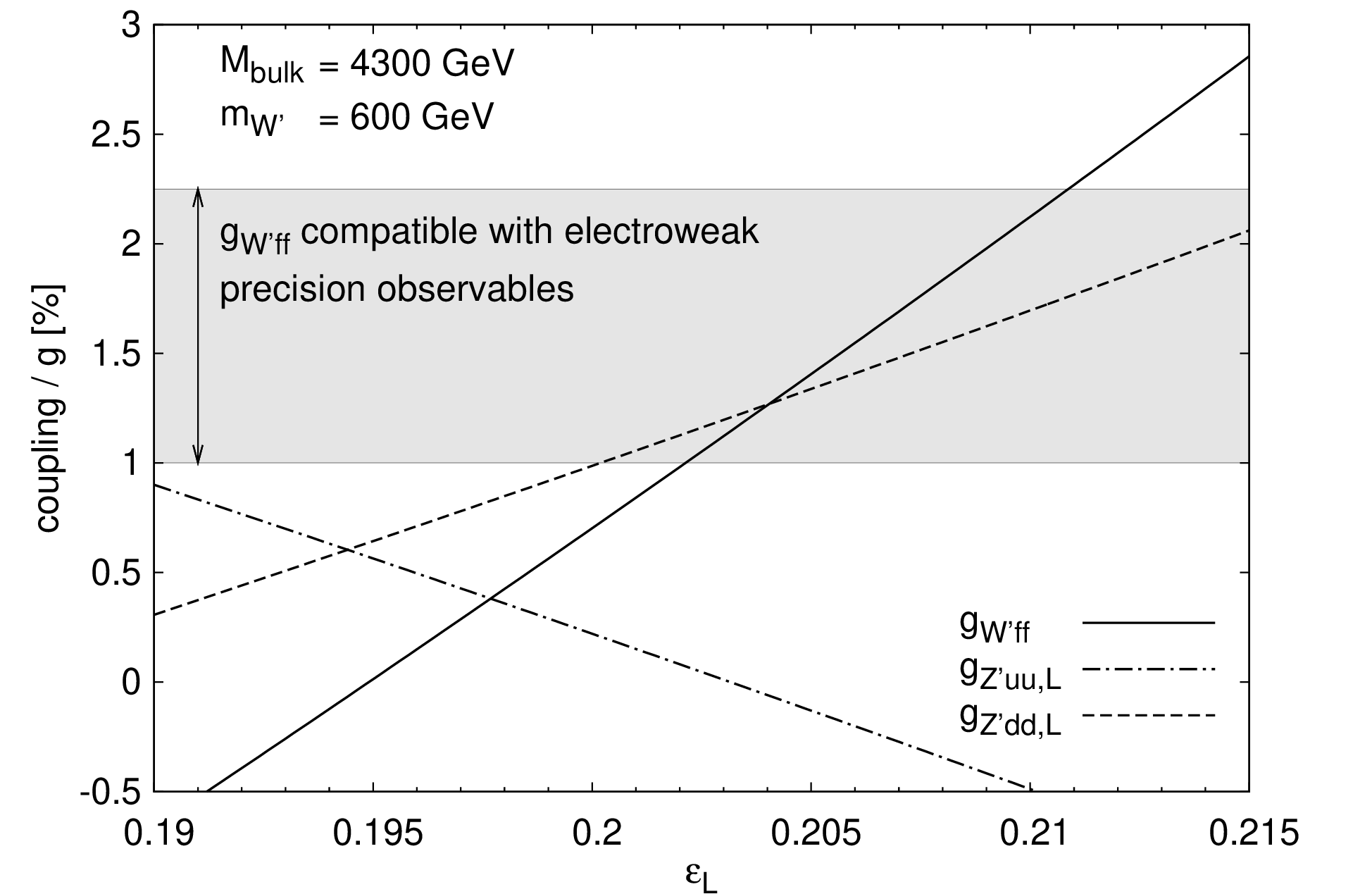}
}
\caption{The same plots as fig.~\ref{fig-6-1-cpl} for $m_{W^\prime}=\unit[380]{GeV}$,
$\unit[600]{GeV}$.}
\label{fig-6-7-cpl}
\end{figure}

In the case of $W^\prime$ production we have the three different final states $jjl\nu$, $jjll$ and
$lll\nu$ which we can use to look for the resonance peak. The left column of fig.~\ref{fig-6-4-whist} shows
the resulting invariant mass distributions for these final states, again for all three
$m_{W^\prime}$ values \eqref{equ-5-2-pspoints} under consideration with $\epsilon_L$ chosen to yield
roughly the maximum allowed value of $g_{W^\prime ff}$ (c.f. fig.~\ref{fig-6-1-cpl} and fig.
\ref{fig-6-7-cpl}). The peaks are clearly visible in all three final states, with the total number
of events contained in them dropping in the order (biggest to smallest)
\[ jjl\nu \quad\longrightarrow\quad jjll \quad\longrightarrow\quad lll\nu \]
due to the declining branching ratio (as can be easily seen by simply counting the number of final
states).

However, going from $jjl\nu$ to $jjll$, the need for the reconstruction of the neutrino
momentum with the associated doubling of the background events also goes away, and the possibility of
cutting on the invariant mass of the $W$ enhances the ratio of signal to background.
Furthermore, the hadronic background is completely removed when going to the fully leptonic final
state $lll\nu$, leading to the much cleaner signal visible in fig.~\ref{fig-6-4-whist} in spite of
the even smaller branching ratio.

\begin{figure}[!b]
\centerline{\begin{tabular}{c||c||c||}
& $m_{W^\prime}=\unit[380]{GeV}$ & $m_{W^\prime}=\unit[600]{GeV}$
\\\hline\hline
\vs{5cm}\parbox{2ex}{\rotatebox{90}{$pp\rightarrow jjl\nu$}} &
\parbox{\dbltableplotheight}{
\includegraphics[height=\dbltableplotheight,angle=270]{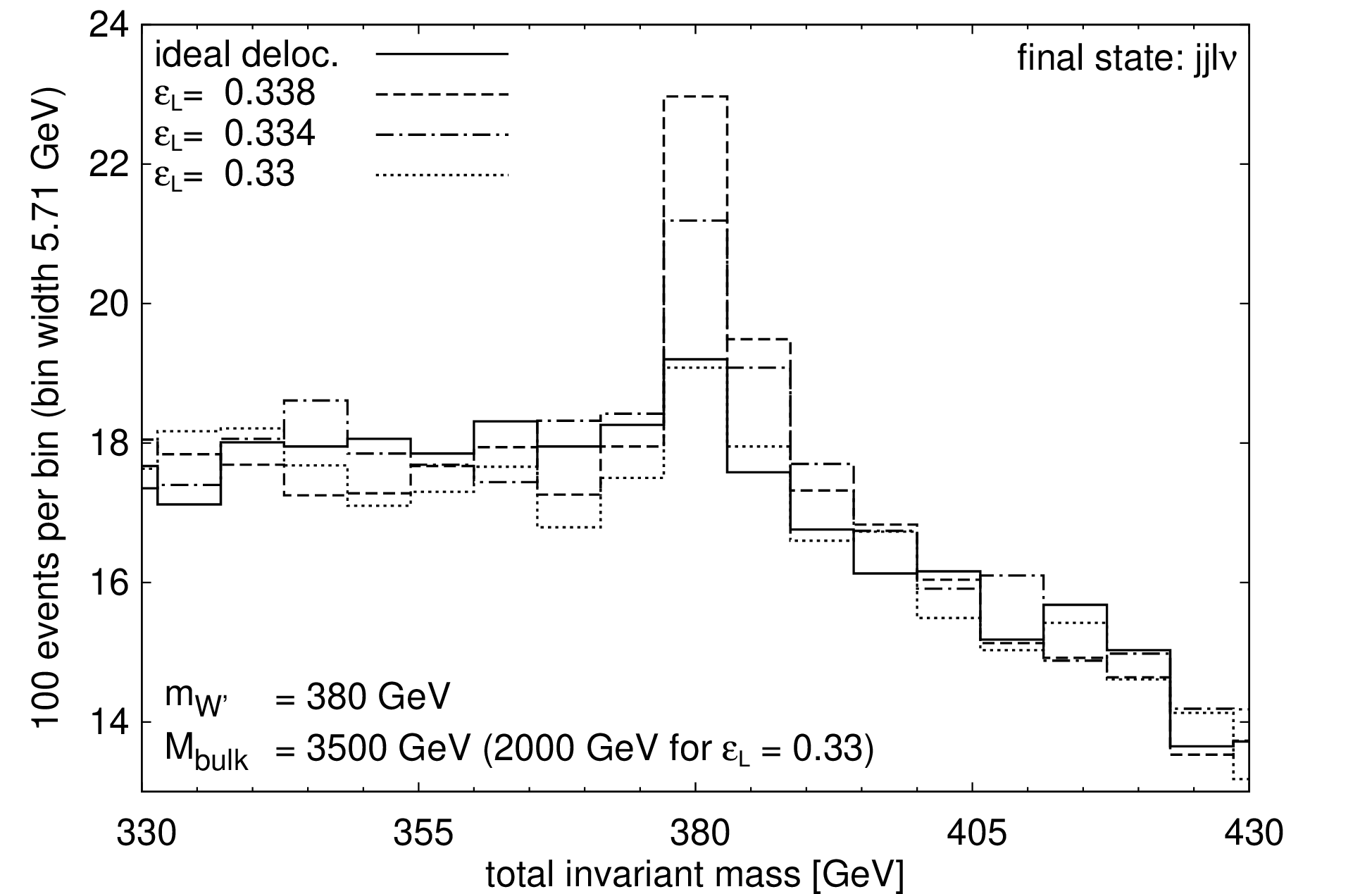}} &
\parbox{\dbltableplotheight}{
\includegraphics[height=\dbltableplotheight,angle=270]{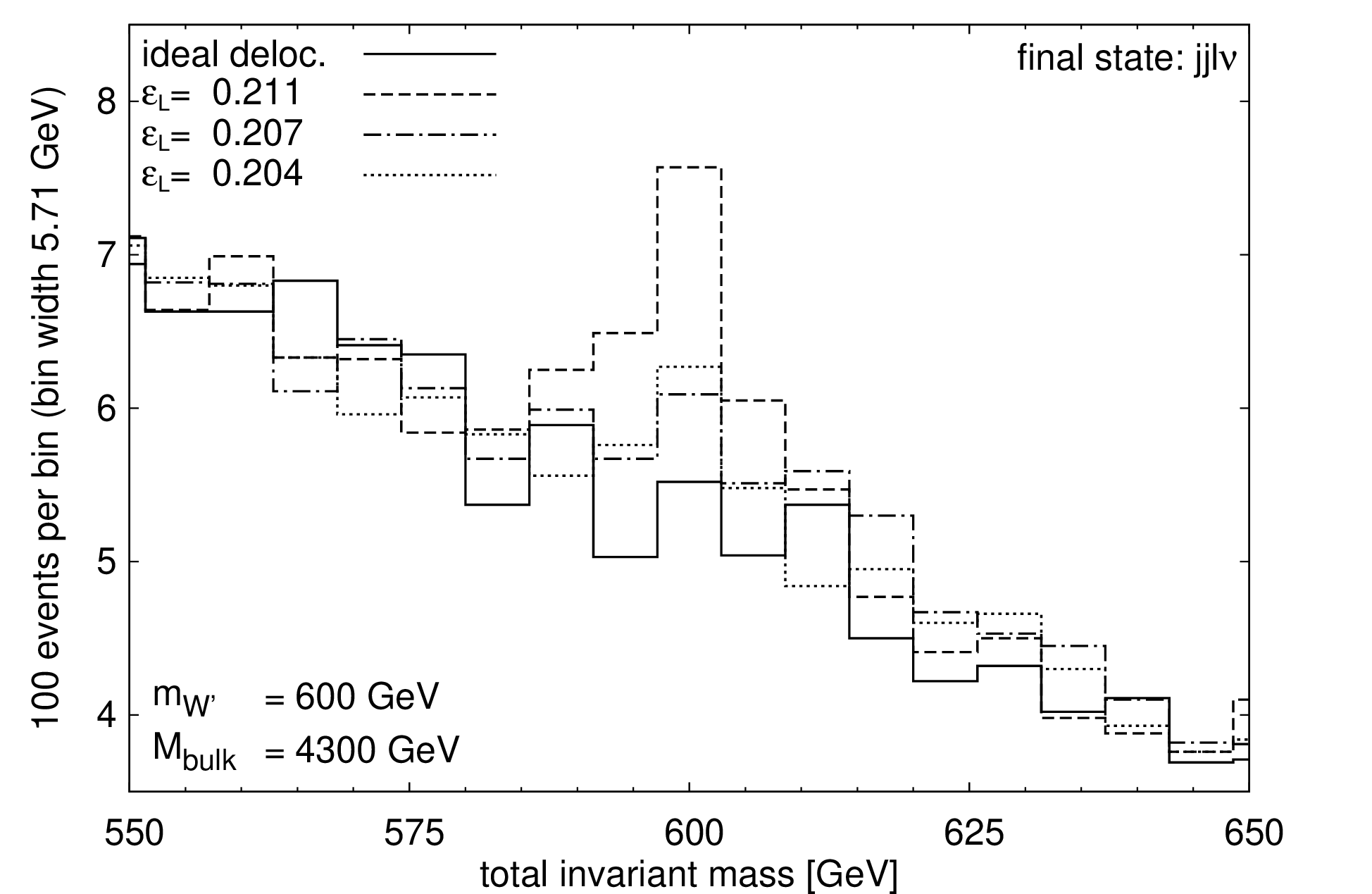}}
\\\hline\hline
\vs{5cm}\parbox{2ex}{\rotatebox{90}{$pp\rightarrow jjll$}} &
\parbox{\dbltableplotheight}{
\includegraphics[height=\dbltableplotheight,angle=270]{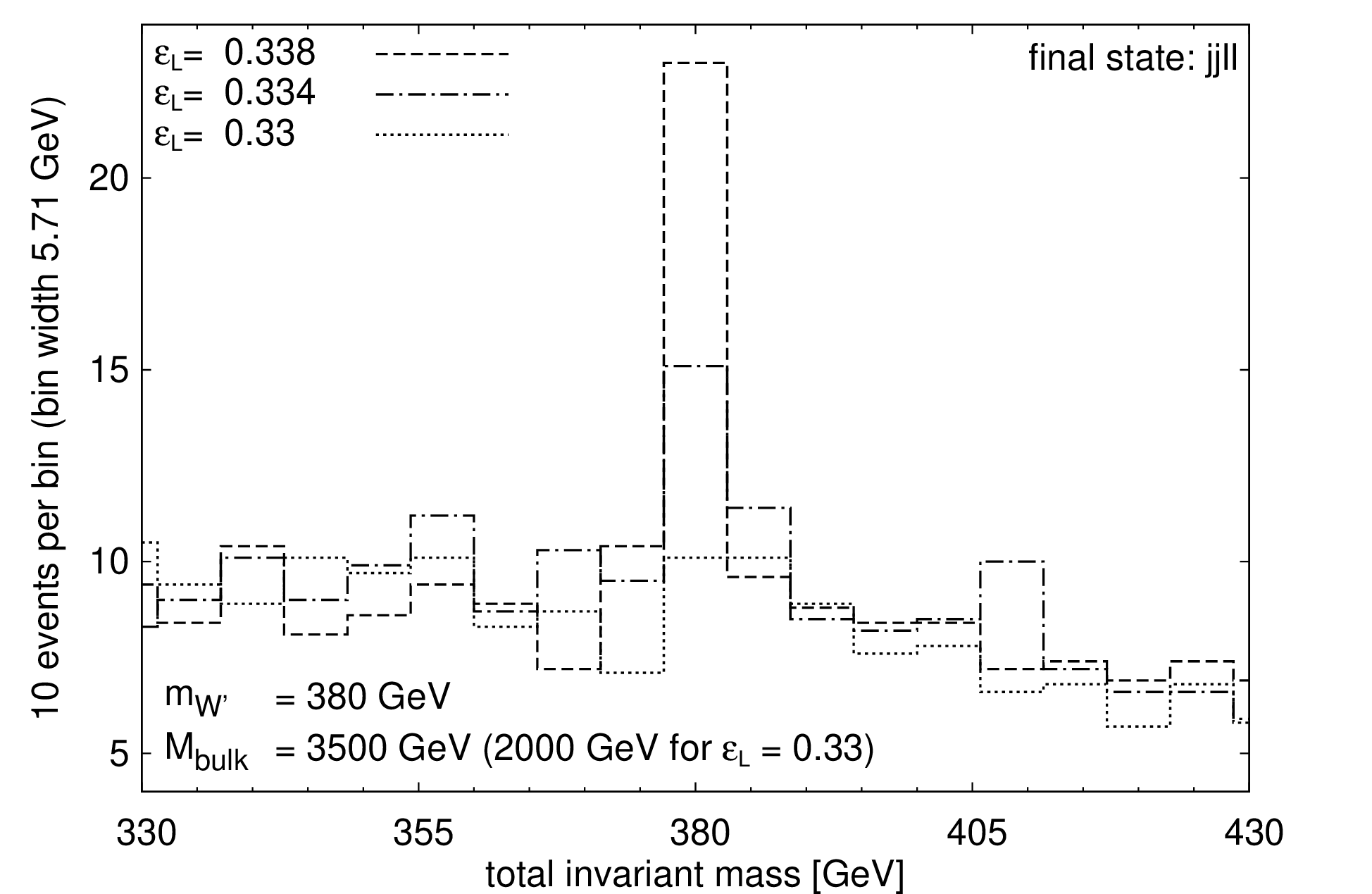}} &
\parbox{\dbltableplotheight}{
\includegraphics[height=\dbltableplotheight,angle=270]{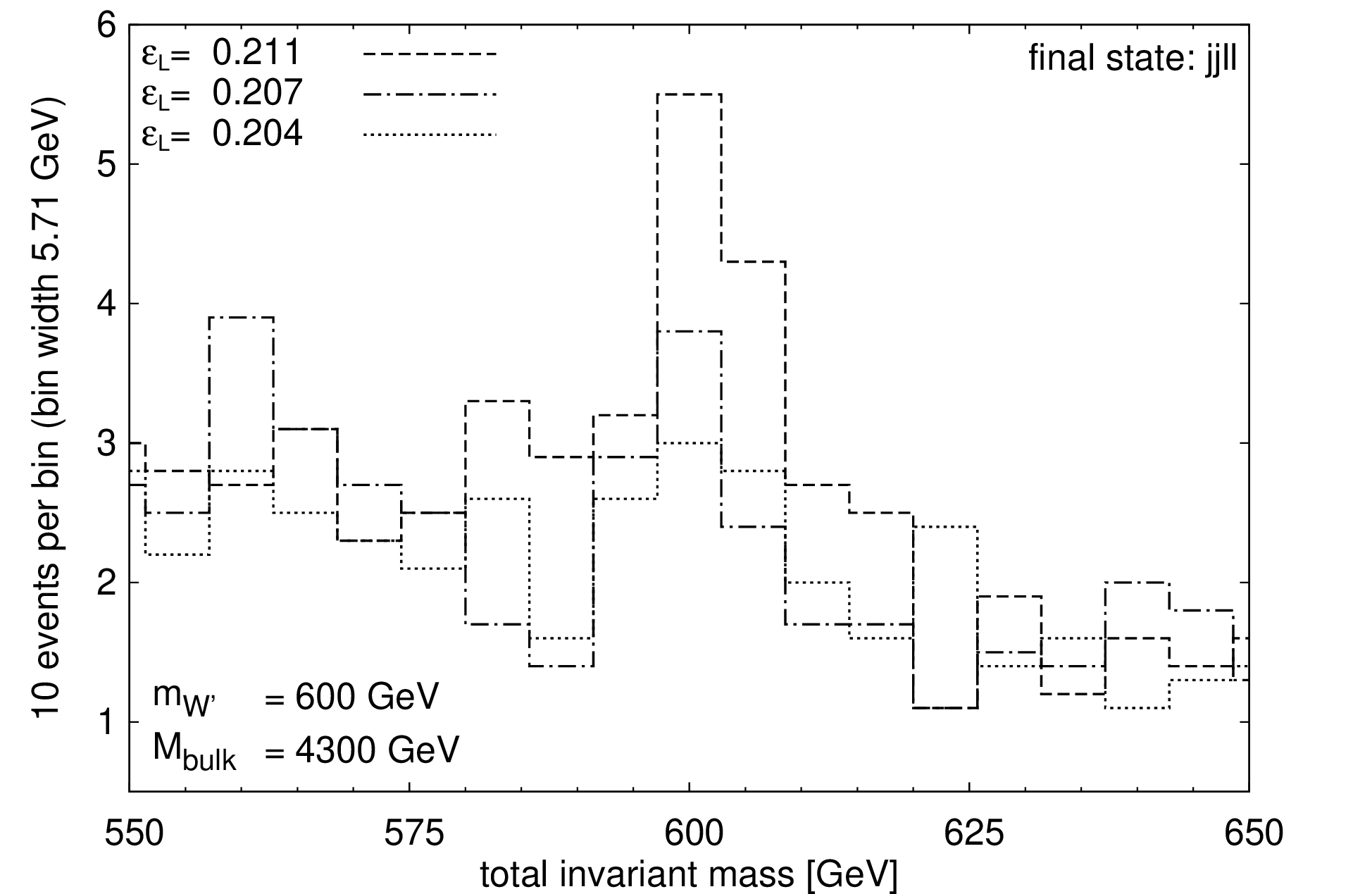}}
\\\hline\hline
\vs{5cm}\parbox{2ex}{\rotatebox{90}{$pp\rightarrow lll\nu$}} &
\parbox{\dbltableplotheight}{
\includegraphics[height=\dbltableplotheight,angle=270]{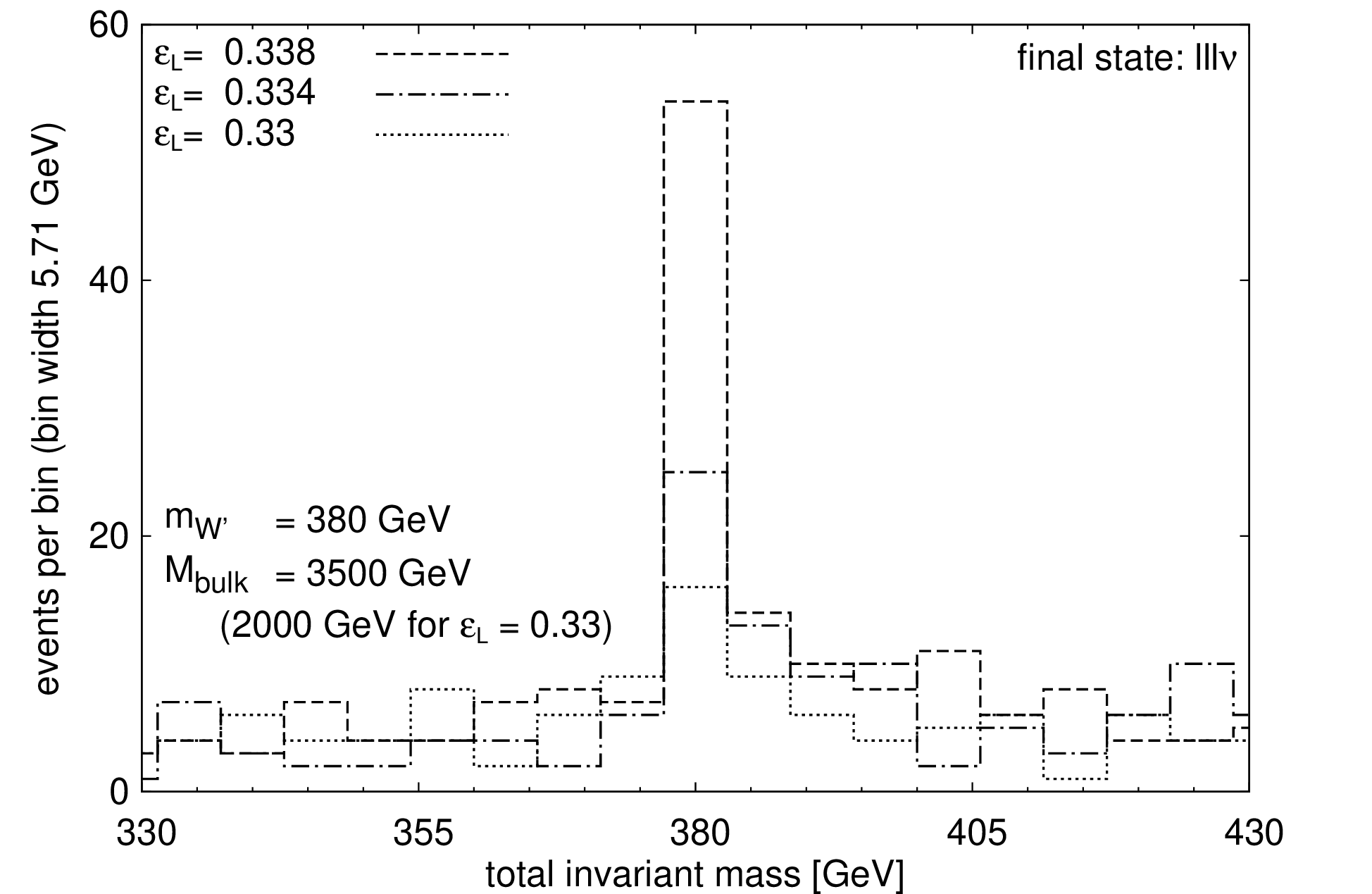}} &
\parbox{\dbltableplotheight}{
\includegraphics[height=\dbltableplotheight,angle=270]{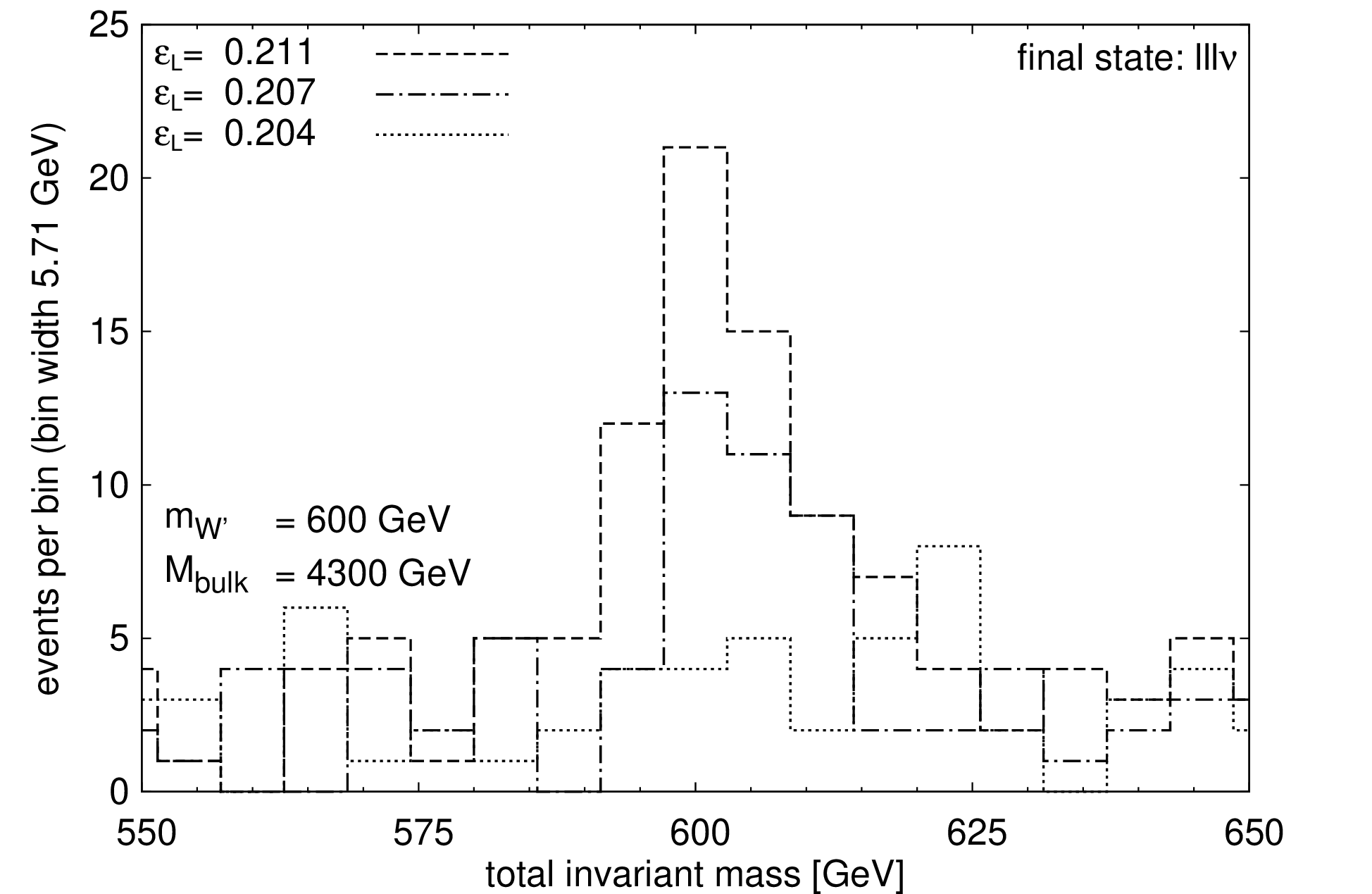}}
\\\hline\hline
\end{tabular}}
\caption{
The $W^\prime$ resonance peaks not covered by fig.~\ref{fig-6-4-whist} for the different final states
under consideration. Different values of $m_{W^\prime}$ and $\epsilon_L$.
}
\label{fig-6-7-whist}
\end{figure}

Contrary to the $Z^\prime$ case, the signal in the $W^\prime$ production process must be
proportional to the square of the $g_{W^\prime ff}$ coupling and is therefore highly sensitive to
the fermion delocalization parameter $\epsilon_L$. The corresponding dramatic 
change in the $m_{W^\prime}=\unit[500]{GeV}$
resonance induced by changing $\epsilon_L$ within the allowed part of parameter space is shown in
the right column of fig.~\ref{fig-6-4-whist}, the change in the peaks for the other $W^\prime$
masses in fig.~\ref{fig-6-7-whist}. From these distributions, it is evident that the peak can be
nearly completely removed within the allowed region of parameter space by choosing a suitable value
for $\epsilon_L$.

As would be expected, the resonance completely
vanishes in the (albeit forbidden) case of ideal delocalization. The histograms show that
this is also true for the $jjl\nu$
final state(with the exception of $m_{W^\prime}=\unit[380]{GeV}$ in which case a small peak from
misidentified $Z^\prime$ events remains), demonstrating that the $\pm\unit[5]{GeV}$ cut around the
$W$ mass is sufficient to
separate $W^\prime$ and $Z^\prime$ contributions to this final state at least at parton level
if the smearing cause by measurement errors is not take into account.

\begin{figure}[!tb]
\centerline{\includegraphics[width=\singleplotwidth,angle=270]{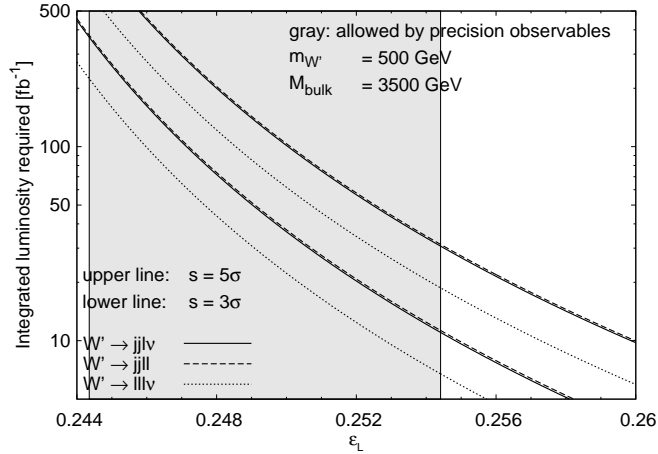}}
\caption{The integrated luminosities required for a $3\sigma$ resp. $5\sigma$ discovery of the
$W^\prime$ in the $s$ channel as a function of $\epsilon_L$. Different final states,
$m_{W^\prime}=\unit[500]{GeV}$.}
\label{fig-6-4-sig}
\end{figure}
\begin{figure}[!tb]
\centerline{
\includegraphics[width=\doubleplotwidth,angle=270]{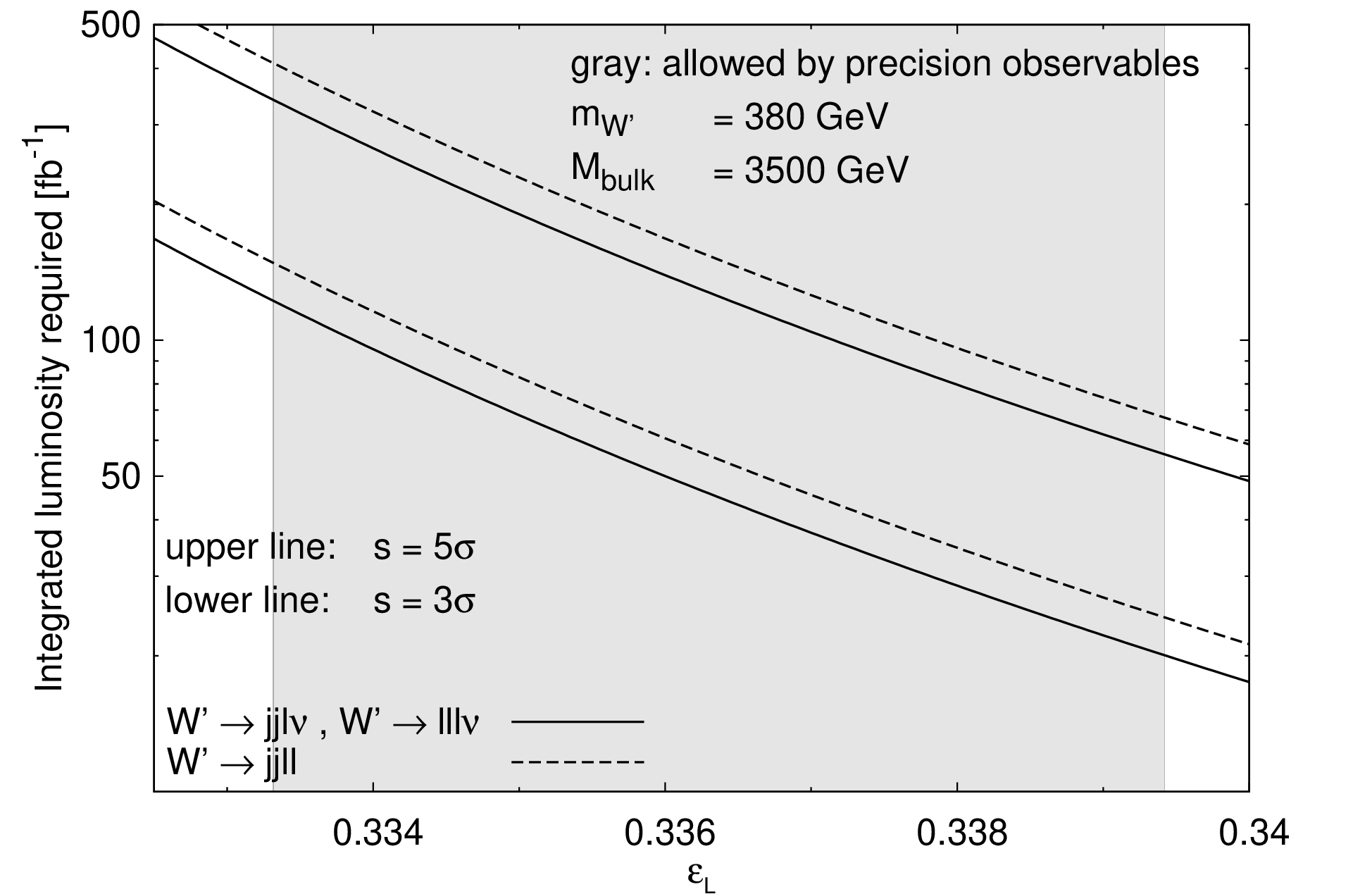}
\includegraphics[width=\doubleplotwidth,angle=270]{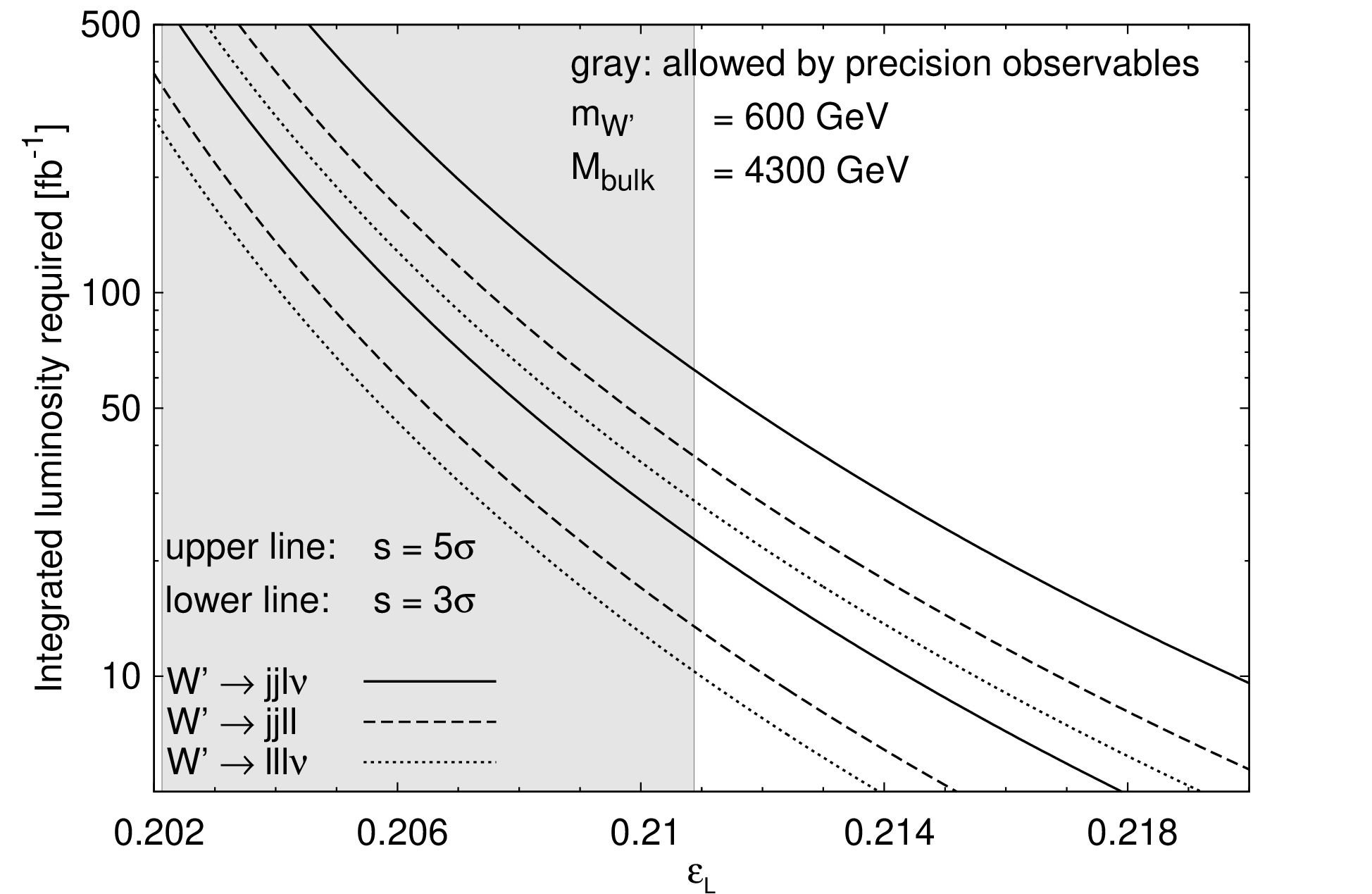}
}
\caption{The same as fig.~\ref{fig-6-4-sig} for $m_{W^\prime}=\unit[380]{GeV}$, $\unit[600]{GeV}$.}
\label{fig-6-7-sig}
\end{figure}

Defining the significance of the signal in exactly the same way\footnote%
{
As the total number of events is much lower in the $jjll$ resp. $lll\nu$ final states when compared
to $jjl\nu$, we generated the Standard Model backround for these by downscaling simulations for
$\ilum=\unit[1000]{fb^{-1}}$ resp. $\ilum=\unit[5000]{fb^{-1}}$.
}
as in \ref{chap-6-3} for
$Z^\prime$ production (without the additional factor $\sqrt{2}$ for $jjll$), we can exploit the
quadratic dependence of the signal on $g_{W^\prime ff}$ in order to obtain an estimate of the integrated
luminosity required to yield a signal with a given significance.
The $5\sigma$ resp. $3\sigma$ result is shown in
fig.~\ref{fig-6-4-sig} as a function of $\epsilon_L$ for $m_{W^\prime}=\unit[500]{GeV}$; the corresponding
plots for the other two masses are shown in fig.~\ref{fig-6-7-sig}.

From these curves it is clear that, while the actual result varies a bit over parameter space,
the performance of the different final states is fairly comparable at the parton level. However,
once detector effects are included, it is likely that the fully leptonic final state is
preferred due to the more precise measurement of the lepton momentum and the absence of complicated
QCD backgrounds --- an assertion which is confirmed by the detector simulation performed in
\cite{fabian:master}.

As far as the prospects of discovering the $W^\prime$ in this process are concerned, fig.
\ref{fig-6-4-sig} and \ref{fig-6-7-sig} clearly demonstrate that there are regions in parameter
space easily accessible in the first $10-\unit[20]{fb^{-1}}$, while it would take
more than $\unit[500]{fb^{-1}}$ for a $5\sigma$ discovery in other regions which is most likely
out of question at the LHC.

\section{Disentangling the $jjl\nu$ final states}
\label{chap-6-5}

Since flavor tagging is impossible for light quark flavors, we have to
rely on invariant mass cuts on the jet pairs in order to be able to separate
the case of the two jets in $jjl\nu$ coming from the decay of a $W$
in $Z^\prime$ production from that of the jets being produced by a
decaying $Z$ in $W^\prime$ production.
However, it may very well be impossible to obtain a resolution of order
$\pm\unit[5]{GeV}$ in the jet invariant mass from experimental data.
In the following, we model the effect of the measurement error on the
$W^\prime/Z^\prime$ separation with a gaussian smearing of the invariant mass of the jets.

In the ideal case of exact $m_{jj}$ measurement, events coming from the decay of a
intermediary $W$/$Z$ are distributed according to a Breit-Wigner distribution
\[ p_b(x,m,\Gamma)\:dx =
\frac{n_b(m,\Gamma)^{-1}}{\left(x^2-m^2\right)^2+\Gamma^2 m^2}\:dx \]
with the normalization factor
\[ n_b(m,\Gamma) = \frac{\pi}{4m^3}\left(1+\frac{\Gamma^2}{m^2}\right)^{-\frac{3}{4}}
\sin^{-1}\left(\frac{1}{2}\atan\frac{\Gamma}{m}\right) \]
Emulating the measurement error in the jet mass by convoluting $p_\text{bw}$ with a
gaussian of standard deviation $\sigma$
\[ 
p_\text{g}(x,\sigma)\:dx = \frac{1}{\sqrt{2\pi}\sigma}e^{-\frac{x^2}{2\sigma^2}}\:dx
\]
we obtain the smeared distribution
\[ p_\text{sm}(x,m,\Gamma,\sigma)\: dx = \int_0^\infty dy\:p_b(y,m,\Gamma)
p_\text{g}(x-y,\sigma)
\]
\begin{figure}[!tb]
\centerline{\includegraphics[width=\singleplotwidth,angle=270]{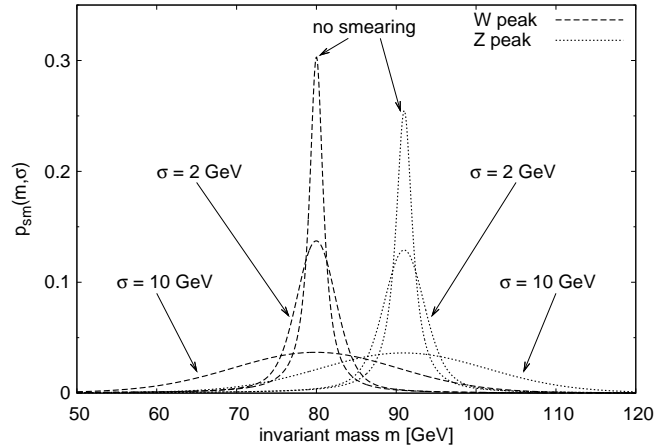}}
\caption{The effect of a gaussian smearing on the Breit-Wigner shape of the $W$ and $Z$
resonances for various widths $\sigma$ of the gaussian.}
\label{fig-6-5-smearedpeaks}
\end{figure}

Figure \ref{fig-6-5-smearedpeaks} shows the effect of this smearing on the Breit-Wigner peaks of
the $Z$ and the $W$. Turning on the smearing and increasing $\sigma$ causes the sharp
Breit-Wigner peaks to decay rapidly, and for $\sigma=\unit[10]{GeV}$, only two very broad
bumps are left. The consequence is that, if a cross section has one contribution which stems
from the decays of a virtual $Z$ and one coming from a virtual $W$, any attempt to isolate
the $Z$ contribution by cutting on the resonance will inevitably also select events coming
from the $W$ decay contaminating the sample (and vice versa). Therefore, our analysis of
the $jjl\nu$ final state will show a $W^\prime$ peak even in the case of ideal
delocalization which is caused by jet pairs from a decaying $W$ misidentified as a $Z$.

If we try to isolate the $W$ peak with a cut on the invariant mass $m_{jj}$
\[ L_W \le m_{jj} \le U_W \]
and the $Z$ peak with a cut
\[ L_Z \le m_{jj} \le U_Z \]
then the resulting event counts $\widetilde{N}_W, \widetilde{N}_Z$ can be
calculated from the true event counts $N_W, N_Z$  coming from a decaying $W$ or $Z$
via a matrix $T$ as
\[ \begin{pmatrix} \widetilde{N}_W \\ \widetilde{N}_Z \end{pmatrix} =
\begin{pmatrix} T_{WW} & T_{WZ} \\ T_{ZW} & T_{ZZ} \end{pmatrix}
\begin{pmatrix} N_W \\ N_Z \end{pmatrix} \]
with entries
\[ T_{ij} = \int_{L_i}^{U_i}dm\:p_\text{sm}(m,m_j,\Gamma_j,\sigma) \]
Inverting $T$ we can calculate the event counts $N_W$ and $N_Z$
\begin{equation} \begin{pmatrix} N_W \\ N_Z \end{pmatrix} = T^{-1}
\begin{pmatrix} \widetilde{N}_W \\ \widetilde{N}_Z \end{pmatrix}
\label{equ-6-5-trans-mat}\end{equation}
The entries of $T$ give the probability of (mis)identifying an event and can be readily
calculated numerically; for example, choosing cuts
\[ L_W=\unit[60]{GeV} \quad,\quad U_W=\unit[85]{GeV} \quad,\quad
L_Z=\unit[86]{GeV} \quad,\quad U_Z=\unit[111]{GeV} \]
yields
\[ T \approx \begin{pmatrix} 0.64 & 0.27 \\ 0.29 & 0.62 \end{pmatrix}
\quad,\quad
T^{-1} \approx \begin{pmatrix} 1.9 & -0.85 \\ -0.89 & 2.0 \end{pmatrix} \]
This way, we can in principle use $T$ to disentangle the
contributions from the $W$ and $Z$ resonances
to the signal in the presence of a measurement error which causes the Breit-Wigner
peaks to lose their shape. However, in order to apply this to actual data, it is vital
to separate the signal from both the reducible and the irreducible
backgrounds as these don't obey a Breit-Wigner distribution.

\begin{figure}[!tb]
\centerline{
\includegraphics[width=\doubleplotwidth,angle=270]{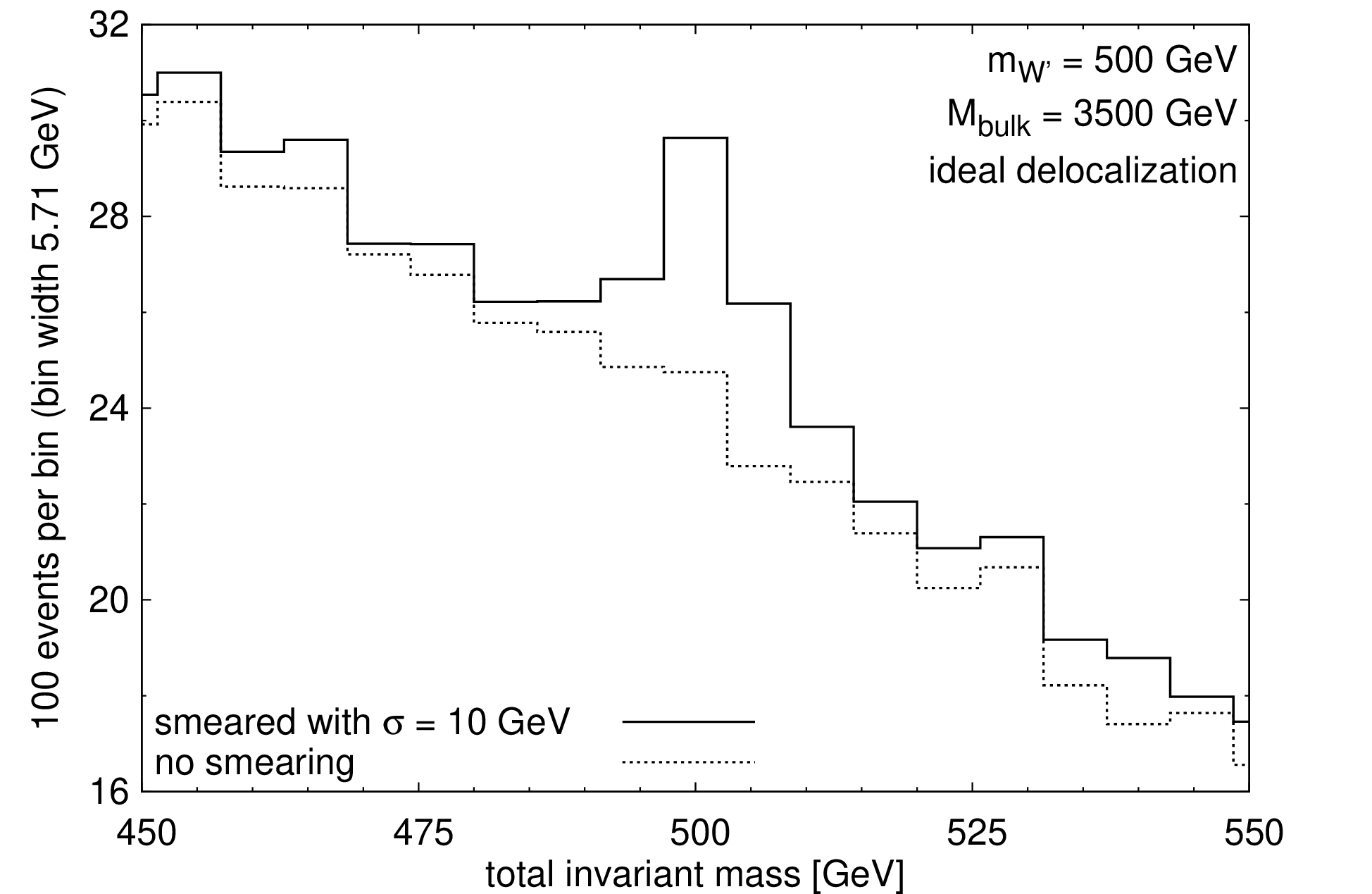}
\includegraphics[width=\doubleplotwidth,angle=270]{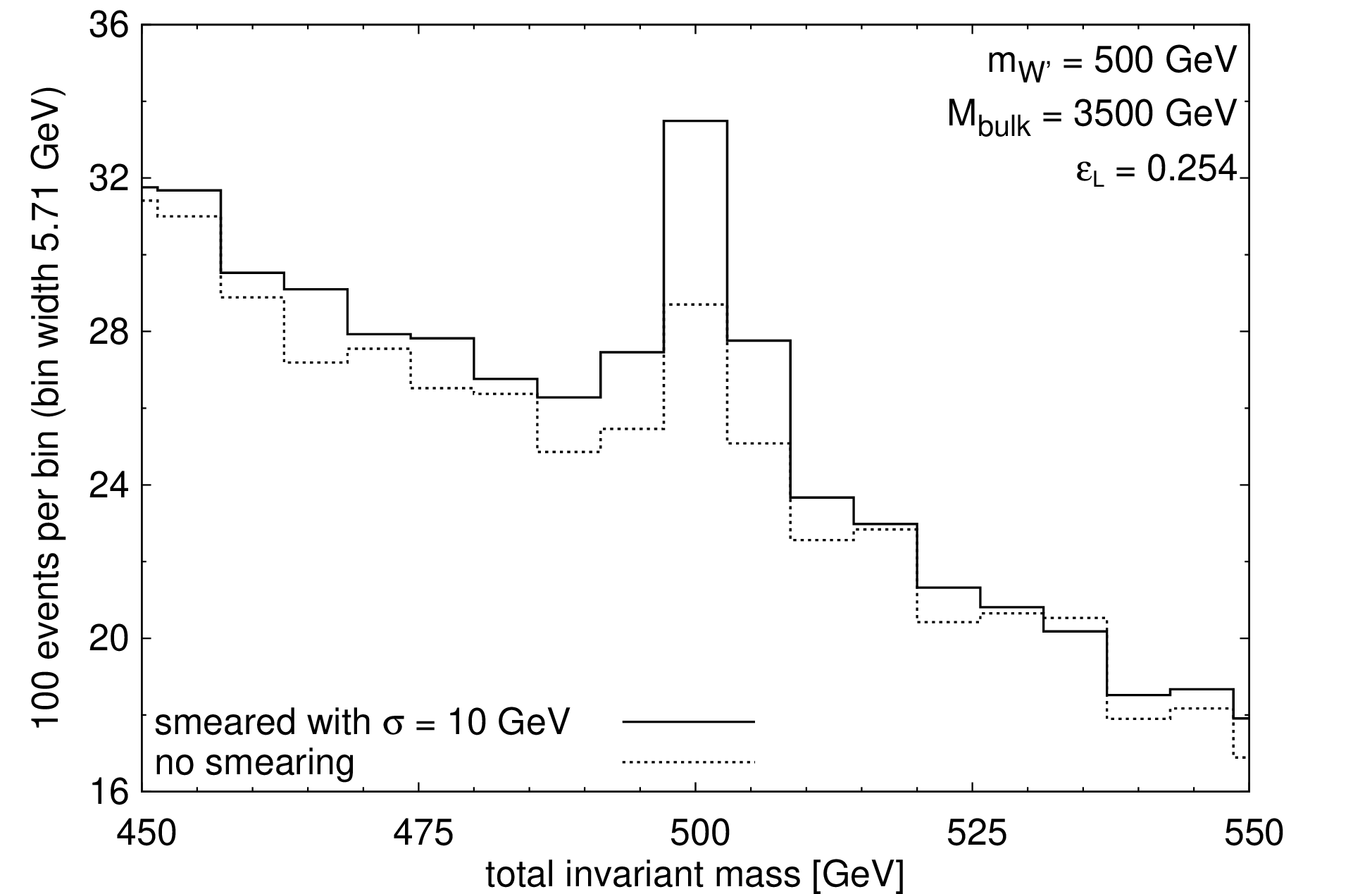}
}
\caption{
\emph{Left}: Signal in the $W^\prime$ detection channel for the case of ideal
delocalization smeared with a gaussian error.
\newline
\emph{Right}: The same for the case of nonzero $g_{W'ff}$
}
\label{fig-6-5-wsmear}
\end{figure}
In order to estimate the significance of a signal obtained this way, we calculate the
standard deviation $\sigma_{N_i}$ of $N_i$ according to
\begin{equation}\label{equ-6-5-recerr}
\sigma_{N_i} = \sqrt{\sum_{j\in W,Z}\left(T^{-1}_{ij}\right)^2\sigma_{\widetilde{N}_j}^2}
\end{equation}
Calculating the fluctuations of $\widetilde{N}_j$ is a bit tricky as the neutrino momentum
reconstruction adds statistically correlated pairs of events to the sample.
According to the analysis
performed in the last two sections \ref{chap-6-3} and \ref{chap-6-4}, we obtain the number of events
$\widetilde{N}$ in the peak by subtracting the background (aka Standard Model expectation) $N_b$
from the total number of events $N_\text{tot}$
\[ \widetilde{N} = N_\text{tot} - N_b \]

The Standard Model value $N_b$ comes from a theoretical calculation and is by definition free of
statistical fluctuations\footnote%
{
Of course, this is a simplification which is not quite true in our analysis where we obtain $N_b$
from a Monte-Carlo simulation. However, due to our generating and downscaling a surplus of events,
the error on $N_b$ is
smaller than that on $N_\text{tot}$, and in a more serious analysis of experimental data,
$\sigma_{N_b}$ could be made arbitrarily small by investing enough time into its determination.
}.
The total number of events $N_\text{tot}$ contains $\widetilde{N}$ events originating from the
resonance which we idealize to be free of double-counting and $N_b$ background events which contain the
correlated double counts. Putting the pieces together, we have
\[
\sigma_{\widetilde N} = \sigma_{N_\text{tot}} = \sqrt{\widetilde{N} + 2 N_b} =
\sqrt {N_\text{tot} + N_b}
\]
which we can now plug into \eqref{equ-6-5-recerr} in order to calculate the statistical fluctuation
we have to expect for the reconstructed event counts and finally arrive at
\begin{equation}
\sigma_{N_i} = \sqrt{\sum_{j\in W,Z}\left(T^{-1}_{ij}\right)^2\left(N_{t,j} + N_{b,j}\right)}
\label{equ-6-5-sigma-after-transfer}\end{equation}

For a simulation of the effect of the measurement error our analysis we have
randomly distributed the invariant mass of the jet pairs within a gaussian with width
$\sigma=\unit[10]{GeV}$ centered around the correct value calculated from Monte Carlo data. We
then performed the same analysis as in the last two sections \ref{chap-6-3} and \ref{chap-6-4}
with $m_{W^\prime}=\unit[500]{GeV}$ and $m_\text{bulk}=\unit[3.5]{TeV}$ both for
$\epsilon_L=0.254$ and for the ideally delocalized scenario. The only difference to the
previous analysis are the cuts on $m_{jj}$ which we enlarged to
\[ \unit[60]{GeV}\le m_{jj}\le\unit[85]{GeV} \quad\text{resp.}\quad
\unit[86]{GeV}\le m_{jj} \le\unit[111]{GeV} \]

Fig.~\ref{fig-6-5-smearedpeaks} shows the resulting effect on the $W^\prime$ peak for the cases
of ideal delocalization (left) and for $\epsilon_L=0.254$ (right). In both cases a peak is
clearly visible, which in the ideally delocalized scenario is only composed of events
with jets coming from a decaying $W$ misidentified as a $Z$.

The number of signal events
$\widetilde{N}_{W/Z}$ after smearing, the significance $s_{W/Z}$ of these
calculated similarly to the last sections, $N_{W/Z}$ obtained from applying the transfer
matrix $T^{-1}$ (\ref{equ-6-5-trans-mat}) and the resulting significance
$N_i/\sigma_{N_i}$ obtained from \eqref{equ-6-5-sigma-after-transfer} are shown in
table \ref{tab-6-5-wzsep}. All peaks are significant with $s>5\sigma$; however, after
applying the transfer matrix, the $W^\prime$ peak vanishes within one standard deviation
for ideal delocalization\footnote%
{
The negative event count in the $W^\prime$ peak is an artifact of the multiplication with $T^{-1}$.
},
while in the case of $\epsilon_L=0.254$ a residue as big as $2\sigma$ remains. The
$Z^\prime$ peak remains significant after applying the transfer matrix although, however, the
significance is reduced because the transfer matrix enlarges the error.
\begin{table}
\centerline{
\begin{tabular}{|c||c|c|c|c|}
\hline\multicolumn{5}{|c|}{ideal delocalization}\\\hline\hline
\vs{4.5ex} & $\widetilde{N}_i$ & $s_i$ & $N_i$ & $\frac{N_i}{\sigma_{N_i}}$ \\\hline\hline
$i=W$ & $3193$ & $17$ & $5126$ & $13$ \\\hline
$i=Z$ & $1371$ & $7.5$ & $-96.10$ & $0.24$ \\\hline
\end{tabular}
\hspace{1cm}
\begin{tabular}{|c||c|c|c|c|}
\hline\multicolumn{5}{|c|}{$\epsilon_L=0.254$}\\\hline\hline
\vs{4.5ex} & $\widetilde{N}_i$ & $s_i$ & $N_i$ & $\frac{N_i}{\sigma_{N_i}}$ \\\hline\hline
$i=W$ & $3767$ & $21$ & $5628$ & $14$ \\\hline
$i=Z$ & $2083$ & $11$ & $811.6$ & $2.0$ \\\hline
\end{tabular}
}
\caption{Comparison of the signals $\widetilde{N}_{W/Z}$ obtained with an gaussian
smearing of the invariant mass of the jets with $\sigma=\unit[10]{GeV}$ to the ``true''
signals $N_{W/Z}$ calculated from the measured ones via the transfer matrix $T^{-1}$.}
\label{tab-6-5-wzsep}
\end{table}

\section{Conclusions}
\label{chap-6-6}

What is the message to be taken from the simulations presented in this chapter? At the beginning,
we have argued that $W^\prime / Z^\prime$ production in the $s$ channel is a very sensitive probe of
the fermion sector in the Three-Site Model and in particular of the delicate interplay of fermion
and KK gauge boson wavefunction which allows the model to evade the precision constraints.

In the simulations presented here we have shown that the heavy gauge bosons, albeit fermiophobic,
can indeed lead to observable resonances in the $s$ channel at the LHC. In the case of the
$Z^\prime$, we have found that the corresponding resonance shows only little dependence on the
fermion delocalization parameter $\epsilon_L$ and should be expected to be visible within the first
$10-\unit[20]{fb^{-1}}$ in the $jjl\nu$ channel for any point in parameter space. At the LHC, such a
signal together with the absence (or near-absence) of a corresponding signal in the dilepton
and Drell-Yan channels would be a strong sign for a fermiophobic, heavy neutral vector resonance like the
Three-Site $Z^\prime$.

As far as the $W^\prime$ is concerned, the simulations have also shown that the resonance might be
accessible at the LHC. However, the magnitude of the signal is highly dependent on
$\epsilon_L$, and there are regions in parameter space in which no signal would be observed at the
LHC. Even in this case though, the existence of the $W^\prime$ could be established via the
strahlung process presented in the last chapter (which is essentially independent of
$\epsilon_L$), and the absence of a corresponding peak in the $s$ channel processes would be a clear sign of its
fermiophobic nature.

We have found three different possible discovery channels for the $W^\prime$, all of which show a
comparable discovery potential at the parton level. However, the $jjl\nu$ type final states contain
contributions from both $W^\prime$ and $Z^\prime$, and if ATLAS and CMS turn out to be unable to resolve
the jet momenta with the accuracy required for separating the resonances
via cuts on the invariant jet mass, some trick like
the one presented in \ref{chap-6-5} needs to be applied if this final state is to be exploited for
$W^\prime$ production. Otherwise, only the combined resonance of $Z^\prime$ and $W^\prime$ can be
observed in this channel which, luckily, at least wouldn't hurt the $Z^\prime$
detection much as the signal is stronger
than the $W^\prime$ one anyways, and the amount of $W^\prime$ events contained in the peak could be
inferred from the other two $W^\prime$ final states available.

All simulations presented in this thesis have been performed at
the parton level only with no detector effects taken into account. In the case of those
presented in this chapter, an ATLAS detector simulation was added in the master's thesis of F.
Bach \cite{fabian:master}. This also includes additional backgrounds which were purposefully not
taken into account in the parton level simulations as they depend on details of the detector and of
the jet
identification algorithm, e.g. the background coming from $t\bar{t}$ production with two jets
escaping down beam pipe.

The result shows that the $Z^\prime$ signature is rather robust and
persists with a $5\sigma$ effect still being visible in the first $5-\unit[30]{fb^{-1}}$ even if
detector effects are taken into account. The $W^\prime$ resonance turns out to be more fragile, with
the separation trick for $jjl\nu$ not working well after the detector simulation and the signal in
$jjll$ being diminished by the measurement error on the jet momenta and additional backgrounds. In
light of this result, $lll\nu$ becomes the preferred final state for this process at the LHC, and
the detector simulation suggests that the parton level results carry over to the actual experiment
in this case. For more details on these issues see \cite{fabian:master}.
\newpage

\section{Additional plots}
\label{chap-6-7}

\chapter{Heavy Fermion Production}
\label{chap-7}

\begin{quote}\itshape
Oliphaunt am I,\\
Biggest of all.
\end{quote}
\hfill\begin{minipage}{0.7\textwidth}\small\raggedleft
(Excerpt from the ``Oliphaunt song'' in J.R.R. Tolkien's ``Lord of the Rings'')
\end{minipage}
\\[5mm]
After discussing at length the prospects of discovering the KK partners of the $Z$ and $W$ as well
as the possibility of getting a handle on their fermiophobic nature,
the only aspect of the model which we have not addressed in our simulations so far are the heavy fermion
partners. With masses above $\unit[1.8]{TeV}$ and widths of several $\unit[100]{GeV}$ (c.f. chapter
\ref{chap-3}), these particles certainly are near the limit of what may be accessible as a resonance at
the LHC. Nevertheless, we will see in this chapter that this may be feasible at least in some
portions of parameter space.

As in chapters \ref{chap-5} and \ref{chap-6}, we will first discuss the different processes that
could be exploited for probing for this kind of new physics in \ref{chap-7-1}. Subsequently, the
details of the simulations we performed will be discussed in \ref{chap-7-2}, while the results are
presented and discussed in \ref{chap-7-3}.

\section{Processes}
\label{chap-7-1}

Besides the fermiophobic $W^\prime$ and $Z^\prime$, the main feature of the
Three-Site Model are the new heavy fermion resonances. As we have seen in chapter \ref{chap-1},
their introduction was required in the Three-Site setup in
order to allow for the tuning of the couplings between the Standard Model fermions and the KK gauge
bosons which in turn is necessary for evading the electroweak precision constraints.
Therefore, regardless of their large mass scale $M_\text{bulk}>\unit[1.8]{TeV}$,
these particles are an integral part of the phenomenology of the Three-Site Model, and if other
experimental results would hint at such a kind of new physics, a way to access these particles would
be desirable.

The production mechanism which might first come to mind are simple Drell-Yan like processes of the
type
\[
\parbox{35mm}{\begin{fmfgraph}(35,17)
\fmfleft{i2,i1}\fmfright{o2,o1}
\fmf{fermion}{i1,v1,i2}\fmf{wiggly}{v1,v2}\fmf{heavy}{o1,v2}\fmf{fermion}{v2,o2}
\fmfdot{v1,v2}
\end{fmfgraph}}
\quad+\quad
\parbox{35mm}{\begin{fmfgraph}(35,17)
\fmfleft{i2,i1}\fmfright{o2,o1}
\fmf{fermion}{i1,v1,i2}\fmf{dbl_wiggly}{v1,v2}\fmf{heavy}{o1,v2}\fmf{fermion}{v2,o2}
\fmfdot{v1,v2}
\end{fmfgraph}}
\]
However, such processes always involve a quark-antiquark pair in the initial state which is very
unlikely to deliver the large amount of energy necessary for the creation of a heavy fermion at a
$pp$ collider like LHC,
and simulating $pp\rightarrow e^+{e^-}^\prime$ with WHIZARD / O'Mega indeed reveals a cross section
of the order of $\unit[0.5]{ab}$ at the LHC.
Clearly, even if we multiply this with with an optimistic factor of $42$ in order to take all Standard Model
fermions visible to the detector into account\footnote%
{
$\D
\left(6\;\text{quarks} \quad\times\quad 3\;\text{colors} \quad+\quad 3\;\text{leptons} \right)
\quad\times\quad2\;\text{(fermions + antifermions)} \quad=\quad 42
$
},
the resulting cross section is still pathetic, and detecting the heavy fermions this way is out
of question.

\begin{figure}
\centerline{\includegraphics[width=\singleplotwidth,angle=270]{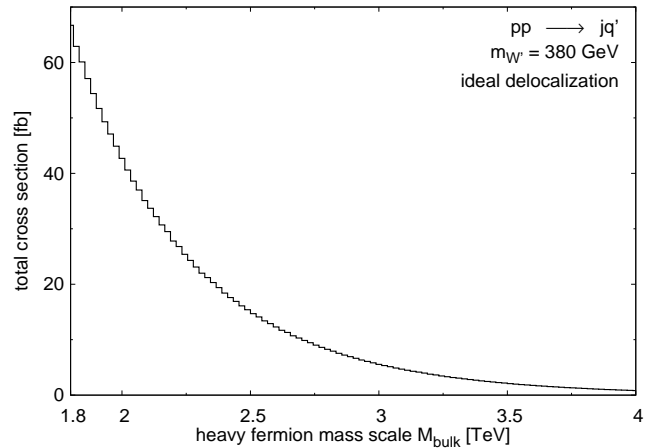}}
\caption{Total cross section for $pp\rightarrow jq^\prime$ as a function of $M_\text{bulk}$.}
\label{fig-7-1-hqprod222}
\end{figure}
However, there is another kind of potential production processes which are mediated by $t$ channel
interactions
\[
\parbox{22mm}{\begin{fmfgraph}(22,25)
\fmfleft{i2,i1}\fmfright{o2,o1}\fmf{fermion}{i1,v1}\fmf{heavy}{v1,o1}\fmf{fermion}{i2,v2,o2}
\fmf{wiggly}{v1,v2}\fmfdot{v1,v2}
\end{fmfgraph}}
\quad+\quad
\parbox{22mm}{\begin{fmfgraph}(22,25)
\fmfleft{i2,i1}\fmfright{o2,o1}\fmf{fermion}{i1,v1}\fmf{heavy}{v1,o1}\fmf{fermion}{i2,v2,o2}
\fmf{dbl_wiggly}{v1,v2}\fmfdot{v1,v2}
\end{fmfgraph}}
\]
This kind of process can contain two valence quarks in the initial state with much more
energy being typically available in the partonic CMS than in the above case of a valence and a sea
quark. The resulting cross section for $pp\rightarrow
jq^\prime$ ($q^\prime$ being any of $u^\prime,\bar{u}^\prime,d^\prime,\bar{d}^\prime$)
is shown in fig.~\ref{fig-7-1-hqprod222} as a function of $M_\text{bulk}$, exploiting the fact that the heavy
partners of the light Standard Model fermions are degenerate (c.f. chapter \ref{chap-3-1}) by summing
over fermion flavors in order to enhance the signal.
Evidently, the available energy is much higher, and the resulting cross section can be as big as
$\approx\unit[66]{fb}$ for the smallest $M_\text{bulk}=\unit[1.8]{TeV}$, dropping only
moderately with nearly $\unit[10]{fb}$ being still available for $M_\text{bulk}=\unit[3]{TeV}$. These
numbers are much more encouraging than those in the Drell-Yan like case, and we can indeed hope that
it might be possible to dig out a heavy quark resonance this way if $M_\text{bulk}$ is not too high.

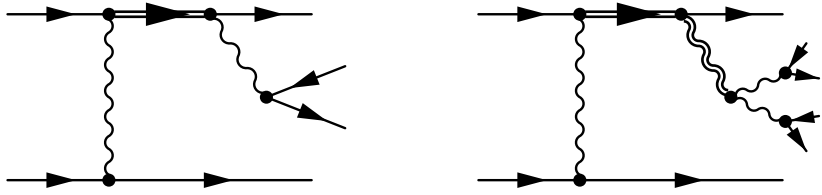
\begin{figure}
\centerline{
\parbox[b]{50mm}{
\begin{fmfgraph}(50,22)
\fmfleft{i2,i1}\fmfright{o4,o3,o2,o1}
\fmf{phantom}{i1,v1,v2,o1}\fmf{phantom}{i2,v3,d1,o4}\fmffreeze
\fmf{phantom}{v1,d2,v3}\fmffreeze\fmf{phantom}{d2,v4}\fmf{fermion}{o2,v4,o3}\fmffreeze
\fmf{wiggly}{v1,v3}\fmf{fermion}{i1,v1}\fmf{heavy}{v1,v2}\fmf{fermion}{v2,o1}
\fmf{fermion}{i2,v3,o4}\fmf{wiggly}{v2,v4}
\fmfdot{v1,v2,v3,v4}
\end{fmfgraph}
}
\hspace{1cm}
\parbox[b]{50mm}{
\begin{fmfgraph}(50,22)
\fmfleft{i2,i1}\fmfright{o6,o5,o4,o3,o2,o1}
\fmf{phantom}{i1,v1,v2,o1}\fmf{phantom}{i2,v3,d1,o6}\fmffreeze
\fmf{phantom}{v1,d2,v3}\fmffreeze\fmf{phantom}{d2,v4}\fmf{phantom}{o2,v4,o5}\fmffreeze
\fmf{wiggly}{v1,v3}\fmf{fermion}{i1,v1}\fmf{heavy}{v1,v2}\fmf{fermion}{v2,o1}
\fmf{fermion}{i2,v3,o6}\fmf{dbl_wiggly}{v2,v4}\fmf{wiggly}{v5,v4,v6}
\fmf{fermion}{o2,v5,o3}\fmf{fermion}{o4,v6,o5}
\fmfdot{v1,v2,v3,v4,v5,v6}
\end{fmfgraph}
\label{fig-7-1-diags}
}}
\caption{Signal diagrams for $t$ channel induced production of heavy fermions with the two
final states as obtained from the different decay channels of the heavy fermions.
}
\end{figure}
As discussed in chapter \ref{chap-3-4}, the heavy fermions decay into a Standard Model fermion and a
gauge boson, the ratio between decays into light and heavy gauge bosons being roughly $1$:$1$.
The Standard Model $W$ and $Z$ then decay into two fermions, leading to four particle final states like
fig.~\ref{fig-7-1-diags} left,
while the $W^\prime$ and $Z^\prime$ cascade into four fermions by means of two intermediary gauge
bosons, resulting in a six particle final state similar to fig.~\ref{fig-7-1-diags} right.

The six particle final states come with the advantage that a cut on the invariant mass of the heavy
gauge boson can be leveraged in order to reduce the number of background events (given that the
$W^\prime / Z^\prime$ mass has been previously determined). However, 
a preliminary study of the six particle final state containing four jets turned out to be quite
involved due to the large number of background processes already present at the parton level, and
the resulting event count in the signal channels was not big enough to leave much hope for a
significant discovery potential in this process. The investigation of six particle final states of
the form $jjlll\nu$ is still an ongoing project, but as processes leading to this final state are
suppressed relative to those with four particles in the final state by
another gauge boson branching factor, it is unlikely that they will offer any advantage over these.

Therefore, we only present results for the four fermion processes in this work. Excluding the
case of four jets due to the large QCD backgrounds, we remain with $jjl\nu$ and $jjll$ as potential
final states. However, as the branching ratio into $Z\rightarrow ll$ is significantly smaller than that of
$W\rightarrow l\nu$, we should expect the $jjl\nu$ final state to perform much better than $jjll$,
provided that we can cope with the missing neutrino momentum.

\begin{figure}
\centerline{\includegraphics[width=\singleplotwidth,angle=270]{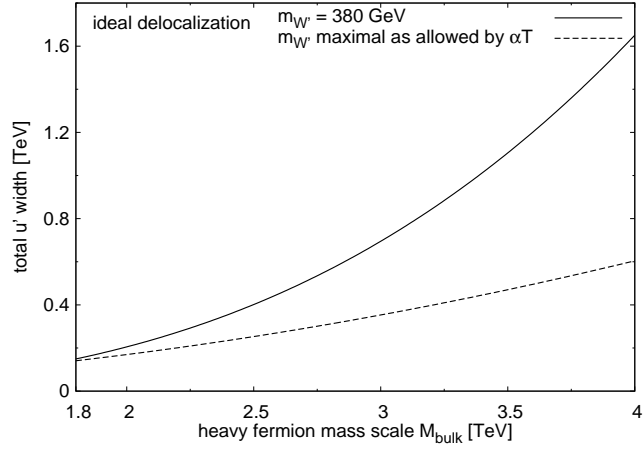}}
\caption{The total $u^\prime$ width as a function of $M_\text{bulk}$ for the minimal value of the
$W^\prime$ mass $m_{W^\prime}=\unit[380]{GeV}$ and for the highest $m_{W^\prime}$ allowed by the
constraint $\alpha T\le 0.005$.}
\label{fig-7-1-wdu-minmax}
\end{figure}
Apart from the mass and couplings of the heavy fermions, the prospects for discovering these
particles as a resonance also depend on their total width. As the amount of phasespace
available for the decay of a particle depends on the mass of the decay products, we should expect
that the total width exhibits a strong dependence on $m_{W^\prime}$. Fig.~\ref{fig-7-1-wdu-minmax}
shows the total $u^\prime$ width as a function of $M_\text{bulk}$ both for
$m_{W^\prime}=\unit[380]{GeV}$ (which is the smallest value allowed by the LEP constraint on the
triple gauge boson couplings) and for the highest $W^\prime$ mass which is still in accordance with
the condition $\alpha T\le0.005$ (c.f. chapter \ref{chap-3-2}). This plot demonstrates that the
range of potential values for the width grows significantly with increasing $M_\text{bulk}$, and for
$M_\text{bulk}=\unit[3]{TeV}$, the difference is more than $\unit[200]{GeV}$. Clearly, this
variation of the width with $m_{W^\prime}$ is big enough to have an influence on the significance of
the resonance.

An additional dependence on $m_{W^\prime}$ comes from the $Wf^\prime f$ / $Zf^\prime f$ couplings
which scale with $x$ (c.f. chapter \ref{chap-3-3}). Looking at \eqref{equ-3-2-x}, it
is obvious that,
while the resonance becomes more narrow and easier to detect with growing $W^\prime$ mass, the
couplings responsible for the production of the heavy quark in fact go down, and it is not clear a priori
which effect dominates the significance of the signal in the end.

In contrast to the $s$ channel production of heavy gauge bosons discussed in the last chapter, we
have no reason to expect a large influence of the delocalization parameter $\epsilon_L$ on this
process. The only possible source of such a dependence ore processes involving a $W^\prime$ in
the $t$ channel, the relative contribution to the amplitude of which would be
\[
\frac{g_{W^\prime qq^\prime,L}\;g_{W^\prime qq,L}}{g_{Wqq^\prime,L}\;g_{Wqq,L}}
\]
(the right-handed components of massless fermions sit completely at the right brane with their
partners residing in the bulk and therefore, the right-handed couplings $g_{W^\prime q^\prime q,R}$ vanish).
This turns out to be less than $10\%$ over the whole parameter space
and therefore, we are free to perform the simulation in the ideally delocalized scenario with the
effect of nonideal delocalization on the result being small and negligible when compared to other
uncertainties.

\section{Simulation Details}
\label{chap-7-2}

In order to examine whether it might indeed be possible to discover the heavy fermions at the LHC,
we have performed full parton level simulations of the processes $pp\rightarrow jjll$ and
$pp\rightarrow jjl\nu$ for an integrated luminosity of $\ilum=\unit[400]{fb^{-1}}$ and a
total energy of $\unit[14]{TeV}$, again with the definitions \eqref{equ-5-2-passign} and the same parton
distributions as in chapters \ref{chap-5} and \ref{chap-6}.
In the case of the final state containing a neutrino, we have
applied the reconstruction method already used in chapter \ref{chap-6} and presented in
\ref{chap-6-2}, again double counting the event if both solutions pass the cuts. To get a
sufficiently smooth estimate of the background, additional simulations in the Standard Model were
performed for an integrated luminosity of $\ilum=\unit[1600]{fb^{-1}}$ and the result then
downscaled by a factor of $4$.

In order to suppress the backgrounds and identify the jet which originates from the heavy fermion,
we concentrate on events in which the total energy available in the parton CMS is just barely enough
to create the KK particle which is therefore only weakly boosted in this frame. In the CMS, this implies
that the jet coming from the decaying heavy quark is emitted isotropically in all directions and is
approximately back-to-back to the $W/Z$ coming from the
decay, while the jet originating from the proton will be aligned with the beam axis. Note that these
kinematics are very different from those exhibited by gauge boson fusion processes like
\[
\begin{fmfgraph}(40,22)
\fmfleft{i2,i1}\fmfright{o4,o3,o2,o1}
\fmf{fermion}{i1,v1,o1}\fmf{fermion}{i2,v2,o4}\fmffreeze
\fmf{wiggly}{v1,v3,v2}\fmf{wiggly}{v3,v4}\fmf{fermion}{o2,v4,o3}
\fmfdot{v1,v2,v3,v4}
\end{fmfgraph}
\]
which feature two high rapidity jets aligned with the beam axis and which can be expected to
constitute the largest part of the irreducible background.

To facilitate the selection of events whose kinematics match these criteria we applied the following cuts in the
parton CMS
\\[2ex]
{\renewcommand{\tabcolsep}{0ex}\renewcommand{\arraystretch}{1.5}
\centerline{\begin{tabular}{|p{1ex}p{0.5\textwidth}p{1ex}|p{1ex}rclp{1ex}|}\hline
&polar angle of jet $j^\prime$ originating from heavy quark &&
&\raisebox{-1.1ex}[-1.1ex]{$0\le$} & \raisebox{-1.1ex}[-1.1ex]{$\;\abs{\cos\theta_{j^\prime}}\;$} &
\raisebox{-1.1ex}[-1.1ex]{$\le 0.7$}&
\\\hline
&polar angle of jet $j$ originating from proton &&
&\raisebox{-1.1ex}[-1.1ex]{$0.95\le$} & \raisebox{-1.1ex}[-1.1ex]{$\abs{\cos\theta_j}$} &
\raisebox{-1.1ex}[-1.1ex]{$\le 1$}&
\\\hline
&angle between $j^\prime$ and $W/Z$&&
&$-1\le$ & $\cos\theta_{jW}$ & $\le -0.8$&
\\\hline
&$p_T$ of heavy quark &&
&$0\le$ & $p_{T,q^\prime}$ & $\le\unit[200]{GeV}$&
\\\hline
&$p_T$ of $j^\prime$ and $W$ &&
&$\unit[500]{GeV}\le$ & $\;p_{T,j^\prime}\;$,\parbox{0mm}{$\;p_{T,W}$} &&
\\\hline
\end{tabular}}}
\\[2ex]
We also applied a $p_T$ cut on all observable momenta (and also on $p_{T,\text{miss}}$ in
the case of $pp\rightarrow jjl\nu$)
\[ p_T \ge \unit[50]{GeV} \]
as well as a cut on the total invariant mass
\[ m_\text{tot} \ge \unit[1]{TeV} \]
for additional background suppression. In order to avoid infrared divergences and for simulating the
acceptance region of the detector, we demanded
\[ -0.99 \le \cos\theta \le 0.99 \]
for the polar and intermediate angles of all observable particles in the hadron CMS, and the same small $x$ cut
already used in the last two chapters \ref{chap-6} -- \ref{chap-7}
\[ x \ge 1.4\cdot10^{-4} \]
was applied on the incoming partons (also in the hadron CMS). In the case of the $jjll$ final state,
we also applied an identification cut on the invariant mass of the dilepton system
\[ \unit[81]{GeV} \le m_{ll} \le \unit[101]{GeV} \]
\begin{table}
{\renewcommand{\arraystretch}{1.2}
\centerline{\begin{tabular}{|c|c||c|c||c|}
\hline
\multicolumn{2}{|c||}{parameters} & \multicolumn{2}{c||}{heavy quark properties} &
point label
\\\hline
$M_\text{bulk}\quad\left[\unit{TeV}\right]$ & $m_{W^\prime}\quad\left[\unit{GeV}\right]$
& $m\quad\left[\unit{TeV}\right]$ & $\Gamma\quad\left[\unit{GeV}\right]$ &
\\\hline\hline
$1.8$ & $380$ & $1.90$ & $150$ & I \\\hline
$2.0$ & $380$ & $2.10$ & $205$ & II \\\hline
& $380$ & $2.31$ & $273$ & IIIa \\\cline{2-5}
\raisebox{1.5ex}[1.5ex]{$2.2$} & $430$ & $2.29$ & $204$ & IIIb\\\hline
& $380$ & $2.63$ & $401$ & IVa \\\cline{2-5}
\raisebox{1.5ex}[1.5ex]{$2.5$} & $460$ & $2.58$ & $257$ & IVb \\\hline
& $380$ & $2.84$ & $506$ & Va \\\cline{2-5}
\raisebox{1.5ex}[1.5ex]{$2.7$} & $480$ & $2.78$ & $294$ & Vb \\\hline
& $380$ & $3.05$ & $628$ & VIa \\\cline{2-5}
\raisebox{1.5ex}[1.5ex]{$2.9$} & $500$ & $2.98$ & $333$ & VIb \\\hline
\end{tabular}}}
\caption{The parameter space points at which the simulations of heavy quark production were
performed. The first two columns define the parameter space point (we always assume ideal
delocalization, c.f. section \ref{chap-7-1}), the second two give mass and width of the KK
partners to the light Standard Model quarks, and the labels used to identify the
points in the analysis are given in the last column.}
\label{tab-7-2-points}
\end{table}
In order to identify the part of parameter space in which the heavy quarks might be visible at the LHC,
we have conducted simulations of the $jjl\nu$ final state at ten different points in parameter space
as shown in tab.~\ref{tab-7-2-points}. For each value of $M_\text{bulk}$, a simulation at
$m_{W^\prime}=\unit[380]{GeV}$ (which is the lower limit given by the LEP / LEP-II constraint on
the triple gauge boson couplings, c.f. chapter \ref{chap-3-2}) was performed. In addition, for
$M_\text{bulk}>\unit[2]{TeV}$, another simulation near the upper limit on $m_{W^\prime}$ (see fig.
\ref{fig-7-1-wdu-minmax}) was conducted in order to probe the dependence of the result on the
$W^\prime$ mass. As the simulations of $pp\rightarrow jjll$ at the points I and II already show
that $jjl\nu$ yields a much better signal, no simulations of $jjll$ were performed at IIIa/b --
VIa/b.

\section{Simulation Results}

\begin{figure}
\centerline{
\includegraphics[width=\doubleplotwidth,angle=270]{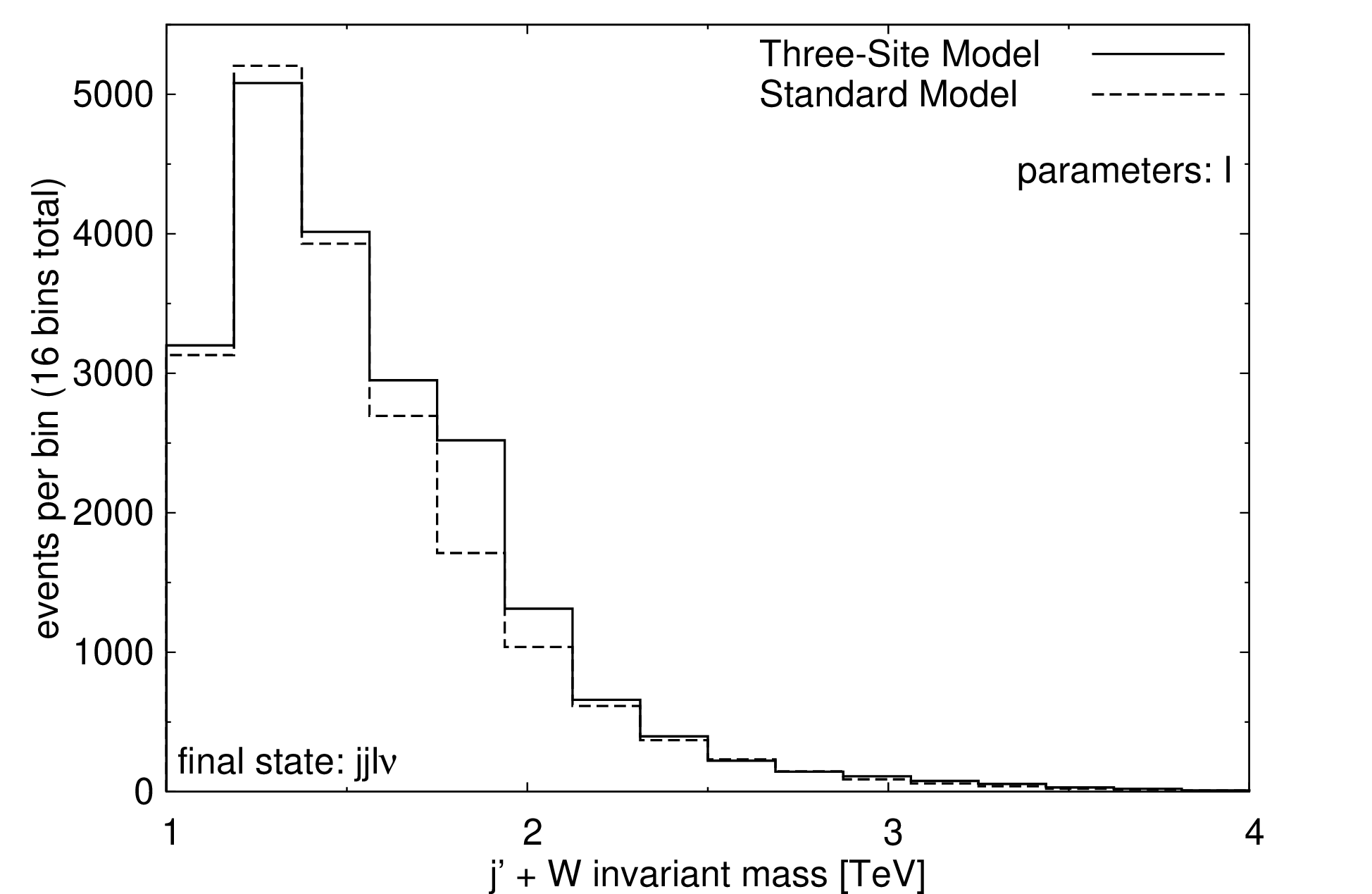}
\includegraphics[width=\doubleplotwidth,angle=270]{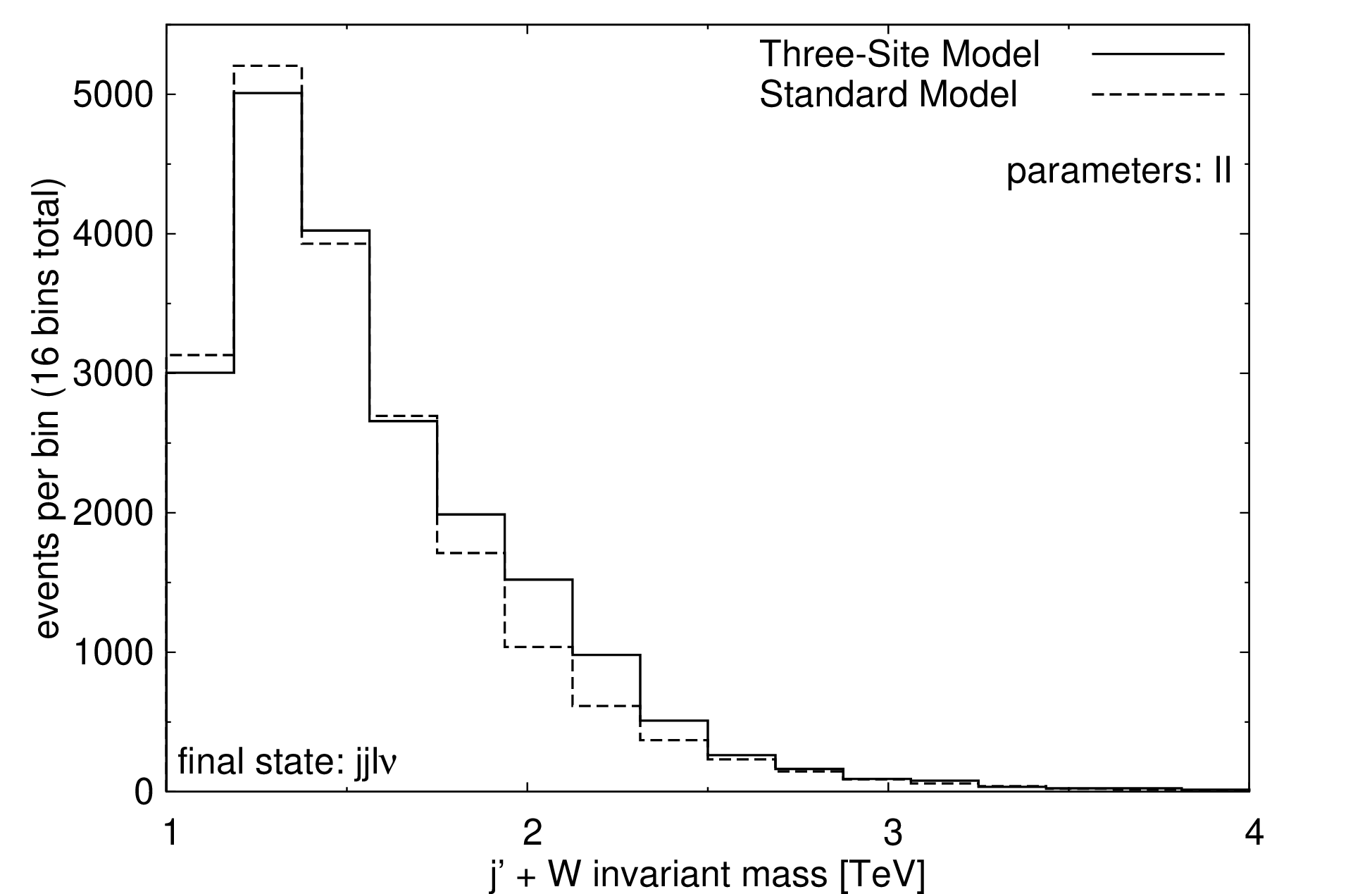}
}
\caption{The heavy quark resonance in the $jjl\nu$ final state for the parameter space points I
(\emph{left}) and II (\emph{right}), visible as an excess over the Standard Model background.}
\label{fig-7-3-wfull}
\end{figure}
Let's start our discussion of the simulation results with the $jjl\nu$ final state.
Fig.~\ref{fig-7-3-wfull} shows the invariant mass of the combination of the jet $j^\prime$ potentially
originating from a heavy quark (as identified by our cuts, see the preceding section) and the
$W$ for the parameter space points I and II (see tab.~\ref{tab-7-2-points}). While the
background is considerable, an excess in the histogram at the invariant mass of the heavy quark over
the Standard Model background is clearly visible (the peak in the distribution around
$\unit[1.2]{TeV}$ is not related to the new physics, but a kinematical artifact induced by the cuts).

\begin{figure}
\centerline{
\includegraphics[width=\doubleplotwidth,angle=270]{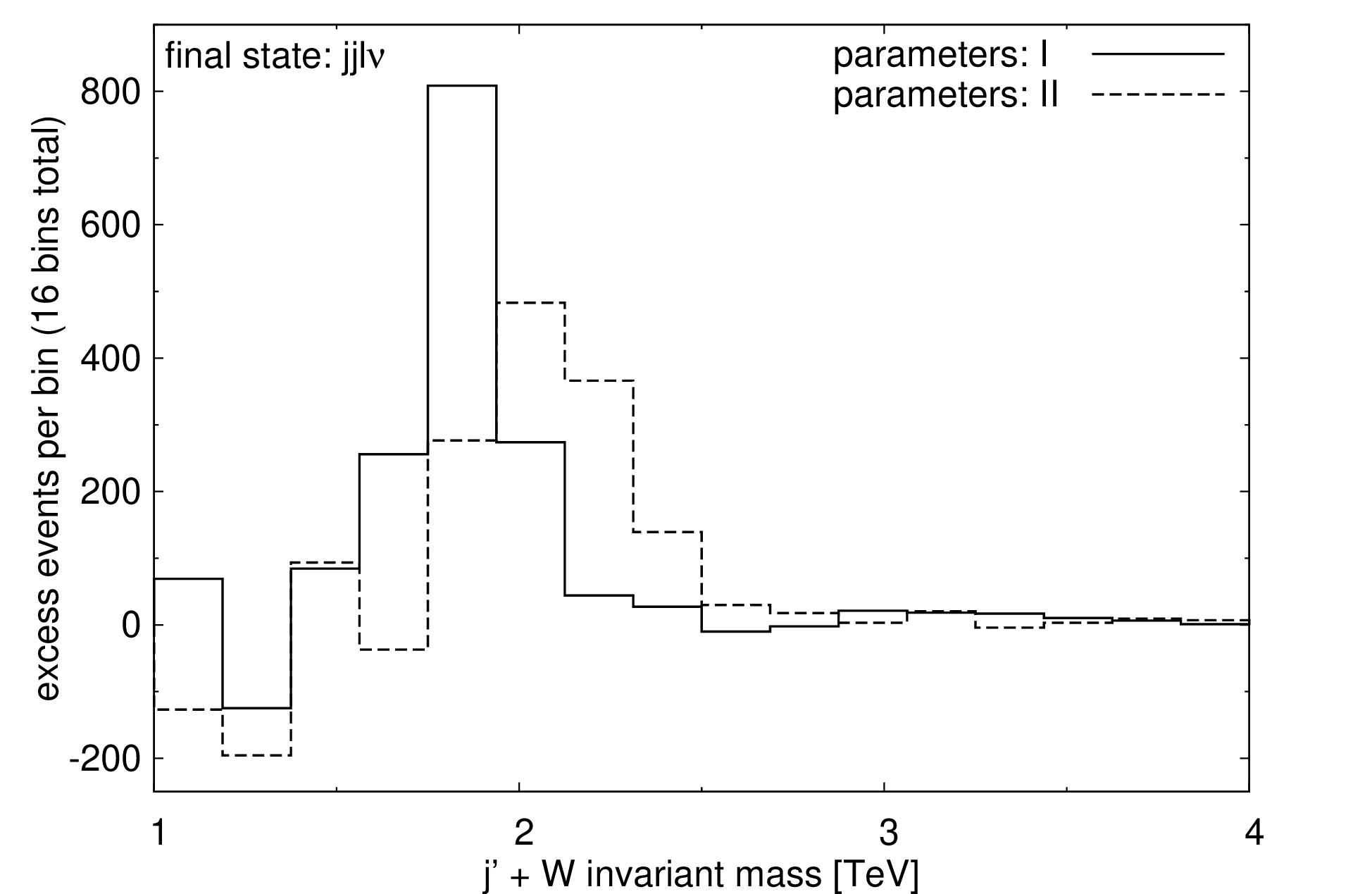}
\includegraphics[width=\doubleplotwidth,angle=270]{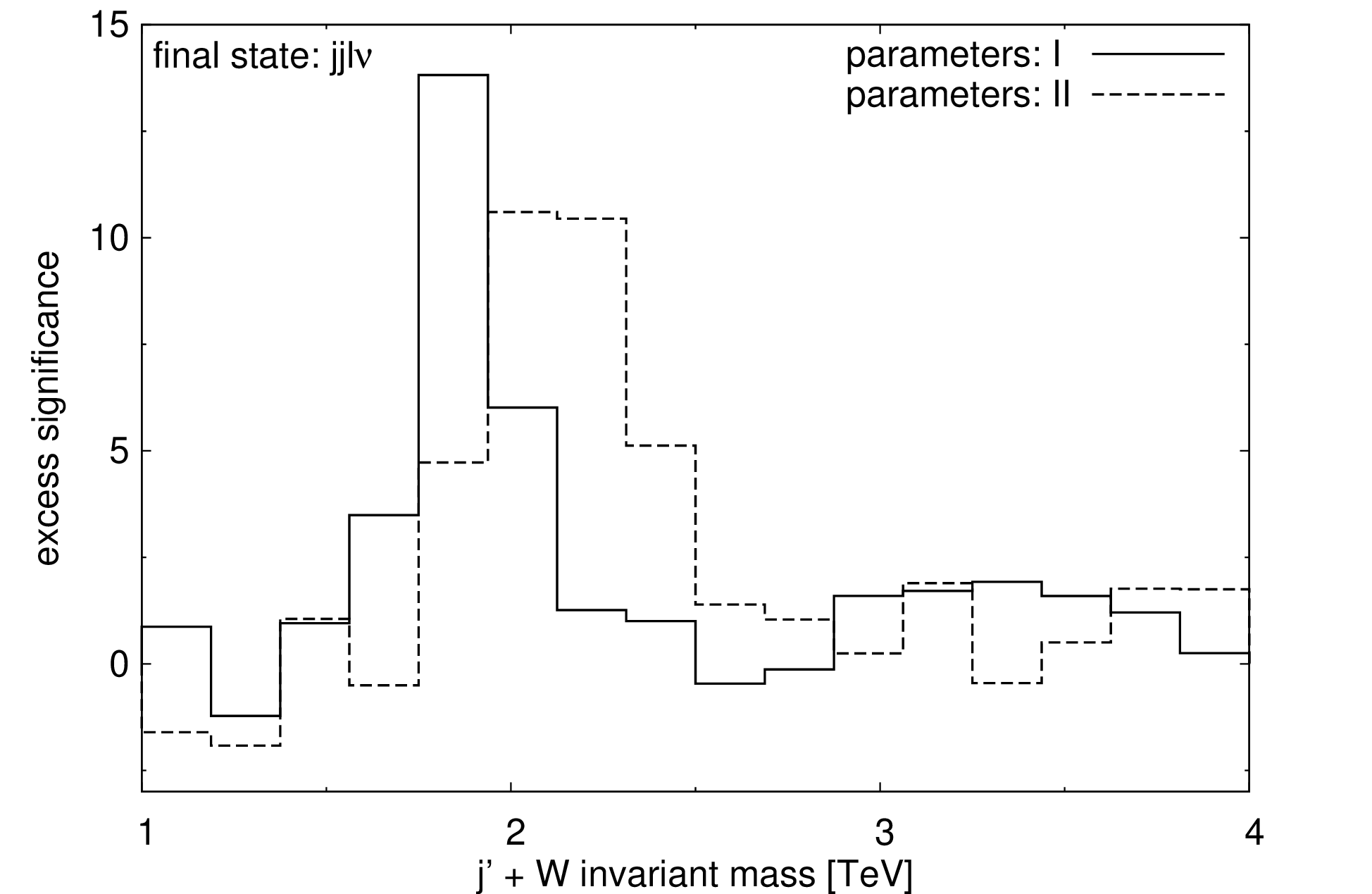}
}
\caption{
\emph{Left}: The excess in the $j^\prime + W$ invariant mass with the heavy quark resonance.
\emph{Right}: The per-bin significance of the excess calculated as described in the text.
}
\label{fig-7-3-diff}
\end{figure}
Calculating the difference between the event counts in the Three-Site Model and those in the
Standard Model as shown in fig.~\ref{fig-7-3-diff} left for points I and II, we find that the excess
is quite big, containing about $1000$ events, and indeed has the shape of a broad resonance peak. In
order to calculate its significance, we follow the same reasoning as in chapter
\ref{chap-6-3} and assume that the double counting performed in the neutrino momentum reconstruction
step simply doubles the event count and define the significance of the excess by
\begin{equation} s_i = \frac{N_{s,i} - N_{b,i}}{\sqrt{2N_{b,i}}} \label{equ-7-3-sig}\end{equation}
where the $N_{s,i}$ are the event counts in the Three-Site Model and the $N_{b,i}$ are the counts in the Standard
Model. The resulting significance is shown on a per-bin basis in fig.~\ref{fig-7-3-diff} right,
demonstrating the excess in the bins lying within the peak to be significant with $s>10$.
In addition, while the
``naked'' difference shown in fig.~\ref{fig-7-3-diff} left also exhibits a noticeable difference in
the leftmost bins around $\unit[1]{TeV}$, fig.~\ref{fig-7-3-diff} right reveals this to be not
significant and in fact just a statistical fluctuation, the conspicuous magnitude of which is
solely due to the number of background events in this energy range.

\begin{figure}
\centerline{\includegraphics[width=\singleplotwidth,angle=270]{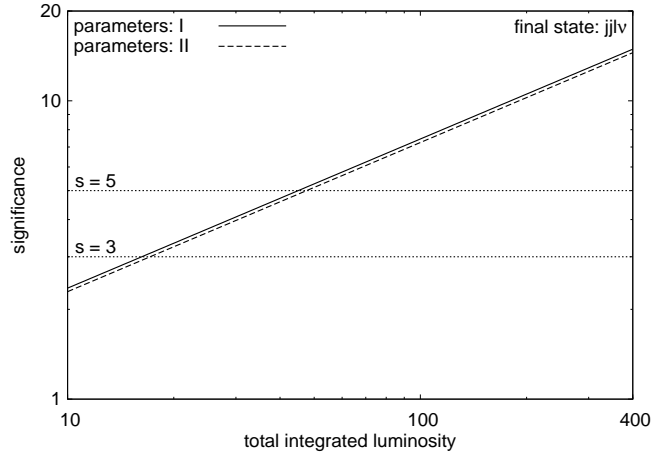}}
\caption{The significance of the heavy quark resonance as defined in the text at the points I and II as
a function of the integrated luminosity together with the $3\sigma$ and $5\sigma$ discovery
thresholds.}
\label{fig-7-3-wsig-I-II}
\end{figure}
In order to get a quantitative estimate of the significance of the whole resonance, we define the signal
$N_s$ to be the number of events within the $\pm\Gamma$ range around the invariant mass\footnote%
{
Choosing a wider interval gives a slightly higher significance of the signal at parameter
space points with low $\Gamma$, but actually turns out degrade the signal at points with high width
like Va and VIa.
}
of the heavy quarks.
From this, we calculate the significance according to \eqref{equ-7-3-sig} just as we did for the
individual bins. The result for the
points I and II is shown in fig.~\ref{fig-7-3-wsig-I-II} as a function of the integrated luminosity
together with the $3\sigma$ and $5\sigma$ discovery thresholds. At both points, the signal exhibits
nearly the same significance of $\approx 15\sigma$ for $\ilum=\unit[400]{fb^{-1}}$ and intersects
the $5\sigma$ threshold already around $\unit[45]{fb^{-1}}$.

\begin{figure}
\centerline{
\begin{tabular}{cc}
\includegraphics[width=\doubleplotwidth,angle=270]{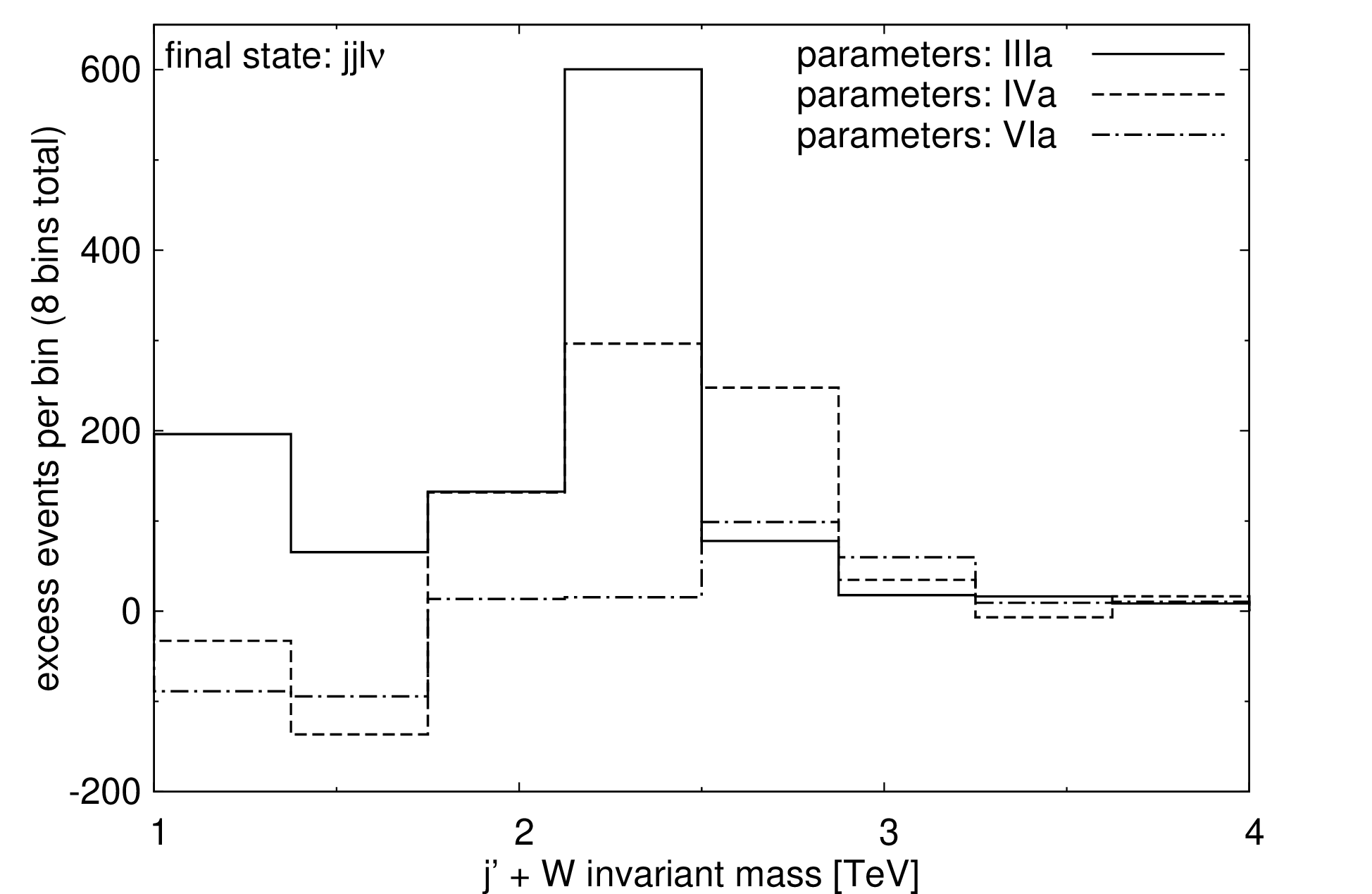} &
\includegraphics[width=\doubleplotwidth,angle=270]{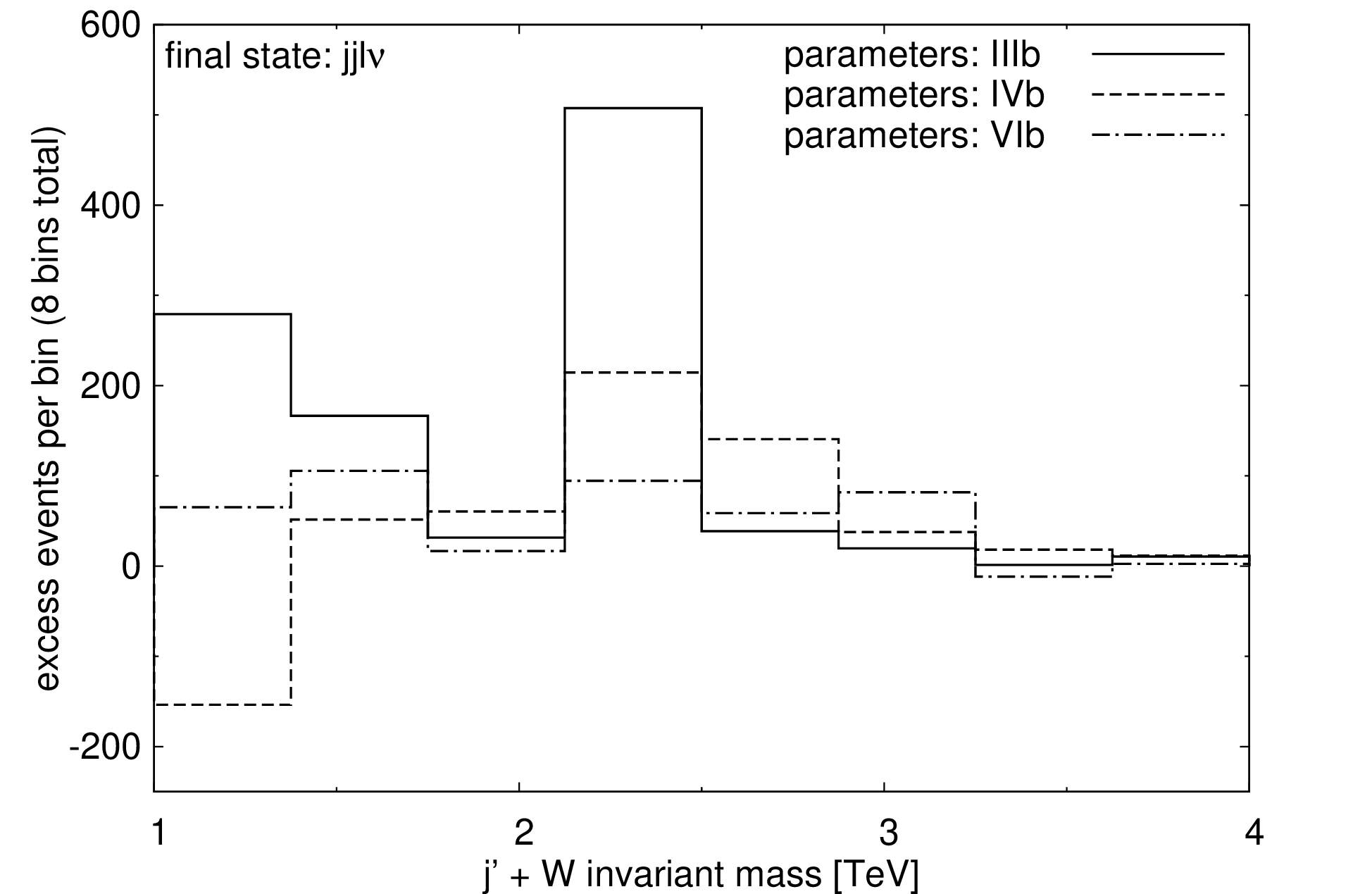}
\\
\includegraphics[width=\doubleplotwidth,angle=270]{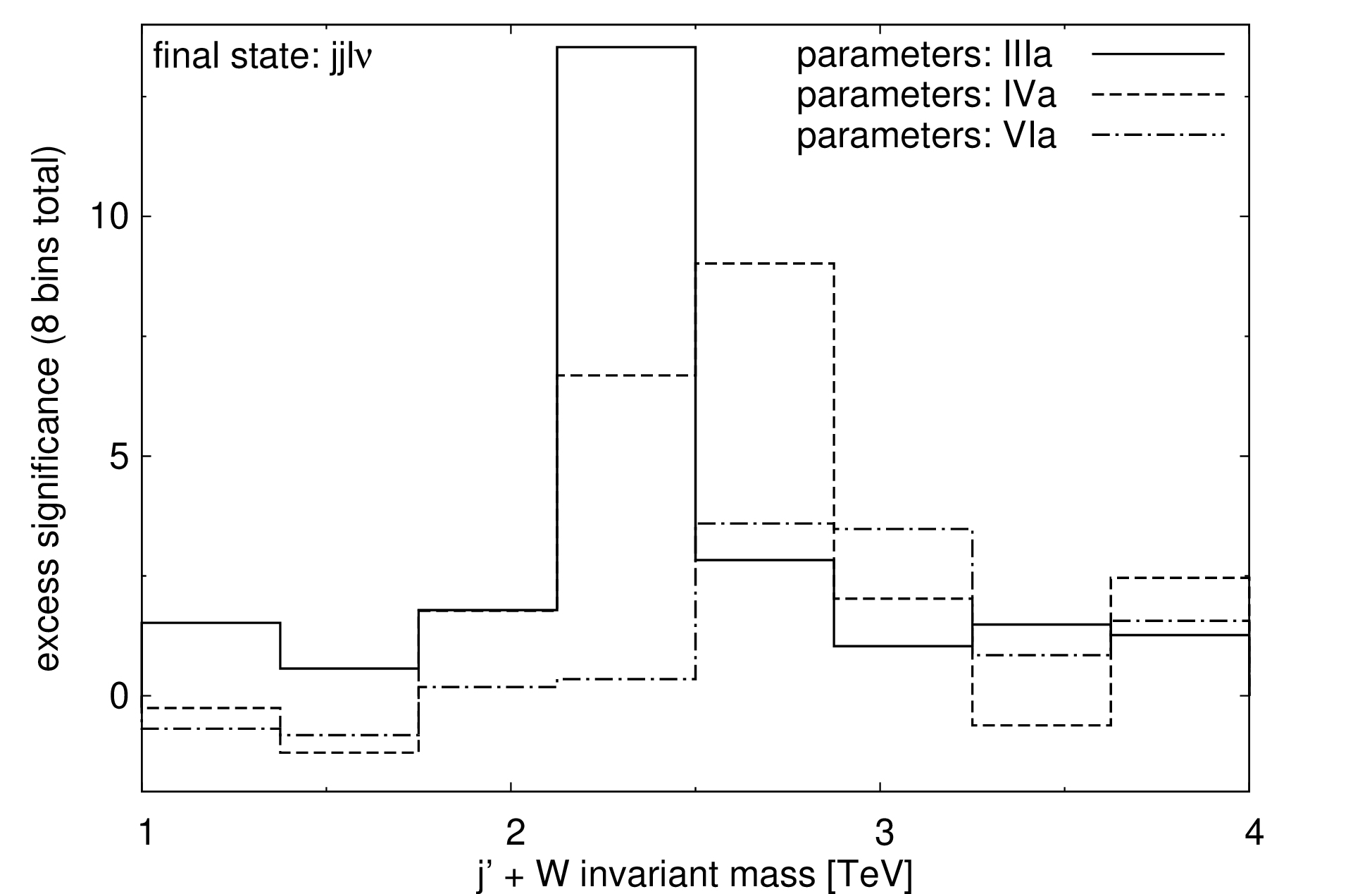} &
\includegraphics[width=\doubleplotwidth,angle=270]{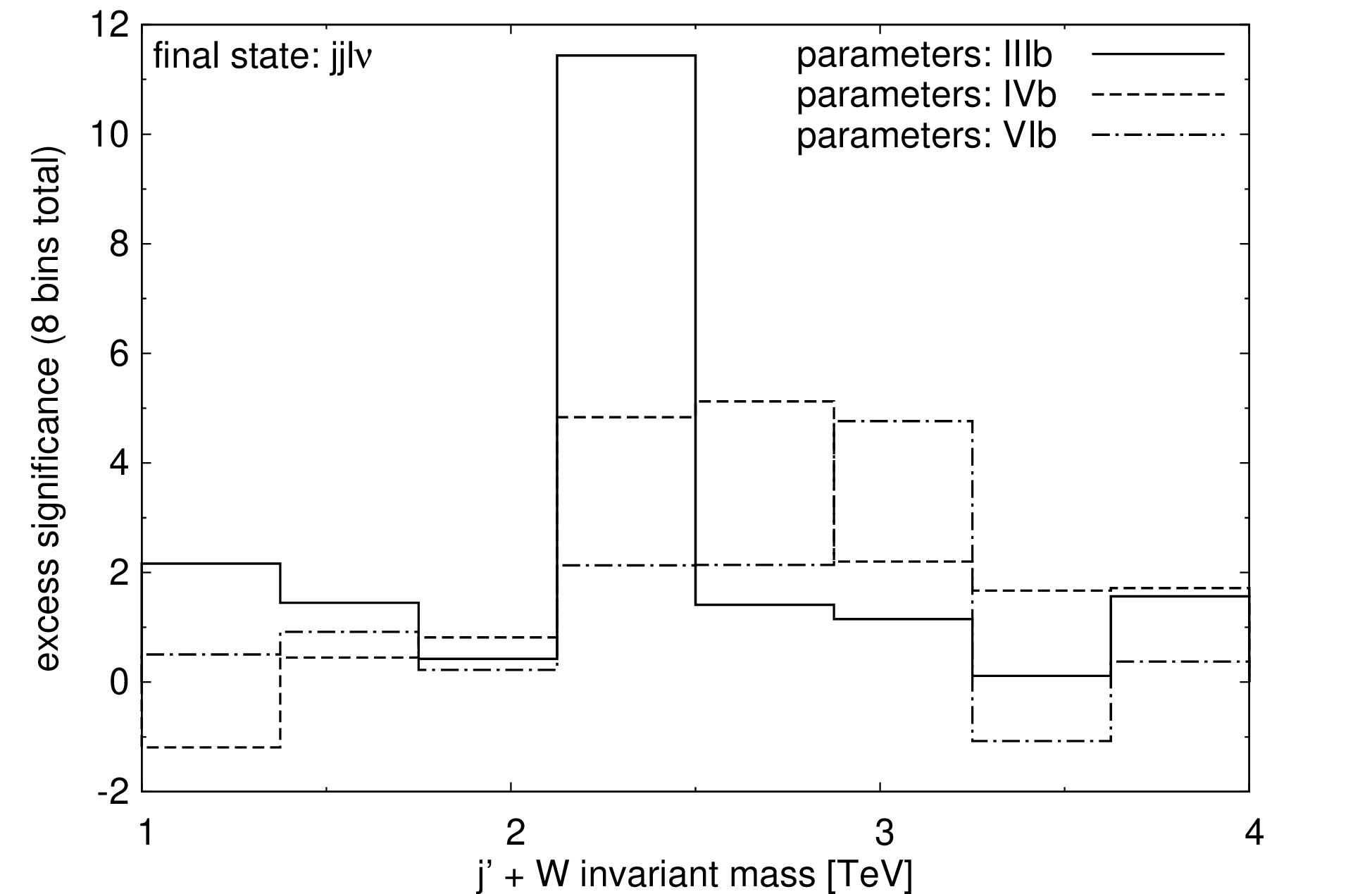}
\\
\includegraphics[width=\doubleplotwidth,angle=270]{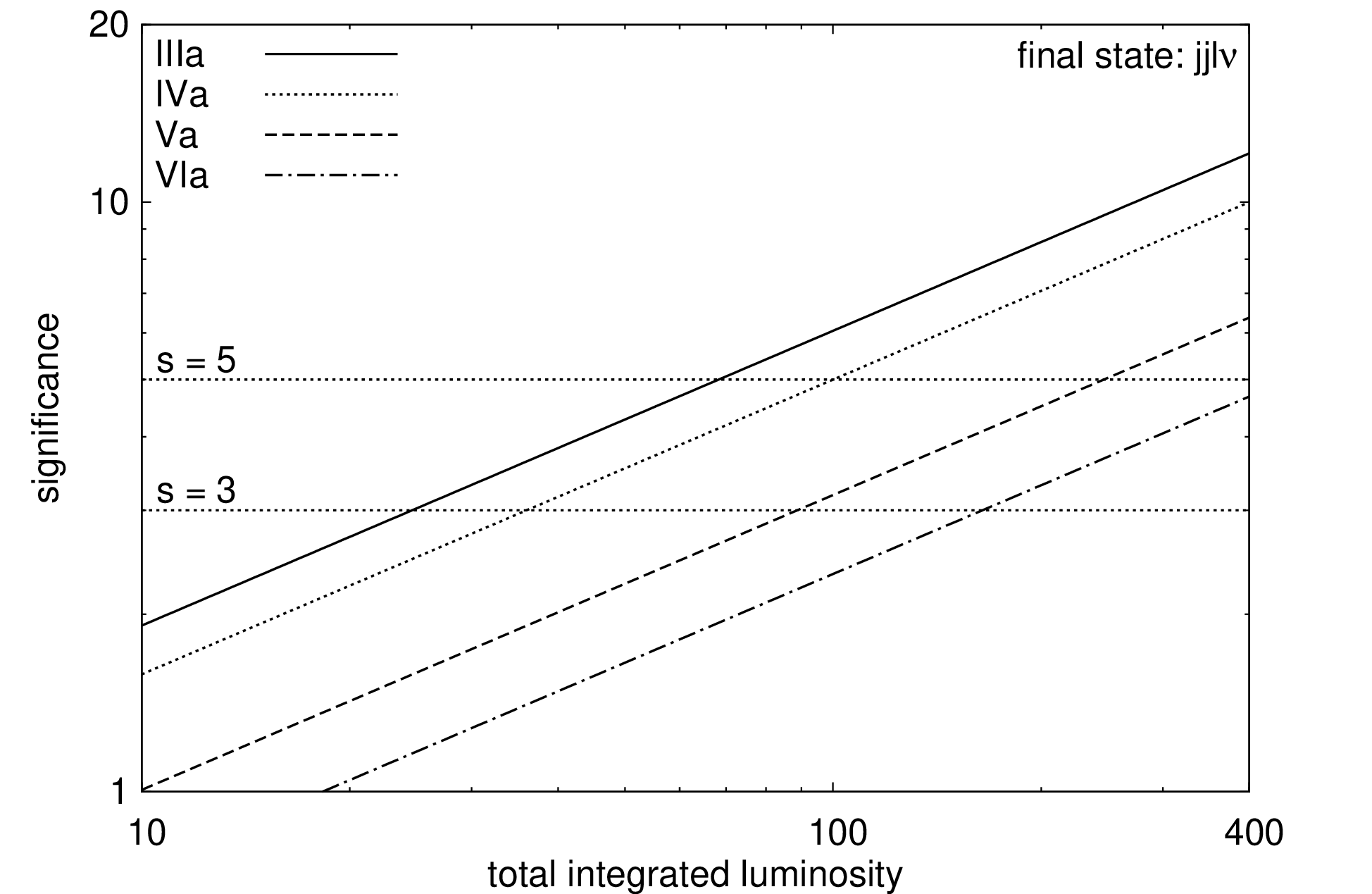} &
\includegraphics[width=\doubleplotwidth,angle=270]{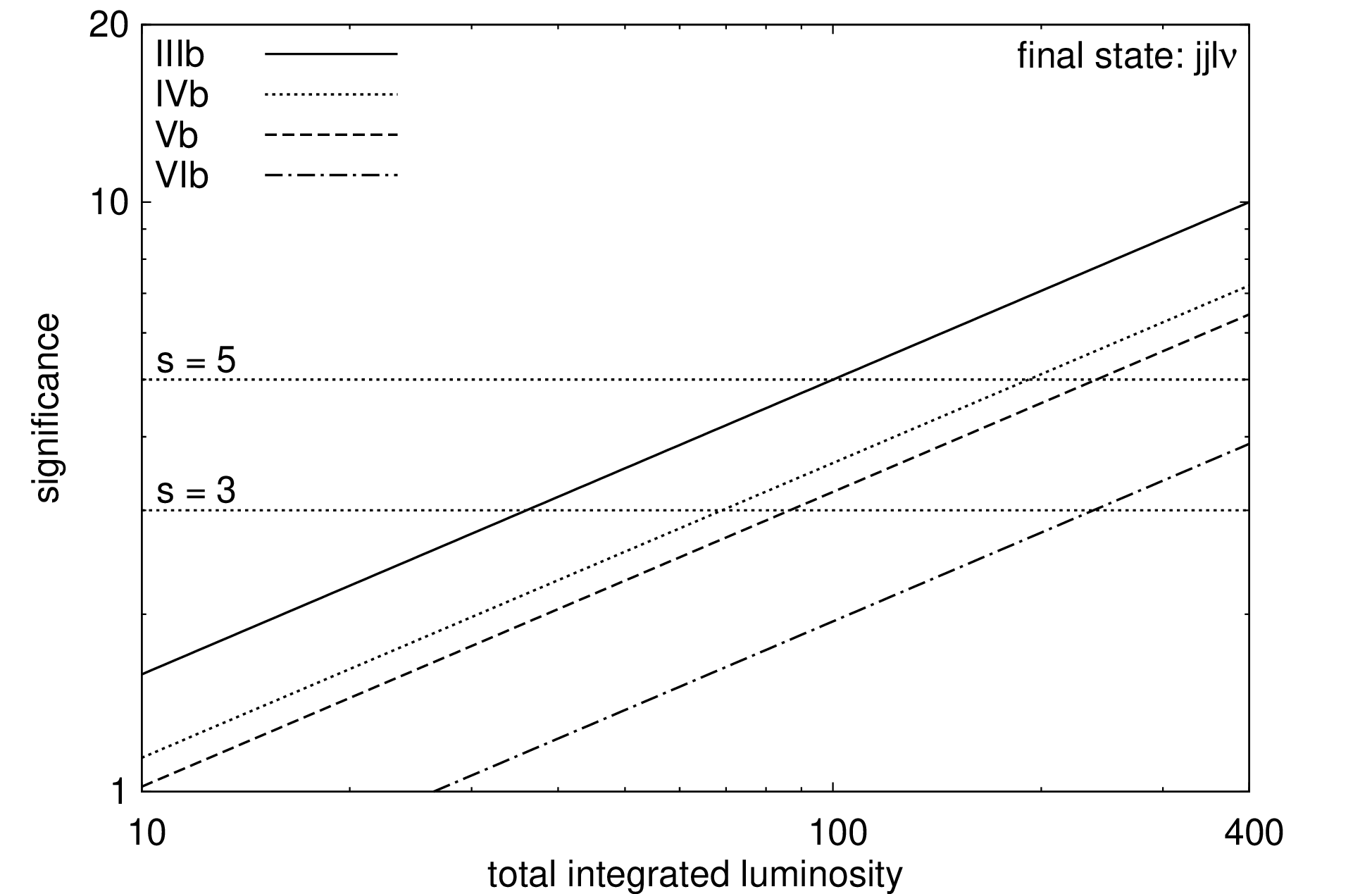}
\end{tabular}
}
\caption{
\emph{Top row}: Similar to fig.~\ref{fig-7-3-diff} left, but for the points III/IV/VIa and
\mbox{III}/\mbox{IV}/\mbox{VIb}.
\emph{Middle row}: Similar to fig.~\ref{fig-7-3-diff} right, but for III/IV/VIa/b.
\emph{Bottom row}: Like fig.~\ref{fig-7-3-wsig-I-II}, but for IIIa/b -- VIa/b.
}
\label{fig-7-3-rest}
\end{figure}
The upper two rows of fig.~\ref{fig-7-3-rest} show the difference in the $j^\prime+W$ invariant mass
between the Three-Site Model and the Standard Model for the parameter space points III/IV/VIa and
III/IV/VIb (Va/b has been excluded in order to make the figures more legible) with the left column giving the
raw event count (similar to fig.~\ref{fig-7-3-diff} left) and the right column again the significance of
the excess (like fig.~\ref{fig-7-3-diff}). For both the points with minimal and maximal
$m_{W^\prime}$, the resonance remains visible while broadening as it moves to higher invariant mass values.

Similar to fig.~\ref{fig-7-3-wsig-I-II}, the bottom row of fig.~\ref{fig-7-3-rest} shows the
significance of the resonance peaks as a function of $\ilum$ for the points IIIa/b -- VIa/b. While the
significance crosses the $5\sigma$ threshold below $\unit[400]{fb^{-1}}$ for IIIa/b -- Va/b,
the significance is just barely $5\sigma$
for VIa and even below that for VIb at $\unit[400]{fb^{-1}}$. For nearly all values of $M_\text{bulk}$,
the significance is worse for the maximum value of $m_{W^\prime}$ even though the resonances are
considerably broader for $m_{W^\prime}=\unit[380]{GeV}$ (c.f. tab.~\ref{tab-7-2-points}). The only
exceptions are Va/b ($M_\text{bulk}=2700$) for which Vb gives a slightly better result than Va.
However, as the determination of the significance comes with an uncertainty of about
$1\sigma$ at $\ilum=\unit[400]{fb^{-1}}$, this may well be a statistical fluctuation.

\begin{figure}[!p]
\centerline{
\includegraphics[width=\doubleplotwidth,angle=270]{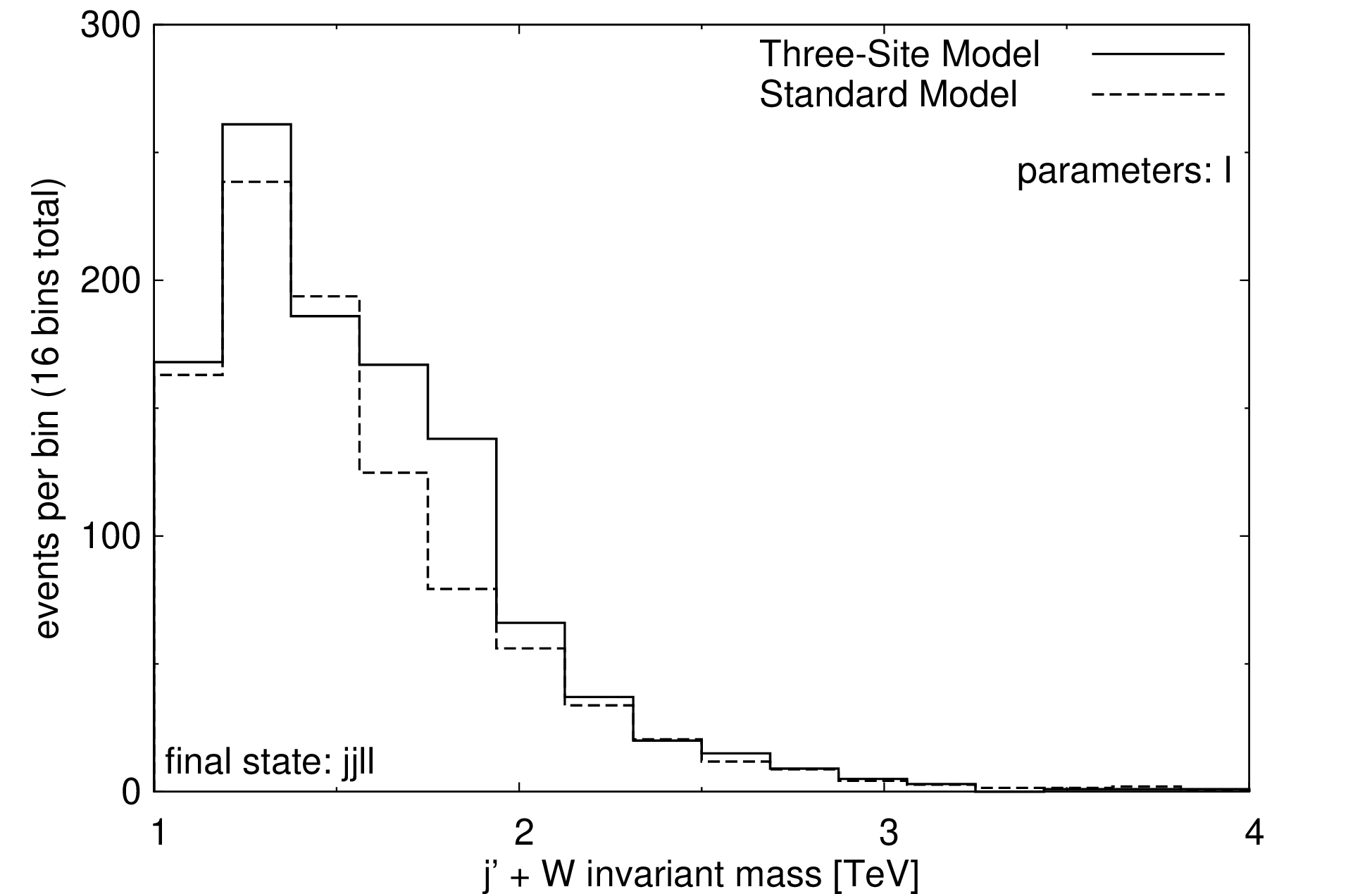}
\includegraphics[width=\doubleplotwidth,angle=270]{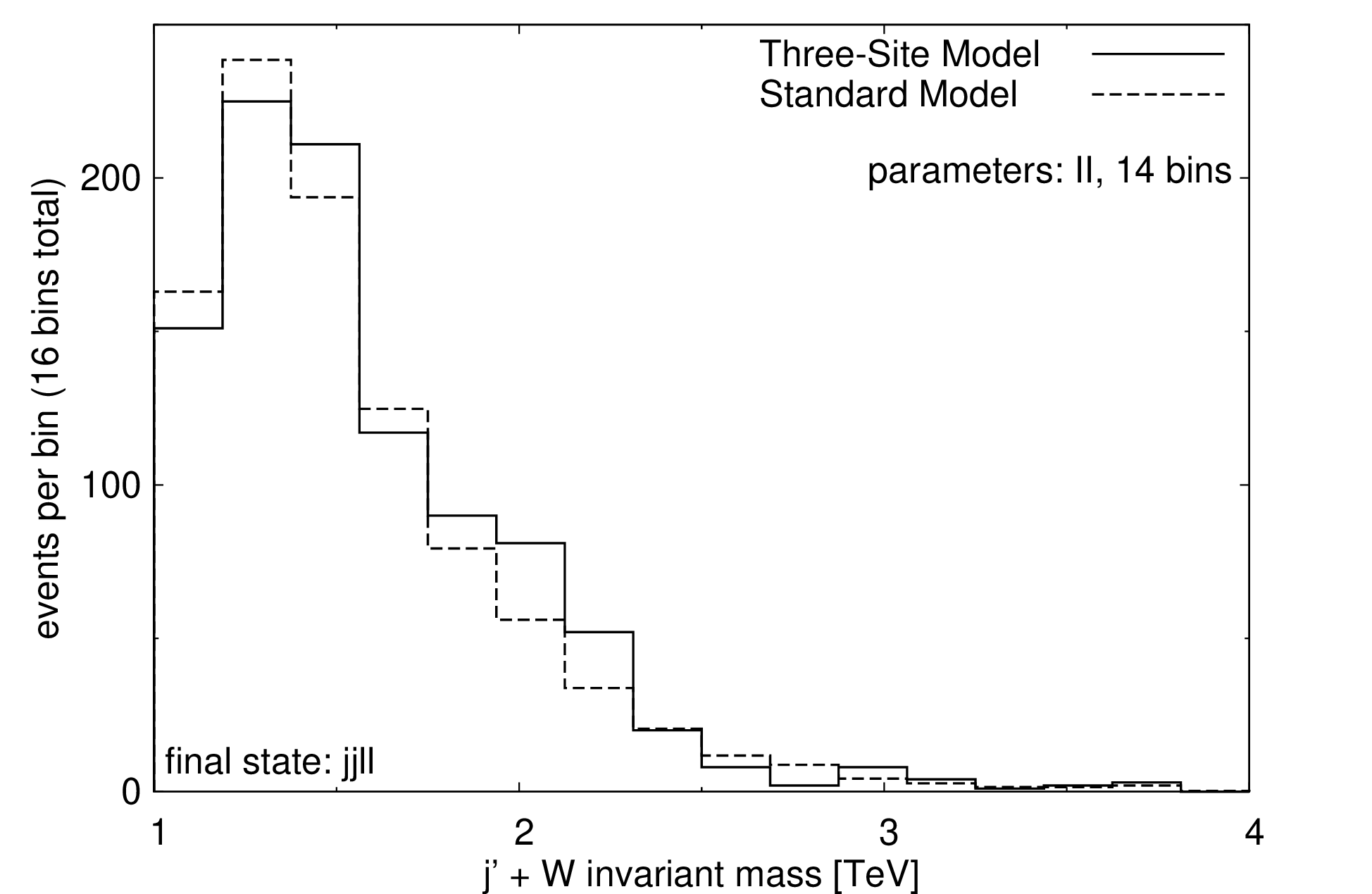}
}
\caption{Like fig.~\ref{fig-7-3-wfull}, but for the $jjll$ final state.}
\label{fig-7-3-zfull}
\end{figure}
\begin{figure}[!p]
\centerline{
\includegraphics[width=\doubleplotwidth,angle=270]{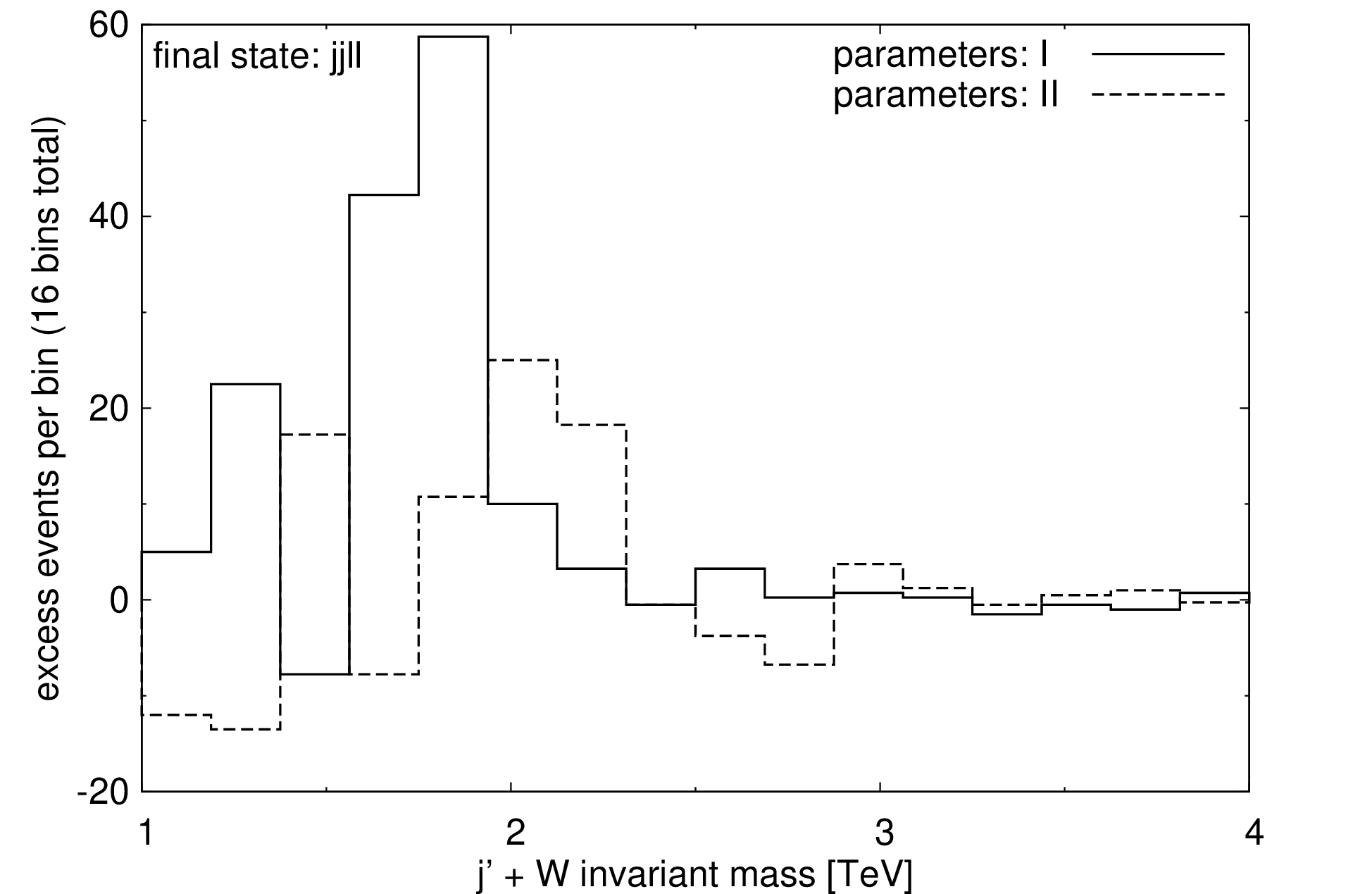}
\includegraphics[width=\doubleplotwidth,angle=270]{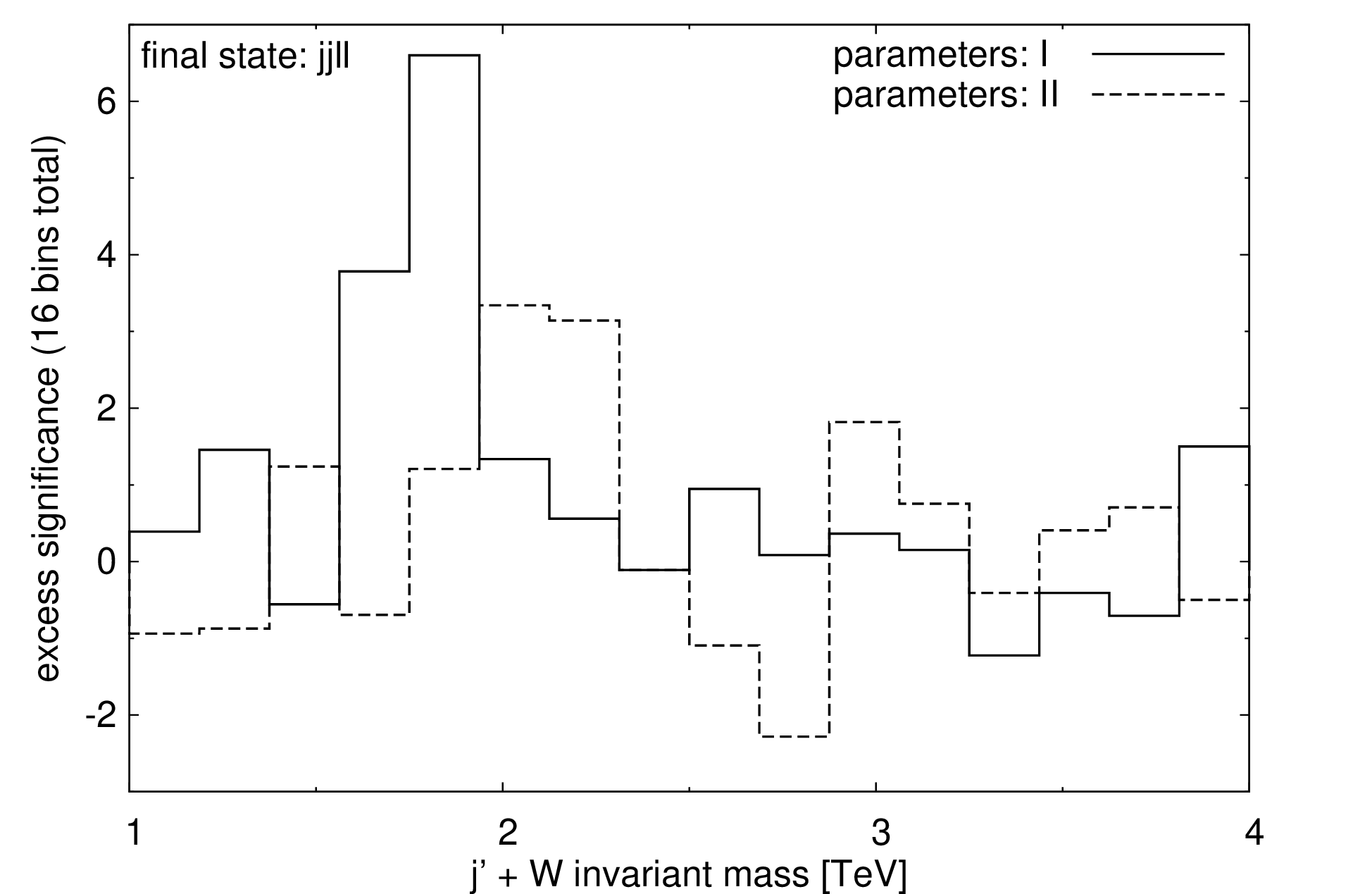}
}
\caption{
Like fig.~\ref{fig-7-3-diff}, but for the $jjll$ final state.
}
\label{fig-7-3-diff-z}
\end{figure}

Let us now move on to the $jjll$ final state. For this process, fig.~\ref{fig-7-3-wfull} shows the
$j^\prime+W$ invariant mass in the Three-Site Model versus the Standard Model expectation in a
fashion similar to fig.~\ref{fig-7-3-zfull} for the parameter space points I and II. Again, the
resonance is clearly visible as an excess over the Standard Model background, but the total event
count is much (more than an order of magnitude) lower than in the $jjl\nu$ case. The corresponding
difference of the Three-Site Model and the Standard Model is shown in fig.~\ref{fig-7-3-diff-z} left,
making the resonance peak manifest, but the fluctuations are much higher than in the $jjl\nu$ case
due to the lower total event count.

\begin{figure}
\centerline{\includegraphics[width=\singleplotwidth,angle=270]{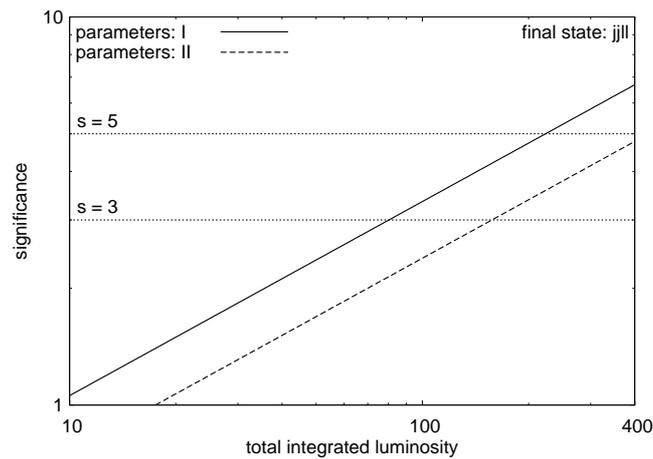}}
\caption{Like fig.~\ref{fig-7-3-wsig-I-II}, but for the $jjll$ final state.}
\label{fig-7-3-zsig-I-II}
\end{figure}

This is emphasized by the significance of the excess showcased
in fig.~\ref{fig-7-3-diff-z} right (with the significance again being calculated via
\eqref{equ-7-3-sig}, but now without the factor $\sqrt{2}$ in the denominator due to the absence of
double counting). Already in this histogram, the significance of the signal turns out to be
much lower than that
observed in fig.~\ref{fig-7-3-diff} right for the $jjl\nu$ case. The total significance of the
signal is shown in fig.~\ref{fig-7-3-zsig-I-II} as a function of $\ilum$. For $\unit[400]{fb^{-1}}$,
it is only about $\approx 6.5\sigma$ at point I and even below $5\sigma$ for II, which
is clearly much worse than what we observed for $jjl\nu$ in fig.~\ref{fig-7-3-wsig-I-II}. As
evidently $jjl\nu$ is much better suited for the detection of the heavy quarks in this process,
with a $5\sigma$ discovery already being difficult for $M_\text{bulk}=\unit[2]{TeV}$ in the $jjll$
final state, we didn't perform any more simulations for $pp\rightarrow jjll$.


\label{chap-7-3}

\clearpage

\section{Conclusions}

While the detection of the heavy fermions at the LHC is challenging due to their large mass and
considerable width, our parton level results show that this should indeed be possible in the
$jjl\nu$ (and to a lesser extend in $jjll$) final states at least in the domain of the parameter space with
$M_\text{bulk}$ smaller than $\unit[2.9]{TeV}$. While the signal
significance depends heavily on $M_\text{bulk}$, the dependence on $m_{W^\prime}$ turns out to be
less than might be expected form the associated change in the width of the resonance which indeed turns
out to be
(over)compensated by the decrease of the relevant couplings. In addition, as argued in the beginning
of this chapter, the effect of changing the delocalization parameter $\epsilon_L$ should be small
and not more than $20\%$ in the worst case.

However, even in the accessible part of parameter space, several caveats remain.
First, this kind of processes is restricted to the heavy quarks, and no information on the heavy
leptons or neutrinos can be obtained this way. Second, due to the
near-degeneracy of the heavy fermions masses,  only the heavy quarks as a whole (and, if small
contributions from flavor mixing are ignored, of these only the $u^\prime$ and $d^\prime$) are
accessible. For a resolution of individual flavors, tagging would be required which is impossible
for the light flavors. Third,
the background is rather large, reducing the broad resonances to an excess in the histograms, from
which the Standard Model background must be subtracted in order to be able to resolve the
resonance, and it is not clear a priori whether a sufficiently good background estimate will be available.

On the pro side, we can expect that the inclusion of hadronization and detector effects into our
parton-level results would not harm the outcome overmuch. The main effect of these effects would
supposedly be a smearing of the invariant mass distribution due to the finite jet resolution, but as
the resonances are very broad anyway, no big harm should come from that. More dangerous is the
potential increase in the reducible background coming from other final states which we did not take
into account in our parton level analysis; this should be studied carefully if such an analysis of the
experimental date should ever be performed.

So, can the heavy fermions be seen at the LHC? In the light of the results of our simulations, the
answer to this question is a definitive ``maybe'', with the final verdict depending on the
actual value of $M_\text{bulk}$, the accuracy of our understanding of the Standard Model background,
the amount of data the LHC will be able to collect and, also important, the actual energy the
collider will reach --- our simulations were performed for $\unit[14]{TeV}$, and less energy would
diminish the result considerably.

\chapter{Conclusions}
\label{chap-8}

\begin{quote}\itshape
The physics is theoretical, but the fun is real.
\end{quote}
\hfill\begin{minipage}{0.7\textwidth}\small\raggedleft
(Tagline of Sheldon's ``Research Lab'' game from ``The Big Bang Theory --- The Guitarist
Amplification'')
\end{minipage}
\\[5mm]
While it certainly does not solve most of the issues the Standard Model is usually criticized for
(apart from the hierarchy problem which obviously is void in Higgsless model), we have argued in
chapter \ref{chap-1} that a scenario like the Three-Site Model does arise in a rather natural
fashion once we try to replace the Higgs with a set of $\sun{2}$ vector bosons as agents for the
preservation of perturbative unitarity at the $\unit{TeV}$ scale. Discussing the properties of the
new structure in the model and the consequences of the experimental constraints in chapter
\ref{chap-3}, we have found that these constraints imply that the new physics is well hidden from
discovery by forcing the $W^\prime / Z^\prime$ to be essentially fermiophobic and the KK partners of
the fermions to be very heavy.

The goal of this thesis was the implementation of the model into the WHIZARD / O'Mega framework and
the application of this implementation to studying the phenomenology exhibited by the model
at the LHC. The
implementation as presented in chapter \ref{chap-4} is complete and has been cross-checked against
two independent implementations in CalcHep and FeynRules. In addition, the FeynRules $\rightarrow$
WHIZARD driver also developed in the context of this thesis and presented in chapter \ref{chap-4-4} will
greatly simplify the future implementation of other models of new physics into WHIZARD / O'Mega.

Due to the constraints on the masses and couplings of the new particles, the LHC phenomenology of
the Three-Site Model turns out to be quite interesting even though the new structure cannot compete
with other extension of the Standard Model (e.g. SUSY) in terms of complexity. Using our WHIZARD /
O'Mega implementation, we have performed a study of the discovery prospects of the new heavy particles
in three different types of processes, the results of which we presented in chapters \ref{chap-5} --
\ref{chap-7}.

For the heavy gauge bosons, essentially two classes of production mechanisms exist with the
$W^\prime/Z^\prime$ coupling either to a Standard Model gauge boson or to a Standard Model fermion.
As a representative of the first class, we have showcased a simulation of the $W^\prime$ strahlung
process in chapter \ref{chap-5}. We find that this process indeed has the potential of uncovering
the $W^\prime$ at the LHC. However, while a $5\sigma$ discovery of the $W^\prime$
should be possible in this process
with considerably less than $\unit[100]{fb^{-1}}$ for $W^\prime$ masses of
between $\unit[380]{GeV}$ and $\unit[500]{GeV}$ (the first value being the lower bound on
$m_{W^\prime}$ from existing experimental data), the integrated luminosity necessary for a $5\sigma$
discovery rises to $\unit[400]{fb^{-1}}$ for $m_{W^\prime}=\unit[600]{GeV}$ (which is the $W^\prime$
mass where the heavy fermions move above the UV cutoff). Considering that no detector effects are
taken into account in these numbers, $m_{W^\prime}=\unit[600]{GeV}$ should present the upper limit
on the reach of the LHC in this process.

As the $W^\prime$ strahlung process is essentially mediated by the gauge sector and does not
contain any information on the fermiophobic couplings of the heavy gauge bosons to the Standard
Model fermions, we have performed another set of simulations presented in chapter \ref{chap-6} where
we studied the production of the $W^\prime / Z^\prime$ in the $s$ channel. For the $Z^\prime$, the
only viable discovery channel leads to the final state $jjl\nu$ which our simulations show to be
suitable for the
discovery of the $Z^\prime$ with the first $10-\unit[20]{fb^{-1}}$ over the whole parameter space.
For the $W^\prime$, three potential discovery channels exists with the final state $jjl\nu$, $jjll$
and $lll\nu$. However, the separation of the $W^\prime$ resonance from
the $Z^\prime$ peak in $jjl\nu$ turns out to be difficult due to the finite experimental resolution
of the jet momenta,
leaving $jjll$ and $lll\nu$ as the most promising channels. We find that the discovery
potential of these ranges from ``possible within the first $\unit[100]{fb^{-1}}$'' to ``utterly
impossible'' for all $W^\prime$ masses depending on the exact value of the delocalization parameter
$\epsilon_L$. As the previously discussed $W^\prime$ strahlung process allows for a discovery of the
$W^\prime$ independent of the delocalization parameter, the combination of these two processes might
not only facilitate the discovery of the $W^\prime$ but even give some hints about its fermiophobic
nature.

Apart from the heavy gauge bosons, the second major component of the new structure in the Three-Site
Model are the heavy partners to the Standard Model fermions. In chapter \ref{chap-7}, a simulation
of the $t$ channel induced production of heavy quarks leading to either a $jjll$ oder a $jjl\nu$ final
state was discussed. The result suggests that, while the mass and width of the heavy fermions makes
them hard to be detected as a resonance, such a measurement might indeed be possible at least for
values of $M_\text{bulk}$ below $\unit[2.9]{TeV}$. However, the success of this in the ``real
world'' would be highly dependent on the exact value of $M_\text{bulk}$ and the accuracy of our
understanding of the expected Standard Model background which must be subtracted in order to establish
the resonance. Still, the situation is not hopeless, and if physics reminiscent of the Three-Site
Model $W^\prime / Z^\prime$ were to be found at the LHC, the heavy fermions could in principle be
probed for.

Combining the results of all our simulations, it is safe to say that the LHC would be capable of
probing for all major components of the new physics in the Three-Site Model. However, there
are areas in parameter space which would remain partially dark at the LHC, and while the
$W^\prime$ and $Z^\prime$ should eventually turn up, the heavy fermions might remain completely
invisible.

Of course, the Three-Site Model is just one example of Higgsless new physics, and there
is no reason why nature should choose this particular scenario (or why nature should be
Higgsless anyway). Still, even in the likely case of it not being a valid description of nature, the
the Three-Site Model remains an interesting example of a model of new physics which is quite
successful at (partially) hiding its new structure from quick discovery. In fact, as all Higgsless
models in which heavy vectors are responsible for the delay of unitarity violation must manage to
pass the precision constraints, phenomena like the fermiophobic couplings of the heavy gauge bosons
or additional heavy fermions occur in other such models as well, with similar difficulties
concerning the detection at the LHC.

\begin{appendix}

\chapter{Notation and Conventions}

\label{chap-conv}

All covariant derivatives appearing in this work are defined with a minus sign
\begin{lequation}{conv-codev}
D^\mu = \partial^\mu - igA^\mu
\end{lequation}
In several places, the generators of the fundamental representation of
$\sun{2}$ are referred to which are defined as
\begin{lequation}{conv-su2gen}
\tau_i = \frac{\sigma_i}{2}
\end{lequation}
with the Pauli matrices $\sigma_i$. Left- and right-handed fermions are defined via the chirality
projectors $\Pi_\pm$ as
\begin{equation} \Psi_{L/R} = \Pi_\pm\Psi = \frac{1\pm\gamma^5}{2}\Psi \label{conv-chiral}\end{equation}

In the introduction to 5D gauge theories given in chapter \ref{chap-2},
a convention is used of Greek Lorentz indices
going from $0$ to $3$ (thus labeling the 4D part of 5D vectors), while Roman Lorenz indices go from
$0$ to $4$.
The 5th coordinate is labeled
either as $x^5$, $y$ or (implicately in sums) $x^4$.
For the metric, the mostly minus convention is used
\begin{lequation}{conv-metric}
g^{\mu\nu} = \diag\left(1,-1,-1,-1\right)^{\mu\nu} \quad,\quad
g^{ab} = \diag\left(1,-1,-1,-1,-1\right)^{ab}
\end{lequation}

It is also necessary to establish a convention on the nomenclature regarding the Standard Model particles
and their heavy partners. If not explicitly stated differently in the text, we refer to the
Standard Model particles as ``light'' or ``KK light'' and to their partners as ``heavy'',
``KK heavy'' or ``KK particles''.

\chapter{A sample spectrum}
\label{app-3}

The following is a sample spectrum generated by running the \verb?spektrum? program described in
appendix \ref{app-5-1}. It is calculated for ideal delocalization $m_W=\unit[500]{GeV}$ and
$M_\text{bulk}=\unit[3.5]{TeV}$ and gives all masses, wavefunctions, widths and couplings (excluding
QED and QCD). The heavy quark widths include the one loop QCD correction (c.f. chapter
\ref{chap-3-4} and appendix \ref{app-4}.

Note that the scale \verb?v? given in the first few lines
must be divided by $\sqrt{2}$ to obtain the symmetry breaking scale $v$
defined in chapter \ref{chap-2-3} due to a difference in convention. For similar reasons, the
parameter \verb?lambda? must be multiplied by $2$ to get the bulk Yukawa coupling $\tilde{\lambda}$.
{\tiny
\begin{verbatim}
 
g0 =  0.6915048     g1 =  2.0914001     g2 =  0.3565523
     -> x =  0.3306421     t =  0.5156179     e =  0.3133290
     -> eps_L =  0.2369703
v =     235.7870519     lambda =     10.4962241
 
 
electron neutrino : light lepton with isospin +1/2 belonging to generation 0
 eps_r:       0.0000000
 wavefunction L :     -0.9730523           0.2305845
 wavefunction R :      0.0000000           1.0000000
 mass:       0.0000000
 width:      0.0000000
 
heavy electron neutrino : heavy lepton with isospin +1/2 belonging to generation 0
 eps_r:       0.0000000
 wavefunction L :      0.2305845           0.9730523
 wavefunction R :      1.0000000           0.0000000
 mass:    3596.9289707
 width:    628.3857980
 
up quark : light quark with isospin +1/2 belonging to generation 0
 eps_r:       0.0000000
 wavefunction L :     -0.9730523           0.2305845
 wavefunction R :      0.0000000           1.0000000
 mass:       0.0000000
 width:      0.0000000
 
heavy up quark : heavy quark with isospin +1/2 belonging to generation 0
 eps_r:       0.0000000
 wavefunction L :      0.2305845           0.9730523
 wavefunction R :      1.0000000           0.0000000
 mass:    3596.9289707
 width:    585.6279797
 
electron : light lepton with isospin -1/2 belonging to generation 0
 eps_r:       0.0000000
 wavefunction L :     -0.9730523           0.2305845
 wavefunction R :      0.0000000           1.0000000
 mass:       0.0000000
 width:      0.0000000
 
heavy electron : heavy lepton with isospin -1/2 belonging to generation 0
 eps_r:       0.0000000
 wavefunction L :      0.2305845           0.9730523
 wavefunction R :      1.0000000           0.0000000
 mass:    3596.9289707
 width:    628.3857980
 
down quark : light quark with isospin -1/2 belonging to generation 0
 eps_r:       0.0000000
 wavefunction L :     -0.9730523           0.2305845
 wavefunction R :      0.0000000           1.0000000
 mass:       0.0000000
 width:      0.0000000
 
heavy down quark : heavy quark with isospin -1/2 belonging to generation 0
 eps_r:       0.0000000
 wavefunction L :      0.2305845           0.9730523
 wavefunction R :      1.0000000           0.0000000
 mass:    3596.9289707
 width:    585.6279797
 
muon neutrino : light lepton with isospin +1/2 belonging to generation 1
 eps_r:       0.0000000
 wavefunction L :     -0.9730523           0.2305845
 wavefunction R :      0.0000000           1.0000000
 mass:       0.0000000
 width:      0.0000000
 
heavy muon neutrino : heavy lepton with isospin +1/2 belonging to generation 1
 eps_r:       0.0000000
 wavefunction L :      0.2305845           0.9730523
 wavefunction R :      1.0000000           0.0000000
 mass:    3596.9289707
 width:    628.3859271
 
charm quark : light quark with isospin +1/2 belonging to generation 1
 eps_r:       0.0015489
 wavefunction L :     -0.9730524           0.2305840
 wavefunction R :     -0.0014665           0.9999989
 mass:       1.2500000
 width:      0.0000000
 
heavy charm quark : heavy quark with isospin +1/2 belonging to generation 1
 eps_r:       0.0015489
 wavefunction L :      0.2305840           0.9730524
 wavefunction R :      0.9999989           0.0014665
 mass:    3596.9328385
 width:    585.6454626
 
muon : light lepton with isospin -1/2 belonging to generation 1
 eps_r:       0.0001313
 wavefunction L :     -0.9730523           0.2305845
 wavefunction R :     -0.0001244           1.0000000
 mass:       0.1060000
 width:      0.0000000
 
heavy muon : heavy lepton with isospin -1/2 belonging to generation 1
 eps_r:       0.0001313
 wavefunction L :      0.2305845           0.9730523
 wavefunction R :      1.0000000           0.0001244
 mass:    3596.9289985
 width:    628.3858577
 
strange quark : light quark with isospin -1/2 belonging to generation 1
 eps_r:       0.0011771
 wavefunction L :     -0.9730524           0.2305842
 wavefunction R :     -0.0011145           0.9999994
 mass:       0.9500000
 width:      0.0000000
 
heavy strange quark : heavy quark with isospin -1/2 belonging to generation 1
 eps_r:       0.0011771
 wavefunction L :      0.2305842           0.9730524
 wavefunction R :      0.9999994           0.0011145
 mass:    3596.9312047
 width:    585.6492771
 
tauon neutrino : light lepton with isospin +1/2 belonging to generation 2
 eps_r:       0.0000000
 wavefunction L :     -0.9730523           0.2305845
 wavefunction R :      0.0000000           1.0000000
 mass:       0.0000000
 width:      0.0000000
 
heavy tauon neutrino : heavy lepton with isospin +1/2 belonging to generation 2
 eps_r:       0.0000000
 wavefunction L :      0.2305845           0.9730523
 wavefunction R :      1.0000000           0.0000000
 mass:    3596.9289707
 width:    628.4222032
 
top quark : light quark with isospin +1/2 belonging to generation 2
 eps_r:       0.2202501
 wavefunction L :     -0.9752869           0.2209422
 wavefunction R :     -0.2046066           0.9788443
 mass:     174.0000000
 width:      1.5230000
 
heavy top quark : heavy quark with isospin +1/2 belonging to generation 2
 eps_r:       0.2202501
 wavefunction L :      0.2209422           0.9752869
 wavefunction R :      0.9788443           0.2046066
 mass:    3674.4892058
 width:    739.9590965
 
tauon : light lepton with isospin -1/2 belonging to generation 2
 eps_r:       0.0022056
 wavefunction L :     -0.9730525           0.2305835
 wavefunction R :     -0.0020883           0.9999978
 mass:       1.7800000
 width:      0.0000000
 
heavy tauon : heavy lepton with isospin -1/2 belonging to generation 2
 eps_r:       0.0022056
 wavefunction L :      0.2305835           0.9730525
 wavefunction R :      0.9999978           0.0020883
 mass:    3596.9368138
 width:    628.4026301
 
bottom quark : light quark with isospin -1/2 belonging to generation 2
 eps_r:       0.0052042
 wavefunction L :     -0.9730536           0.2305789
 wavefunction R :     -0.0049275           0.9999879
 mass:       4.2000000
 width:      0.0000000
 
heavy bottom quark : heavy quark with isospin -1/2 belonging to generation 2
 eps_r:       0.0052042
 wavefunction L :      0.2305789           0.9730536
 wavefunction R :      0.9999879           0.0049275
 mass:    3596.9726381
 width:    910.9435197
 
W boson : light W boson
 wavefuntion:       0.9858825           0.1674384
 mass:      80.4030000
 width:      2.0480000
 
heavy W boson : heavy W boson
 wavefuntion:      -0.1674384           0.9858825
 mass:     500.0000000
 width:      5.0007454
 
Z boson : light Z boson
 wavefuntion:      -0.8759133          -0.1084404           0.4701241
 mass:      91.1880000
 width:      2.4430000
 
heavy Z boson : heavy Z boson
 wavefuntion:       0.1657276          -0.9827488           0.0820918
 mass:     501.6804843
 width:      4.7681804
 
-------------------------------------------
W boson:                light
isospin +1/2 lepton:    light
isospin -1/2 lepton:    light
 
left-handed coupling matrix (rows <-> iso-up,  columns <-> iso-down):
 
           0.6641137           0.0000000           0.0000000
           0.0000000           0.6641137           0.0000000
           0.0000000           0.0000000           0.6641137
 
right-handed coupling matrix (rows <-> iso-up,  columns <-> iso-down):
 
           0.0000000           0.0000000           0.0000000
           0.0000000           0.0000000           0.0000000
           0.0000000           0.0000000           0.0000000
 
-------------------------------------------
 
 
-------------------------------------------
W boson:                light
isospin +1/2 quark:     light
isospin -1/2 quark:     light
 
left-handed coupling matrix (rows <-> iso-up,  columns <-> iso-down):
 
           0.6641137           0.0000000           0.0000000
           0.0000000           0.6641137           0.0000000
           0.0000000           0.0000000           0.6648179
 
right-handed coupling matrix (rows <-> iso-up,  columns <-> iso-down):
 
           0.0000000           0.0000000           0.0000000
           0.0000000           0.0000006           0.0000000
           0.0000000           0.0000000           0.0003530
 
-------------------------------------------
 
 
-------------------------------------------
W boson:                light
isospin +1/2 lepton:    light
isospin -1/2 lepton:    heavy
 
left-handed coupling matrix (rows <-> iso-up,  columns <-> iso-down):
 
          -0.0743928           0.0000000           0.0000000
           0.0000000          -0.0743928           0.0000000
           0.0000000           0.0000000          -0.0743921
 
right-handed coupling matrix (rows <-> iso-up,  columns <-> iso-down):
 
           0.0000000           0.0000000           0.0000000
           0.0000000           0.0000000           0.0000000
           0.0000000           0.0000000           0.0000000
 
-------------------------------------------
 
 
-------------------------------------------
W boson:                light
isospin +1/2 quark:     light
isospin -1/2 quark:     heavy
 
left-handed coupling matrix (rows <-> iso-up,  columns <-> iso-down):
 
          -0.0743928           0.0000000           0.0000000
           0.0000000          -0.0743928           0.0000000
           0.0000000           0.0000000          -0.0780258
 
right-handed coupling matrix (rows <-> iso-up,  columns <-> iso-down):
 
           0.0000000           0.0000000           0.0000000
           0.0000000          -0.0005135           0.0000000
           0.0000000           0.0000000          -0.0716484
 
-------------------------------------------
 
 
-------------------------------------------
W boson:                light
isospin +1/2 lepton:    heavy
isospin -1/2 lepton:    light
 
left-handed coupling matrix (rows <-> iso-up,  columns <-> iso-down):
 
          -0.0743928           0.0000000           0.0000000
           0.0000000          -0.0743928           0.0000000
           0.0000000           0.0000000          -0.0743932
 
right-handed coupling matrix (rows <-> iso-up,  columns <-> iso-down):
 
           0.0000000           0.0000000           0.0000000
           0.0000000          -0.0000435           0.0000000
           0.0000000           0.0000000          -0.0007313
 
-------------------------------------------
 
 
-------------------------------------------
W boson:                light
isospin +1/2 quark:     heavy
isospin -1/2 quark:     light
 
left-handed coupling matrix (rows <-> iso-up,  columns <-> iso-down):
 
          -0.0743928           0.0000000           0.0000000
           0.0000000          -0.0743926           0.0000000
           0.0000000           0.0000000          -0.0678180
 
right-handed coupling matrix (rows <-> iso-up,  columns <-> iso-down):
 
           0.0000000           0.0000000           0.0000000
           0.0000000          -0.0003903           0.0000000
           0.0000000           0.0000000          -0.0016890
 
-------------------------------------------
 
 
-------------------------------------------
W boson:                light
isospin +1/2 lepton:    heavy
isospin -1/2 lepton:    heavy
 
left-handed coupling matrix (rows <-> iso-up,  columns <-> iso-down):
 
           0.3678096           0.0000000           0.0000000
           0.0000000           0.3678096           0.0000000
           0.0000000           0.0000000           0.3678095
 
right-handed coupling matrix (rows <-> iso-up,  columns <-> iso-down):
 
           0.3501807           0.0000000           0.0000000
           0.0000000           0.3501807           0.0000000
           0.0000000           0.0000000           0.3501799
 
-------------------------------------------
 
 
-------------------------------------------
W boson:                light
isospin +1/2 quark:     heavy
isospin -1/2 quark:     heavy
 
left-handed coupling matrix (rows <-> iso-up,  columns <-> iso-down):
 
           0.3678096           0.0000000           0.0000000
           0.0000000           0.3678095           0.0000000
           0.0000000           0.0000000           0.3670548
 
right-handed coupling matrix (rows <-> iso-up,  columns <-> iso-down):
 
           0.3501807           0.0000000           0.0000000
           0.0000000           0.3501801           0.0000000
           0.0000000           0.0000000           0.3427682
 
-------------------------------------------
 
 
-------------------------------------------
W boson:                heavy
isospin +1/2 lepton:    light
isospin -1/2 lepton:    light
 
left-handed coupling matrix (rows <-> iso-up,  columns <-> iso-down):
 
           0.0000000           0.0000000           0.0000000
           0.0000000           0.0000000           0.0000000
           0.0000000           0.0000000          -0.0000005
 
right-handed coupling matrix (rows <-> iso-up,  columns <-> iso-down):
 
           0.0000000           0.0000000           0.0000000
           0.0000000           0.0000000           0.0000000
           0.0000000           0.0000000           0.0000000
 
-------------------------------------------
 
 
-------------------------------------------
W boson:                heavy
isospin +1/2 quark:     light
isospin -1/2 quark:     light
 
left-handed coupling matrix (rows <-> iso-up,  columns <-> iso-down):
 
           0.0000000           0.0000000           0.0000000
           0.0000000          -0.0000004           0.0000000
           0.0000000           0.0000000          -0.0048388
 
right-handed coupling matrix (rows <-> iso-up,  columns <-> iso-down):
 
           0.0000000           0.0000000           0.0000000
           0.0000000           0.0000034           0.0000000
           0.0000000           0.0000000           0.0020788
 
-------------------------------------------
 
 
-------------------------------------------
W boson:                heavy
isospin +1/2 lepton:    light
isospin -1/2 lepton:    heavy
 
left-handed coupling matrix (rows <-> iso-up,  columns <-> iso-down):
 
           0.4886032           0.0000000           0.0000000
           0.0000000           0.4886032           0.0000000
           0.0000000           0.0000000           0.4886032
 
right-handed coupling matrix (rows <-> iso-up,  columns <-> iso-down):
 
           0.0000000           0.0000000           0.0000000
           0.0000000           0.0000000           0.0000000
           0.0000000           0.0000000           0.0000000
 
-------------------------------------------
 
 
-------------------------------------------
W boson:                heavy
isospin +1/2 quark:     light
isospin -1/2 quark:     heavy
 
left-handed coupling matrix (rows <-> iso-up,  columns <-> iso-down):
 
           0.4886032           0.0000000           0.0000000
           0.0000000           0.4886022           0.0000000
           0.0000000           0.0000000           0.4693173
 
right-handed coupling matrix (rows <-> iso-up,  columns <-> iso-down):
 
           0.0000000           0.0000000           0.0000000
           0.0000000          -0.0030238           0.0000000
           0.0000000           0.0000000          -0.4218681
 
-------------------------------------------
 
 
-------------------------------------------
W boson:                heavy
isospin +1/2 lepton:    heavy
isospin -1/2 lepton:    light
 
left-handed coupling matrix (rows <-> iso-up,  columns <-> iso-down):
 
           0.4886032           0.0000000           0.0000000
           0.0000000           0.4886032           0.0000000
           0.0000000           0.0000000           0.4886012
 
right-handed coupling matrix (rows <-> iso-up,  columns <-> iso-down):
 
           0.0000000           0.0000000           0.0000000
           0.0000000          -0.0002564           0.0000000
           0.0000000           0.0000000          -0.0043058
 
-------------------------------------------
 
 
-------------------------------------------
W boson:                heavy
isospin +1/2 quark:     heavy
isospin -1/2 quark:     light
 
left-handed coupling matrix (rows <-> iso-up,  columns <-> iso-down):
 
           0.4886032           0.0000000           0.0000000
           0.0000000           0.4886026           0.0000000
           0.0000000           0.0000000           0.4885680
 
right-handed coupling matrix (rows <-> iso-up,  columns <-> iso-down):
 
           0.0000000           0.0000000           0.0000000
           0.0000000          -0.0022981           0.0000000
           0.0000000           0.0000000          -0.0099449
 
-------------------------------------------
 
 
-------------------------------------------
W boson:                heavy
isospin +1/2 lepton:    heavy
isospin -1/2 lepton:    heavy
 
left-handed coupling matrix (rows <-> iso-up,  columns <-> iso-down):
 
           1.9460904           0.0000000           0.0000000
           0.0000000           1.9460904           0.0000000
           0.0000000           0.0000000           1.9460909
 
right-handed coupling matrix (rows <-> iso-up,  columns <-> iso-down):
 
           2.0618749           0.0000000           0.0000000
           0.0000000           2.0618749           0.0000000
           0.0000000           0.0000000           2.0618704
 
-------------------------------------------
 
 
-------------------------------------------
W boson:                heavy
isospin +1/2 quark:     heavy
isospin -1/2 quark:     heavy
 
left-handed coupling matrix (rows <-> iso-up,  columns <-> iso-down):
 
           1.9460904           0.0000000           0.0000000
           0.0000000           1.9460908           0.0000000
           0.0000000           0.0000000           1.9508340
 
right-handed coupling matrix (rows <-> iso-up,  columns <-> iso-down):
 
           2.0618749           0.0000000           0.0000000
           0.0000000           2.0618714           0.0000000
           0.0000000           0.0000000           2.0182299
 
-------------------------------------------
 
 
couplings to the light Z
------------------------
2x electron neutrino:
 L:      -0.3765880     R:       0.0000000
 
2x up quark:
 L:      -0.2648388     R:       0.1117492
 
2x heavy electron neutrino:
 L:      -0.2072812     R:      -0.1972081
 
2x heavy up quark:
 L:      -0.0955319     R:      -0.0854589
 
electron neutrino , heavy electron neutrino:
 L:       0.0425077     R:       0.0000000
 
up quark , heavy up quark:
 L:       0.0425077     R:       0.0000000
 
2x electron:
 L:       0.2089642     R:      -0.1676238
 
2x down quark:
 L:       0.3207134     R:      -0.0558746
 
2x heavy electron:
 L:       0.0396573     R:       0.0295843
 
2x heavy down quark:
 L:       0.1514065     R:       0.1413335
 
electron , heavy electron:
 L:      -0.0425077     R:       0.0000000
 
down quark , heavy down quark:
 L:      -0.0425077     R:       0.0000000
 
2x muon neutrino:
 L:      -0.3765880     R:       0.0000000
 
2x charm quark:
 L:      -0.2648388     R:       0.1117488
 
2x heavy muon neutrino:
 L:      -0.2072812     R:      -0.1972081
 
2x heavy charm quark:
 L:      -0.0955319     R:      -0.0854585
 
muon neutrino , heavy muon neutrino:
 L:       0.0425077     R:       0.0000000
 
charm quark , heavy charm quark:
 L:       0.0425076     R:       0.0002892
 
2x muon:
 L:       0.2089642     R:      -0.1676238
 
2x strange quark:
 L:       0.3207134     R:      -0.0558744
 
2x heavy muon:
 L:       0.0396573     R:       0.0295843
 
2x heavy strange quark:
 L:       0.1514065     R:       0.1413332
 
muon , heavy muon:
 L:      -0.0425077     R:      -0.0000245
 
strange quark , heavy strange quark:
 L:      -0.0425077     R:      -0.0002198
 
2x tauon neutrino:
 L:      -0.3765880     R:       0.0000000
 
2x top quark:
 L:      -0.2656636     R:       0.1034933
 
2x heavy tauon neutrino:
 L:      -0.2072812     R:      -0.1972081
 
2x heavy top quark:
 L:      -0.0947071     R:      -0.0772030
 
tauon neutrino , heavy tauon neutrino:
 L:       0.0425077     R:       0.0000000
 
top quark , heavy top quark:
 L:       0.0408237     R:       0.0394964
 
2x tauon:
 L:       0.2089642     R:      -0.1676229
 
2x bottom quark:
 L:       0.3207138     R:      -0.0558698
 
2x heavy tauon:
 L:       0.0396573     R:       0.0295834
 
2x heavy bottom quark:
 L:       0.1514061     R:       0.1413287
 
tauon , heavy tauon:
 L:      -0.0425075     R:      -0.0004118
 
bottom quark , heavy bottom quark:
 L:      -0.0425067     R:      -0.0009717
 
couplings to the heavy Z
------------------------
2x electron neutrino:
 L:      -0.0150208     R:       0.0000000
 
2x up quark:
 L:       0.0044925     R:       0.0195133
 
2x heavy electron neutrino:
 L:      -0.9846090     R:      -1.0422955
 
2x heavy up quark:
 L:      -0.9650956     R:      -1.0227822
 
electron neutrino , heavy electron neutrino:
 L:      -0.2434336     R:       0.0000000
 
up quark , heavy up quark:
 L:      -0.2434336     R:       0.0000000
 
2x electron:
 L:      -0.0142492     R:      -0.0292700
 
2x down quark:
 L:       0.0052642     R:      -0.0097567
 
2x heavy electron:
 L:       0.9553390     R:       1.0130255
 
2x heavy down quark:
 L:       0.9748523     R:       1.0325389
 
electron , heavy electron:
 L:       0.2434336     R:       0.0000000
 
down quark , heavy down quark:
 L:       0.2434336     R:       0.0000000
 
2x muon neutrino:
 L:      -0.0150208     R:       0.0000000
 
2x charm quark:
 L:       0.0044928     R:       0.0195111
 
2x heavy muon neutrino:
 L:      -0.9846090     R:      -1.0422955
 
2x heavy charm quark:
 L:      -0.9650959     R:      -1.0227800
 
muon neutrino , heavy muon neutrino:
 L:      -0.2434336     R:       0.0000000
 
charm quark , heavy charm quark:
 L:      -0.2434331     R:       0.0015285
 
2x muon:
 L:      -0.0142492     R:      -0.0292700
 
2x strange quark:
 L:       0.0052640     R:      -0.0097554
 
2x heavy muon:
 L:       0.9553390     R:       1.0130255
 
2x heavy strange quark:
 L:       0.9748525     R:       1.0325376
 
muon , heavy muon:
 L:       0.2434336     R:      -0.0001296
 
strange quark , heavy strange quark:
 L:       0.2434333     R:      -0.0011617
 
2x tauon neutrino:
 L:      -0.0150208     R:       0.0000000
 
2x top quark:
 L:       0.0092162     R:      -0.0241212
 
2x heavy tauon neutrino:
 L:      -0.9846090     R:      -1.0422955
 
2x heavy top quark:
 L:      -0.9698193     R:      -0.9791477
 
tauon neutrino , heavy tauon neutrino:
 L:      -0.2434336     R:       0.0000000
 
top quark , heavy top quark:
 L:      -0.2337897     R:       0.2087489
 
2x tauon:
 L:      -0.0142497     R:      -0.0292655
 
2x bottom quark:
 L:       0.0052614     R:      -0.0097314
 
2x heavy tauon:
 L:       0.9553395     R:       1.0130210
 
2x heavy bottom quark:
 L:       0.9748551     R:       1.0325136
 
tauon , heavy tauon:
 L:       0.2434326     R:      -0.0021766
 
bottom quark , heavy bottom quark:
 L:       0.2434281     R:      -0.0051358
 
 
couplings between gauge bosons
------------------------------
 
2xW boson , Z boson :      -0.5950754
 
2xW boson , heavy Z boson :       0.0537663
 
2xheavy W boson , Z boson :      -0.2374152
 
2xheavy W boson , heavy Z boson :      -1.9944859
 
W boson , heavy W boson , Z boson :       0.0625477
 
W boson , heavy W boson , heavy Z boson :      -0.3581991
 
4x W boson , 0x heavy W boson :       0.4551806
 
3x W boson , 1x heavy W boson :      -0.0564797
 
2x W boson , 2x heavy W boson :       0.1322188
 
1x W boson , 3x heavy W boson :       0.6995732
 
0x W boson , 4x heavy W boson :       4.1325152
 
2x W boson , Z boson , photon :      -0.1864544
 
2x W boson , heavy Z boson , photon :       0.0168466
 
2x heavy W boson , Z boson , photon :      -0.0743891
 
2x heavy W boson , heavy Z boson , photon :      -0.6249303
 
W boson , heavy W boson , Z boson , photon :       0.0195980
 
W boson , heavy W boson , heavy Z boson , photon :      -0.1122342
 
2x W boson , 2x Z boson :       0.3580269
 
2x W boson , 2x heavy Z boson :       0.1311974
 
2x W boson , Z boson , heavy Z boson :      -0.0543996
 
2x heavy W boson , 2x Z boson :       0.0602782
 
2x heavy W boson , 2x heavy Z boson :       4.1062808
 
2x heavy W boson , Z boson , heavy Z boson :       0.4511168
 
W boson , heavy W boson , 2x Z boson :      -0.0520704
 
W boson , heavy W boson , 2x heavy Z boson :       0.6951640
 
W boson , heavy W boson , Z boson , heavy Z boson :       0.0884049
 
 
\end{verbatim}}

\chapter{Two-body decays --- Analytical Results}
\label{app-4}

In this chapter the analytical results for the decay widths presented in chapter \ref{chap-3-4} are
given.

\section{Tree level}
\label{app-4-1}

Consider a heavy particle with momentum $p_1$ and mass $m_1$ decaying into two particles with
momenta $p_2$ / $p_3$ and masses $m_2$ / $m_3$.
\[
\fmfframe(3,3)(3,3){\begin{fmfgraph*}(20,15)
\fmfleft{i}\fmfright{o2,o1}\fmf{fermion}{i,v}\fmf{fermion}{v,o1}
\fmf{fermion}{v,o2}\fmfblob{8thick}{v}
\fmfv{la=$p_1$}{i}\fmfv{la=$p_2$}{o1}\fmfv{la=$p_3$}{o2}
\end{fmfgraph*}}
\]
In the center-of-mass system, we can parameterize the momenta as
\[
p_1^\mu = \left(m_1,0,0,0\right)^\mu \quad,\quad p_2^\mu = \left(E_2,\vec{p}\right)^\mu
\quad,\quad p_3^\mu = \left(E_3,-\vec{p}\right)
\]
Furthermore, momentum conservation and the mass shell conditions can be leveraged to
express the energies and the modulus of the three-momentum as
\begin{multline}\label{equ-a4-1-mshell}
E_2 = \frac{m_1^2 + m_2^2 - m_3^2}{2m_1} \quad,\quad
E_3 = \frac{m_1^2 + m_3^2 - m_2^2}{2m_1} \\
\abs{\vec{p}}^2 = \frac{1}{4m_1^2}\left(m_1^2-\left(m_2+m_3\right)^2\right)
\left(m_1^2-\left(m_2-m_3\right)^2\right)
\end{multline}

The total width of the decaying particle can be expressed as the product of a phasespace factor and
the square of the transition matrix element $\abs{\MM}^2$ (see e.g. \cite{mandlshaw})
\[
\Gamma = \frac{1}{16\pi}\frac{\abs{\vec{p}}}{m_1^2}\left(\prod_f 2m_f\right)
\int_{-1}^1d\cos\theta\:\abs{\MM}^2\]
with the product $\prod_f 2m_f$ running over all external fermions. After performing the spin sum
(including an averaging factor $\frac{1}{N}$ in front of $\Gamma$),
Lorentz invariance requires $\abs{\MM}^2$ to be independent of $\theta$. Inserting
\eqref{equ-a4-1-mshell} and adopting the convention of absorbing the fermion factor into the squared
matrix element, we arrive at the result
\[
\Gamma = \frac{1}{N}\frac{1}{16\pi}\frac{1}{m_1^3}\sqrt{\left(m_1^2-\left(m_2+m_3\right)^2\right)
\left(m_1^2-\left(m_2-m_3\right)^2\right)}\abs{\MM}^2
\]
which is completely expressed in terms of the particle masses, the squared matrix element
$\abs{\MM}^2$ and the averaging factor $\frac{1}{N}$ which is determined from the spin of the
initial state particle\footnote%
{
An additional symmetry factor of $\frac{1}{2}$ could arise if there were identical particles in the
final state. However, this does not occur for the decays relevant to the calculation presented here.
}.

The remaining quantity that has to be determined is the squared matrix element $\abs{\MM}^2$ which
is given in the following for the different spin combinations that are relevant for calculating the widths of
the KK particles in the Three-Site Model.

\subsubsection*{Spin $1$ $\longrightarrow$ Spin $1$ + Spin $1$}

\begin{itemize}
\item Coupling: standard triple gauge boson coupling
\item Mass assignment:
\[
\fmfframe(3,3)(3,3){\begin{fmfgraph*}(20,15)
\fmfleft{i}\fmfright{o2,o1}\fmf{wiggly}{i,v}\fmf{wiggly}{v,o1}
\fmf{wiggly}{v,o2}\fmfdot{v}
\fmfv{la=$m_1$}{i}\fmfv{la=$m_2$}{o1}\fmfv{la=$m_3$}{o2}
\end{fmfgraph*}}
\]
\item Amplitude:
\begin{multline*} \sum_\text{spin}\abs{\MM}^2 = g^2\left(\vs{8ex}
-8\left(m_1^2 + m_2^2 + m_3^2\right)
+2\left(\frac{m_2^4 + m_3^4}{m_1^2} + \frac{m_1^4 + m_3^4}{m_2^2} + 
\frac{m_1^4 + m_2^4}{m_3^2}\right)\right.\\\left.
 -\frac{9}{2}\left(\frac{m_2^2 m_3^2}{m_1^2} + \frac{m_1^2 m_3^2}{m_2^2} +
\frac{m_1^2 m_2^2}{m_3^2}\right)
 +\frac{1}{4}\left(\frac{m_1^6}{m_2^2 m_3^2} + \frac{m_2^6}{m_1^2 m_3^2} +
\frac{m_3^6}{m_1^2 m_2^2}\right)\vs{8ex}\right)
\end{multline*}
\end{itemize}

\subsubsection*{Spin $1$ $\longrightarrow$ Spin $\frac{1}{2}$ + Spin $\frac{1}{2}$}

\begin{itemize}
\item Coupling: $g_V - g_A\gamma_5$
\item Mass assignment:
\[
\fmfframe(3,3)(3,3){\begin{fmfgraph*}(20,15)
\fmfleft{i}\fmfright{o2,o1}\fmf{wiggly}{i,v}\fmf{fermion}{o2,v,o1}\fmfdot{v}
\fmfv{la=$m_1$}{i}\fmfv{la=$m_2$}{o1}\fmfv{la=$m_3$}{o2}
\end{fmfgraph*}}
\]
\item Amplitude:
\begin{multline*}
\sum_\text{spins}\abs{\MM}^2 =
\left(g_V^2 + g_A^2\right)\left(2\left(2m_1^2 - m_2^2 - m_3^2\right) -
2\frac{m_2^4 + m_3^4}{m_1^2} + 4\frac{m_2^2 m_3^2}{m_1^2}\right) \\
+ 12\left(g_V^2 - g_A^2\right)m_1m_3
\end{multline*}
\end{itemize}

\subsubsection*{Spin $\frac{1}{2}$ $\longrightarrow$ Spin $\frac{1}{2}$ + Spin $1$}

\begin{itemize}
\item Coupling: $g_V - g_A\gamma_5$
\item Mass assignment:
\[
\fmfframe(3,3)(3,3){\begin{fmfgraph*}(20,15)
\fmfleft{i}\fmfright{o2,o1}\fmf{fermion}{i,v}\fmf{fermion}{v,o2}\fmf{wiggly}{v,o1}\fmfdot{v}
\fmfv{la=$m_1$}{i}\fmfv{la=$m_2$}{o1}\fmfv{la=$m_3$}{o2}
\end{fmfgraph*}}
\]
\item Amplitude:
\begin{multline}\label{equ-a4-1-fwf}
\sum_\text{spins}\abs{\MM}^2 =
\left(g_V^2 + g_A^2\right) \left(2\left(m_1^2 + m_3^2 - 2m_2^2\right) +
2\frac{m_1^4+m_3^4}{m_2^2} - 4\frac{m_1^2 m_3^2}{m_2^2}\right) \\
- 12\left(g_V^2 - g_A^2\right)m_1m_3
\end{multline}
\end{itemize}

\section{$\order{\alpha_s}$ corrections to the heavy quark widths}
\label{app-4-2}

According to \eqref{equ-3-4-nlo}, the calculation of the $\order{\alpha_s}$ QCD corrections to the
heavy quark widths can be split into
the calculation of the virtual correction and that of the real ones. In order to subtract the
divergences arising in the loop integrals, we have to specify a renormalization scheme. For this
calculation, we use a modified $\overline{\text{MS}}$ scheme where the heavy particle masses are renormalized
on-shell such that the masses obtained by diagonalizing the Lagrangian are treated as pole masses.

The consistency of the result has been checked by confirming infrared finiteness. In addition, it
has been cross-checked against the calculation of the $\order{\alpha_s}$ correction to the top decay
in \cite{Liu:1990py}.

\subsubsection*{Virtual corrections}

For computing the corrections which arise from the exchange of virtual gluons, we decompose the
squared and spin summed two body decay matrix element as
\begin{equation}\label{equ-a4-2-vdec}
\sum_\text{spin}\abs{\MM}^2 = M_0 + \alpha_s \left(M_{1,1\text{PI}} + M_{1,\delta Z}\right) +
\order{\alpha_s^2}
\end{equation}
The first term in the decomposition is just the tree level matrix element
\[
M_0 =
\abs{\parbox{20mm}{\begin{fmfgraph}(20,15)
\fmfset{curly_len}{2mm}
\fmfleft{i1}\fmfright{o2,o1}\fmf{heavy}{i1,v}\fmf{fermion}{v,o2}\fmf{wiggly}{v,o1}\fmfdot{v}
\end{fmfgraph}}}^2
\]
given by \eqref{equ-a4-1-fwf}. The second term is the interference between the tree level matrix
element and the QCD vertex correction
\[
M_{1,1\text{PI}} =
2\Re\left(\parbox{20mm}{
\begin{fmfgraph}(20,15)
\fmfset{curly_len}{2mm}
\fmfleft{i1}\fmfright{o2,o1}\fmf{heavy}{i1,v}\fmf{fermion}{v,o2}\fmf{wiggly}{v,o1}\fmfdot{v}
\end{fmfgraph}}\right)\times\left(\parbox{20mm}{
\begin{fmfgraph}(20,15)
\fmfset{curly_len}{2mm}
\fmfleft{i1}\fmfright{o2,o1}\fmf{heavy}{i1,v}\fmf{fermion}{v,o2}\fmf{wiggly}{v,o1}\fmfdot{v}
\fmffreeze\fmf{phantom}{i1,v1,v,v2,o2}\fmffreeze
\fmf{gluon,right,te=0.5}{v1,v2}
\end{fmfgraph}}\right)^*
\]

The third term $M_{1,\delta Z}$ arises from the wave function renormalization
\[
M_{1,\delta Z}=
2\Re\left(\parbox{20mm}{
\begin{fmfgraph}(20,15)
\fmfset{curly_len}{2mm}
\fmfleft{i1}\fmfright{o2,o1}\fmf{heavy}{i1,v}\fmf{fermion}{v,o2}\fmf{wiggly}{v,o1}\fmfdot{v}
\end{fmfgraph}}\right)\times\\
\left(\parbox{20mm}{\begin{fmfgraph}(20,15)
\fmfset{curly_len}{2mm}
\fmfleft{i1}\fmfright{o2,o1}\fmf{heavy}{i1,v}\fmf{fermion}{v,o2}\fmfdot{v}\fmf{wiggly}{v,o1}
\fmffreeze\fmf{phantom}{i1,v1,v2,v}\fmffreeze
\fmf{gluon,right}{v1,v2}
\end{fmfgraph}} +\;
\parbox{20mm}{\begin{fmfgraph}(20,15)
\fmfset{curly_len}{2mm}
\fmfleft{i1}\fmfright{o2,o1}\fmf{heavy,te=2}{i1,v}\fmf{fermion}{v,o2}\fmf{wiggly}{v,o1}\fmfdot{v}
\fmffreeze\fmf{phantom}{v,v1,v2,o2}\fmffreeze
\fmf{gluon,right}{v1,v2}
\end{fmfgraph}}
\right)^*
\]
In order to better understand the origin of this term and how to calculate it, consider the full two point
function $\hat\Sigma(p^2)$ of a scalar\footnote%
{
Choosing a scalar is convenient to make the argument more transparent. Of course, the
same reasoning applies to fields of arbitrary spin.}
\[
\hat\Sigma(p^2) \qquad=\qquad \parbox{30mm}{\begin{fmfgraph}(30,10)
\fmfleft{i}\fmfright{o}\fmf{plain}{i,v,o}
\fmfv{de.sh=circle,de.si=10.1thick,de.fi=hatched}{v}
\end{fmfgraph}}
\]
In the vicinity of the mass pole, $\hat\Sigma$ has a Laurent expansion
\[ \hat\Sigma(p^2) = \frac{Zi}{p^2 - m^2} + \order{1} \]
with the pole mass $m$ and residual $Z$. At tree level, the residual is $Z=1$. However, loop
corrections give rise to a nontrivial value of $Z$ which depends on the renormalization
scheme chosen.
If we apply LSZ reduction (see e.g. \cite{itzykson}) and
amputate legs from a Green's function to obtain a transition matrix element, we have to keep track of
the potentially nontrivial residual
\begin{equation}\label{equ-a4-2-lsz}
\MM \;=\; \frac{-i}{\sqrt{Z}} \lim_{p^2\rightarrow m^2}(p^2-m^2)\; \parbox{45mm}{
\fmfframe(0,0)(0,0){\begin{fmfgraph*}(40,20)
\fmfleft{i}\fmfright{o3,o2,o1}
\fmf{phantom,te=1.5}{i,v2}\fmf{plain}{o1,v2,o2}\fmf{plain}{v2,o3}\fmffreeze
\fmf{plain,te=1.5}{i,v1}\fmf{plain}{v1,v2}
\fmfv{de.sh=circle,de.si=10.1thick,de.fi=hatched}{v1}\fmfblob{15thick}{v2}
\fmfv{la=$p$,la.an=45}{i}
\end{fmfgraph*}}}
\!\!\!= \;\;
\sqrt{Z}\;\parbox{30mm}{\begin{fmfgraph*}(30,20)
\fmfleft{i}\fmfright{o3,o2,o1}
\fmf{plain,te=3}{i,v}\fmf{plain}{o1,v,o2}\fmf{plain}{o3,v}
\fmfblob{15thick}{v}\fmfv{la=$p$,la.an=45}{i}
\end{fmfgraph*}}
\end{equation}
\eqref{equ-a4-2-lsz} shows that we have to multiply the amputated Greens function with the proper
residual $\sqrt{Z}$ for each external leg in order to correctly obtain the transition matrix
element. This is precisely the origin of $M_{1,\delta Z}$. If we work in perturbation theory, then
the residual can be expanded as
\begin{equation}\label{equ-a4-2-res}
Z = 1 + \alpha_s \Delta Z_1 + \order{\alpha_s^2}
\end{equation}
and $M_{1,\delta Z}$ is correctly calculated as
\[ M_{1,\delta Z} = M_0\left(\Delta Z_{1,f^\prime} + \Delta Z_{1,f}\right) \]
with the residuals of the two fermions $f$ and $f^\prime$.

For the actual calculation of the residual, decompose the one particle irreducible (1PI) two point
function as
\begin{equation}\label{equ-a4-2-decpi}
i\Pi(p^2) \quad=
\quad\parbox{30mm}{\begin{fmfgraph*}(30,10)\fmfleft{i}\fmfright{o}\fmf{fermion}{i,v,o}
\fmfblob{10thick}{v}\end{fmfgraph*}}\quad
=\quad i\left(\slashed p\Delta(p^2) + m\Sigma(p^2)\right)
\end{equation}
with scalar functions $\Delta$ and $\Sigma$. The full two point function is then obtained by
Dyson resummation of the 1PI function
\[
\hat\Sigma = \frac{i}{\slashed{p} - m} + \frac{i}{\slashed{p} - m}\left(i\Pi\right)
\frac{i}{\slashed{p} - m} + \ldots =
\frac{i}{\slashed{p} - m}\frac{1}{1+\frac{\Pi}{\slashed{p} - m}} =
\frac{i}{\slashed{p} - m + \Pi}
\]
Using the decomposition \eqref{equ-a4-2-decpi}, this can be more properly written as
\[
\hat{\Sigma} = \frac{i}{1+\Delta}\cdot\frac{\slashed{p}+m\frac{1-\Sigma}{1+\Delta}}
{p^2-m^2\left(\frac{1-\Sigma}{1+\Delta}\right)^2}
\]
As we have chosen to renormalize the mass on-shell\footnote%
{
If the calculation is performed in a different renormalization scheme, then we can always perform a
finite renormalization of the mass
\[ m \rightarrow m - m\delta m \quad,\quad \Sigma \rightarrow \Sigma + \delta m + \order{\alpha_s^2}\]
such that the on-shell
condition is fulfilled. As the final $\order{\alpha_s}$ result \eqref{equ-a4-2-dz} only depends on the
derivative $\Sigma^\prime$ which is
independent on the finite renormalization $\delta m$ to order $\order{\alpha_s}$,
it is in fact valid for any
renormalization scheme (provided that we set $m$ to the pole mass in \eqref{equ-a4-2-dz}).
}, the self energy $\hat\Sigma$ must have a pole at $p^2 = m^2$, and we can perform Laurent
expansions around $m$
\[
\frac{1-\Sigma(p^2)}{1+\Delta(p^2)} = 1+ \order{p^2-m^2}
\]\begin{equation}\label{equ-a4-2-laurent}
\frac{i}{B(p^2)} = \frac{1}{1+\Delta}\cdot\frac{i}{p^2-m^2\left(\frac{1-\Sigma}{1+\Delta}\right)^2} =
\frac{iZ}{p^2-m^2} + \order{1}
\end{equation}
\eqref{equ-a4-2-laurent} implies that the self energy can be expanded around the pole as
\[ \hat\Sigma = \frac{iZ}{\slashed{p}-m} + \order{1} \]
with $Z$ being the residual we are looking for and which therefore can be calculated as
\[ \frac{1}{Z} =\frac{d}{dp^2}B(p^2)\left.\vs{4ex}\right|_{p^2=m^2} \]
Plugging in \eqref{equ-a4-2-laurent} and expanding in $\alpha_s$, we finally obtain
\begin{equation}\label{equ-a4-2-dz}
\Delta Z = - \Delta(m^2) - 2m^2\left(\Delta^\prime(m^2) + \Sigma^\prime(m^2)\right)
\end{equation}

The actual calculation of $M_{1,1\text{PI}}$ and $\Delta Z$ was carried out using FeynArts \cite{Hahn:2000kx}
and FormCalc \cite{Hahn:2004rf}. The result for $M_{1,1\text{PI}}$ is
{
\newcommand{\linea}[1]{\shoveleft{\hspace{4ex}#1\hfill}\\}
\newcommand{\lineb}[1]{\shoveleft{\hspace{8ex}#1\hfill}\\}
\newcommand{\linec}[1]{\shoveleft{\hspace{12ex}#1\hfill}\\}
\begin{multline}\label{equ-a4-2-nlores}
\shoveleft{M_{1,1\text{PI}}=\hfill}\\
\linea{-\frac{2}{3\pi} M_0}
\linea{+\frac{4g_V}{3\pi m_2^2}\left(\left(m_1 - m_3\right)^2 - m_2^2\right)\left(\vs{5ex}\right.}
\lineb{+B_0\left({m_1}^2+2 {m_3} {m_1}+2 {m_2}^2+{m_3}^2\right)}
\lineb{+C_0\left({m_1}^4+2 {m_3} {m_1}^3+\left({m_2}^2+2 {m_3}^2\right) {m_1}^2+
	\left(2{m_3}^3-2 {m_2}^2 {m_3}\right) {m_1}\right.}
\linec{-\left.2 {m_2}^4+{m_3}^4+{m_2}^2 {m_3}^2\right)}
\lineb{+C_1\left({m_1}^4+2 {m_3} {m_1}^3+\left({m_2}^2+2 {m_3}^2\right) {m_1}^2+{m_3}\left({m_2}^2+2
	{m_3}^2\right){m_1}\right.}
\linec{\left.-2 {m_2}^4+{m_3}^4+4 {m_2}^2 {m_3}^2 \right)}
\lineb{+C_2\left({m_1}^4+2 {m_3} {m_1}^3+2 \left(2 {m_2}^2+{m_3}^2\right) {m_1}^2+{m_3}
	\left({m_2}^2+2 {m_3}^2\right) {m_1}\right.}
\linec{\left.-2 {m_2}^4+{m_3}^4+{m_2}^2 {m_3}^2\right)}
\lineb{-C_{00}\left(2 \left({m_1}^2+2 {m_3} {m_1}+2 {m_2}^2+{m_3}^2\right)\right)}
\lineb{-C_{11}\left({m_3} ({m_1}+{m_3}) \left({m_1}^2+2 {m_3} {m_1}-{m_2}^2+{m_3}^2\right)\right)}
\lineb{-C_{12}\left(({m_1}+{m_3})^2 \left({m_1}^2+2 {m_3} {m_1}-{m_2}^2+{m_3}^2\right)\right)}
\lineb{-C_{22}\left({m_1} ({m_1}+{m_3}) \left({m_1}^2+2 {m_3}
	{m_1}-{m_2}^2+{m_3}^2\right)\right)\left.\vs{5ex}\right)}
\linea{+\frac{4g_A}{3\pi m_2^2}\left(\left(m_1 + m_3\right)^2 - m_2^2\right)\left(\vs{5ex}\right.}
\lineb{+B_0\left({m_1}^2-2 {m_3} {m_1}+2 {m_2}^2+{m_3}^2\right)}
\lineb{+C_0\left({m_1}^4-2 {m_3} {m_1}^3+\left({m_2}^2+2 {m_3}^2\right) {m_1}^2+2 {m_3}
	\left({m_2}^2-{m_3}^2\right) {m_1}\right.}
\linec{\left.-2 {m_2}^4+{m_3}^4+{m_2}^2 {m_3}^2\right)}
\lineb{+C_1\left({m_1}^4-2 {m_3} {m_1}^3+\left({m_2}^2+2 {m_3}^2\right) {m_1}^2-{m_3}
	\left({m_2}^2+2 {m_3}^2\right) {m_1}\right.}
\linec{\left.-2 {m_2}^4+{m_3}^4+4 {m_2}^2 {m_3}^2\right)}
\lineb{+C_2\left({m_1}^4-2 {m_3} {m_1}^3+2 \left(2 {m_2}^2+{m_3}^2\right) {m_1}^2-{m_3}
	\left({m_2}^2+2 {m_3}^2\right) {m_1}\right.}
\linec{\left.-2 {m_2}^4+{m_3}^4+{m_2}^2 {m_3}^2\displaybreak\right)}
\lineb{-C_{00}\left(2 \left({m_1}^2-2 {m_3} {m_1}+2 {m_2}^2+{m_3}^2\right)\right)}
\lineb{-C_{11}\left(({m_3}-{m_1}) {m_3} \left({m_1}^2-2 {m_3} {m_1}-{m_2}^2+{m_3}^2\right)\right)}
\lineb{-C_{12}\left(({m_1}-{m_3})^2 \left({m_1}^2-2 {m_3} {m_1}-{m_2}^2+{m_3}^2\right)\right)}
\shoveleft{\hspace{8ex}-C_{22}\left({m_1} ({m_1}-{m_3}) \left({m_1}^2-2 {m_3}
	{m_1}-{m_2}^2+{m_3}^2\right)\right)\left.\vs{5ex}\right)\hfill}
\end{multline}
}
where the same assignments as in \eqref{equ-a4-1-fwf} are used for the masses of the particles.
$B_i$, $C_i$ and $C_{ij}$ appearing in \eqref{equ-a4-2-nlores} are the scalar and tensor integrals
defined in \cite{Denner:1991kt} with
\begin{align*}
B_i &= B_i\left(m_2^2,m_3^2,m_1^2\right) \\
C_i &= C_i\left(m_3^2,m_2^2,m_1^2,0,m_3^2,m_1^2\right) \\
C_{ij} &= C_{ij}\left(m_3^2,m_2^2,m_1^2,0,m_3^2,m_1^2\right)
\end{align*}

For the $\order{\alpha_s}$ correction to the the residual \eqref{equ-a4-2-res}, we obtain the result
\begin{multline}\label{equ-a4-2-dzres}
\Delta Z = -\frac{1}{3\pi} + \frac{2}{3\pi}
\left(\vs{5ex}B_0\left(m^2,0,m^2\right) + B_1\left(m^2,0,m^2\right) -\right.\\
\left. 2m^2\frac{\partial}{\partial p^2}\left.\left(B_0\left(p^2,0,m^2\right) -
B_1\left(p^2,0,m^2\right)\right)\vs{5ex}\right|_{p^2= m^2}\right)
\end{multline}
with the fermion mass $m$.

Both \eqref{equ-a4-2-nlores} and \eqref{equ-a4-2-dzres} contain a infrared divergence which
originates
from the vanishing gluon mass and which is understood to be regularized by a
small gluon mass $\omega$. The divergence in $\omega$ drops out once the contributions of gluon
radiation are added, allowing for a finite limit $\omega\rightarrow 0 $.
By the virtue of gauge invariance, the dependence on the renormalization scale
$\mu$ drops out of the NLO result \eqref{equ-a4-2-vdec}, leaving the running of $\alpha_s$
as the only source of the scale uncertainty.

\subsubsection*{Real corrections}

The remaining piece to calculate is the decay width of the heavy quark going into a quark, a
heavy gauge boson and an on-shell gluon
\[
\Gamma_\text{real} \propto
\left|
\parbox{20mm }{\begin{fmfgraph}(20,15)
\fmfset{curly_len}{2mm}
\fmfleft{i1}\fmfright{o4,o3,o2,o1}\fmf{heavy}{i1,v}\fmf{fermion}{v,o4}\fmf{wiggly}{v,o1}
\fmffreeze\fmf{phantom}{v,v1,o4}\fmffreeze\fmf{gluon}{v1,o2}\fmfdot{v}
\end{fmfgraph}}+
\parbox{20mm }{\begin{fmfgraph}(20,15)
\fmfset{curly_len}{2mm}
\fmftop{t4,t3,t2,t1}\fmfleft{i1}\fmfright{o2,o1}\fmf{heavy}{i1,v}\fmf{fermion}{v,o2}\fmf{wiggly}{v,o1}
\fmffreeze\fmf{phantom}{i1,v1,v}\fmffreeze\fmf{gluon}{v1,t2}\fmfdot{v}
\end{fmfgraph}}
\right|^2
\]

For the calculation, we choose a mass and momentum assignment analogous to \eqref{equ-a4-1-fwf}
\[
\fmfframe(3,3)(3,3){\begin{fmfgraph*}(25,15)
\fmfset{curly_len}{2mm}
\fmfleft{i}\fmfright{o3,o2,o1}\fmf{heavy,te=2}{i,v}\fmf{fermion}{v,o3}\fmf{wiggly}{v,o2}
\fmf{gluon}{v,o1}\fmfblob{8thick}{v}
\fmfv{la=$p_1,,m_1$}{i}\fmfv{la=$p_2,,m_2$}{o2}\fmfv{la=$p_3,,m_3$}{o3}
\fmfv{la=$q,,\omega$}{o1}
\end{fmfgraph*}}
\]
and label the phasespace integrals appearing in the calculation in accordance with
\cite{Denner:1991kt} as
\begin{equation}\label{equ-a4-2-phsint}
I_{i_1\ldots i_n}^{j_1\ldots i_m} = \frac{1}{\pi^2}\int
\frac{d^3p_2}{2p_2^0}\frac{d^3p_3}{2p_3^0}\frac{d^3q}{2q^0}\delta\left(p_1+p_2+p_3-q\right)
\frac{\left(\pm2qp_{j_1}\right)\ldots\left(\pm2qp_{j_m}\right)}
{\left(\pm2qp_{i_1}\right)\ldots\left(\pm2qp_{i_n}\right)}
\end{equation}
with the minus signs belonging to $p_1$ and the plus signs to $p_{2/3}$. Analytical expressions for the
integrals \eqref{equ-a4-2-phsint} can be found in \cite{Denner:1991kt}.

The calculation has been carried out with FORM \cite{Vermaseren:2000nd} and we choose to decompose
the results as 
\begin{equation}\label{equ-a4-2-rrres}
\Gamma_\text{real} = -\frac{\alpha_s}{3\pi^2 m_2^2 m_1}\left(g_V A_V + g_A A_A\right)
\end{equation}
$A_V$ is defined in terms of the integrals \eqref{equ-a4-2-phsint} as
\begin{multline*}
\shoveleft{A_V = \left(m_3 - m_1\right)^2 I +\hfill}\\
\shoveleft{\hspace{8ex}\left(\left(m_3 - m_1\right)^2 - m_2^2\right)
	\left(\left(m_3 + m_1\right)^2 + 2m_2^2\right)\times\hfill}\\
\shoveright{\left(I_1+ I_3 + m_1^2 I_{11} + \left(m_1^2 - m_2^2 + m_3^2\right)I_{13} +
	m_3^2I_{33}\right)+\hspace{8ex}}\\
\shoveright{\hfill
\left(m_2^2 + \frac{1}{2}\left(m_3 - m_1\right)^2\right)\left(I_3^1 + I_1^3\right)}
\end{multline*}
and $A_A$ can be obtained from $A_V$ by swapping the sign of $m_1$ (just in the polynomials, not in
the integrals \eqref{equ-a4-2-phsint}).

\chapter{Code}
\label{app-5}

This chapter is intended to give an overview over some of the code written in the course of this
thesis. However, it does not intended to cover the whole source which can be downloaded from
\\[2mm]
\centerline\myurl

\section{Coupling Library}
\label{app-5-1}

\lstset{language={[95]FORTRAN},tabsize=3,columns=flexible,%
	basicstyle={\small\ttfamily},stringstyle={\sffamily}}

The coupling library is split into four core modules and a number of support modules and programs.
The core consists of:
\begin{itemize}
\item\lstinline?module tdefs?: Global definitions.
\item\lstinline?module threeshl?: Actual implementation of the model; calculates masses, couplings
and widths.
\item\lstinline?module nlowidth?: Contains the numerical code for the calculation of the
$\order{\alpha_s}$ QCD corrections, c.f. app.~\ref{app-4-2}.
\item\lstinline?module tglue?: Definitions and functions for access from the O'Mega-generated
amplitudes.
\end{itemize}

The support components are
\begin{itemize}
\item\lstinline?module tscript?: Allows to access to masses, widths, couplings etc. by text
identifiers like e.g.
{\small\verb?lhcoupling(T b W)?} for the left-handed $t^\prime bW^\prime$ coupling.
\item\lstinline?program spektrum?: Dumps spectrum and couplings for a specific point in parameter
space (c.f. app.~\ref{app-3}).
\item\lstinline?program threeshleval?: Simple command-line tool which evaluates a quantity as
specified by a \lstinline?tscript? function.
\item\lstinline?program paraplot?: Generates a density plot of a \lstinline?tscript? function
over parameter space, taking a list of constraints into account.
\item\lstinline?scan?: Bash script for creating a ``catalogue'' of density plots.
\item\lstinline?plotter.pl?: Perl frontend to \lstinline?threeshleval? for plotting quantities over
$m_{W^\prime}$ or $M_\text{bulk}$.
\item\lstinline?threeshl_mlink?: Mathematica wrapper around the libraries. Allows to access quantities via
\lstinline?tscript? functions.
\end{itemize}

In this section, some pieces of code will be presented to illustrate organization, implementation
and usage of
the package. The interested reader should consult the woven documentation generated from the noweb
source which can be downloaded from the URL quoted at the beginning of this chapter which
is more detailed and also gives the analytic formulae employed by the module.

\subsubsection*{Definitions: \lstinline?module tdefs?}

The module \lstinline?tdefs? defines some basic constants
\begin{lstlisting}
integer, parameter :: double=selected_real_kind &
	(precision(1.)+1, range(1.)+1)
integer, parameter :: slength = 256
real(kind=double), parameter :: &
	pi=3.1415926535897932385_double
\end{lstlisting}
\lstinline?slength? defines a standard length for character variables, and the other two constants
are self-explanatory. In addition, the module defines the \lstinline?output_unit? and
\lstinline?error_onit? either via the \lstinline?iso_fortran_env? (a FORTRAN 2003 feature) or as
an ordinary constant in case this module is not supported by the compiler.

\subsubsection*{The actual calculations: \lstinline?module threeshl?}

The first duty of this module is to provide the necessary infrastructure to store the masses, widths,
wavefunctions and couplings.
To this end, each particle is assigned a number between $0$ and $64$ whose bits encode different quantum
numbers
\begin{table}[!h]
\centerline{\begin{tabular}{|c|c|}
\hline bits & meaning \\ \hline
0 & 0: Light state, 1: Heavy state \\ \hline
1 & 0: Fermion, 1: Boson \\ \hline
2 & 0: Lepton, 1: Quark (0: W, 1: Z) \\ \hline
3 & 0: Isospin $\frac{1}{2}$, 1: Isospin $-\frac{1}{2}$ \\ \hline
4-5 & Fermion generation \\ \hline \end{tabular}}
\end{table}
\begin{lstlisting}
integer, parameter, public :: e_bcd=B'001000', nue_bcd=B'000000', &
	mu_bcd=B'011000', numu_bcd=B'010000', tau_bcd=B'101000', &
	nutau_bcd=B'100000', he_bcd=B'001001',hnue_bcd=B'000001', &
	hmu_bcd=B'011001', hnumu_bcd=B'010001', htau_bcd=B'101001', &
	hnutau_bcd=B'100001', u_bcd=B'000100', d_bcd=B'001100', &
	c_bcd=B'010100', s_bcd=B'011100', t_bcd=B'100100', &
	b_bcd=B'101100', hu_bcd=B'000101', hd_bcd=B'001101', &
	hc_bcd=B'010101', hs_bcd=B'011101', ht_bcd=B'100101', &
	hb_bcd=B'101101', w_bcd=B'010', hw_bcd=B'011', z_bcd=B'110', &
	hz_bcd=B'111', a_bcd=63
\end{lstlisting}
(although it doesn't fit into the quantum number scheme, the photon snaps nicely into its place).
For the indexing of the coupling arrays, index ranges are defined which are nontrivial and disjunct
in order to allow for better compile-time and run-time consistency checks
\begin{table}[!h]\centerline{\begin{tabular}{|c|c|}\hline
specifier & meaning \\ \hline\hline
\lstinline?l_chir?, \lstinline?r_chir? & chirality \\ \hline
\lstinline?l_mode?, \lstinline?h_mode?, \lstinline?lh_mode? & KK modes / mode combinations \\ \hline
\lstinline?iso_up?, \lstinline?iso_down? & isospin \\ \hline
\lstinline?lat_0?, \lstinline?lat_1?, \lstinline?lat_2? & lattice sites \\ \hline
\lstinline?gen_0?, \lstinline?gen_1?, \lstinline?gen_3? & fermion generations \\ \hline
\lstinline?ftype_l?, \lstinline?ftype_q? & lepton / quark \\ \hline
\lstinline?ptype_b?, \lstinline?ptype_f? & boson / fermion \\ \hline
\lstinline?btype_w?, \lstinline?btype_z?, \lstinline?btype_a? & $W$ / $Z$ \\ \hline
\end{tabular}}\end{table}
\begin{lstlisting}
integer, parameter, public :: l_chir=100, r_chir=101, &
	l_mode=110, h_mode=111, lh_mode=112, iso_up=120, &
	iso_down=121, lat_0=130, lat_1=131, lat_2=132, &
	gen_0=140, gen_1=141, gen_2=142, ftype_l=150, &
	ftype_q=151, ptype_b=160, ptype_f=161, btype_w=170, &
	btype_z=171, btype_a=172
\end{lstlisting}

The masses and widths of the particles are then stored in arrays indexed by the integer assigned to
the particle
\begin{lstlisting}
real(kind=double), public, target :: &
	mass_array(0:63)=0., width_array(0:63)=0.
\end{lstlisting}
while the couplings are stored in arrays which are indexed by the quantum numbers of the lines
meeting at the vertex
\begin{lstlisting}
real(kind=double), public, target :: &
	g_w_lep(l_mode:h_mode, l_mode:h_mode, gen_0:gen_2, &
		l_mode:h_mode, gen_0:gen_2, l_chir:r_chir)= 0., &
	g_w_quark(l_mode:h_mode, l_mode:h_mode, gen_0:gen_2, &
	l_mode:h_mode, gen_0:gen_2, l_chir:r_chir)= 0.
real(kind=double), public, target :: &
	g_z_lep(l_mode:h_mode, l_mode:lh_mode, gen_0:gen_2, &
		iso_up:iso_down, l_chir:r_chir)= 0.,&
	g_z_quark(l_mode:h_mode, l_mode:lh_mode, gen_0:gen_2, &
		iso_up:iso_down, l_chir:r_chir)= 0.
real(kind=double), public, target :: &
	g_wwz(l_mode:lh_mode, l_mode:h_mode)= 0.
real(kind=double), public, target :: &
	g_wwww(0:4)=(/(0.,i=1,5)/), &
	g_wwzz(l_mode:lh_mode, l_mode:lh_mode)= 0., &
	g_wwza(l_mode:lh_mode, l_mode:h_mode)= 0.
\end{lstlisting}
(the coupling between four $W^{(\prime)}$ is an exception and is enumerated by the number of
$W^\prime$ at the vertex). All other couplings are fixed by electromagnetic or $\sun{3}_C$ gauge
invariance. For convenience when calculating the couplings, the wavefunctions are organized in a
similar way
\begin{lstlisting}
real(kind=double) :: &
	wfunct_w(l_mode:h_mode, lat_0:lat_1)= 0., &
	wfunct_z(l_mode:h_mode, lat_0:lat_2)= 0.
real(kind=double) :: &
	wfunct_lep_l(l_mode:h_mode, gen_0:gen_2, &
		iso_up:iso_down, lat_0:lat_1)= 0.,&
	wfunct_lep_r(l_mode:h_mode, gen_0:gen_2, &
		iso_up:iso_down, lat_1:lat_2)= 0.,&
	wfunct_quark_l(l_mode:h_mode, gen_0:gen_2, &
		iso_up:iso_down, lat_0:lat_1)= 0.,&
	wfunct_quark_r(l_mode:h_mode, gen_0:gen_2, &
		iso_up:iso_down, lat_1:lat_2)= 0.
\end{lstlisting}

The parameters of the Three-Site Lagrangian are defined as
\begin{lstlisting}
real(kind=double), target :: sigma_vev=0., g0=0., g1=0., &
	g2=0., x=0., lambda=0., t=0., eps_l, &
	eps_r(ftype_l:ftype_q, gen_0:gen_2, iso_up:iso_down)= 0., e
\end{lstlisting}
where $t$ is redundant and defined for convenience only. \lstinline?sigma_vev? must be divided by
$\sqrt{2}$ to obtain the symmetry breaking scale $v$ defined in chapter \ref{chap-2-3}, and
\lstinline?lambda? has to be multiplied by $2$ to obtain $\tilde\lambda$ (due to a difference in
convention).

The Standard Model parameters are defined as variables which have to be set prior to calling the
diagonalization routines
\begin{lstlisting}
real(kind=double), public :: &
	me_pdg=0._double, mmu_pdg=0.106_double, mtau_pdg=1.78_double, &
	muq_pdg=0._double, mdq_pdg=0._double, mcq_pdg=1.25_double, &
	msq_pdg=0.95_double, mtq_pdg=174._double, mbq_pdg=4.2_double, &
	mw_pdg=80.403_double, mz_pdg=91.188_double, &
	e_pdg=0.313329_double, ww_pdg=2.048_double, &
	wz_pdg=2.443_double, wt_pdg=1.523_double
\end{lstlisting}

In addition, there is an error reporting system built into the module which traps invalid
function calls or points in parameter space at which the analytic formulae contain complex roots.
Each function pushes its name via a call to \lstinline?estack_push? on a stack when it begins
execution and removes it again via \lstinline?estack_pop? at the end.
A subprogram \lstinline?panic? is called if an error occurs, which then prints an error message and
the contents of the error stack to help localizing the origin of the failure. A version of the
square root \lstinline?msqrt? is provided which traps complex roots.

Apart from the Standard Model parameters and the mass / width arrays, most variables, subprograms
and functions are declared as
\lstinline?private? and prefixed with \lstinline?threeshl_? for export (the variables are exported
by pointers).

Prior to using the library, the subprogram \lstinline?init? must be called
\begin{lstlisting}
subroutine init
#	ifdef __NLOW__
	call nlow_init
#	endif
	call set_names
	call init_pointers
end subroutine init
\end{lstlisting}
which initializes a couple of internal quantities (if the $\order{\alpha_s}$ corrections have been
enabled at compile time, the corresponding library is also initialized).
Afterwards, the Standard Model variables can be redefined if desired and the Model is then
initialized by a call to
\begin{lstlisting}
subroutine pdg_init_wgap_bmass (mhw, bmass, el)
real(kind=double), intent(in) :: mhw, bmass
real(kind=double), optional, intent(in) :: el
character(len=slength), parameter :: fname="pdb_init_wgap_bmass"
	call errstack_push(fname)
	if ((mhw < mw_pdg) .or. (bmass < 0.)) &
		call panic(err_invalid_parameters, 0)
	call gauge_cpl_from_sm_wgap(mw_pdg, mz_pdg, e_pdg, mhw)
	if (present (el)) then
		eps_l = el
	else
		eps_l = eps_l_of_x(x)
	end if
	lambda = bmass / sqrt(2._double) / sigma_vev
	call translate_fermion_masses( (/me_pdg, mmu_pdg, mtau_pdg/), &
		(/muq_pdg, mcq_pdg, mtq_pdg/), (/mdq_pdg, msq_pdg, mbq_pdg/))
	call diagonalize
	call errstack_pop
end subroutine pdg_init_wgap_bmass
\end{lstlisting}
The call to \lstinline?gauge_cpl_from_sm_wgap? first calculates the gauge couplings from the
$W,W^\prime$ and $Z$ masses and the electromagnetic gauge coupling $e$. If $\epsilon_L$ is not
given, ideal delocalization is assumed and $\epsilon_L$ is calculated from $x$. $\tilde\lambda$ is
then set from $M_\text{bulk}$ and \lstinline?translate_fermion_masses? is called to calculate the
$\epsilon_R$ from the SM fermion masses. Finally, \lstinline?diagonalize? is called.

For more information on the guts of \lstinline?diagonalize? refer to the woven source. In a
nutshell, this subprogram performs the following steps
\begin{itemize}
\item Calculate the gauge boson wavefunctions and masses.
\item Calculate the fermion wavefunctions and masses.
\item Calculate the couplings.
\end{itemize}
In particular, the masses of the Standard Model particles are not taken from the input parameters,
but are also calculated from the mass matrices in order to provide a simple consistency check.

An error during the calculation usually triggers the termination of the program. This behavior can
be modified by setting the flag \lstinline?threeshl_quit_on_panic? to \lstinline?.false.?, in which
case the library returns to the caller,
printing an error message to STDOUT and setting the flag
\lstinline?threeshl_error?.

After the couplings have been calculated, the user can call the subprogram
\lstinline?calculate_widths? in order to calculate the widths and fill the width array (only the
widths of the KK particles are calculated, the $t, W$ and $Z$ widths are taken from the Standard
Model parameters, and the other Standard Model particles are set to zero width). If activated at
compile time, the $\order{\alpha_s}$ corrections to the heavy quark widths can be included. This can
be controlled via the global flag \lstinline?threeshl_use_nlow? which is set to \lstinline?.true.?
per default. In the calculation, $\alpha_s$ is evaluated at the heavy quark mass.

It is also possible to initialize the model in the massless limit described in chapter
\ref{chap-4-3}. To this end, the subprogram \lstinline?init_ward? must be called
\begin{lstlisting}
subroutine init_ward (mx, ct, ph)
real(kind=double), intent(in) :: mx, ct, ph
integer :: gen, iso
character(len=slength), parameter :: fname = "init_ward"
	call errstack_push(fname)
	if ( (ct .le. -1._double) .or. &
			(ct .ge. 1._double) .or. (mx == 0.)) &
		call panic (err_invalid_parameters, 0)
	mass_array = 0.
	t = msqrt(1._double/ct**2 - 1._double)
	x = mx
	e = e_pdg
	g0 = e * msqrt( 1._double + x**2 + 1._double/t**2 )
	g1 = g0 / x
	g2 = g0 *  t
	sigma_vev = 0.
	lambda = 0.
	eps_l = 0.
	eps_r = 0.
	wfunct_w(l_mode, :) = (/cos(ph), sin(ph)/)
	wfunct_w(h_mode, :) = (/-sin(ph), cos(ph)/)
	wfunct_z(l_mode, :) = (/-g2/2._double/g1 - g1/g2 , &
		g2/2._double/g0 - g0/2._double/g2 , g1/g0 + g0/2._double/g1 /)
	wfunct_z(l_mode, :) = wfunct_z(l_mode, :) / &
		msqrt(wfunct_z(l_mode, lat_0)**2 + &
			wfunct_z(l_mode, lat_1)**2 + &
			wfunct_z(l_mode, lat_2)**2)
	wfunct_z(h_mode, :) = &
		(/-g0/2._double, g1, -g2/2._double/) / &
		msqrt(g0**2/4._double + g1**2 + g2**2/4._double)
	do gen = gen_0, gen_2 ; do iso = iso_up, iso_down
		wfunct_lep_l(l_mode, gen, iso, :) = (/cos(ph), sin(ph)/)
		wfunct_lep_l(h_mode, gen, iso, :) = (/-sin(ph), cos(ph)/)
		wfunct_lep_r(l_mode, gen, iso, :) = (/sin(ph), cos(ph)/)
		wfunct_lep_r(h_mode, gen ,iso, :) = (/cos(ph), -sin(ph)/)
		wfunct_quark_l(l_mode, gen, iso, :) = (/cos(ph), sin(ph)/)
		wfunct_quark_l(h_mode, gen, iso, :) = (/-sin(ph), cos(ph)/)
		wfunct_quark_r(l_mode, gen, iso, :) = (/sin(ph), cos(ph)/)
		wfunct_quark_r(h_mode, gen ,iso, :) = (/cos(ph),-sin(ph)/)
	end do; end do
	call calculate_couplings
	call errstack_pop
end subroutine init_ward
\end{lstlisting}
This first sets all masses to zero and calculates the wavefunctions and parameters according to
chapter \ref{chap-4-3}. Then, the couplings are calculated via the overlap of the wavefunctions just
like in the physical case.

After using the library, the function \lstinline?finalize? should be called.

\subsubsection*{$\order{\alpha_s}$ corrections to the fermion widths: \lstinline?module nlowidth?}

The module \lstinline?nlowidth? implements the numerical calculation of the analytical expressions
given in app.~\ref{app-4-2}. As this might be useful in other contexts than the Three-Site Model,
this module is kept separate from the noweb source of the rest of the package and doesn't contain
any dependencies on the other packages. The analytical
expressions are included from external files which have been autogenerated from Mathematica, and a
FORTRAN 90 wrapper around the LoopTools library (which is FORTRAN 77 code) is supplied.
As there is no woven documentation for this module, it is completely reprinted and documented here

\begin{lstlisting}
module nlowidth
use ltglue
\end{lstlisting}
The LoopTools library outputs status and error messages to STDOUT. Unfortunately, there is no way to
change this behavior which disturbs the operation of programs like \lstinline?spektrum? or
\lstinline?threeshl?. To this end, the module calls \lstinline?libc? functions to redirect these
messages to STDERR (a trick shamelessly stolen from the Mathematica wrapper around LoopTools). This
relies on FORTRAN 2003 features and can be deactivated at compile time via the preprocessor variable
\lstinline?__NO__SILENCER__?.
\begin{lstlisting}
#ifndef __NO_SILENCER__
use, intrinsic :: iso_c_binding
use, intrinsic :: iso_fortran_env
#endif

implicit none
private
\end{lstlisting}
All \lstinline?public? qualifiers are prefixed with \lstinline?nlow_?, everything else is
\lstinline?private?.
\begin{lstlisting}
public :: nlow_dz, nlow_nlo, nlow_rrad, nlow_lo, nlow_width, &
	nlow_wrel, nlow_init, nlow_finalize, nlow_alfas, nlow_alfas_mz, &
	nlow_mt, nlow_mb, nlow_mc, nlow_mz, nlow_b2mode, &
	nlow_b2mode_series, nlow_b2mode_full, nlow_b2mode_auto, &
	nlow_b2auto_thresh
\end{lstlisting}
The parameter $\beta_2$ used in \cite{Denner:1991kt} for the definition of the three-particle
phasespace integrals turns out to be numerically unstable in the massless limit for the final state
fermion. Therefore, the module can either use the full analytic expression for $\beta_2$ or an
expansion in
\[ \epsilon = \frac{m_0 m_2}{m_0^2 - m_1^2} \]
to order $\order{\epsilon^{19}}$ (using the naming conventions from app.~\ref{app-4}).
This choice can be controlled via \lstinline?nlow_b2mode? with the
default setting \lstinline?nlow_b2mode_auto? using the full expression for $\beta_2$ above
\lstinline?nlow_b2auto_thresh? and the expansion below it. The remaining public parameters which can
be modified are the $c,b,t$ and $Z$ masses and the value of $\alpha_s$ at the $Z$ pole
which are required for the evolution of the coupling.
\begin{lstlisting}
integer, parameter :: nlow_b2mode_series=0, nlow_b2mode_full=1, &
   nlow_b2mode_auto=2
integer, parameter :: double=selected_real_kind&
	(precision(1.) + 1, range(1.) + 1)
real(kind=double), parameter :: pi=3.1415926535897932385_double
real(kind=double) :: nlow_alfas_mz=0.1176_double, &
	nlow_mt=174._double, nlow_mb=4.2, nlow_mc=1.25, &
	nlow_mz=91.188_double, nlow_b2auto_thresh=0.1_double
integer :: stdout_copy, nlow_b2mode=nlow_b2mode_auto
\end{lstlisting}
The following functions are overloaded and can be called with or without $m_2$. If $m_2$ is
committed, then the limit $m_2\rightarrow0$ is performed (see the actual implementation below for
details); otherwise, the treatment of $m_2$ is
determined from \lstinline?nlow_b2mode?.

\lstinline?nlow_rrad? calculates the
$1\rightarrow 3$ decay width an external gluon, \lstinline?nlow_width? the full $\order{\alpha_s}$
width (including virtual + real corrections) and \lstinline?nlow_wrel? returns the
$\order{\alpha_s}$ width normalized to the leading order result.
\begin{lstlisting}
interface nlow_rrad
   module procedure rrad_nobs
   module procedure rrad_bs
end interface nlow_rrad

interface nlow_width
   module procedure width_nobs
   module procedure width_bs
end interface nlow_width

interface nlow_wrel
   module procedure wrel_nobs
   module procedure wrel_bs
end interface nlow_wrel
\end{lstlisting}
This is the C interface necessary for redirecting the LoopTools output.
\begin{lstlisting}
#ifndef __NO_SILENCER__
interface
	function dup (s) result (t) bind (c, name="dup")
		import :: c_int
		integer (kind=c_int), value :: s
		integer (kind=c_int) :: t
	end function dup

	function dup2 (s, t) result (r) bind (c, name="dup2")
		import :: c_int
		integer (kind=c_int), value :: s, t
		integer (kind=c_int) :: r
	end function dup2
end interface
#endif

contains
\end{lstlisting}
\lstinline?redirect? and \lstinline?end_redirect? are responsible for the actual redirection.
\begin{lstlisting}
subroutine redirect
integer :: dummy
#  ifndef __NO_SILENCER__
   flush (output_unit, iostat=dummy)
   flush (error_unit, iostat=dummy)
   dummy = dup2 (2, 1)
#  endif
end subroutine redirect

subroutine end_redirect
integer :: dummy
#  ifndef __NO_SILENCER__
   flush (output_unit, iostat=dummy)
   flush (error_unit, iostat=dummy)
   dummy = dup2 (stdout_copy, 1)
#  endif
end subroutine end_redirect
\end{lstlisting}
The following two functions should be called before and after using the library.
\begin{lstlisting}
subroutine nlow_init
#  ifndef __NO_SILENCER__
   stdout_copy = dup (1)
#  endif
   call redirect
   call ffini
   call end_redirect
end subroutine nlow_init

subroutine nlow_finalize
   call redirect
   call ffexi
   call end_redirect
end subroutine nlow_finalize
\end{lstlisting}
Calculates the normalization factor for $1 \rightarrow 2$. \lstinline?mt?, \lstinline?mb? and
\lstinline?mw? are $m_0$, $m_2$ and $m_1$ (adhering to the conventions of app.~\ref{app-4}).
\begin{lstlisting}
function wdnorm (mt, mb, mw) result (res)
real(kind=double), intent(in) :: mt, mb, mw
real(kind=double) :: res
   call redirect
   if ((min (mt, mb, mw) < 0.) .or. (mt < mb + mw)) &
      print *, "WARNING: calculating broken width!"
   res = sqrt((mt**2 - (mb + mw)**2)*(mt**2 - (mb - mw)**2)) / &
		mt**3 / 32._double / pi
   call end_redirect
end function wdnorm
\end{lstlisting}
This function calculates the $\order{\alpha_s}$ contribution to the residual of
the quark propagator \eqref{equ-a4-2-dzres}. The actual code for $\Delta Z$ is included from {\small\verb?dz.inc?}.
\lstinline?m? is the fermion mass, and \lstinline?l? the infinitesimal gluon mass $\omega$
\begin{lstlisting}
function nlow_dz (m, l) result (res)
real(kind=double), intent(in) :: m, l
real(kind=double) :: res
complex(kind=double) :: dz
   call redirect
   call setlambda (l**2)
#  include "dz.inc"
   if (aimag (dz) /= 0._double) print *, &
      "WARNING: dz not strictly real! ", aimag (dz)
   res = real (dz)
   call end_redirect
end function nlow_dz
\end{lstlisting}
Calculates the $\order{\alpha_s}$ 1PI correction to the $1\rightarrow 2$ width.
Again, the actual code is included from an extra file.
\lstinline?gv? and \lstinline?ga? are the vector and axial couplings and \lstinline?l? is the gluon
mass $\omega$.
\begin{lstlisting}
function nlow_nlo (mt, mb, mw, gv, ga, l) result (res)
real(kind=double), intent(in) :: mt, mb, mw, gv, ga, l
real(kind=double) :: res
complex(kind=double) :: nlo
   call redirect
   call setlambda (l**2)
#  include "nlo.inc"
   if (aimag (nlo) /= 0._double) print *, &
      "WARNING: nlo not strictly real! ", aimag (nlo)
   res = real (nlo) * wdnorm (mt, mb, mw)
   call end_redirect
end function nlow_nlo
\end{lstlisting}
The next two \lstinline?private? functions are the $m_2\ne0$ resp. $m_2=0$ branches of the
\lstinline?public? \lstinline?nlow_rrad? and calculate the $1\rightarrow3$ decay width for real
gluon radiation.

In the first case, the code for the analytical expressions of the phasespace integrals
\eqref{equ-a4-2-phsint} is taken from \cite{Denner:1991kt} and included from external files.
Depending on \lstinline?nlow_b2mode? and $m_2$, the series expansion of $\beta_2$ (see above) is
used. 
The analytical expression for the $1\rightarrow3$ in terms of the phase space integrals is read in
from {\small\verb?rrad.inc?}.

In the second case, an infrared singularity arises in the limit $m_2\rightarrow0$ which cancels
between virtual and real corrections. The function sets $m_2=\unit[10^{-50}]{GeV}$ in order to
regularize this divergence and then evaluates a version of the $2\rightarrow3$ width where only the
logarithms in $m_2$ are retained.
\begin{lstlisting}
function rrad_nobs (mt, mb, mw, gv, ga, l) result (res)
real(kind=double), intent(in) :: mt, mb, mw, gv, ga, l
real(kind=double) :: res, ep
complex(kind=double) :: rrad, k, b0, b1, b2, int0, int101, int10, &
	int11, int200, int201, int211, int110, b2_series
   call redirect
#  include "k.inc"
#  include "b0.inc"
#  include "b1.inc"
   ep = mb*mt/(mt**2 - mw**2)
   select case (nlow_b2mode)
      case (nlow_b2mode_full)
#        include "b2.inc"
      case (nlow_b2mode_series)
#        include "b2_series.inc"
         b2 = b2_series
      case (nlow_b2mode_auto)
         if (ep < nlow_b2auto_thresh) then
#           include "b2_series.inc"
            b2 = b2_series
         else
#           include "b2.inc"
         end if
      case default
         print *, "WARNING: invalid nlow_b2mode; using nlow_b2mode_full!"
#        include "b2.inc"
   end select
#  include "int0.inc"
#  include "int101.inc"
#  include "int10.inc"
#  include "int11.inc"
#  include "int200.inc"
#  include "int201.inc"
#  include "int211.inc"
#  include "int110.inc"
#  include "rrad.inc"
   if (aimag (rrad) /= 0._double) print *, &
      "WARNING: rrad not strictly real! ", aimag(rrad)
   res = real (rrad)
   call end_redirect
end function rrad_nobs

function rrad_bs (mmt, mmw, gv, ga, l) result (res)
real(kind=double), intent(in) :: mmt, mmw, gv, ga, l
real(kind=double) :: res
complex(kind=double) :: rrad_bsoft, mb, mt, mw
   mt = mmt; mw = mmw
   mb = (1.E-50_double, 0._double)
   call redirect
#  include "rrad_bsoft.inc"
   if (aimag (rrad_bsoft) /= 0._double) print *, &
      "WARNING: rrad_bsoft not strictly real! ", aimag (rrad_bsoft)
   res = real (rrad_bsoft)
   call end_redirect
end function rrad_bs
\end{lstlisting}
\lstinline?nlow_lo? calculates the leading order $1\rightarrow2$ width.
\begin{lstlisting}
function nlow_lo (mt, mb, mw, gv, ga) result (res)
real(kind=double), intent(in) :: mt, mb, mw, gv, ga
real(kind=double) :: res, lo
#  include "lo.inc"
   res = lo * wdnorm (mt, mb, mw)
end function nlow_lo
\end{lstlisting}
The following two functions are drivers which assemble the NLO width from the leading order result
plus the real and virtual corrections. In the $m_2=0$ case, $m_2$ is set to the infrared regulator
$m_2=\unit[10^{-50}]{GeV}$ in order to regularize the divergence which cancels between virtual and real
corrections. \lstinline?alfas? is the value for $\alpha_s$ to be used for the calculation (see below
for a function which evolves $\alpha_s$ to the desired scale).
\begin{lstlisting}
function width_bs (mt, mw, gv, ga, alfas, l) result (res)
real(kind=double), intent(in) :: mt, mw, gv, ga, l, alfas
real(kind=double) :: res, low, mb
   mb = 1.E-50_double
   low = nlow_lo(mt, mb, mw, gv, ga)
   res = low + alfas * &
         ((nlow_dz (mt, l) + nlow_dz (mb, l)) * low + &
         nlow_nlo (mt, mb, mw, gv, ga, l) + &
			nlow_rrad (mt, mw, gv, ga, l))
end function width_bs

function width_nobs (mt, mb, mw, gv, ga, alfas, l) result (res)
real(kind=double), intent(in) :: mt, mb, mw, gv, ga, l, alfas
real(kind=double) :: res, low
   low = nlow_lo(mt, mb, mw, gv, ga)
   res = low + alfas * &
         ((nlow_dz (mt, l) + nlow_dz (mb, l)) * low + &
         nlow_nlo (mt, mb, mw, gv, ga, l) + &
			nlow_rrad (mt, mb, mw, gv, ga, l))
end function width_nobs
\end{lstlisting}
These two function return the NLO with normalized to the LO result.
\begin{lstlisting}
function wrel_nobs (mt, mb, mw, gv, ga, alfas, l) &
	result (res)
real(kind=double), intent(in) :: mt, mb, mw, gv, ga, l, alfas
real(kind=double) :: res
   res = nlow_width (mt, mb, mw, gv, ga, alfas, l) &
      / nlow_lo (mt, mb, mw, gv, ga)
end function wrel_nobs

function wrel_bs (mt, mw, gv, ga, alfas, l) result (res)
real(kind=double), intent(in) :: mt, mw, gv, ga, l, alfas
real(kind=double) :: res
   res = nlow_width (mt, mw, gv, ga, alfas, l) &
      / nlow_lo (mt, 1.E-50_double, mw, gv, ga)
end function wrel_bs
\end{lstlisting}
\lstinline?nlow_alfas? evolves $\alpha_s$ to a given scale. The mass thresholds and the value
$\alpha_s$ at the $Z$ pole are global parameters which can be adjusted (see above).
\begin{lstlisting}
function nlow_alfas (l) result (res)
real(kind=double), intent(in) :: l
real(kind=double) :: res, a
   call redirect
   a = nlow_alfas_mz
   if (l < 0) then
      print *, "WARNING: not evolving alfas to a negative scale!"
   elseif (l > nlow_mt) then
      a = evolve (nlow_mt, nlow_mz, a, 5)
      a = evolve (l, nlow_mt, a, 6)
   elseif (l > nlow_mb) then
      a = evolve (l, nlow_mz, a, 5)
   elseif ( l > nlow_mc) then
      a = evolve (nlow_mb, nlow_mz, a ,5)
      a = evolve (l, nlow_mb, a, 4)
   else
      a = evolve (nlow_mb, nlow_mz, a, 5)
      a = evolve (nlow_mc, nlow_mb, a, 4)
      a = evolve (l, nlow_mc, a, 3)
   end if
   res = a
   call end_redirect

contains

function evolve (l1, l0, a0, n) result (res1)
real(kind=double), intent(in) :: l1, l0, a0
integer, intent(in) :: n
real(kind=double) :: res1
   res1 = 1._double / &
      ( (11._double - 2._double*real(n, kind=double)/3._double) &
      / 2._double / pi * log (l1 / l0) + 1._double / a0)
end function evolve
end function nlow_alfas

end module nlowidth
\end{lstlisting}

\subsubsection*{Interfacing O'Mega: \lstinline?module tglue?}

The \lstinline?tglue? module provides the interface to the matrix element code generated by O'Mega.
This is necessary for several reasons:
\begin{itemize}
\item Several couplings to the photon and to the gluons are not defined in \lstinline?threeshl? but
are required for the matrix element code to function.
\item The couplings between three gauge bosons must be defined with an additional factor $i$ for
usage in O'Mega.
\item The $W^{(\prime)}ff$ type couplings calculated in \lstinline?threeshl? are taken w.r.t. to
the $W_{1/2}^{(\prime)}$ states and need to be divided by an additional factor of $\sqrt{2}$ in
order to obtain the couplings to the ${W^\pm}^{(\prime)}$.
\item The couplings between gauge bosons and fermions are decomposed w.r.t. to the left- and
right-handed projectors, while the O'Mega implementation uses VA-type couplings.
\end{itemize}
To this end, additional couplings are defined in this module
\begin{lstlisting}
complex(kind=double), public :: &
	g_a_lep, g_a_quark (iso_up:iso_down), g_aaww, &
	ig_aww, ig_wwz (l_mode:lh_mode, l_mode:h_mode)
complex(kind=double), public :: &
	g_w_lep_va (1:2, l_mode:h_mode, l_mode:h_mode, &
		gen_0:gen_2, l_mode:h_mode, gen_0:gen_2), &
	g_w_quark_va (1:2, l_mode:h_mode, l_mode:h_mode, &
		gen_0:gen_2, l_mode:h_mode, gen_0:gen_2), &
	g_z_lep_va (1:2, l_mode:h_mode, l_mode:lh_mode, &
		gen_0:gen_2, iso_up:iso_down), &
	g_z_quark_va (1:2, l_mode:h_mode, l_mode:lh_mode, &
		gen_0:gen_2, iso_up:iso_down)
real(kind=double), public :: g_s=1.218_double
complex(kind=double), public :: ig_s_norm, g_s_norm, g_s_norm2
\end{lstlisting}
{\sloppy
In order to initialize these couplings, the functions \lstinline?tglue_init? (and
\lstinline?tglue_init_ward? for the massless limit) are provided which boast the same interfaces as
\lstinline?threeshl_pdg_init_wgap_bmass? and \lstinline?threeshl_ward? (see above). The
\lstinline?tglue? initialization automatically initializes the \lstinline?threeshl? module (and the
\lstinline?tscript? module, see below). Similarly, a \lstinline?tglue_finalize? is provided which
calls all other initialization routines.

}

\subsubsection*{An example of how to use \lstinline?tscript?: \lstinline?program threeshleval?}

Instead of discussing the \lstinline?tscript? module in detail, the source \lstinline?treeshleval?
is reprinted, a small program which takes $m_{W^\prime}$ $M_\text{bulk}$ and optionally $\epsilon_L$
(for nonideal delocalization) as well as a tscript function definition as command line arguments and returns the
evaluated function.

\begin{lstlisting}
program threeshleval
use threeshl
use tdefs
use tscript
use tglue
implicit none
real(kind=double) :: mhw=-1., mbulk=-1., eps_l
real(kind=double), pointer :: value
logical :: el_present = .false.
character(len=slength) :: fun
\end{lstlisting}
The \lstinline?tscript? module defines the \lstinline?tscript_tokenize_object? data type which is initialized by a call to
\lstinline?tscript_create_tobject? with the character string which is to be interpreted as a
function definition.
\begin{lstlisting}
type(tscript_tokenize_object) :: tobject
\end{lstlisting}
First, the program checks if it has been correctly called. If this is not the case, it prints the
syntax and quits (see below). Otherwise, the variables are set from the command line arguments.
\begin{lstlisting}
if (.not. ((iargc () >= 3) .and. (iargc () <= 5))) call print_usage
call getarg (1, fun); mhw = str_to_double (fun)
call getarg (2, fun); mbulk = str_to_double (fun)
call getarg (3, fun)
if (iargc() == 4) then
	if (trim (fun) == "no_nlow") then
		threeshl_use_nlow = .false.
	else
		eps_l = str_to_double (fun)
		el_present = .true.
	end if
	call getarg (4, fun)
end if
if (iargc () == 5) then
	eps_l = str_to_double (fun)
	call getarg (4, fun)
	if (trim (fun) == "no_nlow") then
		threeshl_use_nlow = .false.
	else
		call print_usage
	end if
	call getarg (5, fun)
end if
\end{lstlisting}
In order to parse the function string, a \lstinline?tscript_tokenize_object? must be created.
This is then decoded by a call to \lstinline?tscript_decode_fspec? in order to obtain a
pointer to the corresponding quantity.
If an error occurred during the parsing, the syntax is printed (can be switched off via the flag
\lstinline?tscript_show_syntax?) and the program is terminated if \lstinline?threesh_quit_on_panic?
is set.
\begin{lstlisting}
tobject = tscript_create_tobject(fun)
value => tscript_decode_fspec(tobject)
\end{lstlisting}
The model is initialized by a call to \lstinline?tglue_init? and the function value is printed to
STDOUT. After calling \lstinline?tglue_finalize?, the program quits.
\begin{lstlisting}
if (el_present) then
	call tglue_init (mhw, mbulk, el=eps_l)
else
	call tglue_init (mhw, mbulk)
end if
print '(F25.15)', value
call tglue_finalize

contains
\end{lstlisting}
This subprogram prints the syntax and then in turn calls \lstinline?tscript_print_syntax? which shows
the \lstinline?tscript? syntax and quits if \lstinline?threeshl_quit_on_panic? is set.
\begin{lstlisting}
subroutine print_usage
	print *, "usage: threeshleval mhw mbulk [eps_l] [no_nlow] function"
	print *
	print *, "mhw and mbulk are the mass of the heavy W and the bulk mass;"
	print *, "if eps_l is not present, it is determined via ideal"
	print *, "delocalization. As for the function specifier:"
	print *
	call tscript_print_syntax
end subroutine print_usage
\end{lstlisting}
This tries to interpret a character string as a floating point number.
\begin{lstlisting}
function str_to_double (str) result(res)
character(len=slength), intent(in) :: str
real(kind=double) :: res
integer :: errstat
	read (unit=str, fmt='(F50.50)', iostat=errstat), res
	if (errstat .ne. 0) call print_usage
end function str_to_double

end program threeshleval
\end{lstlisting}

\section{O'Mega Module}
\label{app-5-2}

{
\newcommand{\camlval}[1]{$\ocwlowerid{#1}$}
\originalTeX
%%%%%%%%%%%%%%%%%%%%%%%%%%%%%%%%%%%%%%%%%%%%%%%%%%%%%%%%%%%%%%%%%
%% This file has been automatically generated with the command
%% ocamlweb -s --header --no-preamble --no-index -o includes/camlcode.tex includes/camlcode.ml 
%%%%%%%%%%%%%%%%%%%%%%%%%%%%%%%%%%%%%%%%%%%%%%%%%%%%%%%%%%%%%%%%%
\label{includes/camlcode.ml:0}%
In this section, only a part of the O'Caml source of the model module is presented. The reader
interested in studying the full code is referred to the commented ocamlweb source.

\subsubsection*{Options}

In order to allow for an easy setup of the various options discussed \ref{chap-4-3}, the O'Mega
module is implemented as a functor which maps two option modules to the final model module.
The signatures of these modules are

\ocweol
\label{includes/camlcode.ml:437}%
\medskip
\ocwbegincode{}\ocwindent{0.00em}
\ocwkw{module}~\ocwkw{type}~$\ocwupperid{Threeshl\_options}~=$\ocweol
\ocwindent{4.00em}
\ocwkw{sig}\ocweol
\ocwindent{8.00em}
\ocwkw{val}~$\ocwlowerid{include\_ckm}:~$\ocwbt{bool}\ocweol
\ocwindent{8.00em}
\ocwkw{val}~$\ocwlowerid{include\_hf}:~$\ocwbt{bool}\ocweol
\ocwindent{8.00em}
\ocwkw{val}~$\ocwlowerid{diet}:~$\ocwbt{bool}\ocweol
\ocwindent{4.00em}
\ocwkw{end}\medskip

\label{includes/camlcode.ml:543}%
\ocwindent{0.00em}
\ocwkw{type}~$\ocwlowerid{qcd\_implementation}~=~\ocwupperid{Disabled}~\mid{}~\ocwupperid{Colflow}$\medskip

\label{includes/camlcode.ml:589}%
\ocwindent{0.00em}
\ocwkw{module}~\ocwkw{type}~$\ocwupperid{Threeshl\_colopt}~=$\ocweol
\ocwindent{4.00em}
\ocwkw{sig}\ocweol
\ocwindent{8.00em}
\ocwkw{val}~$\ocwlowerid{o}:~\ocwlowerid{qcd\_implementation}$\ocweol
\ocwindent{4.00em}
\ocwkw{end}\medskip

\ocwendcode{}\ocwindent{0.00em}
The meaning of the different options is:
\begin{itemize}
\item \camlval{include\_ckm}: Include flavor violating couplings.
\item \camlval{include\_hf}: Include the heavy fermions.
\item \camlval{diet}: Setting this to \camlval{true} discards all couplings between the $W^\prime$
and the leptons or the first two quark generations (only implemented in the case of
\camlval{include\_ckm}=\camlval{false}).
\item \camlval{Theeshl\_colopt.o}: Method for the treatment of color. Setting this to
\camlval{Colflow} includes the gluons and the QCD couplings with the sign conventions choosen
correctly for the colorizer module to work, \camlval{Disabled} excludes the gluons.
\end{itemize}
The functor is then defined as

\ocweol
\label{includes/camlcode.ml:1380}%
\medskip
\ocwbegincode{}\ocwindent{0.00em}
\ocwkw{module}~$\ocwupperid{Threeshl'}~(\ocwupperid{Module\_options}:~\ocwupperid{Threeshl\_options})$\ocweol
\ocwindent{4.00em}
$(\ocwupperid{Module\_colopt}:~\ocwupperid{Threeshl\_colopt})~=~$\medskip

\ocwendcode{}\ocwindent{0.00em}
Other options that the user might want to control can be set via command line options when the
binary is invoked

\ocweol
\label{includes/camlcode.ml:1613}%
\medskip
\ocwbegincode{}\ocwindent{0.00em}
\ocwkw{let}~$\ocwlowerid{default\_width}~=~$\ocwbt{ref}~$\ocwupperid{Timelike}$\ocweol
\ocwindent{0.00em}
\ocwkw{let}~$\ocwlowerid{all\_feynman}~=~$\ocwbt{ref}~\ocwkw{false}\medskip

\label{includes/camlcode.ml:1675}%
\ocwindent{0.00em}
\ocwkw{let}~$\ocwlowerid{options}~=~\ocwupperid{Options.}\ocwlowerid{create}~[$\ocweol
\ocwindent{4.00em}
\ocwstring{"constant\_width"},~$\ocwupperid{Arg.Unit}~($\ocwkw{fun}~$\ocwlowerid{\_}~\rightarrow{}~\ocwlowerid{default\_width}~:=~\ocwupperid{Constant}),$\ocweol
\ocwindent{8.00em}
\ocwstring{"use\ocwvspace{}constant\ocwvspace{}width\ocwvspace{}(also\ocwvspace{}in\ocwvspace{}t\symbol{45}channel)"};\ocweol
\ocwindent{4.00em}
\ocwstring{"custom\_width"},~$\ocwupperid{Arg.String}~($\ocwkw{fun}~$\ocwlowerid{x}~\rightarrow{}~\ocwlowerid{default\_width}~:=~\ocwupperid{Custom}~\ocwlowerid{x}),$\ocweol
\ocwindent{8.00em}
\ocwstring{"use\ocwvspace{}custom\ocwvspace{}width"};\ocweol
\ocwindent{4.00em}
\ocwstring{"cancel\_widths"},~$\ocwupperid{Arg.Unit}~($\ocwkw{fun}~$\ocwlowerid{\_}~\rightarrow{}~\ocwlowerid{default\_width}~:=~\ocwupperid{Vanishing}),$\ocweol
\ocwindent{8.00em}
\ocwstring{"use\ocwvspace{}vanishing\ocwvspace{}width"};\ocweol
\ocwindent{4.00em}
\ocwstring{"all\_feynman"},~$\ocwupperid{Arg.Unit}~($\ocwkw{fun}~$\ocwlowerid{\_}~\rightarrow{}~\ocwlowerid{all\_feynman}~:=~$\ocwkw{true}$),$\ocweol
\ocwindent{8.00em}
\ocwstring{"assign\ocwvspace{}feynman\ocwvspace{}gauge\ocwvspace{}propagators\ocwvspace{}to\ocwvspace{}all\ocwvspace{}gauge\ocwvspace{}bosons\symbol{92}n"}\ocweol
\ocwindent{8.00em}
\^{}~\ocwstring{"\symbol{92}t(for\ocwvspace{}checking\ocwvspace{}the\ocwvspace{}ward\ocwvspace{}identities);"}\ocweol
\ocwindent{8.00em}
\^{}~\ocwstring{"use\ocwvspace{}only\ocwvspace{}if\ocwvspace{}you\ocwvspace{}*really*\ocwvspace{}know\symbol{92}n"}\ocweol
\ocwindent{8.00em}
\^{}~\ocwstring{"\symbol{92}twhat\ocwvspace{}you\ocwvspace{}are\ocwvspace{}doing"}$]$\medskip

\ocwendcode{}\ocwindent{0.00em}
\subsubsection*{Flavors}

The first duty of the module is the definition of a type \camlval{flavor} which enumerates the
different particles. This is done by first defining quantum numbers that encode KK mode,
generation, ``charge'' (differentiates between particles / antiparticles and is also defined for
neutrinos) and isospin and then defining \camlval{flavor} through suitable constructors.

\ocweol
\label{includes/camlcode.ml:2626}%
\medskip
\ocwbegincode{}\ocwindent{0.00em}
\ocwkw{type}~$\ocwlowerid{kkmode}~=~\ocwupperid{Light}~\mid{}~\ocwupperid{Heavy}$\ocweol
\ocwindent{0.00em}
\ocwkw{type}~$\ocwlowerid{generation}~=~\ocwupperid{Gen0}~\mid{}~\ocwupperid{Gen1}~\mid{}~\ocwupperid{Gen2}$\ocweol
\ocwindent{0.00em}
\ocwkw{type}~$\ocwlowerid{csign}~=~\ocwupperid{Pos}~\mid{}~\ocwupperid{Neg}$\ocweol
\ocwindent{0.00em}
\ocwkw{type}~$\ocwlowerid{isospin}~=~\ocwupperid{Iso\_up}~\mid{}~\ocwupperid{Iso\_down}$\medskip

\label{includes/camlcode.ml:2748}%
\ocwindent{0.00em}
\ocwkw{type}~$\ocwlowerid{fermion}~=~$\ocweol
\ocwindent{4.00em}
$\mid{}~\ocwupperid{Lepton}~$\ocwkw{of}~$(\ocwlowerid{kkmode}~\times{}~\ocwlowerid{csign}~\times{}~\ocwlowerid{generation}~\times{}~\ocwlowerid{isospin})$\ocweol
\ocwindent{4.00em}
$\mid{}~\ocwupperid{Quark}~$\ocwkw{of}~$(\ocwlowerid{kkmode}~\times{}~\ocwlowerid{csign}~\times{}~\ocwlowerid{generation}~\times{}~\ocwlowerid{isospin})$\medskip

\label{includes/camlcode.ml:2870}%
\ocwindent{0.00em}
\ocwkw{type}~$\ocwlowerid{boson}~=$\ocweol
\ocwindent{4.00em}
$\mid{}~\ocwupperid{W}~$\ocwkw{of}~$(\ocwlowerid{kkmode}~\times{}~\ocwlowerid{csign})$\ocweol
\ocwindent{4.00em}
$\mid{}~\ocwupperid{Z}~$\ocwkw{of}~$\ocwlowerid{kkmode}$\ocweol
\ocwindent{4.00em}
$\mid{}~\ocwupperid{A}$\ocweol
\ocwindent{4.00em}
$\mid{}~\ocwupperid{G}$\medskip

\label{includes/camlcode.ml:2934}%
\ocwindent{4.00em}
\ocwkw{type}~$\ocwlowerid{flavor}~=~\ocwupperid{Fermion}~$\ocwkw{of}~$\ocwlowerid{fermion}~\mid{}~\ocwupperid{Boson}~$\ocwkw{of}~$\ocwlowerid{boson}$\medskip

\ocwendcode{}\ocwindent{0.00em}
\mbox{}

Two functions \camlval{revmap} and \camlval{revmap2} are defined which apply a list of functions to a
value or resp. each element of a list of values and return the result as a flat list.

\ocweol
\label{includes/camlcode.ml:3190}%
\medskip
\ocwbegincode{}\ocwindent{0.00em}
\ocwkw{let}~$\ocwlowerid{revmap}~\ocwlowerid{funs}~\ocwlowerid{v}~=~\ocwupperid{List.}\ocwlowerid{map}~($\ocwkw{fun}~$\ocwlowerid{x}~\rightarrow{}~\ocwlowerid{x}~\ocwlowerid{v})~\ocwlowerid{funs}$\ocweol
\ocwindent{0.00em}
\ocwkw{let}~$\ocwlowerid{revmap2}~\ocwlowerid{funs}~\ocwlowerid{vals}~=~\ocwupperid{ThoList.}\ocwlowerid{flatmap}~(\ocwlowerid{revmap}~\ocwlowerid{funs})~\ocwlowerid{vals}$\medskip

\ocwendcode{}\ocwindent{0.00em}
Together with a couple of functions that map to the constructors

\ocweol
\label{includes/camlcode.ml:3371}%
\medskip
\ocwbegincode{}\ocwindent{0.00em}
\ocwkw{let}~$\ocwlowerid{lepton}~\ocwlowerid{kk}~\ocwlowerid{cs}~\ocwlowerid{gen}~\ocwlowerid{iso}~=~\ocwupperid{Lepton}~(\ocwlowerid{kk},~\ocwlowerid{cs},~\ocwlowerid{gen},~\ocwlowerid{iso})$\ocweol
\ocwindent{0.00em}
\ocwkw{let}~$\ocwlowerid{quark}~\ocwlowerid{kk}~\ocwlowerid{cs}~\ocwlowerid{gen}~\ocwlowerid{iso}~=~\ocwupperid{Quark}~(\ocwlowerid{kk},~\ocwlowerid{cs},~\ocwlowerid{gen},~\ocwlowerid{iso})$\ocweol
\ocwindent{0.00em}
\ocwkw{let}~$\ocwlowerid{w}~\ocwlowerid{kk}~\ocwlowerid{cs}~=~\ocwupperid{W}~(\ocwlowerid{kk},~\ocwlowerid{cs})$\ocweol
\ocwindent{0.00em}
\ocwkw{let}~$\ocwlowerid{z}~\ocwlowerid{kk}~=~\ocwupperid{Z}~\ocwlowerid{kk}$\ocweol
\ocwindent{0.00em}
\ocwkw{let}~$\ocwlowerid{flavor\_of\_f}~\ocwlowerid{x}~=~\ocwupperid{Fermion}~\ocwlowerid{x}$\ocweol
\ocwindent{0.00em}
\ocwkw{let}~$\ocwlowerid{flavor\_of\_b}~\ocwlowerid{x}~=~\ocwupperid{Boson}~\ocwlowerid{x}$\medskip

\ocwendcode{}\ocwindent{0.00em}
and several functions which loop a list of functions over the quantum numbers

\ocweol
\label{includes/camlcode.ml:3660}%
\medskip
\ocwbegincode{}\ocwindent{0.00em}
\ocwkw{let}~$\ocwlowerid{loop\_kk}~\ocwlowerid{flist}~=~\ocwlowerid{revmap2}~\ocwlowerid{flist}~[\ocwupperid{Light};~\ocwupperid{Heavy}]$\ocweol
\ocwindent{0.00em}
\ocwkw{let}~$\ocwlowerid{loop\_cs}~\ocwlowerid{flist}~=~\ocwlowerid{revmap2}~\ocwlowerid{flist}~[\ocwupperid{Pos};~\ocwupperid{Neg}]$\ocweol
\ocwindent{0.00em}
\ocwkw{let}~$\ocwlowerid{loop\_gen}~\ocwlowerid{flist}~=~\ocwlowerid{revmap2}~\ocwlowerid{flist}~[\ocwupperid{Gen0};~\ocwupperid{Gen1};~\ocwupperid{Gen2}]$\ocweol
\ocwindent{0.00em}
\ocwkw{let}~$\ocwlowerid{loop\_iso}~\ocwlowerid{flist}~=~\ocwlowerid{revmap2}~\ocwlowerid{flist}~[\ocwupperid{Iso\_up};~\ocwupperid{Iso\_down}]$\ocweol
\ocwindent{0.00em}
\ocwkw{let}~$\ocwlowerid{cloop\_kk}~\ocwlowerid{flist}~=~$\ocwkw{match}~$\ocwupperid{Module\_options.}\ocwlowerid{include\_hf}~$\ocwkw{with}\ocweol
\ocwindent{4.00em}
$\mid{}~$\ocwkw{true}~$\rightarrow{}~\ocwlowerid{loop\_kk}~\ocwlowerid{flist}$\ocweol
\ocwindent{4.00em}
$\mid{}~$\ocwkw{false}~$\rightarrow{}~\ocwlowerid{revmap}~\ocwlowerid{flist}~\ocwupperid{Light}$\medskip

\ocwendcode{}\ocwindent{0.00em}
these two functions allow for the easy creation of a list of all particles which is required by
O'Mega
\ocweol
\label{includes/camlcode.ml:4086}%
\medskip
\ocwbegincode{}\ocwindent{0.00em}
\ocwkw{let}~$\ocwlowerid{all\_leptons}~=~\ocwlowerid{loop\_iso}~(\ocwlowerid{loop\_gen}~(\ocwlowerid{loop\_cs}~(\ocwlowerid{cloop\_kk}~[\ocwlowerid{lepton}]~)))$\ocweol
\ocwindent{0.00em}
\ocwkw{let}~$\ocwlowerid{all\_quarks}~=~\ocwlowerid{loop\_iso}(~\ocwlowerid{loop\_gen}~(\ocwlowerid{loop\_cs}~(\ocwlowerid{cloop\_kk}~[\ocwlowerid{quark}]~)))$\ocweol
\ocwindent{0.00em}
\ocwkw{let}~$\ocwlowerid{all\_bosons}~=~(\ocwlowerid{loop\_cs}~(\ocwlowerid{loop\_kk}~[\ocwlowerid{w}]~))~@~[\ocwupperid{Z}~\ocwupperid{Light};~\ocwupperid{Z}~\ocwupperid{Heavy}]~@~[\ocwupperid{A}]~@$\ocweol
\ocwindent{4.00em}
$($\ocwkw{match}~$\ocwupperid{Module\_colopt.}\ocwlowerid{o}~$\ocwkw{with}~$\ocwupperid{Colflow}~\rightarrow{}~[\ocwupperid{G}]~\mid{}~\ocwlowerid{\_}~\rightarrow{}~[\,])$\medskip

\label{includes/camlcode.ml:4349}%
\ocwindent{0.00em}
\ocwkw{let}~$\ocwlowerid{flavors}~()~=~(\ocwupperid{List.}\ocwlowerid{map}~\ocwlowerid{flavor\_of\_f}~(\ocwlowerid{all\_leptons}~@~\ocwlowerid{all\_quarks}))~@$\ocweol
\ocwindent{4.00em}
$(\ocwupperid{List.}\ocwlowerid{map}~\ocwlowerid{flavor\_of\_b}~\ocwlowerid{all\_bosons})$\medskip

\ocwendcode{}\ocwindent{0.00em}
(using \camlval{clopp\_kk} instead of \camlval{loop\_kk} automatically implements the option
\camlval{include\_hf} for the exclusion of the heavy fermions).

\subsubsection*{Helpers}

Apart the enumeration of the particles, the module is required to provide some query functions
which O'Mega can use in order to obtain information on the properties of the particles. The enumeration of
the particles by quantum number allows to do this in a rather convenient way as demonstrated by the examples
of $\sun{3}_C$ and Lorentz representation.

\ocweol
\label{includes/camlcode.ml:4999}%
\medskip
\ocwbegincode{}\ocwindent{0.00em}
\ocwkw{let}~$\ocwlowerid{color}~=$\ocweol
\ocwindent{0.00em}
\ocwkw{let}~$\ocwlowerid{quarkrep}~=~$\ocwkw{function}\ocweol
\ocwindent{4.00em}
$\mid{}~(\ocwlowerid{\_},~\ocwupperid{Pos},~\ocwlowerid{\_},~\ocwlowerid{\_})~\rightarrow{}~\ocwupperid{Color.SUN}~3$\ocweol
\ocwindent{4.00em}
$\mid{}~(\ocwlowerid{\_},~\ocwupperid{Neg},~\ocwlowerid{\_},~\ocwlowerid{\_})~\rightarrow{}~\ocwupperid{Color.SUN}~(-3)$\ocweol
\ocwindent{0.00em}
\ocwkw{in}~\ocwkw{function}\ocweol
\ocwindent{4.00em}
$\mid{}~\ocwupperid{Fermion}~(\ocwupperid{Quark}~\ocwlowerid{x})~\rightarrow{}~\ocwlowerid{quarkrep}~\ocwlowerid{x}$\ocweol
\ocwindent{4.00em}
$\mid{}~\ocwupperid{Boson}~\ocwupperid{G}~\rightarrow{}~\ocwupperid{Color.AdjSUN}~3$\ocweol
\ocwindent{4.00em}
$\mid{}~\ocwlowerid{\_}~\rightarrow{}~\ocwupperid{Color.Singlet}$\medskip

\label{includes/camlcode.ml:5203}%
\ocwindent{0.00em}
\ocwkw{let}~$\ocwlowerid{lorentz}~=~$\ocweol
\ocwindent{0.00em}
\ocwkw{let}~$\ocwlowerid{spinor}~=~$\ocwkw{function}\ocweol
\ocwindent{4.00em}
$\mid{}~(\ocwlowerid{\_},~\ocwupperid{Pos},~\ocwlowerid{\_},~\ocwlowerid{\_})~\rightarrow{}~\ocwupperid{Spinor}$\ocweol
\ocwindent{4.00em}
$\mid{}~(\ocwlowerid{\_},~\ocwupperid{Neg},~\ocwlowerid{\_},~\ocwlowerid{\_})~\rightarrow{}~\ocwupperid{ConjSpinor}$\ocweol
\ocwindent{0.00em}
\ocwkw{in}~\ocwkw{function}\ocweol
\ocwindent{4.00em}
$\mid{}~\ocwupperid{Fermion}~(\ocwupperid{Lepton}~\ocwlowerid{x})~\mid{}~\ocwupperid{Fermion}~(\ocwupperid{Quark}~\ocwlowerid{x})~\rightarrow{}~\ocwlowerid{spinor}~\ocwlowerid{x}$\ocweol
\ocwindent{4.00em}
$\mid{}~\ocwupperid{Boson}~(\ocwupperid{W}~\ocwlowerid{\_})~\mid{}~\ocwupperid{Boson}~(\ocwupperid{Z}~\ocwlowerid{\_})~\rightarrow{}~\ocwupperid{Massive\_Vector}$\ocweol
\ocwindent{4.00em}
$\mid{}~\ocwupperid{Boson}~\ocwupperid{A}~\rightarrow{}~\ocwupperid{Vector}$\ocweol
\ocwindent{4.00em}
$\mid{}~\ocwupperid{Boson}~\ocwupperid{G}~\rightarrow{}~\ocwupperid{Vector}$\medskip

\ocwendcode{}\ocwindent{0.00em}
\mbox{}

Another type of helper which is very important for the interaction with WHIZARD is the function
\camlval{pdg} which returns the PDG code assigned to a particle.

\ocweol
\label{includes/camlcode.ml:5633}%
\medskip
\ocwbegincode{}\ocwindent{0.00em}
\ocwkw{let}~$\ocwlowerid{int\_of\_csign}~=~$\ocwkw{function}~$\ocwupperid{Pos}~\rightarrow{}~1~\mid{}~\ocwupperid{Neg}~\rightarrow{}~-1$\ocweol
\ocwindent{0.00em}
\ocwkw{let}~$\ocwlowerid{int\_of\_gen}~=~$\ocwkw{function}~$\ocwupperid{Gen0}~\rightarrow{}~1~\mid{}~\ocwupperid{Gen1}~\rightarrow{}~2~\mid{}~\ocwupperid{Gen2}~\rightarrow{}~3$\medskip

\label{includes/camlcode.ml:5743}%
\ocwindent{0.00em}
\ocwkw{let}~$\ocwlowerid{pdg}~=$\ocweol
\ocwindent{0.00em}
\ocwkw{let}~$\ocwlowerid{iso\_delta}~=~$\ocwkw{function}~$\ocwupperid{Iso\_down}~\rightarrow{}~0~\mid{}~\ocwupperid{Iso\_up}~\rightarrow{}~1$\ocweol
\ocwindent{0.00em}
\ocwkw{in}~\ocwkw{let}~$\ocwlowerid{gen\_delta}~=~$\ocwkw{function}~$\ocwupperid{Gen0}~\rightarrow{}~0~\mid{}~\ocwupperid{Gen1}~\rightarrow{}~2~\mid{}~\ocwupperid{Gen2}~\rightarrow{}~4$\ocweol
\ocwindent{0.00em}
\ocwkw{in}~\ocwkw{let}~$\ocwlowerid{kk\_delta}~=~$\ocwkw{function}~$\ocwupperid{Light}~\rightarrow{}~0~\mid{}~\ocwupperid{Heavy}~\rightarrow{}~9900$\ocweol
\ocwindent{0.00em}
\ocwkw{in}~\ocwkw{function}\ocweol
\ocwindent{4.00em}
$\mid{}~\ocwupperid{Fermion}~(~\ocwupperid{Lepton}~(\ocwlowerid{kk},~\ocwlowerid{cs},~\ocwlowerid{gen},~\ocwlowerid{iso}))~\rightarrow{}$\ocweol
\ocwindent{8.00em}
$(\ocwlowerid{int\_of\_csign}~\ocwlowerid{cs})~\times{}~(11~+~(\ocwlowerid{gen\_delta}~\ocwlowerid{gen})~+~(\ocwlowerid{iso\_delta}~\ocwlowerid{iso})~+~(\ocwlowerid{kk\_delta}~\ocwlowerid{kk}))$\ocweol
\ocwindent{4.00em}
$\mid{}~\ocwupperid{Fermion}~(~\ocwupperid{Quark}~(\ocwlowerid{kk},~\ocwlowerid{cs},~\ocwlowerid{gen},~\ocwlowerid{iso}))~\rightarrow{}~$\ocweol
\ocwindent{8.00em}
$(\ocwlowerid{int\_of\_csign}~\ocwlowerid{cs})~\times{}~(1~+~(\ocwlowerid{gen\_delta}~\ocwlowerid{gen})~+~(\ocwlowerid{iso\_delta}~\ocwlowerid{iso})+~(\ocwlowerid{kk\_delta}~\ocwlowerid{kk}))$\ocweol
\ocwindent{4.00em}
$\mid{}~\ocwupperid{Boson}~(\ocwupperid{W}~(\ocwlowerid{kk},~\ocwlowerid{cs}))~\rightarrow{}~(\ocwlowerid{int\_of\_csign}~\ocwlowerid{cs})~\times{}~(24~+~(\ocwlowerid{kk\_delta}~\ocwlowerid{kk}))$\ocweol
\ocwindent{4.00em}
$\mid{}~\ocwupperid{Boson}~(\ocwupperid{Z}~\ocwlowerid{kk})~\rightarrow{}~23~+~(\ocwlowerid{kk\_delta}~\ocwlowerid{kk})$\ocweol
\ocwindent{4.00em}
$\mid{}~\ocwupperid{Boson}~\ocwupperid{A}~\rightarrow{}~22$\ocweol
\ocwindent{4.00em}
$\mid{}~\ocwupperid{Boson}~\ocwupperid{G}~\rightarrow{}~21$\medskip

\ocwendcode{}\ocwindent{0.00em}
This function is designed such that it adheres to the Monte Carlo numbering scheme for the Standard
Model particles and assigns the PDG code of their partner prefixed with $99$ to their KK partners,
e.g. $11/-11$ to the $e^-/e^+$ and $9911 / -9911$ to the ${e^-}^\prime / {e^+}^\prime$.

Although not included in the interface, a very important function is the translation of a flavor into
the constant identifying it in the FORTRAN module (see app. \ref{app-5-1}).

\ocweol
\label{includes/camlcode.ml:6788}%
\medskip
\ocwbegincode{}\ocwindent{0.00em}
\ocwkw{let}~$\ocwlowerid{bcdi\_of\_flavor}~=~$\ocweol
\ocwindent{0.00em}
\ocwkw{let}~$\ocwlowerid{prefix}~=~$\ocwkw{function}\ocweol
\ocwindent{4.00em}
$\mid{}~\ocwupperid{Fermion}~(\ocwupperid{Lepton}~(\ocwupperid{Heavy},~\ocwlowerid{\_},~\ocwlowerid{\_},~\ocwlowerid{\_}))~\mid{}~\ocwupperid{Fermion}~(\ocwupperid{Quark}~(\ocwupperid{Heavy},~\ocwlowerid{\_},~\ocwlowerid{\_},~\ocwlowerid{\_}))$\ocweol
\ocwindent{4.00em}
$\mid{}~\ocwupperid{Boson}~(\ocwupperid{W}~(\ocwupperid{Heavy},~\ocwlowerid{\_}))~\mid{}~\ocwupperid{Boson}~(\ocwupperid{Z}~\ocwupperid{Heavy})~\rightarrow{}~$\ocwstring{"h"}\ocweol
\ocwindent{4.00em}
$\mid{}~\ocwlowerid{\_}~\rightarrow{}~$\ocwstring{""}\ocweol
\ocwindent{0.00em}
\ocwkw{in}~\ocwkw{let}~$\ocwlowerid{rump}~=~$\ocwkw{function}\ocweol
\ocwindent{4.00em}
$\mid{}~\ocwupperid{Fermion}~(\ocwupperid{Lepton}~\ocwlowerid{spec})~\rightarrow{}~($\ocwkw{match}~$\ocwlowerid{spec}~$\ocwkw{with}\ocweol
\ocwindent{8.00em}
$\mid{}~(\ocwlowerid{\_},~\ocwlowerid{\_},~\ocwupperid{Gen0},~\ocwupperid{Iso\_up})~\rightarrow{}~$\ocwstring{"nue"}\ocweol
\ocwindent{8.00em}
$\mid{}~(\ocwlowerid{\_},~\ocwlowerid{\_},~\ocwupperid{Gen0},~\ocwupperid{Iso\_down})~\rightarrow{}~$\ocwstring{"e"}\ocweol
\ocwindent{8.00em}
$\mid{}~(\ocwlowerid{\_},~\ocwlowerid{\_},~\ocwupperid{Gen1},~\ocwupperid{Iso\_up})~\rightarrow{}~$\ocwstring{"numu"}\ocweol
\ocwindent{8.00em}
$\mid{}~(\ocwlowerid{\_},~\ocwlowerid{\_},~\ocwupperid{Gen1},~\ocwupperid{Iso\_down})~\rightarrow{}~$\ocwstring{"mu"}\ocweol
\ocwindent{8.00em}
$\mid{}~(\ocwlowerid{\_},~\ocwlowerid{\_},~\ocwupperid{Gen2},~\ocwupperid{Iso\_up})~\rightarrow{}~$\ocwstring{"nutau"}\ocweol
\ocwindent{8.00em}
$\mid{}~(\ocwlowerid{\_},~\ocwlowerid{\_},~\ocwupperid{Gen2},~\ocwupperid{Iso\_down})~\rightarrow{}~$\ocwstring{"tau"}$)$\ocweol
\ocwindent{4.00em}
$\mid{}~\ocwupperid{Fermion}~(\ocwupperid{Quark}~\ocwlowerid{spec})~\rightarrow{}~($\ocwkw{match}~$\ocwlowerid{spec}~$\ocwkw{with}\ocweol
\ocwindent{8.00em}
$\mid{}~(\ocwlowerid{\_},~\ocwlowerid{\_},~\ocwupperid{Gen0},~\ocwupperid{Iso\_up})~\rightarrow{}~$\ocwstring{"u"}\ocweol
\ocwindent{8.00em}
$\mid{}~(\ocwlowerid{\_},~\ocwlowerid{\_},~\ocwupperid{Gen0},~\ocwupperid{Iso\_down})~\rightarrow{}~$\ocwstring{"d"}\ocweol
\ocwindent{8.00em}
$\mid{}~(\ocwlowerid{\_},~\ocwlowerid{\_},~\ocwupperid{Gen1},~\ocwupperid{Iso\_up})~\rightarrow{}~$\ocwstring{"c"}\ocweol
\ocwindent{8.00em}
$\mid{}~(\ocwlowerid{\_},~\ocwlowerid{\_},~\ocwupperid{Gen1},~\ocwupperid{Iso\_down})~\rightarrow{}~$\ocwstring{"s"}\ocweol
\ocwindent{8.00em}
$\mid{}~(\ocwlowerid{\_},~\ocwlowerid{\_},~\ocwupperid{Gen2},~\ocwupperid{Iso\_up})~\rightarrow{}~$\ocwstring{"t"}\ocweol
\ocwindent{8.00em}
$\mid{}~(\ocwlowerid{\_},~\ocwlowerid{\_},~\ocwupperid{Gen2},~\ocwupperid{Iso\_down})~\rightarrow{}~$\ocwstring{"b"}$)$\ocweol
\ocwindent{4.00em}
$\mid{}~\ocwupperid{Boson}~(\ocwupperid{W}~\ocwlowerid{\_})~\rightarrow{}~$\ocwstring{"w"}~$\mid{}~\ocwupperid{Boson}~(\ocwupperid{Z}~\ocwlowerid{\_})~\rightarrow{}~$\ocwstring{"z"}\ocweol
\ocwindent{4.00em}
$\mid{}~\ocwupperid{Boson}~\ocwupperid{A}~\rightarrow{}~\ocwlowerid{invalid\_arg}~$\ocwstring{"Csmodels1.bcd\_of\_flavor:\ocwvspace{}no\ocwvspace{}bcd\ocwvspace{}for\ocwvspace{}photon!"}\ocweol
\ocwindent{4.00em}
$\mid{}~\ocwupperid{Boson}~\ocwupperid{G}~\rightarrow{}~\ocwlowerid{invalid\_arg}~$\ocwstring{"Csmodels1.bcd\_of\_flavor:\ocwvspace{}no\ocwvspace{}bcd\ocwvspace{}for\ocwvspace{}gluon!"}\ocweol
\ocwindent{0.00em}
\ocwkw{in}~\ocwkw{function}~$\ocwlowerid{x}~\rightarrow{}~(\ocwlowerid{prefix}~\ocwlowerid{x})~$\^{}~$(\ocwlowerid{rump}~\ocwlowerid{x})~$\^{}~\ocwstring{"\_bcd"}\medskip

\ocwendcode{}\ocwindent{0.00em}
Other functions which are required by O'Mega for example to translate particle flavors in character
strings suitable for the backend or to communicate with the user on the command line are constructed in a
likewise fashion.

\subsubsection*{Couplings and vertices}

In order to be able to represent the coupling constants as closely to the conventions of the
FORTRAN module as possible, another type referring to the combination of KK modes at a vertex
rather than the individual modes is defined.

\ocweol
\label{includes/camlcode.ml:8226}%
\medskip
\ocwbegincode{}\ocwindent{0.00em}
\ocwkw{type}~$\ocwlowerid{kk2}~=~\ocwupperid{Light2}~\mid{}~\ocwupperid{Heavy2}~\mid{}~\ocwupperid{Light\_Heavy}$\ocweol
\ocwindent{0.00em}
\ocwkw{let}~$\ocwlowerid{loop\_kk2}~\ocwlowerid{flist}~=~\ocwlowerid{revmap2}~\ocwlowerid{flist}~[\ocwupperid{Light2};~\ocwupperid{Heavy2};~\ocwupperid{Light\_Heavy}]$\ocweol
\ocwindent{0.00em}
\ocwkw{let}~$\ocwlowerid{cloop\_kk2}~\ocwlowerid{flist}~=~$\ocwkw{match}~$\ocwupperid{Module\_options.}\ocwlowerid{include\_hf}~$\ocwkw{with}\ocweol
\ocwindent{4.00em}
$\mid{}~$\ocwkw{true}~$\rightarrow{}~\ocwlowerid{loop\_kk2}~\ocwlowerid{flist}$\ocweol
\ocwindent{4.00em}
$\mid{}~$\ocwkw{false}~$\rightarrow{}~\ocwlowerid{revmap}~\ocwlowerid{flist}~\ocwupperid{Light2}$\medskip

\ocwendcode{}\ocwindent{0.00em}
The coupling constants are then defined to mimic their FORTRAN counterparts as closely as possible.

\ocweol
\label{includes/camlcode.ml:8557}%
\medskip
\ocwbegincode{}\ocwindent{0.00em}
\ocwkw{type}~$\ocwlowerid{constant}~=$\ocweol
\ocwindent{4.00em}
$\mid{}~\ocwupperid{G\_a\_lep}~\mid{}~\ocwupperid{G\_a\_quark}~$\ocwkw{of}~$\ocwlowerid{isospin}$\ocweol
\ocwindent{4.00em}
$\mid{}~\ocwupperid{G\_aww}~\mid{}~\ocwupperid{G\_aaww}$\ocweol
\ocwindent{4.00em}
$\mid{}~\ocwupperid{G\_w\_lep}~$\ocwkw{of}~$(\ocwlowerid{kkmode}~\times{}~\ocwlowerid{kkmode}~\times{}~\ocwlowerid{generation}~\times{}~\ocwlowerid{kkmode}~\times{}~\ocwlowerid{generation})$\ocweol
\ocwindent{4.00em}
$\mid{}~\ocwupperid{G\_w\_quark}~$\ocwkw{of}~$(\ocwlowerid{kkmode}~\times{}~\ocwlowerid{kkmode}~\times{}~\ocwlowerid{generation}~\times{}~\ocwlowerid{kkmode}~\times{}~\ocwlowerid{generation})$\ocweol
\ocwindent{4.00em}
$\mid{}~\ocwupperid{G\_z\_lep}~$\ocwkw{of}~$(\ocwlowerid{kkmode}~\times{}~\ocwlowerid{kk2}~\times{}~\ocwlowerid{generation}~\times{}~\ocwlowerid{isospin})$\ocweol
\ocwindent{4.00em}
$\mid{}~\ocwupperid{G\_z\_quark}~$\ocwkw{of}~$(\ocwlowerid{kkmode}~\times{}~\ocwlowerid{kk2}~\times{}~\ocwlowerid{generation}~\times{}~\ocwlowerid{isospin})$\ocweol
\ocwindent{4.00em}
$\mid{}~\ocwupperid{G\_wwz}~$\ocwkw{of}~$(\ocwlowerid{kk2}~\times{}~\ocwlowerid{kkmode})$\ocweol
\ocwindent{4.00em}
$\mid{}~\ocwupperid{G\_wwzz}~$\ocwkw{of}~$(\ocwlowerid{kk2}~\times{}~\ocwlowerid{kk2})$\ocweol
\ocwindent{4.00em}
$\mid{}~\ocwupperid{G\_wwza}~$\ocwkw{of}~$(\ocwlowerid{kk2}~\times{}~\ocwlowerid{kkmode})$\ocweol
\ocwindent{4.00em}
$\mid{}~\ocwupperid{G\_wwww}~$\ocwkw{of}~\ocwbt{int}\ocweol
\ocwindent{4.00em}
$\mid{}~\ocwupperid{G\_s}$\ocweol
\ocwindent{4.00em}
$\mid{}~\ocwupperid{IG\_s}$\ocweol
\ocwindent{4.00em}
$\mid{}~\ocwupperid{G\_s2}$\medskip

\ocwendcode{}\ocwindent{0.00em}
\mbox{}

Using the \camlval{loop\_xx} function allows for a compact definition of the vertex lists, e.g. for
the $\gamma ff$ and $Zff$ type vertices

\ocweol
\label{includes/camlcode.ml:9144}%
\medskip
\ocwbegincode{}\ocwindent{0.00em}
\ocwkw{let}~$\ocwlowerid{vertices\_all}~=$\ocweol
\ocwindent{0.00em}
\ocwkw{let}~$\ocwlowerid{vgen}~\ocwlowerid{kk}~\ocwlowerid{gen}~=$\ocweol
\ocwindent{4.00em}
$((\ocwupperid{Fermion}~(\ocwupperid{Lepton}~(\ocwlowerid{kk},~\ocwupperid{Neg},~\ocwlowerid{gen},~\ocwupperid{Iso\_down})),~\ocwupperid{Boson}~\ocwupperid{A},~\ocwupperid{Fermion}~(\ocwupperid{Lepton}~(\ocwlowerid{kk},~\ocwupperid{Pos},~\ocwlowerid{gen},$\ocweol
\ocwindent{8.00em}
$\ocwupperid{Iso\_down}))),~\ocwupperid{FBF}(1,~\ocwupperid{Psibar},~\ocwupperid{V},~\ocwupperid{Psi}),~\ocwupperid{G\_a\_lep})$\ocweol
\ocwindent{0.00em}
\ocwkw{in}~$\ocwlowerid{loop\_gen}~(\ocwlowerid{cloop\_kk}~[\ocwlowerid{vgen}])$\medskip

\label{includes/camlcode.ml:9346}%
\ocwindent{0.00em}
\ocwkw{let}~$\ocwlowerid{vertices\_aqq}~=$\ocweol
\ocwindent{0.00em}
\ocwkw{let}~$\ocwlowerid{vgen}~\ocwlowerid{kk}~\ocwlowerid{gen}~\ocwlowerid{iso}~=$\ocweol
\ocwindent{4.00em}
$((\ocwupperid{Fermion}~(\ocwupperid{Quark}~(\ocwlowerid{kk},~\ocwupperid{Neg},~\ocwlowerid{gen},~\ocwlowerid{iso})),~\ocwupperid{Boson}~\ocwupperid{A},~\ocwupperid{Fermion}~(\ocwupperid{Quark}~(\ocwlowerid{kk},~\ocwupperid{Pos},~\ocwlowerid{gen},$\ocweol
\ocwindent{8.00em}
$\ocwlowerid{iso}))),~\ocwupperid{FBF}(1,~\ocwupperid{Psibar},~\ocwupperid{V},~\ocwupperid{Psi}),~\ocwupperid{G\_a\_quark}~\ocwlowerid{iso})$\ocweol
\ocwindent{0.00em}
\ocwkw{in}~$\ocwlowerid{loop\_iso}~(\ocwlowerid{loop\_gen}~(\ocwlowerid{cloop\_kk}~[\ocwlowerid{vgen}]))$\medskip

\label{includes/camlcode.ml:9557}%
\ocwindent{0.00em}
\ocwkw{let}~$\ocwlowerid{vertices\_zll}~=$\ocweol
\ocwindent{0.00em}
\ocwkw{let}~$\ocwlowerid{vgen}~\ocwlowerid{kkz}~\ocwlowerid{kk\_f}~\ocwlowerid{kk\_fbar}~\ocwlowerid{gen}~\ocwlowerid{iso}~=$\ocweol
\ocwindent{4.00em}
$((\ocwupperid{Fermion}~(\ocwupperid{Lepton}~(\ocwlowerid{kk\_fbar},~\ocwupperid{Neg},~\ocwlowerid{gen},~\ocwlowerid{iso})),~\ocwupperid{Boson}~(\ocwupperid{Z}~\ocwlowerid{kkz}),$\ocweol
\ocwindent{8.00em}
$\ocwupperid{Fermion}~(\ocwupperid{Lepton}~(\ocwlowerid{kk\_f},~\ocwupperid{Pos},~\ocwlowerid{gen},~\ocwlowerid{iso}))),$\ocweol
\ocwindent{8.00em}
$\ocwupperid{FBF}~(1,~\ocwupperid{Psibar},~\ocwupperid{VA2},~\ocwupperid{Psi}),$\ocweol
\ocwindent{8.00em}
$\ocwupperid{G\_z\_lep}~(\ocwlowerid{kkz},~\ocwlowerid{get\_kk2}~(\ocwlowerid{kk\_f},~\ocwlowerid{kk\_fbar}),~\ocwlowerid{gen},~\ocwlowerid{iso}))$\ocweol
\ocwindent{0.00em}
\ocwkw{in}~$\ocwlowerid{loop\_iso}~(\ocwlowerid{loop\_gen}~(\ocwlowerid{cloop\_kk}~(\ocwlowerid{cloop\_kk}~(\ocwlowerid{loop\_kk}~[\ocwlowerid{vgen}]~))))$\medskip

\label{includes/camlcode.ml:9861}%
\ocwindent{0.00em}
\ocwkw{let}~$\ocwlowerid{vertices\_zqq}~=$\ocweol
\ocwindent{0.00em}
\ocwkw{let}~$\ocwlowerid{vgen}~\ocwlowerid{kkz}~\ocwlowerid{kk\_f}~\ocwlowerid{kk\_fbar}~\ocwlowerid{gen}~\ocwlowerid{iso}~=$\ocweol
\ocwindent{4.00em}
$((\ocwupperid{Fermion}~(\ocwupperid{Quark}~(\ocwlowerid{kk\_fbar},~\ocwupperid{Neg},~\ocwlowerid{gen},~\ocwlowerid{iso})),~\ocwupperid{Boson}~(\ocwupperid{Z}~\ocwlowerid{kkz}),$\ocweol
\ocwindent{8.00em}
$\ocwupperid{Fermion}~(\ocwupperid{Quark}~(\ocwlowerid{kk\_f},~\ocwupperid{Pos},~\ocwlowerid{gen},~\ocwlowerid{iso}))),$\ocweol
\ocwindent{8.00em}
$\ocwupperid{FBF}~(1,~\ocwupperid{Psibar},~\ocwupperid{VA2},~\ocwupperid{Psi}),$\ocweol
\ocwindent{8.00em}
$\ocwupperid{G\_z\_quark}~(\ocwlowerid{kkz},~\ocwlowerid{get\_kk2}~(\ocwlowerid{kk\_f},~\ocwlowerid{kk\_fbar}),~\ocwlowerid{gen},~\ocwlowerid{iso}))$\ocweol
\ocwindent{0.00em}
\ocwkw{in}~$\ocwlowerid{loop\_iso}~(\ocwlowerid{loop\_gen}~(\ocwlowerid{cloop\_kk}~(\ocwlowerid{cloop\_kk}~(\ocwlowerid{loop\_kk}~[\ocwlowerid{vgen}]~))))$\medskip

\ocwendcode{}\ocwindent{0.00em}
The $Wff$ vertex lists are defined in different versions in order to implement the different options
, e.g. for $Wqq$ type vertices

\ocweol
\label{includes/camlcode.ml:10304}%
\medskip
\ocwbegincode{}\ocwindent{0.00em}
\ocwkw{let}~$\ocwlowerid{vertices\_wqq\_no\_ckm}~=$\ocweol
\ocwindent{0.00em}
\ocwkw{let}~$\ocwlowerid{vgen}~\ocwlowerid{kkw}~\ocwlowerid{kk\_f}~\ocwlowerid{kk\_fbar}~\ocwlowerid{iso\_f}~\ocwlowerid{gen}~=$\ocweol
\ocwindent{4.00em}
$((\ocwupperid{Fermion}~(\ocwupperid{Quark}~(\ocwlowerid{kk\_fbar},~\ocwupperid{Neg},~\ocwlowerid{gen},~\ocwlowerid{conj\_iso}~\ocwlowerid{iso\_f})),$\ocweol
\ocwindent{8.00em}
$\ocwupperid{Boson}~(\ocwupperid{W}~(\ocwlowerid{kkw},~($\ocwkw{match}~$\ocwlowerid{iso\_f}~$\ocwkw{with}~$\ocwupperid{Iso\_up}~\rightarrow{}~\ocwupperid{Neg}~\mid{}~\ocwlowerid{\_}~\rightarrow{}~\ocwupperid{Pos}))),$\ocweol
\ocwindent{8.00em}
$\ocwupperid{Fermion}~(\ocwupperid{Quark}~(\ocwlowerid{kk\_f},~\ocwupperid{Pos},~\ocwlowerid{gen},~\ocwlowerid{iso\_f}))),$\ocweol
\ocwindent{8.00em}
$\ocwupperid{FBF}~(1,~\ocwupperid{Psibar},~\ocwupperid{VA2},~\ocwupperid{Psi}),$\ocweol
\ocwindent{8.00em}
$\ocwupperid{G\_w\_quark}~(\ocwlowerid{kkw},~($\ocwkw{match}~$\ocwlowerid{iso\_f}~$\ocwkw{with}~$\ocwupperid{Iso\_up}~\rightarrow{}~\ocwlowerid{kk\_f}~\mid{}~\ocwlowerid{\_}~\rightarrow{}~\ocwlowerid{kk\_fbar}),~\ocwlowerid{gen},$\ocweol
\ocwindent{12.00em}
$($\ocwkw{match}~$\ocwlowerid{iso\_f}~$\ocwkw{with}~$\ocwupperid{Iso\_up}~\rightarrow{}~\ocwlowerid{kk\_fbar}~\mid{}~\ocwlowerid{\_}~\rightarrow{}~\ocwlowerid{kk\_f}),~\ocwlowerid{gen})~)$\ocweol
\ocwindent{0.00em}
\ocwkw{in}~$\ocwlowerid{loop\_gen}~(\ocwlowerid{loop\_iso}~(\ocwlowerid{cloop\_kk}~(\ocwlowerid{cloop\_kk}~(\ocwlowerid{loop\_kk}~[\ocwlowerid{vgen}]~))))$\medskip

\label{includes/camlcode.ml:10758}%
\ocwindent{0.00em}
\ocwkw{let}~$\ocwlowerid{vertices\_wqq\_no\_ckm\_diet}~=$\ocweol
\ocwindent{0.00em}
\ocwkw{let}~$\ocwlowerid{filter}~=~$\ocwkw{function}\ocweol
\ocwindent{4.00em}
$\mid{}~((\ocwupperid{Fermion}~(\ocwupperid{Quark}~(\ocwupperid{Light},~\ocwlowerid{\_},~\ocwlowerid{gen},~\ocwlowerid{\_})),~\ocwupperid{Boson}~(\ocwupperid{W}~(\ocwupperid{Heavy},~\ocwlowerid{\_})),$\ocweol
\ocwindent{8.00em}
$\ocwupperid{Fermion}~(\ocwupperid{Quark}~(\ocwupperid{Light},~\ocwlowerid{\_},~\ocwlowerid{\_},~\ocwlowerid{\_}))),~\ocwlowerid{\_},~\ocwlowerid{\_})~\rightarrow{}~$\ocweol
\ocwindent{12.00em}
$($\ocwkw{match}~$\ocwlowerid{gen}~$\ocwkw{with}~$\ocwupperid{Gen2}~\rightarrow{}~$\ocwkw{true}~$\mid{}~\ocwlowerid{\_}~\rightarrow{}~$\ocwkw{false}$)$\ocweol
\ocwindent{4.00em}
$\mid{}~\ocwlowerid{\_}~\rightarrow{}~$\ocwkw{true}\ocweol
\ocwindent{0.00em}
\ocwkw{in}~$\ocwupperid{List.}\ocwlowerid{filter}~\ocwlowerid{filter}~\ocwlowerid{vertices\_wqq\_no\_ckm}$\medskip

\label{includes/camlcode.ml:11023}%
\ocwindent{0.00em}
\ocwkw{let}~$\ocwlowerid{vertices\_wqq}~=$\ocweol
\ocwindent{0.00em}
\ocwkw{let}~$\ocwlowerid{vgen}~\ocwlowerid{kkw}~\ocwlowerid{kk\_f}~\ocwlowerid{gen\_f}~\ocwlowerid{kk\_fbar}~\ocwlowerid{gen\_fbar}~\ocwlowerid{iso\_f}~=$\ocweol
\ocwindent{4.00em}
$((\ocwupperid{Fermion}~(\ocwupperid{Quark}~(\ocwlowerid{kk\_fbar},~\ocwupperid{Neg},~\ocwlowerid{gen\_fbar},~\ocwlowerid{conj\_iso}~\ocwlowerid{iso\_f})),$\ocweol
\ocwindent{8.00em}
$\ocwupperid{Boson}~(\ocwupperid{W}~(\ocwlowerid{kkw},~($\ocwkw{match}~$\ocwlowerid{iso\_f}~$\ocwkw{with}~$\ocwupperid{Iso\_up}~\rightarrow{}~\ocwupperid{Neg}~\mid{}~\ocwlowerid{\_}~\rightarrow{}~\ocwupperid{Pos}))),$\ocweol
\ocwindent{8.00em}
$\ocwupperid{Fermion}~(\ocwupperid{Quark}~(\ocwlowerid{kk\_f},~\ocwupperid{Pos},~\ocwlowerid{gen\_f},~\ocwlowerid{iso\_f}))),$\ocweol
\ocwindent{8.00em}
$\ocwupperid{FBF}~(1,~\ocwupperid{Psibar},~\ocwupperid{VA2},~\ocwupperid{Psi}),$\ocweol
\ocwindent{8.00em}
$\ocwupperid{G\_w\_quark}~($\ocwkw{match}~$\ocwlowerid{iso\_f}~$\ocwkw{with}\ocweol
\ocwindent{12.00em}
$\mid{}~\ocwupperid{Iso\_up}~\rightarrow{}~(\ocwlowerid{kkw},~\ocwlowerid{kk\_f},~\ocwlowerid{gen\_f},~\ocwlowerid{kk\_fbar},~\ocwlowerid{gen\_fbar})$\ocweol
\ocwindent{12.00em}
$\mid{}~\ocwupperid{Iso\_down}~\rightarrow{}~(\ocwlowerid{kkw},~\ocwlowerid{kk\_fbar},~\ocwlowerid{gen\_fbar},~\ocwlowerid{kk\_f},~\ocwlowerid{gen\_f})))$\ocweol
\ocwindent{0.00em}
\ocwkw{in}~$\ocwlowerid{loop\_iso}~(\ocwlowerid{loop\_gen}~(\ocwlowerid{cloop\_kk}~(\ocwlowerid{loop\_gen}~(\ocwlowerid{cloop\_kk}~(\ocwlowerid{loop\_kk}~[\ocwlowerid{vgen}]~)))))$\medskip

\ocwendcode{}\ocwindent{0.00em}
The actual vertex list passed to O'Mega is then assembled according to the selected options.

\ocweol
\label{includes/camlcode.ml:11606}%
\medskip
\ocwbegincode{}\ocwindent{0.00em}
\ocwkw{let}~$\ocwlowerid{vertices}~()~=~(\ocwlowerid{vertices\_all}~@~\ocwlowerid{vertices\_aqq}~@~$\ocweol
\ocwindent{4.00em}
$($\ocwkw{match}~$\ocwupperid{Module\_options.}\ocwlowerid{diet}~$\ocwkw{with}\ocweol
\ocwindent{8.00em}
$\mid{}~$\ocwkw{false}~$\rightarrow{}~\ocwlowerid{vertices\_wll}$\ocweol
\ocwindent{8.00em}
$\mid{}~$\ocwkw{true}~$\rightarrow{}~\ocwlowerid{vertices\_wll\_diet})~@$\ocweol
\ocwindent{4.00em}
$($\ocwkw{match}~$(\ocwupperid{Module\_options.}\ocwlowerid{include\_ckm},~\ocwupperid{Module\_options.}\ocwlowerid{diet})~$\ocwkw{with}\ocweol
\ocwindent{8.00em}
$\mid{}~($\ocwkw{true},~\ocwkw{false}$)~\rightarrow{}~\ocwlowerid{vertices\_wqq}$\ocweol
\ocwindent{8.00em}
$\mid{}~($\ocwkw{false},~\ocwkw{false}$)~\rightarrow{}~\ocwlowerid{vertices\_wqq\_no\_ckm}$\ocweol
\ocwindent{8.00em}
$\mid{}~($\ocwkw{false},~\ocwkw{true}$)~\rightarrow{}~\ocwlowerid{vertices\_wqq\_no\_ckm\_diet}$\ocweol
\ocwindent{8.00em}
$\mid{}~($\ocwkw{true},~\ocwkw{true}$)~\rightarrow{}~\ocwlowerid{raise}~(\ocwupperid{Failure}$\ocweol
\ocwindent{12.00em}
$($\ocwstring{"Modules4.Threeshl.vertices:\ocwvspace{}CKM\ocwvspace{}matrix\ocwvspace{}together\ocwvspace{}with"}~\^{}\ocweol
\ocwindent{16.00em}
\ocwstring{"\ocwvspace{}option\ocwvspace{}diet\ocwvspace{}is\ocwvspace{}not\ocwvspace{}implemented\ocwvspace{}yet!"}$)))~@$\ocweol
\ocwindent{4.00em}
$\ocwlowerid{vertices\_zll}~@~\ocwlowerid{vertices\_zqq}~@~\ocwlowerid{vertices\_aww}~@~\ocwlowerid{vertices\_zww}~@$\ocweol
\ocwindent{4.00em}
$($\ocwkw{match}~$\ocwupperid{Module\_colopt.}\ocwlowerid{o}~$\ocwkw{with}\ocweol
\ocwindent{8.00em}
$\mid{}~\ocwupperid{Colflow}~\rightarrow{}~\ocwlowerid{vertices\_gqq}~@~\ocwlowerid{vertices\_ggg}$\ocweol
\ocwindent{8.00em}
$\mid{}~\ocwlowerid{\_}~\rightarrow{}~[\,]),$\ocweol
\ocwindent{4.00em}
$\ocwlowerid{vertices\_aaww}~@~\ocwlowerid{vertices\_wwzz}~@~\ocwlowerid{vertices\_wwza}~@~\ocwlowerid{vertices\_wwww}~@$\ocweol
\ocwindent{4.00em}
$($\ocwkw{match}~$\ocwupperid{Module\_colopt.}\ocwlowerid{o}~$\ocwkw{with}\ocweol
\ocwindent{8.00em}
$\mid{}~\ocwupperid{Colflow}~\rightarrow{}~\ocwlowerid{vertices\_gggg}$\ocweol
\ocwindent{8.00em}
$\mid{}~\ocwlowerid{\_}~\rightarrow{}~[\,])$\ocweol
\ocwindent{4.00em}
,~$[\,])$\ocweol
\ocweol
\ocwendcode{}
\germanTeX
\selectlanguage{english}
}

\section{WHIZARD Quirks}
\label{app-5-3}

In WHIZARD, models are described by a model file and a piece of FORTRAN glue. The model file defines
free and derived parameters, the particles and a vertex list.
The FORTRAN glue is called prior to the evaluation of the first matrix element and
has to take care of setting up and providing all parameters and couplings that are required by the
matrix element generation code to function.
In principle, writing the model file is a straightforward task, and the FORTRAN glue would be the
perfect place to perform the \lstinline?tglue? initialization. However, in the case
of the Three-Site Model, implementing the masses of the particles in this fashion turns out
to be a problem.

In the particle definition, WHIZARD requires the mass to be either zero or a free / derived
parameter.
In the Three-Site Model however, the masses are complicated functions which are
calculated by the \lstinline?threeshl? module and stored in the \lstinline?mass_array? array. It is
possible to assign the members of this array to derived parameters like\\[1ex]
\centerline{\ttfamily
derived mhz \hspace{10ex} mass\_array(hz\_bcd)
}\\[2ex]
However, the members of this array get only initialized upon calling the \lstinline?tglue?
initialization which, if this would be called from the FORTRAN glue, would be only after the derived
parameters have been evaluated. This way, the correct masses would not propagate into the
phasespace code which therefore would fail.

The only way to get around this without modifying WHIZARD would be to define the masses as free
parameters and calculate them with an external program which writes out a WHIZARD input file.
However, this would be rather cumbersome and error-prone. Therefore, the version of WHIZARD which
contains the Three-Site Model has been modified to generate code which calls \lstinline?tglue_init?
before the derived parameters are calculated. The amplitude is modified not to use the usual
FORTRAN glue (which is unnecessary for the Three-Site Model and therefore omitted) but instead use
\lstinline?threeshl? and \lstinline?tglue?.

Other minor changes include the build system which is modified to properly use the external library
and an external O'Mega tree (WHIZARD includes its own version of O'Mega). However, these are minor
issues which could be avoided, while the problem discussed in the last paragraph can only be solved
cleanly by a change to the WHIZARD infrastructure.

As a result of these issues, the Three-Site Model is currently not included in the official distribution
of the 1.9x branch of WHIZARD but only available in a modified WHIZARD package which can be
downloaded from the URL
quoted at the beginning of this chapter. The upcoming WHIZARD 2.0, however, will remove these
limitations and will contain the implementation as part of the official distribution.

%\section{FeynRules}
%\label{app-5-4}

\bibliography{physics}

\end{appendix}

\pagestyle{plain}
\renewcommand{\thepage}{}
\selectlanguage{german}

\chapter*{Danksagung}

Wohl keine Doktorarbeit dieser Welt entsteht ohne die Unterst"utzung des Autors durch seine
Mitmenschen. Daher sind an dieser Stelle (ohne besondere Reihenfolge oder Anspruch auf
Vollst"andigkeit) diejenigen Menschen aufgef"uhrt, ohne deren Unterst"utzung die vorliegende
Arbeit in dieser Form nicht m"oglich gewesen w"are.

\begin{itemize}
\item Meine Familie, die mich mein gesamtes Studium hindurch in jeder Hinsicht unterst"utzt hat und
ohne deren R"uckhalt diese Doktorarbeit niemals entstanden w"are.
\item Meine Frau Barbara, deren Geduld durch die Physik manchmal schwer strapaziert wird und die
dennoch immer zu mir und meiner Arbeit stand und steht.
\item Prof. Dr. Thorsten Ohl, der diese Arbeit geduldig betreut und mir in zahllosen Diskussionen
immer wieder zu neuen Erkenntnissen verholfen hat.
\item Prof. Dr. R"uckl, der mir die Durchf"uhrung dieser Arbeit an seinem Lehrstuhl erm"oglicht hat.
\item Prof. Dr. Hans Fraas, der in seinen Vorlesungen mein Interesse an der Theoretischen
Physik geweckt hat.
\item Dem Graduiertenkolleg GRK4711 (wenn auch keine Person im eigentlich Sinne), welches durch ein
Stipendium diese Arbeit wesentlich unterst"utzt und mir zudem einen f"unfw"ochigen Aufenthalt
an der MSU erm"oglicht hat.
\item Meinen Mitkollegiaten f"ur eine angenehme und anregende Arbeitsatmosph"are, in der ich auch
das eine oder andere gelernt habe, was nichts mit Teilchenphysik zu tun hat.
\item Brigitte Wehner, die in allen formalen Angelegenheiten immer hilfreich zur Seite stand, und
ohne die ich sicherlich die Erstattung von so mancher Dienstreise verbummelt h"atte.
\item Meinen Freunden, Kollegen und B"urogenossen Lisa Edelh"auser, Alexander Knochel und Thomas
Schutzmeier f"ur viele Diskussionen "uber Physik und noch mehr v"ollig unphysikalischen Spa"s.
\item Fabian Bach f"ur viele erhellende Diskussionen "uber meine Arbeit.
\item Lisa f"ur das Korrekturlesen.
\item Elizabeth Simmons und Sekhar Chivukula, die meinen sechsw"ochigen Aufenthalt an der Michigan
State University erm"oglicht und durch Ihre Gastfreundlichkeit angenehm gemacht haben.
\item Neil Christensen f"ur die angenehme und fruchtbare Zusammenarbeit an dem
FeynRules-WHIZARD-Interface sowie f"ur den ausf"uhrlichen Vergleich der verschiedenen
Implementationen des Modells.
\item Alle meine Freunde, die ich hier zwar leider nicht aufz"ahlen kann, aber ohne deren
Unterst"utzung vieles schwerer und manches unm"oglich w"are.
\item Zahllose Open-Source-Entwickler, ohne deren unerm"udlichen und leider viel zu selten
honorierten Einsatz kaum eines der Programme existieren w"urde, welche die Umsetzung dieser
Arbeit erst m"oglich gemacht haben.
\end{itemize}

\chapter*{Lebenslauf}

\subsubsection{Pers"onliche Daten}

{
\renewcommand{\arraystretch}{1.5}

\begin{tabular}{p{0.3\textwidth}p{0.7\textwidth}}
Name & Christian Sepp Wolfgang Speckner \\
Geburtsdatum & 19.08.1980\\
Geburtsort & W"urzburg \\
Familienstand & verheiratet (04.04.2009), keine Kinder \\
Nationalit"at & Deutsch
\end{tabular}

\subsubsection{Ausbildung}

\begin{tabular}{p{0.3\textwidth}p{0.7\textwidth}}
1986 -- 1990 & Grundschule Mistelbach\\
1990 -- 1999 & Graf-M"unster-Gymnasium Bayreuth\\
Fr"uhjahr 1999 & Abitur\\
Wintersemester 2000 -- Wintersemester 2003 & Studium Mathematik und Physik f"ur das Lehramt
an Gymnasien, Universit"at W"urzburg\\
Sommer 2003 & Staatl. Zwischenpr"ufung f"ur das Lehramt an Gymnasien\\
Wintersemester 2003 -- Wintersemester 2006 & Studium Diplomphysik, Universit"at W"urzburg \\
Januar 2007 & Verleihung des Titels \glqq Diplom-Physiker (Univ.)\grqq\\
Januar 2007 & Beginn der Promotion an der Universit"at W"urzburg\\
Januar 2007 & Aufnahme als Stipendiat in das DFG-Gra"-du"-ier"-ten"-kol"-leg GRK4711
\glqq Theoretische Teilchenphysik und Astrophysik\grqq in W"urzburg
\end{tabular}

}

{\parindent0ex
\chapter*{Versicherung an Eides Statt}

Ich versichere hiermit an Eides statt, da"s ich diese Dissertation eigenst"andig, selbst"andig und
insb. ohne Hife eines kommerziellen Promotionsberaters angefertigt habe und keine anderen als die
angegebenen Quellen und Hilfsmittel benutzt habe.\\[1cm]

{\bfseries W"urzburg, den\hspace{5mm}\underline{\hspace{3cm}}\\[2cm]
(Christian Speckner)}

\chapter*{Erkl"arung}

\thispagestyle{empty}
Ich erkl"are hiermit, da"s die vorliegende Dissertation weder in gleicher noch anderer Form bereits
in einem anderen Pr"ufungsfach vorgelegen hat.\\[1cm]

{\bfseries W"urzburg, den\hspace{5mm}\underline{\hspace{3cm}}\\[2cm]
(Christian Speckner)}
}

\end{fmffile}

\end{document}